%% file: header.tex
\documentclass[a4paper,12pt]{book}

\usepackage[left=2cm,  right=2cm,  top=2cm,  bottom=2cm]{geometry} 

\usepackage[utf8]{inputenc}	
\usepackage[USenglish]{babel}

\usepackage{blindtext}
\usepackage{amsmath}	
\usepackage{bm}
\usepackage{amssymb}	
\usepackage{setspace}	
\usepackage{hyperref}	
\usepackage{titlesec}

\usepackage{indentfirst} 
\usepackage{booktabs}	
\usepackage{placeins}	
\usepackage{flafter}	

\usepackage[nottoc,numbib]{tocbibind}	

\usepackage{mathtools} 
\usepackage{caption}
\usepackage{subcaption}
\usepackage{placeins}	
\usepackage{flafter}	
\usepackage{xcolor}
\usepackage{verbatim}
\usepackage{tensor}
\usepackage{cancel}
\usepackage{textcomp}

\usepackage{empheq} 
\usepackage{mmacells}
\usepackage{fixltx2e}

\usepackage{pgfornament} 
\usepackage{braket}

\usepackage{bibentry}
\usepackage{upgreek}

\usepackage[citation-order, nobysame]{amsrefs}

\BibSpec{article}{%
    +{}  {\PrintAuthors}                {author}
    +{,} { \textit}                     {title}
    +{.} { }                            {part}
    +{:} { \textit}                     {subtitle}
    +{,} { \PrintContributions}         {contribution}
    +{.} { \PrintPartials}              {partial}
    +{,} { }                            {journal}
    +{}  { \textbf}                     {volume}
    +{,} { \PrintConference}            {conference}
    +{}  { \PrintDatePV}                {date}
    +{,} { \issuetext}                  {number}
    +{,} { \eprintpages}                {pages}
    +{,} { }                            {status}
    +{,} { \PrintDOI}                   {doi}
    +{,} { \eprint}                     {eprint} 
    +{}  { \parenthesize}               {language}
    +{}  { \PrintTranslation}           {translation}
    +{;} { \PrintReprint}               {reprint}
    +{.} { }                            {note}
    +{.} {}                             {transition}
    +{}  {\SentenceSpace \PrintReviews} {review}
}

\renewcommand{\eprint}[1]{\href{https://arxiv.org/abs/#1}{arXiv:#1}}

\mmaDefineMathReplacement[≤]{<=}{\leq}
\mmaDefineMathReplacement[≥]{>=}{\geq}
\mmaDefineMathReplacement[≠]{!=}{\neq}
\mmaDefineMathReplacement[→]{->}{\to}[2]
\mmaDefineMathReplacement[⧴]{:>}{:\hspace{-.2em}\to}[2]
\mmaDefineMathReplacement{∉}{\notin}
\mmaDefineMathReplacement{∞}{\infty}
\mmaDefineMathReplacement{𝕕}{\mathbbm{d}}

\newcommand{\be}{\begin{equation}}
\newcommand{\ee}{\end{equation}}

\mmaSet{
  morefv={gobble=2},
  linklocaluri=mma/symbol/definition:#1,
  morecellgraphics={yoffset=1.9ex}
}

\begin{document}
\pagestyle{plain}
\frontmatter

\hypersetup{pageanchor=false}

\newcommand*\varhrulefill[1][0.4pt]{\leavevmode\leaders\hrule height#1\hfill\kern0pt}

\thispagestyle{empty}
    \begin{center}

    \vspace*{.15\textheight}

    \noindent\varhrulefill[0.7mm] \\[0.9cm]

    {\Huge \bfseries Transport coefficients associated to black holes on the brane: analysis of the shear viscosity-to-entropy density ratio \par}\vspace{0.9cm} 
    
    \noindent\varhrulefill[0.7mm] \\[1.5cm]
    
      \textsc{
      {\Large Pedro Henrique Meert Ferreira}}\\[4cm]

    

    {\LARGE 2022}\\[4cm] 

    \vfill
    \end{center}

\thispagestyle{empty}

    \begin{center}

    {\large Federal University of ABC}\\[6cm]

    \LARGE{\bf Transport coefficients associated to black holes on the brane: analysis of the shear viscosity-to-entropy density ratio} \\[16ex] 
\vspace{-3.5cm}
    \Large {\bf PhD Thesis}\\[3ex]
    \large {\bf \underline{Refereeing Committee}\\ Prof. Dr. Henrique Boschi Filho\\Prof. Dr. Nelson Ricardo Freitas Braga\\ Prof. Dr. Horatiu Stefan Nastase\\ Prof. Dr. Mauricio Richartz}\\[2ex]
    \large {\bf Advised by Prof. Dr. Roldão da Rocha}\\[14ex]

    {\Large Pedro Henrique Meert Ferreira }\\[4cm]

    {\large Santo André}\\[0.2cm]
    {\large 2022}
    \end{center}
\thispagestyle{empty}


 \vspace{1.37cm}
\begin{center}
  {\large \scshape \bfseries Universidade Federal do ABC
  \vspace{.5cm}
\thispagestyle{empty}
Centro de Ciências Naturais e Humanas}
\end{center}
\vspace{3.2cm}
\begin{center}
  {\large \scshape \bfseries Pedro Henrique Meert Ferreira}
\end{center}
\vspace{2.7cm}
\begin{center}
  {\LARGE \scshape \bfseries Coeficientes de transporte associados a buracos negros na brana: an\'alise da razão entre viscosidade de cisalhamento e a densidade de entropia}
\end{center}
\vspace{3cm}
{\bfseries
\noindent
Orientador:  Prof  Dr. Roldão da Rocha
\vspace{.25cm}
}

\vspace{1.4cm}
\begin{flushright}
  \begin{minipage}[c]{.6\textwidth}
    \begin{flushleft}
      Tese de Doutorado
      apresentada ao Centro de Ciências Naturais e Humanas para \\ \noindent  obtenção do título de Doutor em Física
    \end{flushleft}
  \end{minipage}
\end{flushright}
\vspace{1.3cm}
{\footnotesize \scshape
Este exemplar corresponde à versão final \\da 
tese
defendida 
pelo aluno \\Pedro Henrique Meert Ferreira
\\
e orientada pelo Prof Dr. Roldão da Rocha
}

\vspace{1.2cm}

\vfill
\begin{center}
  {\small \scshape \bfseries Santo André\\2022}
\end{center}
\thispagestyle{empty}
\vspace{3cm}
\par Throughout the PhD we published the following research papers:
\begin{itemize}
	\item P.~Meert and R.~da Rocha,
	``The emergence of flagpole and flag-dipole fermions in fluid/gravity correspondence,''
	Eur. Phys. J. C \textbf{78} (2018) no.12, 1012.
	\href{https://arxiv.org/abs/1809.01104}{[arXiv:1809.01104 [hep-th]]}.
	\item A.~J.~Ferreira-Martins, P.~Meert and R.~da Rocha,
	``Deformed AdS$_4$ \textendash{}Reissner\textendash{}Nordstr\"om black branes and shear viscosity-to-entropy density ratio,''
	Eur. Phys. J. C \textbf{79} (2019) no.8, 646.
	\href{https://arxiv.org/abs/1904.01093}{[arXiv:1904.01093 [hep-th]]}.
	\item A.~J.~Ferreira\textendash{}Martins, P.~Meert and R.~da Rocha,
	``AdS$_5$-Schwarzschild deformed black branes and hydrodynamic transport coefficients,''
	Nucl. Phys. B \textbf{957} (2020), 115087.
	\href{https://arxiv.org/abs/1912.04837}{[arXiv:1912.04837 [hep-th]]}.
	\item P.~Meert and R.~da Rocha,
	``Probing the minimal geometric deformation with trace and Weyl anomalies,''
	Nucl. Phys. B \textbf{967} (2021), 115420.
	\href{https://arxiv.org/abs/2006.02564}{[arXiv:2006.02564 [gr-qc]]}.
	\item P.~Meert and R.~da Rocha,
	``Gravitational decoupling, hairy black holes and conformal anomalies,''
	Eur. Phys. J. C \textbf{82} (2022) no.2, 175.
	\href{https://arxiv.org/abs/2109.06289}{[arXiv:2109.06289 [hep-th]]}.
\end{itemize}
\par Between September 2021 and March 2022 I was at the Universit\`a di Bologna as a visiting scholar under the CAPES/PrInt grant, working with Prof. Roberto Casadio. The research we conducted during the 6 months is going to be published, and the paper is in preparation. The details of the article below are not definitive:
\begin{itemize}
	\item P.~Meert, R.~da Rocha, R.~Casadio, L.~Tabarroni, W.~Barreto,
	``Informational entropic aspects of quantum description of collapsing balls of dust,''
	(2022), in preparation.
\end{itemize}
\newpage
\section*{CAPES acknowledgment}

This study was financed in part by the Coordenação de Aperfeiçoamento de Pessoal de Nível Superior - Brasil (CAPES) - Finance Code 001, CAPES/PrInt grant 88887.571337/2020-00, and I would also like to thank Universit\'a di Bologna for the Hospitality. 

\newpage
\begin{center}
	{\large \scshape \bfseries Resumo}
\end{center}
Nesta tese aplicamos a correspond\^encia AdS/CFT a espa\c{c}os-tempo oriundos de teorias que s\~ao extens\~oes da Relatividade Geral. Em particular, estamos interessados em calcular o coeficiente de transporte, a viscosidade de cisalhamento, associada \`a teoria de campo efetiva que \'e dual ao espa\c{c}o-tempo de um buraco negro. Tamb\'em \'e poss\'ivel obter quantidades termodin\^amicas via aplica\c{c}\~ao da correspond\^encia AdS/CFT, das quais a entropia tem import\^ancia, principalmente quando \'e tomada a raz\~ao da viscosidade de cisalhamento com a densidade de entropia. Essa raz\~ao \'e conjecturada ter um valor m\'inimo, em unidades naturais é dado por $\frac{1}{4\pi}$. Os principais resultados contidos nessa tese s\~ao os c\'alculos da raz\~ao  para dois casos diferentes, e a utilização da conjectura KSS -- que estabelece o valor m\'inimo para a raz\~ao --, para investiga\c{c}\~ao de propriedades sobre as m\'etricas deformadas. Fazemos uma revis\~ao da Relatividade Geral e Buracos Negros, e tamb\'em introduzimos o formalismo de Branas, obtendo as equa\c{c}\~oes de Einstein efetivas e solu\c{c}\~oes de Buracos Negros na Brana, que podem ser vistos como extens\~oes de solu\c{c}\~oes conhecidas na Relatividade Geral. Discutimos a formula\c{c}\~ao da correspond\^encia AdS/CFT, e como a aplicar para calcular coeficientes de transporte. Aplicamos ent\~ao esse conhecimento para calcular o coeficiente de transporte associado \`as solu\c{c}\~oes que s\~ao extens\~oes da Relatividade Geral. Naturalmente no processo de desenvolvimento desse trabalho outras quest\~oes interessantes chamam aten\c{c}\~ao, assim inclu\'imos dois estudos que s\~ao relacionados a esse desenvolvimento. O primeiro trata de uma investiga\c{c}\~ao sobre coeficientes de transporte calculados utilizando a correspond\^encia quando setores fermi\^onicos s\~ao inclu\'idos no modelo. No segundo \'e proposta uma forma de avaliar a aplica\c{c}\~ao da correspond\^encia AdS/CFT atrav\'es do c\'alculo da anomalia de Weyl.\\
\linebreak
Palavras chave: Coeficientes de transporte, Razão entre viscosidade de cisalhamento e densidade de entropia, AdS/CFT.
\newpage
\begin{center}
	{\large \scshape \bfseries Abstract}
\end{center}
In this thesis, we apply the AdS/CFT correspondence to space-times associated with extensions of General Relativity. We are particularly interested in calculating the transport coefficient, the shear viscosity, associated to the effective field theory whose dual is a black hole space-time in the bulk. The correspondence also allows us to compute thermodynamic quantities, of which the entropy, plays a prominent role when the ratio between the shear viscosity-to-entropy density is taken. This ratio is conjectured to have a minimum value of $\frac{1}{4\pi}$, in natural units. The main results presented in this thesis consist of calculating the ratio  for two different cases, and then applying the KSS conjecture -- which establishes the minimum value for the ratio --, to investigate properties associated with the deformed space-time metrics. A brief review of General Relativity and Black Holes is presented, and the so-called Brane World formalism is introduced, obtaining the equivalent of Einstein’s equations and Black Hole solutions on the brane, which can be seen as extensions of solutions known in General Relativity. We discuss the formulation of AdS/CFT correspondence, and how to apply it to compute transport coefficients. Then applying this knowledge to compute transport coefficients associated to the solutions constituting an extension of General Relativity. Naturally, throughout the process of developing this work other interesting questions arise, therefore we include two studies related to these developments. In one we investigate the computation of transport coefficients when we have a fermionic sector in the model. The second is a proposal on how to evaluate the consistency of AdS/CFT correspondence via the calculation of the Weyl anomaly.\\
\linebreak
Keywords: Transport coefficients, shear viscosity-to-entropy density ratio, AdS/CFT.
\begingroup 
\let\cleardoublepage\clearpage		
\tableofcontents

\onehalfspacing
\input{chapters/preface.tex}
\mainmatter
\input{chapters/branes.tex}

\input{chapters/AdSCFT_Review.tex}

\input{chapters/Transport_coefficients.tex}

\input{chapters/thesis_result.tex}

\input{chapters/conclusions.tex}
\appendix
\input{chapters/app-admtex.tex}
\input{chapters/app-electricweyl.tex}
\input{chapters/Frr0beta.tex}
\backmatter

\newpage
\let\Section\section 
\def\section*#1{\Section{#1}} 
\bibliographystyle{unsrt}
\bibliography{header}
\endgroup
\end{document}

%% file: chapters/preface.tex
\chapter*{Preface}
\par I like to think of this thesis as a filtered compilation of my studying during the past years. Filtered in the sense many ideas were left out, some because they were not developed further, and others would not fit in the context of this thesis. We started this project with one goal, but ended up changing some ideas throughout the process. This happened because we noticed that some aspects of our investigation were showing up to be very different from what we expected. For instance, the second published work lead to an unexpected conclusion: an argument in favor of the uniqueness of a solution, rather than the possibility of extending the results -- which was our expectation.

\par We start by reviewing some fundamental concepts related to General Relativity and Black Holes, to then introduce the idea of Black Holes and the Brane. I would like to emphasize that Branes constitute an entire area of research on its own, and this thesis is not about branes. Instead, we will use the setup, and introduce only the necessary ideas so one can understand where the solutions we are working with -- the black holes on the brane -- are coming from. Obviously, a wide literature on the subject is provided, for none of these were my own ideas.

\par The objective innovation of this study is probably the way in which we use to characterize the viability of the analyzed black branes. We apply AdS/CFT correspondence techniques and analyze the results in the context of the dual theories, and make conclusions about the black branes given the outcomes from the dual theory. This is only possible because of a famous \textit{conjecture}: the Kovtun--Son--Starinets (KSS) lower bound on the shear viscosity-to-entropy density ratio, which we will refer to as the KSS bound. Essentially this conjecture establishes a minimum value for the ratio, and it tends to be looked at in a similar way as the speed of light is to velocity, i.e. one cannot exceed it. We emphasize the word conjecture in the case of KSS bound because this is based on a theoretical prediction. In fact, the experiments reveal this values to be lower, but within the predicted experimental error. As this is obtained analytically using the AdS/CFT correspondence the conjecture proposes that if one was able to perform a measurement accurate enough the value found would be greater than, or equal to $1/4\pi$.

\par Anti-de Sitter black holes gained a lot of interest since the release of the works by Maldacena, Witten, Gubser, Klebanov, and Polyakov \cite{Maldacena:1997re, Witten:1998qj,Gubser:1998bc}, because the duality associates it to thermal states of field theories. The full conjecture would require the gravity theory to be quantum in order to obtain the complete description, but to this day this is unknown. Nonetheless, we are able to obtain an effective description applying the GKPW procedure, which relates the classical gravity action to effective field theory. It was found that the AdS-Schwarzschild black hole can be associated with the so-called Quark-Gluon plasma \cite{Arnold:2000dr,Huot:2006ys,Policastro:2001yc,Policastro:2002se,Karch:2002sh,Janik:2006ft}, and the AdS/CFT correspondence provides a way to compute some properties that are otherwise impossible. A similar association can be made for metals that behave in an unconventional way, called strange metals, in which case the AdS--Reissner--Nordstr\"om metric plays a central role \cite{Hartnoll:2009sz,Hartnoll:2008kx}.

\par The GKPW formula mentioned in the previous paragraph is what enables the link between classical gravity and effective field theory. Technically, the formula provides a way to compute $n$-point functions, but in the specific context where it is applied, these functions are transport coefficients, such as shear viscosity or conductivity. An important point though, is that despite linking black hole spacetimes with properties of fluids and metals, for example, the conjecture itself does not provide information about the microscopic constituents of  the system -- in the examples presented in the text the reader will notice that there is experimental agreement with specific materials, but this is \textit{a posteriori}. 

\par Among all the findings related to the application of AdS/CFT to compute transport coefficients, two are outstanding and deserve some more attention. The KSS bound, which plays an important role in the results to be described in this work, is intriguing because the result arises very naturally from general assumptions. It is a conjecture nonetheless, and one can find arguments supporting the lower bound \cite{Pesci:2009xh,Pesci:2009dc,Lawrence:2021cvt}, as well as situations where it is explicitly violated \cite{Cremonini:2009sy,Rebhan:2011vd,Buchel:2008vz,Brigante:2007nu}. The second result is related to the non-fermi liquids, sometimes called strange metals, these are materials with unconventional properties regarding their conductivity \cite{Hartnoll:2009sz}. Although known before the AdS/CFT conjecture made its first appearance in the literature \cite{Varma:1989zz}, the first analytical results were provided by calculations using the AdS-Reissner-Nordstr\"om black hole in the context of AdS/CFT \cite{SSLee:2009,Liu:2014dva,Hartnoll:2009ns}. Although we do not study this system specifically, it is worth mentioning since it was one of the most striking results provided by the conjecture.

\par Chapters 2, 3, and 4 constitute mostly a review of topics that are relevant to understanding the results on chapter 5. Namely black holes in GR and on the brane, the AdS/CFT correspondence, and linear response theory. The last two combined become a powerful tool to compute transport coefficients. We give concrete examples from the literature on how to obtain quantities such as conductivity and shear viscosity, and also show one way of deriving the KSS bound.

\par Chapter 5, along with Sections \ref{adsschw-def} and \ref{adsrn-def} are the original parts of this work. Although parts of it are not in the exact same form, these works were published \cite{Meert:2018qzk,Ferreira-Martins:2019wym,Ferreira-Martins:2019svk,Meert:2021khi} throughout the years while we were conducting the studies. The decision of splitting part of the results into sections in the first chapter, and most of them in the last chapter is only for easiness of contextualization. Since the metrics arise in the context of gravitation theory, it feels more natural to discuss how one arrives at their expressions in this context, rather than discussing the derivation after talking about AdS/CFT and transport coefficients.

\chapter*{Notation and conventions}
\par Throughout this work we adopt certain conventions that we will expose here. 

\par The index notation for vectors, covectors, and tensors are respectively
\begin{align*}
	\vec{V}&=\sum_{i}V^{i}\hat{e}_{i}\\\tilde{\omega}&=\sum_{i}\omega_{i}dx^{i}\\T_{i}^{\text{ }j}&=\sum_{i}\sum_{j}T_{i}^{\text{ }j}\hat{e}_{j}dx^{i}
\end{align*}
where $V^{i},\omega_{i},T_{i}^{\text{ }j}$ are the components of $\vec{V},\tilde{\omega},\textbf{T}$ with respect to the basis $\hat{e},dx$. We will assume that the reader is familiar with the basis when they are mentioned – for solutions of General Relativity and alternative theories of gravity, for example – and work with the components only. Therefore, when it is said ``a vector$ V^{i}$” we refer to the vector itself, not to a particular component.

\par Indices run through different values. We shall adopt the following convention:
\begin{itemize}
	\item $i,j,k,\ldots$ lowercase latin letter from the middle of the alphabet run from $1\ldots3$, and are associated with the spatial coordinates of space-time.
	\item Lowercase greek letters $\alpha,\beta,\gamma,\ldots,\mu,\nu,\rho,\ldots$ run from $0\ldots4$, and are associated with space-time coordinates.
	\item Uppercase latin letters from the middle of the alphabet $M,N,P,Q,\ldots$ are associated with higher dimensional space-times, and run from $0\ldots5$ or $0\ldots6$. When they appear it is going to be mentioned the dimension.
\end{itemize}
In the examples given here we will use $i,j,k,\ldots$ but these are general notions applied to all indices throughout the work.

\par The so-called Einstein implicit sum for tensor equations is
\begin{equation*}
	\sum_{i}V_{i}U^{i}=V_{i}U^{i}
\end{equation*}
This means that we drop the summation symbol $\sum$ to simplify the notation. Notice, however, that the summation occurs only over repeated indices, so for the expression
\begin{equation*}
	T_{i}^{\text{ }j}V^{i}=U^{j}
\end{equation*}
the left hand-side has only one implicit sum over the index $i$. In the expression above $j$ is referred to as a free index, and therefore indicates that this expression refers to vector components – in this case $U^{j}$.

\par Regarding units, unless it is explicitly stated in the text we shall use units where $c=G=k_{B}=\hbar=1$. The metric signature will always be with the time component negative, therefore in 4 dimensions the Minkowski space-time reads
\begin{equation*}
	ds^{2}=-dt^{2}+dx^{2}+dy^{2}+dz^{2}
\end{equation*}
The so-called AdS radius, $L$, appearing on AdS space-time metrics will be set to unity, unless otherwise stated. 

%% file: chapters/branes.tex
	\newpage
	\chapter{Black Holes} \label{ch-2}
	
	\par We start by reviewing the metric backgrounds that are going to be used throughout this work: black holes. For a long time, these objects were just a matter of theoretical speculation \cite{2009JAHH...12...90M}, but became a reality once the theory of General Relativity was published \cite{Einstein:1915by}. Essentially, a black hole can be regarded as a region of space-time where the gravitational attraction is so strong that even light cannot escape, hence the word ``black" in its name. 
	We should also remark that we are not only concerned with black holes from General Relativity, in fact, most of our results to be presented in the next chapters make use of black holes appearing in theories that are extensions or modifications of Einstein's theory of General Relativity.
	
	\section{Black Holes in General Relativity}
	
	\subsection{Overview of General Relativity}
	\par The General Theory of Relativity is currently the standard description of gravity. It was first presented to the community in 1915 by Albert Einstein \cite{Einstein:1915by} as a generalization of his theory of Special Relativity. In fact, General Relativity (GR) is a theory about space-time rather than gravity itself, whereas gravity emerges as a feature of space-time, namely: curvature. There is a wide range of applications of this theory, from understanding the nature of space-time itself and the gravitational interaction at the classical level, celestial motion, astrophysics, evolution and stability of stars, galaxies, black holes, and cosmology.
	
	
	\par This chapter is an introduction to the basic features of GR, needed to explore the realm of AdS/CFT. It is not intended to be an introduction to GR \textit{per se}, instead, it is presented such that instrumental use of the theory can be made in a consistent manner in future chapters. For an introductory level, there are two references that have shown to be very useful to the writer, these are books from Bernard Schutz \cite{Schutz:1985jx} and Sean Carroll \cite{Carroll:2004st}. Both books cover GR from scratch and require a minimum amount of knowledge to be understood, these are intended for the reader with no familiarity at all with GR. More advanced books on the subject are the ones from Wald \cite{Wald:1984rg}, MTW \cite{Misner:1974qy}, Weinberg \cite{Weinberg:1972kfs} and Hawking \& Ellis \cite{Hawking:1973uf}. These contain a formal approach to the subject and explore deeply the features of the theory. 
	
	
	\subsubsection*{Mathematical preliminaries}
	
	\par This is a very short summary of the mathematics of GR. It is by no means intended to be a formal introduction, but rather a short summary of the basic properties and expressions of tensors that will appear in the next sections. It is instructive to recall some notions of geometry before presenting the action and deriving the equations of motion.
	
	\par Throughout this chapter only pseudo-Riemannian manifolds are considered, the pseudo indicates a minus sign in one component of the metric tensor, which we now proceed to describe. The metric tensor encodes all properties of the geometry, from the metric tensor quantities such as the connection and curvature can be calculated. Formally it is called the First Fundamental Form of the manifold and characterizes the inner product, i.e. defines angles and lengths. For instance, consider the inner product of Special Relativity between four-vectors $A^{\mu}$ and $B^{\mu}$
	\begin{equation} \label{eq:1}
		A_{\mu}B^{\mu}=-A_{0}B^{0}+A_{i}B^{i}\ .
	\end{equation}
	In Eq. \eqref{eq:1} the metric is given by $\eta_{\mu\nu}=\text{diag}\left(-1,+1,+1,+1\right)$, the well known Minkowski metric. GR generalizes this notion and allows the components of the metric to be functions of the coordinates as well as not necessarily diagonal. A generic space-time metric will have its components denoted by $g_{\mu\nu}\equiv g_{\mu\nu}\left(x^{\mu}\right)$ and the following properties hold
	\begin{align}
		\begin{aligned}
			g_{\mu\nu}=g_{\nu\mu} \ , \ \ \ \
			g_{\mu\lambda}g^{\lambda\nu}=\delta_{\mu}^{\nu}\ ,
		\end{aligned}
	\end{align}
	i.e. it is symmetric and invertible.
	
	\par Upon generalization of the metric, one needs to introduce the covariant derivative, $\nabla_{\mu}$, as a replacement for the ordinary derivative $\partial_{\mu}$. It is necessary in order to define parallelism and the transport of vectors from one point to another in the manifold. From an intuitive point of view, one can think about the covariant derivative as an ordinary derivative plus a correction term that accounts for the change in the vector itself due to it being moved from one point to another in the manifold. Denoting a vector by $A^{\mu}$, its covariant derivative is defined by
	\begin{equation} \label{eq:2}
		\nabla_{\nu}A^{\mu}=\partial_{\nu}A^{\mu}+\Gamma_{\nu\lambda}^{\mu}A^{\lambda} \ .
	\end{equation}
	Similarly, let $A_{\mu}$ be a covector and $A^{\mu}_{\nu}$ a mixed tensor, the covariant derivative acts like
	\begin{align}
		\nabla_{\nu}A_{\mu}=\partial_{\nu}A_{\mu}-\Gamma_{\nu\mu}^{\rho}A_{\rho}	\label{eq:3}\ ,			\ \ \ \
		\nabla_{\lambda}A_{\text{ }\nu}^{\mu}=\partial_{\lambda}A_{\text{ }\nu}^{\mu}+\Gamma_{\lambda\rho}^{\mu}A_{\text{ }\nu}	^{\rho}-\Gamma_{\lambda\nu}^{\rho}A_{\text{ }\rho}^{\mu}  \ . 
	\end{align} 
	And so on for different combinations of indices in all sorts of tensors.
	
	\par At this point it should be emphasised that $\Gamma_{\nu\lambda}^{\mu}$ is only a connection that accounts to transporting the vector. Let $a,b\in\mathbb{R}$, $\varphi\equiv\varphi(x^{\mu})$ a scalar field, and $A,B$ be vectors or tensors, so the covariant derivative satisfies the four properties:
	\begin{itemize}
		\item Linearity: $\nabla\left(aA+bB\right)=a\nabla A+b\nabla B$.
		\item Acts as the regular derivative on scalar fields: $\nabla\varphi=\partial\varphi$.
		\item Obeys Leibniz rule: $\nabla\left(AB\right)=\left(\nabla A\right)B+A\left(\nabla B\right)$.
		\item Metric compatibility: $\nabla_{\lambda}g_{\mu\nu}=0$.
	\end{itemize}
	
	\par Consider the commutator of covariant derivatives, as given in Eq. \eqref{eq:2}, acting on a vector field $A^{\mu}$:
	\begin{equation} \label{eq:5}
		\left[\nabla_{\mu},\nabla_{\nu}\right]A^{\lambda}=R_{\text{ }\mu\nu\rho}^{\lambda}A^{\rho}+T_{\text{ }\mu\nu}^{\rho}\nabla_{\rho}V^{\lambda}\ ,
	\end{equation}
	the first term is the Riemann tensor contracted with the vector, whereas the second one is the torsion tensor. This relation enables one to extract all the relevant geometric quantities from knowledge of the covariant derivative operator. Put in other words, given the connection $\Gamma$ all geometric properties of a manifold can be obtained.
	
	\par There are several ways of deriving the GR connection, the so-called Levi-Civita connection, or sometimes Christoffel connection. The argument given here is the simplest one for time's sake, the interested reader is referred to one of the references at the beginning of this chapter for other, more physically intuitive, derivations.
	
	\par There is a theorem \cite{RicciTheoremBook}  that establishes that torsion and curvature are uniquely defined by the connection, this fact enables a significant simplification. It is widely known in GR that torsion plays no role at all in the theory, and all the geometric effects are related to curvature. This is, of course, a choice - a wise one, but nonetheless, a choice -, that simplifies a lot of the expressions. From this theorem, there is one, and only one, connection that accounts for zero torsion in Eq. \eqref{eq:5} and it is given by:
	\begin{equation} \label{eq:6}
		\Gamma_{\mu\nu}^{\lambda}=\frac{1}{2}g^{\lambda\rho}\left(\partial_{\mu}g_{\rho\nu}+\partial_{\nu}g_{\rho\mu}-\partial_{\rho}g_{\mu\nu}\right) \ .
	\end{equation}
	As one can easily check, this connection is symmetric and thus, makes the second term in Eq. \eqref{eq:5} vanish. Eq. \eqref{eq:6} can be derived directly from the metric compatibility condition once it is assumed that $\Gamma_{\mu\nu}^{\lambda}=\Gamma_{\nu\mu}^{\lambda}$.
	
	\par Using Eq. \eqref{eq:6} in Eq. \eqref{eq:5} one can derive the explicit expression for the Riemann tensor (recall that the torsion vanishes identically)
	\begin{align*}
		\begin{aligned}
			\left[\nabla_{\mu},\nabla_{\nu}\right]V^{\lambda}&=\nabla_{\mu}\left(\nabla_{\nu}V^{\lambda}\right)-\nabla_{\nu}\left(\nabla_{\mu}V^{\lambda}\right)\\
			&=\partial_{\mu}\nabla_{\nu}V^{\lambda}-\Gamma_{\mu\nu}^{\rho}\nabla_{\rho}V^{\lambda}+\Gamma_{\mu\rho}^{\lambda}\nabla_{\nu}V^{\rho}-\partial_{\nu}\nabla_{\mu}V^{\lambda}+\Gamma_{\mu\nu}^{\rho}\nabla_{\rho}V^{\lambda}-\Gamma_{\nu\rho}^{\lambda}\nabla_{\mu}V^{\rho}\\
			&=\partial_{\mu}\left(\partial_{\nu}V^{\lambda}+\Gamma_{\nu\theta}^{\lambda}V^{\theta}\right)-\Gamma_{\mu\nu}^{\rho}\left(\partial_{\rho}V^{\lambda}+\Gamma_{\rho\theta}^{\lambda}V^{\theta}\right)+\Gamma_{\mu\rho}^{\lambda}\left(\partial_{\nu}V^{\rho}+\Gamma_{\nu\theta}^{\rho}V^{\theta}\right)\\
			&-\partial_{\nu}\left(\partial_{\mu}V^{\lambda}+\Gamma_{\mu\theta}^{\lambda}V^{\theta}\right)+\Gamma_{\mu\nu}^{\rho}\left(\partial_{\rho}V^{\lambda}+\Gamma_{\rho\theta}^{\lambda}V^{\theta}\right)-\Gamma_{\nu\rho}^{\lambda}\left(\partial_{\mu}V^{\rho}+\Gamma_{\mu\theta}^{\rho}V^{\theta}\right)\\
			&=\partial_{\mu}\Gamma_{\nu\theta}^{\lambda}V^{\theta}-\partial_{\nu}\Gamma_{\mu\theta}^{\lambda}V^{\theta}+\Gamma_{\mu\rho}^{\lambda}\Gamma_{\nu\theta}^{\rho}V^{\theta}-\Gamma_{\nu\rho}^{\lambda}\Gamma_{\mu\theta}^{\rho}V^{\theta}\\
			&=\left(\partial_{\mu}\Gamma_{\nu\theta}^{\lambda}-\partial_{\nu}\Gamma_{\mu\theta}^{\lambda}+\Gamma_{\mu\rho}^{\lambda}\Gamma_{\nu\theta}^{\rho}-\Gamma_{\nu\rho}^{\lambda}\Gamma_{\mu\theta}^{\rho}\right)V^{\theta}\\
			&=R_{\text{ }\theta\mu\nu}^{\lambda}V^{\theta}
		\end{aligned}
	\end{align*}
	i.e. 
	\begin{equation} \label{eq:7}
		R_{\text{ }\theta\mu\nu}^{\lambda}=\partial_{\mu}\Gamma_{\nu\theta}^{\lambda}-\partial_{\nu}\Gamma_{\mu\theta}^{\lambda}+\Gamma_{\mu\rho}^{\lambda}\Gamma_{\nu\theta}^{\rho}-\Gamma_{\nu\rho}^{\lambda}\Gamma_{\mu\theta}^{\rho}\ .
	\end{equation}
	The Ricci tensor is defined from the Riemann tensor by the contraction
	\begin{equation} \label{eq:8}
		R_{\mu\nu}=R_{\text{ }\mu\lambda\nu}^{\lambda}\ .
	\end{equation}
	And taking the trace of the Ricci tensor one obtains Ricci scalar
	\begin{equation} \label{eq:9}
		g^{\mu\nu}R_{\mu\nu}=R\ .
	\end{equation}
	Notice that they can all be obtained from the metric tensor since the connection is given by derivatives of the metric, c.f. Eq. \eqref{eq:6}, and all these tensors are obtained from the connection.
	
	
	
	\subsubsection*{The Einstein-Hilbert action}
	\par The action principle, the method of minimizing the action in order to find the equations of motion, is the quickest way to introduce the Einstein equation. 
	One usually applies the action principle by writing the action functional as
	\begin{equation} \label{eq:10}
		S=\int d^{4}x\sqrt{-g}\mathcal{L}\ ,
	\end{equation}
	where $\mathcal{L}$ is, in fact, a sum of various contributions to the action, i.e. electromagnetic, gravitational, scalar, more general gauge fields, etc. The principle states that the variation of the action \eqref{eq:10}  (which is a minimum)  vanishes, and the solution to this problem are the equations of motion of the system. 
	
	\par Let $\mathcal{L}$ in Eq. \eqref{eq:10} be
	\begin{equation} \label{eq:11}
		\mathcal{L} = \mathcal{L}_{grav}+\mathcal{L}_{matter}\ ,
	\end{equation}
	where the first term comprises all the effects of the gravitational field in the paradigm of GR, which means that it describes how the curvature of the spacetime is modified by the second term, which is everything else.\footnote{In GR any other field that is not the metric is considered matter since it contributes to the energy-momentum tensor.}
	
	\par The gravitational part of the Lagrangian \eqref{eq:11} is given by
	\begin{equation} \label{eq:12}
		\mathcal{L}_{grav}=\frac{8\pi G}{c^4}\left(R+2\Lambda\right)\ ,
	\end{equation}
	where $R$ is the scalar curvature, c.f. Eq. \eqref{eq:9}, and $\Lambda$ the cosmological constant. The action $S=\int d^4x \sqrt{-g}\frac{8\pi G}{c^4}\left(R+2\Lambda\right)$ is usually called Hilbert Action (or Einstein-Hilbert action) after David Hilbert who first introduced it as a way to obtain Einstein's equations. Einstein himself derived his equation using geometric arguments and the conservation of the energy-momentum tensor \cite{Einstein:1915by}.
	
	\par Writing Eq. \eqref{eq:10} using Eqs. \eqref{eq:11} and \eqref{eq:12} one finds
	\begin{equation} \label{eq:13}
		S=\int d^4x\sqrt{-g}\left[\frac{1}{2\kappa}\left(R+2\Lambda\right)+\mathcal{L}_{matter}\right]\ ,
	\end{equation}
	where $\frac{1}{2\kappa}=\frac{8\pi G}{c^4}$. Variation of Eq. \eqref{eq:13} with respect to the metric tensor\footnote{Variations can be taken with respect to any field appearing in the action functional, it is usually written as $\frac{\delta S}{\delta\Phi}=0$, where $\Phi$ is the field. In fact this is just a cumbersome notation, what is meant is $\delta S=\int d^4x\delta \Phi[\ldots]=0$.} leads to the Einstein's equation. Consider the general form of the variation of \eqref{eq:13}:
	\begin{align}
		\begin{aligned} \label{eq:14}
			\delta S&=\int d^{4}x\delta g^{\mu\nu}\left[\frac{1}{2\kappa}\frac{\delta\left(\sqrt{-g}R\right)}{\delta g^{\mu\nu}}+\frac{\delta\left(\sqrt{-g}\mathcal{L}_{matter}\right)}{\delta g^{\mu\nu}}\right]\\
			&=\int d^{4}x\delta g^{\mu\nu}\left[\frac{1}{2\kappa}\left(R\frac{\delta\sqrt{-g}}{\delta g^{\mu\nu}}+\sqrt{-g}\frac{\delta R}{\delta g^{\mu\nu}}\right)+\frac{\delta\left(\sqrt{-g}\mathcal{L}_{matter}\right)}{\delta g^{\mu\nu}}\right]\\
			&=\int d^{4}x\delta g^{\mu\nu}\sqrt{-g}\left[\frac{1}{2\kappa}\left(\frac{R}{\sqrt{-g}}\frac{\delta\sqrt{-g}}{\delta g^{\mu\nu}}+\frac{\delta R}{\delta g^{\mu\nu}}\right)+\frac{1}{\sqrt{-g}}\frac{\delta\left(\sqrt{-g}\mathcal{L}_{matter}\right)}{\delta g^{\mu\nu}}\right] \ .
		\end{aligned}
	\end{align}
	The second term inside the square brackets is defined as the energy-momentum tensor
	\begin{equation} \label{eq:15}
		T_{\mu\nu}=-\frac{2}{\sqrt{-g}}\frac{\delta\left(\sqrt{-g}\mathcal{L}_{matter}\right)}{\delta g^{\mu\nu}}\ .
	\end{equation}
	Since this variation must hold for arbitrary $\delta g^{\mu\nu}$ the geometric part of Einstein's equation is obtained from the term:
	\begin{equation} \label{eq:16}
		\frac{R}{\sqrt{-g}}\frac{\delta\sqrt{-g}}{\delta g^{\mu\nu}}+\frac{\delta R}{\delta g^{\mu\nu}}= G_{\mu\nu}\ ,
	\end{equation}
	as will be shown.
	
	\par Variation of the determinant requires using the following identity
	\begin{equation}\label{eq:17}
		\ln\left(\det M\right)=\text{Tr}\left(\ln M\right)\ ,
	\end{equation}
	which applies to any matrix $M$.
	Varying this identity leads to
	\begin{align}
		\begin{aligned}	\label{eq:18-2}
			\delta\ln\left(\det M\right)&=\delta\text{Tr}\left(\ln M\right)\ , \\
			\frac{1}{\det M}\delta M&=\text{Tr}\left(M^{-1}\delta M\right)\ .
		\end{aligned}
	\end{align}
	Now let $\det M=\det g_{\mu\nu}=g$\, thus
	\begin{align} \label{eq:19}
		\delta g&=gg^{\mu\nu}\delta g_{\mu\nu}\ .
	\end{align}
	Finally, putting all the pieces together to write the variation of $\sqrt{-g}$:
	\begin{align}
		\begin{aligned} \label{eq:20}
			\delta\sqrt{-g}&=-\frac{1}{2\sqrt{-g}}\delta g\\
			&=-\frac{g}{2\sqrt{-g}}g^{\mu\nu}\delta g_{\mu\nu}\\
			&=\frac{g}{2\sqrt{-g}}g_{\mu\nu}\delta g^{\mu\nu}\\
			&=-\frac{\sqrt{-g}}{2}g_{\mu\nu}\delta g^{\mu\nu}\ .
		\end{aligned}
	\end{align}
	Notice that the sign change from the second to third line occurs because $\delta\left(g^{\mu\nu}g_{\mu\nu}\right)=\delta\left[\text{Tr}(\text{Id})\right]=0\implies g_{\mu\nu}\delta g^{\mu\nu}+g^{\mu\nu}\delta g_{\mu\nu}=0$.
	
	\par The second term on the left--hand side of Eq. \eqref{eq:16} requires to look back on the definition of the scalar curvature (c.f. Eq. \eqref{eq:9}) as the trace of the contraction of the Riemann tensor. Thus, by varying the Riemann tensor, which according to Eq. \eqref{eq:7} is
	\begin{equation} \label{eq:21}
		\delta R_{\text{ }\theta\mu\nu}^{\lambda}=\partial_{\mu}\delta\Gamma_{\nu\theta}^{\lambda}-\partial_{\nu}\delta\Gamma_{\mu\theta}^{\lambda}+\Gamma_{\mu\alpha}^{\lambda}\delta\Gamma_{\nu\theta}^{\alpha}+\Gamma_{\mu\alpha}^{\lambda}\delta\Gamma_{\nu\theta}^{\alpha}-\Gamma_{\nu\alpha}^{\lambda}\delta\Gamma_{\mu\theta}^{\alpha}-\Gamma_{\mu\theta}^{\alpha}\delta\Gamma_{\nu\alpha}^{\lambda}+\Gamma_{\nu\theta}^{\alpha}\delta\Gamma_{\mu\alpha}^{\lambda}\ .
	\end{equation}
	The variation of the connection is defined by $\Gamma_{\mu\nu}^{\lambda}\mapsto\Gamma_{\mu\nu}^{\lambda}+\delta\Gamma_{\mu\nu}^{\lambda}$, i.e. $\delta\Gamma_{\mu\nu}^{\lambda}$ is the difference between two connections, therefore it is a tensor. This fact allows one to take the covariant derivative $\nabla_{\alpha}\delta\Gamma_{\mu\nu}^{\lambda}$:
	\begin{equation}\label{eq:22}
		\nabla_{\mu}\delta\Gamma_{\nu\theta}^{\lambda}=\partial_{\mu}\delta\Gamma_{\mu\theta}^{\lambda}+\Gamma_{\mu\alpha}^{\lambda}\delta\Gamma_{\nu\theta}^{\alpha}-\Gamma_{\mu\nu}^{\alpha}\delta\Gamma_{\alpha\theta}^{\lambda}-\Gamma_{\mu\theta}^{\alpha}\delta\Gamma_{\nu\alpha}^{\lambda}\ ,
	\end{equation}
	and similarly for $\nabla_{\nu}\delta\Gamma_{\mu\theta}^{\lambda}$. After a little algebra one sees that Eq. \eqref{eq:21} is equivalent to
	\begin{equation}\label{eq:23}
		\delta R_{\text{ }\theta\mu\nu}^{\lambda}=\nabla_{\mu}\delta\Gamma_{\nu\theta}^{\lambda}-\nabla_{\nu}\delta\Gamma_{\mu\theta}^{\lambda}\ .
	\end{equation}
	Contraction of Eq. \eqref{eq:23} gives the variation of Ricci tensor
	\begin{equation}\label{eq:24}
		\delta R_{\theta\nu}=\delta R_{\text{ }\theta\mu\nu}^{\mu}=\nabla_{\mu}\delta\Gamma_{\nu\theta}^{\mu}-\nabla_{\nu}\delta\Gamma_{\mu\theta}^{\mu}\ .
	\end{equation}
	Multiplying by the inverse metric the variation of scalar curvature is obtained:
	\begin{align}
		\begin{aligned}\label{eq:25}
			\delta R=g^{\mu\nu}\delta R_{\mu\nu}&=g^{\mu\nu}\left(\nabla_{\alpha}\delta\Gamma_{\mu\nu}^{\alpha}-\nabla_{\nu}\delta\Gamma_{\mu\alpha}^{\alpha}\right)\\
			&=\nabla_{\alpha}g^{\mu\nu}\delta\Gamma_{\mu\nu}^{\alpha}-\nabla_{\nu}g^{\mu\nu}\delta\Gamma_{\mu\alpha}^{\alpha}\\
			&=\nabla_{\alpha}\left(g^{\mu\nu}\delta\Gamma_{\mu\nu}^{\alpha}-g^{\mu\alpha}\delta\Gamma_{\mu\nu}^{\nu}\right)\ ,
		\end{aligned}
	\end{align}
	where it is shown that this particular variation can be written as a single covariant derivative, which upon multiplication by $\sqrt{-g}$ becomes a total derivative, so by Stokes theorem, the integral of this term vanishes:
	\begin{equation} \label{eq:26}
		\int d^{4}x\sqrt{-g}\nabla_{\alpha}\left(g^{\mu\nu}\delta\Gamma_{\mu\nu}^{\alpha}-g^{\mu\alpha}\delta\Gamma_{\mu\nu}^{\nu}\right)=0\ .
	\end{equation}
	Finally, notice that from the definition of scalar curvature, Eq. \eqref{eq:9}:
	\begin{equation}\label{eq:27}
		\delta\left(g^{\mu\nu}R_{\mu\nu}\right)=R_{\mu\nu}\delta g^{\mu\nu}-g^{\mu\nu}\delta R_{\mu\nu}\ ,
	\end{equation}
	from Eq. \eqref{eq:26} it is seen that the second term vanishes upon integration, so the action in Eq. \eqref{eq:14} becomes
	\begin{equation} \label{eq:28}
		\delta S=\int d^{4}x\delta g^{\mu\nu}\sqrt{-g}\left[\frac{1}{2\kappa}\left(-\frac{R}{2}g_{\mu\nu}+R_{\mu\nu}\right)-\frac{1}{2}T_{\mu\nu}\right]\ ,
	\end{equation}
	where Eqs. \eqref{eq:15}, \eqref{eq:20} and \eqref{eq:27} were used. The term in the square brackets leads to the Einstein's equations
	\begin{equation} \label{eq:29}
		R_{\mu\nu}-\frac{R}{2}g_{\mu\nu}=\kappa T_{\mu\nu}\ .
	\end{equation}
	
	\par As a final remark for this section, it must be said that the cosmological constant present in Eq. \eqref{eq:13} was not considered but it can be easily seen that its contribution to the Einstein's equation can be derived from Eq. \eqref{eq:20} since that is the only variation contributing. Einstein's equations including the cosmological constant $\Lambda$ are
	\begin{equation} \label{eq:30}
		R_{\mu\nu}-\frac{R}{2}g_{\mu\nu}+\Lambda g_{\mu\nu}=\kappa T_{\mu\nu}\ .
	\end{equation}
	The cosmological constant allows for solutions with positive and negative constant curvature in vacuum, which are of great importance in this work. 
	
	\subsection{Anti-de Sitter space-time} \label{sec:2.1}
	\par There is a class of solutions to Einstein's equations called maximally symmetric spaces. These space-times have all possible symmetries allowed and it can be shown that it depends on the scalar curvature. For a $D-$dimensional manifold there are $\frac{1}{2}D(D+1)$ possible symmetries, which are translations and rotations along all the axes (recall that when the space-time has a Lorentzian signature some rotations are actually boosts, but this does not change the counting of symmetries). For such a manifold the curvature must be the same everywhere, which means that the Riemann tensor doesn't change at all, regardless of symmetry transformations done to its components. In fact, there are only three tensors that are invariant under these conditions: the Kronecker delta, Levi-Civita tensor (totally anti-symmetric tensor), and the metric tensor. Thus the Riemann curvature tensor must be, at most, a combination of these three in such circumstances. From the symmetry properties of the Riemann tensor, one can only find one combination \cite{Wald:1984rg}
	\begin{equation}\label{eq:31}
		R_{\mu\nu\lambda\theta}=k\left(g_{\mu\lambda}g_{\nu\theta}-g_{\mu\theta}g_{\nu\lambda}\right)\ ,
	\end{equation}
	where $k$ is determined by contracting the expression twice with the metric tensor. In doing that, one obtains
	\begin{equation}\label{eq:32}
		k=\frac{R}{D\left(D-1\right)}\ .
	\end{equation}
	
	\par Eqs. \eqref{eq:31} and \eqref{eq:32} show that maximally symmetric spaces are basically described by the scalar curvature and, since it is the same everywhere, there are only three possibilities: positive, negative, and zero. For vanishing scalar curvature it is obvious that the Riemann tensor will vanish as well (c.f.  Eq. \eqref{eq:31}), and that is Minkowski space-time, the space-time of Special Relativity. For positive and negative values they produce the solutions called de Sitter and anti-de Sitter space-times, respectively. This is where the cosmological constant, $\Lambda$ in Eq. \eqref{eq:30}, plays an important role. Consider Einstein's equations in vacuum, i.e. $T_{\mu\nu}=0$ and take its trace, from Eq. \eqref{eq:30}:
	\begin{equation} \label{eq:33}
		R=4\Lambda\ ,
	\end{equation}
	thus one sees that if the cosmological constant is absent the only possibility for the scalar curvature is actually zero. Also, notice that the sign of the cosmological constant determines the sign of the scalar curvature. In the remainder of this subsection the solution for negative cosmological constant is treated, hence negative scalar curvature, which is called the anti-de Sitter space-time.
	
	\par It is usual to rename the constant $k$ in Eq. \eqref{eq:32} as
	\begin{equation} \label{eq:34}
		k=\frac{1}{L^2}\ ,
	\end{equation}
	where $L$ is called radius of curvature, thus
	\begin{align}\label{eq:35}
		R=\frac{D\left(D-1\right)}{L^{2}}\text{, }&\Lambda=-\frac{D\left(D-1\right)}{4L^{2}}\ .
	\end{align}
	
	\par The explicit expression for the metric is obtained by embedding the solution in a higher dimensional space-time, let its dimension be $N=D+1$, and the metric given by the following expression in coordinates $\left\{ \xi^{0},\ldots,\xi^{D}\right\}$
	\begin{equation} \label{eq:36}
		ds^{2}=-\left(d\xi^{0}\right)^{2}-\left(d\xi^{1}\right)^{2}+\left(d\xi^{2}\right)^{2}+\left(d\xi^{3}\right)^{2}+\ldots+\left(d\xi^{n}\right)^{2}\ ,
	\end{equation}
	i.e. $\mathbb{R}^{(n-1,2)}$. Then AdS space-time is defined by the constraint
	\begin{equation}\label{eq:37}
		\xi_{\mu}\xi^{\mu}=-\left(\xi^{0}\right)^{2}-\left(\xi^{1}\right)^{2}+\left(\xi^{2}\right)^{2}+\ldots+\left(\xi^{n}\right)^{2}=-L^{2}\ .
	\end{equation}
	This is a hyperboloid. Considering the case for $D=4$, one parametrization is given by
	\begin{align}
		\begin{aligned}\label{eq:38}
			\xi^{0}=L\sin\left(\frac{t}{L}\right)\cosh\left(\frac{r}{L}\right),\ \ \ \
			\xi^{1}=L\cos\left(\frac{t}{L}\right)\cosh\left(\frac{r}{L}\right),\ \ \ \
			\vec{\xi}=L\sinh\left(\frac{r}{L}\right)\vec{n},
		\end{aligned}
	\end{align}
	where $\vec{\xi}=\left(\xi^2,\xi^3,\xi^4\right)$ and $\vec{n}=\left(\sin\theta\cos\phi,\sin\theta\sin\phi,\cos\theta\right)$, so the metric is
	\begin{equation}\label{eq:39}
		ds^{2}=L^{2}\left(-\cosh^{2}\left(\frac{R}{L}\right)dt^{2}+dr^{2}+\sinh^2\left(\frac{r}{L}\right)d\Omega_{2}^{2}\right)\ .
	\end{equation}
	Notice that $t=const.$ is the metric of hyperbolic space. 
	Two things that should be remarked about Eq. \eqref{eq:39} are: i) it has a singularity along $r=0$, ii) its time coordinate is periodic. Both of these problems are due to the choice of coordinates in Eq. \eqref{eq:38}, in fact, these coordinates are not of much use and another coordinate system is more suitable to describe AdS space-time.
	
	\par Define the coordinates $\left\{ t,r,x,y\right\} =\left\{ t,r,\vec{x}\right\}$ in the following way
	\begin{align}
		\begin{aligned} \label{eq:40}
			\xi^{0}&=\frac{L}{2r}\left(1+\frac{r^{2}}{L^{4}}\left(L^{2}+\vec{x}^{2}-t^{2}\right)\right), \ \ \
			\xi^{1}=\frac{r}{L}t,\ \ \ \
			\xi^{i}=\frac{r}{L}x^{i},\\
			\xi^{D+1}&=\frac{L}{2r}\left(1-\frac{r^{2}}{L^{4}}\left(L^{2}+\vec{x}^{2}-t^{2}\right)\right),
		\end{aligned}
	\end{align} 
	in these coordinates Eq. \eqref{eq:36} becomes
	\begin{equation}\label{eq:41}
		ds^{2}=-\frac{r^{2}}{L^{2}}dt^{2}+\frac{L^{2}}{r^{2}}dr^{2}+\frac{r^{2}}{L^{2}}d\vec{x^{2}}\ .
	\end{equation}
	This is the AdS space-time in the so-called Poincaré Coordinates, it is the most commonly used coordinate system in holography. Notice that in these coordinates $-\infty<t<\infty$, $r>0$. The singularity is located at the point $r=0$, but now it can be seen to be a Killing horizon. Also $r\to\infty$ approaches the boundary of AdS space-time, sometimes called the conformal boundary, for it is in fact Minkowski space-time with conformal factor $r/L$.
	
	\par Now, by defining $u=r/L$ Eq. \eqref{eq:41} reads
	\begin{equation}\label{eq:42}
		ds^{2}=L^{2}\left(\frac{du^{2}}{u^{2}}+u^{2}dx_{\mu}dx^{\mu}\right)\ ,
	\end{equation}
	where $dx_{\mu}dx^{\mu}=-dt^{2}+d\vec{x}^{2}$. Further, defining $z=1/u$, Eq. \eqref{eq:42} is cast in the following form
	\begin{equation} \label{eq:43}
		ds^{2}=\frac{L^{2}}{z^{2}}\left(dz^{2}+dx_{\mu}dx^{\mu}\right)\ .
	\end{equation}
	All forms of AdS space-time metric, Eqs. \eqref{eq:41}-\eqref{eq:43}, exhibit the same properties mentioned below Eq. \eqref{eq:41}, it is just a matter of convenience using one or another by the simple transformation that leads from one to the other. A complete discussion of multiple coordinate systems that can be used for the AdS space-time can be found at \cite{Bayona:2005nq}.
	
	\subsection{Schwarzschild Black Hole}
	\par Apart from the trivial solutions of Einstein's equations (metrics with constant components), probably the simplest one is the so-called Schwarzschild space-time. It is a spherically symmetric static solution, which essentially means that the components of the metric tensor depend only on the distance coordinate from the origin $r$.  It is also a vacuum solution - i.e. exterior part of the body -, so we set the energy-momentum tensor to zero.
	
	\par One starts by plugging the following ansatz in the Einstein's equations
	\begin{equation}\label{eq:2-1}
		ds^{2}=-e^{2a\left(r\right)}dt^{2}+e^{2b\left(r\right)}dr^{2}+r^{2}d\Omega^{2}\ ,
	\end{equation}
	notice that all the components depend only on $r$, as previously stated. Moreover $d\Omega^2=d\theta^2+\sin^2\theta d\phi^2$ is the (squared) solid angle element of a unit sphere. The solution for Einstein's equations are then reduced to solving the following system
	\begin{align}
		\begin{aligned} \label{eq:2-2}
			R_{tt}+R_{rr}&=&0 , \\
			R_{\theta \theta} &=& 0 .
		\end{aligned}
	\end{align}
	A simple computation of the Ricci tensor for metric \eqref{eq:2-1} gives the following
	\begin{align}
		\begin{aligned} \label{eq:2-3}
			\frac{d}{dr}\left(re^{2a\left(r\right)}\right)&=1\ , \\
			a(r)&=-b(r)+const.\ .
		\end{aligned}
	\end{align}
	Where we identify the second equation as a constraint. The solution for the differential equation reads
	\begin{equation}\label{eq:2-4}
		e^{2a}=1-\frac{2R_S}{r}\ ,
	\end{equation}
	when substituted in \eqref{eq:2-1} gives the Schwarzschild metric
	\begin{equation}\label{eq:2-5}
		ds^{2}=-\left(1-\frac{2R_S}{r}\right)dt^{2}+\left(1-\frac{2R_S}{r}\right)^{-1}dr^{2}+r^{2}d\Omega^{2}\ .
	\end{equation}
	In these expressions, $R_S$ is a constant of integration, which can be identified with $R_S=2GM$, where $M$ is the total mass of the body, which in this case is treated as point-like since we are concerned about the vacuum region, i.e. exterior to the mass distribution. This is called the Schwarzschild radius, and it is this region that defines what a black hole is.

	
	\subsection{Reissner-Nordstr\"om Black Hole} \label{rnBH}
	\par Having knowledge of the Schwarzschild solution, one can study what happens when, aside from mass, one adds an electric charge to the body associated with the curvature in space-time. This of course adds a non-zero energy-momentum tensor to Einstein's equations and therefore gives a new solution, even when keeping all the other assumptions unchanged. Notice, however, that the non-zero energy-momentum tensor is due to the electric field produced by the body, the region of interest still does not have matter content.
	
	\par To couple the gravitational to electromagnetic field we simply add the Maxwell Lagrangian to the Einstein-Hilbert action, c.f. Eq. \eqref{eq:13}:
	\begin{equation} \label{eq:2-6}
		\mathcal{L}_{matter}=-\frac{1}{4}F_{\alpha\beta}F^{\alpha\beta}\ ,
	\end{equation}
	where $F_{\mu\nu}$ is the Maxwell tensor, defined from the gauge potential $A_{\mu}$ by $F_{\mu\nu}=2\partial_{[\mu}A_{\nu ]}$. The energy momentum tensor, c.f. Eq. \eqref{eq:15}, can be written as
	\begin{equation} \label{eq:2-6,5}
		T_{\mu\nu}=-2\frac{\delta\mathcal{L}_{matter}}{\delta g^{\mu\nu}}+g_{\mu\nu}\mathcal{L}_{matter} = F_{\mu\alpha}F_{\nu}^{\text{ }\alpha}-\frac{1}{4}g_{\mu\nu}F_{\alpha\beta}F^{\alpha\beta}\ .
	\end{equation}
	
	\par The coupled system is formed by Einstein's, c.f. \eqref{eq:29}, and Maxwell's equations
	\begin{equation} \label{eq:2-7}
		\partial_{\mu}\left(\sqrt{-g}F^{\mu\nu}\right)=0\ .
	\end{equation}
	We will consider an electrostatic configuration, along with spherical symmetry, which allows us to identify the only non-zero component of $F_{\mu\nu}$
	\begin{equation} \label{eq:2-8}
		E_r=F_{tr}=f(r)\ .
	\end{equation}
	Formally one should consider time dependency, but as a consequence of the equations of motion \eqref{eq:2-7}, one can show that such dependency is not present for the configuration we are considering. 
	
	\par For the sake of brevity, the full calculations are not carried over. The components $T_{tt}$ and $T_{rr}$ are equal and have opposite sign, thus the combination $R_{tt}+R_{rr}=0$ holds and the second result in Eq. \eqref{eq:2-3} is obtained. Solving the Maxwell equation for the electric field, the $t$-component of Eq. \eqref{eq:2-7} will be, after all, substitutions
	\begin{align}
		\begin{aligned}\label{eq:2-9}
			\partial_{r}\left(r^{2}F^{rt}\right)=\partial_{r}\left(r^{2}f\right)=0
			\implies f\left(r\right)=\frac{const.}{r^{2}}\ .
		\end{aligned}
	\end{align}
	The constant is identified with the electric charge, $Q$, requiring that this assumes the form of Coulomb's law of electrostatics.
	
	\par Now there is only the $\theta\theta$-component of Einstein's equations to be solved, the differential equation is slightly different from Eq. \eqref{eq:2-3}, in fact, it reads
	\begin{equation}\label{eq:2-10}
		\frac{d}{dr}\left(re^{2a\left(r\right)}\right)=1-\frac{Q^{2}}{r^{2}}\ ,
	\end{equation}
	and the solution is
	\begin{equation}
		e^{2a\left(r\right)}=1-\frac{R_{S}}{r}+\frac{Q^{2}}{r^{2}}\ .
	\end{equation}
	From our ansatz and considerations made here, we find the Reissner-Nortstr\"om metric
	\begin{equation}\label{eq:2-11}
		ds^{2}=-\left(1-\frac{R_{S}}{r}+\frac{Q^{2}}{r^{2}}\right)dt^{2}+\left(1-\frac{R_{S}}{r}+\frac{Q^{2}}{r^{2}}\right)^{-1}dr^{2}+r^{2}d\Omega^{2}\ .
	\end{equation}
	
	\par Differently from Schwarzschild space-time we can readily see that there are 2 singular surfaces for the $r$ coordinate, these are usually denoted by $r_{\pm}$, and one can verify that it is dependent on the mass and charge of the black hole:
	\begin{equation}\label{eq:2-12}
		r_{\pm}=M\pm\sqrt{M^{2}-Q^{2}}\ .
	\end{equation}
These two are in fact null hypersurfaces, meaning they are indeed event horizons of the space-time. The fact that components of the metric vanish or diverge at these values has to do with the coordinate system being used to describe the space-time, the only point that shows an unavoidable divergence is $r=0$, which can be seen by the evaluation of curvature scalars such as the Kretschmann scalar.
	
	\section{Thermodynamics of black holes}\label{sec:3}
	
	\par In the late '60s and during the '70s it was found that black holes obey a set of laws that resembles the laws of thermodynamics, the works of Carter, Bekenstein, Hawking, and many others \cite{Bardeen:1973gs, Bekenstein:1973ur, Hawking:1974rv} are seminal references to the topic which lives to this day, and in fact constitutes a research area on its own. It all started as an analogy, originating from the similarity between equations obtained for black holes and the already known laws of thermodynamics, but far from stopping at this point, it is actually shown formally that the laws of black hole mechanics hold in their own and the term thermodynamics of black holes is used merely as terminology.
	
	\par Here the laws of black hole mechanics are stated and some of its most important features are highlighted since the subject is very technical and would demand the introduction of a large mathematical machinery we refer to \cite{Hawking:1973uf, Wald:1995yp} for a complete discussion.
	\par The zeroth law of thermodynamics defines temperature using equilibrium states. For black holes, the analog of an equilibrium state is a stationary space-time, which is the fate of a black hole. It can be shown that a quantity called surface gravity is constant everywhere at the event horizon of a stationary black hole \cite{Wald:1995yp}, when measured by an observer at asymptotic infinity. It is defined in purely geometrical terms as \cite{Wald:1984rg}
	\begin{equation}\label{eq:64}
		\kappa^{2}=\lim_{r\to r_{0}}\left[\frac{\left(\xi^{\mu}\nabla_{\mu}\xi^{\lambda}\right)\left(\xi^{\nu}\nabla_{\nu}\xi_{\lambda}\right)}{\xi^{\rho}\xi_{\rho}}\right]\ ,
	\end{equation}
	where $\xi$ is the time like Killing vector,\footnote{ \label{footnote}
	The Killing equation in local coordinates reads \begin{equation*}
		\nabla_{\mu}X_{\nu}+\nabla_{\nu}X_{\mu}=0\ .
	\end{equation*} The vector fields satisfying these equations are called Killing vector fields. They are related to the symmetries of the space-time, in fact, the Killing equation is obtained in the current form from demanding that the vector $X_{\mu}$ preserves the metric.  Killing vector fields are infinitesimal generators of isometries of the metric tensor. }
	and we are considering the case for a static space-time -- for a more general, stationary solution, the expression is similar but the Killing vector is a linear combination of the Killing vectors associated with time translations and rotations --. For static space-times the Killing vector is simply $\xi_{\mu}=\partial_{t}$.\footnote{It is worth mentioning that \eqref{eq:64} will only work when the coordinates are regular at the horizon, otherwise it will lead to a meaningless expression.}
	
	\par Identifying Eq. \eqref{eq:64} with the actual surface gravity requires some effort, the proof is found in \cite{Wald:1984rg}, after the calculations, it can be verified that
	\begin{equation}
		\kappa=\lim_{r\to r_{0}}(Va)\ ,
	\end{equation}
	where $V$ is the redshift factor and $a$ the magnitude of the four acceleration. Essentially the surface gravity measures the force necessary to keep a particle from falling into the event horizon, such force is measured by an observer at infinity. The formal relation between the surface gravity and temperature is given by
	\begin{equation}\label{eq:66}
		T=\frac{\kappa}{2\pi}\ ,
	\end{equation}
	it was derived first by Hawking \cite{Hawking:1974rv} in his work where he also introduced the so-called Hawking radiation, which is the cornerstone for the development of thermodynamics of black holes - it was this work that started to establish the laws of black hole mechanics, before that it was considered an analogy with classical thermodynamic equations.
	
	\par Next consider a more general black hole, it is called Kerr-Newman black hole \cite{Newman:1965my}. The metric in Boyer-Lindquist coordinates is
	\begin{equation}
	ds^{2}=-\left(dt-a\sin^{2}\theta d\phi\right)^{2}\frac{\Delta}{\rho^{2}}+\left(\frac{dr^{2}}{\Delta}+d\theta^{2}\right)\rho^{2}+\left[\left(r^{2}+a^{2}\right)d\phi-acdt\right]^{2}\frac{\sin^{2}\theta}{\rho}\ ,
	\end{equation}
	where
	\begin{align}
	a&=\frac{J}{M}\\\rho^{2}&=r^{2}+a^{2}\cos^{2}\theta\\\Delta&=r^{2}-R_{S}r+a^{2}+R_{Q}^{2}
	\end{align}
	with $R_{S}=2M$ and $R_{Q}^{2}=\frac{Q^{2}}{4\pi}$.	This black hole has all the properties allowed by the no-hair theorem: mass, charge, and angular momentum. It is a stationary and axisymmetric space-time, and its derivation, as well as properties, go far beyond those of Schwarzschild and Reissner-Nordstr\"om.
	
	\par In \cite{Smarr:1973zz} it is shown that the surface area of the horizon of the Kerr--Newman black hole is given by the following
	\begin{equation}\label{eq:67}
		A=4\pi\left[2M^{2}-Q^{2}+2M^{2}\left(1-\frac{Q^{2}}{M^{2}}-\frac{J^{2}}{M^{4}}\right)^{1/2}\right]\ ,
	\end{equation}
	where $J$ is the angular momentum. In Eq. \eqref{eq:67} it is considered that $Q^2<M^2$ and $J^2<M^4$. From Eq. \eqref{eq:67} one can derive the following relation
	\begin{equation}\label{eq:68}
		dM=\frac{1}{8\pi}\kappa dA+\Omega dJ+\Phi dQ\ ,
	\end{equation}
	where $\frac{1}{8\pi}\kappa=\frac{\partial M}{\partial A}, \Omega=\frac{\partial J}{\partial A}$, etc.  Eq. \eqref{eq:68} is called the first law of black hole mechanics, for it resembles the first law of thermodynamics. This law plays a crucial role in proving that Kerr-Newman is, in fact, the only black hole that can exist in an asymptotically flat space-time, governed by Einstein's equations \cite{Hawking:1973uf, Wald:1993ki}. Moreover, it is important in the study of matter accretion in black holes and galaxies, which is subject of astrophysics \cite{Raine:2005bs}.
	
	\par The first term in Eq. \eqref{eq:68} gives a hint for the second law of black hole mechanics. Remember that $dE=TdS+\ldots$ is the first term of the first law of thermodynamics. One would be inclined to write $\kappa dA\sim TdS$, and it is actually correct to identify the entropy with the area of the black hole.  In fact, the horizon area has a similar property to entropy, which is: that it can only increase. It can be shown \cite{Wald:1995yp} that for Einstein--Hilbert action, c.f. Eq. \eqref{eq:13}, the entropy of a black hole is given by
	\begin{equation}\label{eq:69}
		S=\frac{c^{3}A}{4G\hbar}\ ,
	\end{equation}
	where the constants are displayed on purpose, to highlight the fact that quantum mechanics plays an important role in calculating this entropy, hence the appearance of $\hbar$. Notice that Eq. \eqref{eq:69} depends only on the surface area of the black hole event horizon, $A$. This is taken to be a hint towards holography since a quantity defined \textit{a priori} on a $D$-dimensional manifold depends on a $(D-1)$-dimensional quantity. 
	
	\par Defining entropy for a black hole requires one to reformulate the second law of thermodynamics, in what is called the generalized second law of thermodynamics \cite{Bekenstein:1973ur}. Since the event horizon and beyond are literally invisible areas, once an object (presumably possessing a given amount of entropy) falls inside a black hole, its entropy would vanish for an external observer - thus decreasing, which is clearly against the second law of thermodynamics. Hence the generalized second law, states that the entropy for an ordinary system that falls into the black hole is compensated:
	\begin{equation}\label{eq:70}
		\Delta S_{0}+\Delta S_{BH}\geq0\ ,
	\end{equation}
	with $\Delta S_{0}$ the change in entropy of the ordinary (exterior to the black hole) system.
	
	\par Finally, the third law of thermodynamics (also called Nerst law) is still an object of debate when it comes to the black hole scenario. In one of its forms, Nerst-Simon's statement of the third law, asserts that ``the entropy of a system at absolute zero temperature either vanishes or becomes independent of its intensive thermodynamic parameters". It can be shown that for the Kerr-Newman black hole this is, at least theoretically, not true. In the extremal case, where the sum of angular momentum density and electric charge are equal to the mass, one finds zero temperature without vanishing entropy. Moreover, the entropy depends on the angular momentum density, which from the first law, c.f. Eq. \eqref{eq:68}, can be identified with an intensive variable of the system. Theory aside, experimental tests have shown that the statement of the third law is unattainable by black holes because the extremal condition is never met in real astrophysical scenarios.
	
	\subsubsection*{Thermodynamics of Reissner-Nordstr\"om black hole}
	\par To end this section some explicit calculations regarding the thermodynamics of RN black holes are shown. This is not only instructive but will be important in later chapters, hence it also accounts for future reference. Notice that the RN black hole is the simplest case in which an extremal black hole can be defined, thus one can verify explicitly the statement regarding the third law, made in the last paragraph.
	
	\par To avoid the need of introducing other coordinate systems one can make use of the first law of black hole mechanics, Eq. \eqref{eq:68}, to evaluate the temperature. To use such an expression one needs the entropy, which, under the adoption of natural units $c=G=\hbar=1$ is seen to be proportional to the area of the black hole. The RN black hole has spherical horizons, and as seen previously the outermost is $r_+$, c.f. Eq. \eqref{eq:2-12}. The area is simply
	\begin{equation}\label{eq:71}
		A=4\pi r_+^2=4\pi\left(M+\sqrt{M^2-Q^2}\right)^2\ ,
	\end{equation}
	so the entropy of the RN black hole is
	\begin{equation}\label{eq:72}
		S=\pi r_+^2=\pi\left(M+\sqrt{M^2-Q^2}\right)^2\ .
	\end{equation}
	Notice how $S=\pi Q^2$ on the extremal case, i.e. non-zero.
	
	\par Writing the first law in a slightly different way
	\begin{equation}\label{eq:73}
		TdS=dM-\Phi dQ\ ,
	\end{equation}
	where usage of Eqs. \eqref{eq:66} and \eqref{eq:69} has been made. Varying the entropy given by Eq. \eqref{eq:72}
	\begin{equation}\label{eq:74}
		dS=\left(\frac{\partial S}{\partial M}\right)_{Q}dM+\left(\frac{\partial S}{\partial Q}\right)_{M}dQ\ ,
	\end{equation}
	where standard notation of thermodynamics is used. It is straightforward to check that
	\begin{align}
		\begin{aligned}\label{eq:75}
			\frac{\partial S}{\partial M}&=2\pi\frac{\left(M+\sqrt{M^{2}-Q^{2}}\right)^{2}}{\sqrt{M^{2}-Q^{2}}}\\
			\frac{\partial S}{\partial Q}&=-2\pi Q\frac{\left(M+\sqrt{M^{2}-Q^{2}}\right)}{\sqrt{M^{2}-Q^{2}}}
		\end{aligned}
	\end{align}
	Comparing Eqs. \eqref{eq:74} with \eqref{eq:73} and using Eq. \eqref{eq:75} one finds that
	\begin{equation}\label{eq:76}
		T=\frac{\sqrt{M^{2}-Q^{2}}}{2\pi\left(M+\sqrt{M^{2}-Q^{2}}\right)^{2}}\ .
	\end{equation}
	The surface gravity can be read off immediately by multiplying Eq. \eqref{eq:76} by $2\pi$. In doing this procedure one also finds the electric potential
	\begin{equation}\label{eq:77}
		\Phi=\frac{Q}{M+\sqrt{M^2-Q^2}}=\frac{Q}{r_+}\ ,
	\end{equation}
	which was a result already known from solving Maxwell's equations.
	
	\par An intriguing feature emerges from this calculation when the extremal case, i.e. $M=Q$, is considered, which is a non-zero entropy for vanishing temperature. In the context of holography such property gained a lot of attention in recent years, and the theory dual to the RN black hole in the AdS background is interesting on its own and shall be discussed in the next chapters. For now, we just notice this as a matter of curiosity.
	
	\par There are other ways of computing the temperature, namely introducing the imaginary time and demanding it be regular (this can be found \cite{Nastase:2017cxp}), and by direct application of Eq. \eqref{eq:66} once the coordinates are changed to the so-called Eddington--Finkelstein coordinate system, which makes the radial and temporal coordinates regular at the horizon (for those interested in this route, it is done at \cite{Raine:2005bs}). Both of these methods arrive at the same expression, Eq. \eqref{eq:76}, but would require the introduction of more mathematical machinery.
	
	\section{AdS Black Holes} \label{sec-2.3}
	\par As our intention is to apply the AdS/CFT correspondence to black hole geometries, we need backgrounds that are asymptotically anti-de Sitter. So far, both black holes presented are asymptotically flat, meaning that on the $r\to\infty$ limit the metrics become Minkowski space-time. 
	
	\subsection{AdS-Schwarzschild Black Hole}
	
	\par Let us consider the simplest case of a black hole in AdS, the so-called AdS-Schwarzschild black hole, it is derived from action 
	\begin{equation}\label{eq:2-13}
		S=\frac{1}{2\kappa_{d+1}^{2}}\int d^{d+1}x\sqrt{-g}\left[R+\frac{d\left(d-1\right)}{L^{2}}\right]+\frac{1}{2\kappa_{d+1}^{2}}\int d^{d}x\sqrt{-\gamma}\left[2K+\frac{2\left(d-1\right)}{L^{2}}\right]\ ,
	\end{equation}
	the second term is necessary so that the holographic problem is well defined, i.e. it removes ambiguities from the boundary so that the variational problem for the equations of motion is well defined. $K$ is the extrinsic curvature, and $\gamma$ the determinant of the induced metric. Without going into technical details (the steps are similar to those presented in previous sections, where the solutions for Schwarzschild and Reissner-Nordstr\"om black holes were obtained), the solution for the equations of motion is the following metric
	\begin{align}
		\begin{aligned}\label{eq:2-14}
			ds^{2}&=-f\left(r\right)dt^{2}+f^{-1}\left(r\right)dr^{2}+r^{2}h_{ij}dx^{i}dx^{j}\ , \\
			f\left(r\right)&=k-\frac{M}{r^{d-2}}+\frac{r^{2}}{L^{2}}\ .
		\end{aligned}
	\end{align}
	$h_{ij}$ is the horizon metric and is a function of $\{x_i\}$ coordinates only. This solution is a space-time known as Einstein space \cite{Hawking:1973uf} provided the horizon metric satisfied the following condition
	\begin{equation}\label{eq:2-15}
		R_{ij}=\left(d-2\right)kh_{ij}\ ,
	\end{equation}
	where $k$ can assume values of $-1,0,+1$. This means that the horizon can have different geometries and, moreover, Eq. \eqref{eq:2-14} is, in fact, a family of solutions determined by $k$. The most relevant scenario for AdS/CFT correspondence is when $k=0$, in which case it is a black hole with a planar (or flat) horizon - this is often called a black brane. For technical details, as well as solutions with other horizon geometries one is referred to \cite{Aminneborg:1996iz,Birmingham:1998nr,Gibbons:1976ue,Hawking:1982dh,Mann:1996gj}.
	
	\par In fact, for $k=0$, one can check that Eq. \eqref{eq:2-14} has a horizon at $r_h=(ML^2)^{\frac{1}{d}}$, with toroidal topology \cite{Aminneborg:1996iz}. Also, for $r\to\infty$, Eq. \eqref{eq:2-14} becomes the AdS space-time metric in Poincarè coordinates, c.f. Eq. \eqref{eq:41}.

	\section{Black Holes from Alternative Gravitation Theories}
	\par In this work, we are going to focus on the application of AdS/CFT correspondence to black hole space-times obtained from theories that are modifications (or extensions) of General Relativity. In this section, we will briefly describe some of these theories and present the solutions that will be used in future chapters, where we obtain the main results of this thesis.
	
	\par The background geometries to be introduced in this chapter all have a thing in common: they are the type of solution called black branes. As it was briefly described in the end of the previous section, the black brane is a special case within the family of solutions of the asymptotically AdS black holes, namely, when the horizon is planar. All of our solutions will have the general form
	\begin{equation} \label{eq:18}
		ds^{2}=-\frac{r^2}{r_0^2} f\left(r\right)dt^{2}+\frac{r_0^2}{r^2} g\left(r\right)dr^{2}+r^{2}h_{ij}dx^{i}dx^{j}\ ,
	\end{equation}
	for $f(r)\neq g(r)$, and $i,j$ going from $0$ to $3$ or $0$ to $4$, depending on the particular solution we are working with.
	
	\subsection{Effective Einstein's equation on the brane} \label{blackbranes}
	\par There's a wide literature about the so-called brane world scenario, which covers the historical perspective and the first models \cite{Maartens:2010ar,Brax:2001qd,Arkani-Hamed:1998jmv,Shiromizu:1999wj,Shiromizu:2001jm,Shiromizu:2001ve,Sasaki:1999mi,Rubakov:2001kp,Randall:1999ee,Maeda:2003ar,Hawking:2000kj,Gubser:1999vj,Garriga:1999yh,Garriga:1999bq,Germani:2001du,Dvali:2000hr,Dadhich:2000am,Chamblin:2000ra,Chamblin:1999by,Cavaglia:2002si,Neves:2021dqx,Abdalla:2006qj}. Here we will discuss it in a direct way, describing how to obtain the equations of motion and in which situation they recover General Relativity, therefore constituting an extension of the familiar theory of gravitation.
	
	\par In the brane world scenario, our world (4-dimensional space-time manifold) is described by a domain wall, called 3-brane, in 5-dimensional space-time. All the matter fields are localized in the 3-brane and only gravity at high energies is able to propagate throughout the 5-dimensional bulk. In essence, this description pictures our universe as a (hyper) surface embedded in a higher-dimensional bulk, and we describe our universe from the bulk perspective \cite{Maartens:2010ar}. In Appendix \ref{adm-app} we obtain Einstein's equations in the ADM formalism, the process of foliating the space-time involves similar techniques.
	
	\par In order to fulfill the criteria outlined in the previous paragraph, one has to describe the evolution of the brane, that is, the (equivalent of) Einstein's Fields Equations for the brane. Since we are considering the brane as an embedded surface on a higher dimensional space-time, we will need to relate quantities defined on the bulk with quantities on the brane. The Einstein equation for general dimensions read
	\begin{equation} \label{1.3.1-1}
		R_{MN}-\frac{1}{d}Rg_{MN}=k T_{MN}\ ,
	\end{equation}
	where we write $k$ for simplicity in the proportionality constant. The cosmological constant can be introduced in the matter action to give $T_{AB}=-\Lambda g_{AB}$. We will use the equation like this for now for simplicity since our interest is to relate quantities in high dimension with one less dimension - i.e. project them onto the brane.
	
	
	\par From now on we denote quantities on the brane by $\left(M,q_{\mu\nu}\right)$, whereas $\left(V,g_{\mu\nu}\right)$ refers to the codimension-1 bulk. Let $n^{M}$ be an unit normal vector to $M$, such that $q_{\alpha \beta}=q_{MN}=g_{MN}-n_{M}n_{N}$ is the induced metric on the brane. The notation for indices is as follows: $M,N,\ldots$ runs from $0,\ldots,d$, whereas $\alpha, \beta,\ldots$ from $0,\ldots,(d-1)$ - the coordinate singled out by the projection from the bulk to the brane is the coordinate defining the bulk. Suppose a general coordinate system $x^0,\ldots,x^d$, by the definition of normal vector we have $g_{x^d x^d}-n_{x^d}n_{x^d}=0$, in this sense $q_{\alpha\beta}$ has $d-1$ dimensions.
	
	The Gauss equation reads
	\begin{equation} \label{1.3.1-2}
		^{\left(4\right)}R_{\text{ }\beta\gamma\delta}^{\alpha}=^{\left(5\right)}R_{\text{ }NPQ}^{M}q_{M}^{\text{ }\alpha}q_{\beta}^{\text{ }N}q_{\gamma}^{\text{ }P}q_{\delta}^{\text{ }Q}+K_{\text{ }\gamma}^{\alpha}K_{\beta\delta}-K_{\text{ }\delta}^{\alpha}K_{\beta\gamma}\ .
	\end{equation}
	This equation simply relates the Riemann tensors in 5 and 4 dimensions via de extrinsic curvature. The Riemann tensor itself has the usual definition, c.f. \eqref{eq:7}. The Codazzi equation
	\begin{equation} \label{1.3.1-3}
		D_{\nu}K_{\alpha}^{\text{ }\nu}-D_{\alpha}K=^{\left(5\right)}R_{MN}n^{M}q_{\alpha}^{\text{ }N}\ ,
	\end{equation}
	relates the extrinsic curvature to the codimension-1 Ricci tensor. In Eq. \eqref{1.3.1-3} $K_{\alpha\beta}=q_{\alpha}^{\text{ }M}q_{\beta}^{\text{ }N}\nabla_{M}n_{N}$ is the extrinsic curvature on $M$, $K=K_{\mu}^{\text{ }\mu}$ its trace, and $D_{\mu}$ is the covariant derivative with respect to $q_{\mu\nu}$.
	
	\par We now to proceed to derive the equation describing the brane geometry from the 5 dimensional Einstein's Equations. Contracting Eq. \eqref{1.3.1-2} on indices $\alpha$ and $\gamma$
	\begin{equation} \label{1.3.1-4}
		^{\left(4\right)}R_{\beta\delta}=^{\left(5\right)}R_{MN}q_{\beta}^{\text{ }M}q_{\delta}^{\text{ }M}-\tilde{\Psi}_{\beta\delta}+KK_{\beta\delta}-K_{\text{ }\delta}^{\alpha}K_{\beta\alpha}\ ,
	\end{equation}
	where we defined 
	\begin{equation}\label{1.3.1-5}
		\tilde{\Psi}_{\beta\delta}=\text{ }^{\left(5\right)}R_{MNP}^{\text{ }\text{ }\text{ }\text{ }Q}n_{Q}n^{N}q_{\beta}^{\text{ }M}q_{\delta}^{\text{ }P}\ .
	\end{equation}
	Contracting again we have\footnote{The first term is written with $M, N$ indices for generality, because one term in that contraction obviously vanishes, since $q_{x^d x^d}=0$.}
	\begin{equation}\label{1.3.1-6}
		^{\left(4\right)}R=\text{ }^{\left(5\right)}R_{MN}q^{MN}-\tilde{\Psi}_{\mu\nu}q^{\mu\nu}+K^{2}-K_{\mu\nu}K^{\mu\nu}\ .
	\end{equation}
	This allows us to write the Einstein tensor associated with the brane using quantities defined in the bulk
	\begin{align}
		\begin{aligned}\label{1.3.1-7}
			^{\left(4\right)}G_{\alpha\beta}&=\left(^{\left(5\right)}R_{MN}-\frac{1}{2}\text{ }^{\left(5\right)}Rg_{MN}\right)q_{\alpha}^{\text{ }M}q_{\beta}^{\text{ }N}+\text{ }^{\left(5\right)}R_{MN}n^{M}n^{N}q_{\alpha\beta}\\&\text{ }\text{ }\text{ }+K_{\alpha\beta}K-K_{\mu\beta}K_{\alpha}^{\text{ }\mu}-\tilde{\Psi}_{\alpha\beta}-\frac{q_{\alpha\beta}}{2}\left(K^{2}-K_{\mu\nu}K^{\mu\nu}\right)\ .
		\end{aligned}
	\end{align}
	Thus we are able to identify parts that come from the brane world description, for instance, the first term between parenthesis is the 5D energy-momentum tensor, defined by the 5D Einstein field equation, projected onto the brane. Notice, however, that these remaining terms are geometric quantities, and we would like to describe them in terms of the more familiar energy-momentum tensor, or something similar that has a clearer physical interpretation.
	
	\par Taking the trace of the 5D Einstein's equations
	\begin{align}
		\begin{aligned}\label{1.3.1-8}
			^{\left(5\right)}R_{MN}g^{MN}-\frac{1}{2}Rg_{MN}g^{MN}&=8\pi G_{5}T_{MN}g^{MN}\ ,	\\
			^{\left(5\right)}R-\frac{5}{2}\text{ }^{\left(5\right)}R&=8\pi G_{5}T\ ,		\\
			^{\left(5\right)}R&=-\frac{16\pi G_{5}}{3}T\ .
		\end{aligned}
	\end{align}
	Allowing us to write the Ricci tensor as
	\begin{equation}\label{1.3.1-9}
		^{\left(5\right)}R_{MN}=8\pi G_{5}\left(T_{MN}-\frac{T}{3}g_{MN}\right)\ .
	\end{equation}
	Thus, in 4 dimensions the Einstein's equations now read
	\begin{align}
		\begin{aligned}\label{1.3.1-10}
			^{\left(4\right)}G_{\alpha\beta}&=8\pi G_{5}\left[T_{MN}q_{\alpha}^{\text{ }M}q_{\beta}^{\text{ }N}+q_{\alpha\beta}\left(T_{MN}n^{M}n^{N}-\frac{T}{3}\right)\right]\\&\text{ }\text{ }\text{ }+K_{\alpha\beta}K-K_{\mu\beta}K_{\alpha}^{\text{ }\mu}-\tilde{\Psi}_{\alpha\beta}-\frac{q_{\alpha\beta}}{2}\left(K^{2}-K_{\mu\nu}K^{\mu\nu}\right)\ .
		\end{aligned}	
	\end{align}
	From our previous definition, c.f. \eqref{1.3.1-5}, we can see that
	\begin{equation}\label{1.3.1-11}
		\tilde{\Psi}_{\alpha\beta}=\Psi_{\alpha\beta}+\frac{8\pi G_{5}}{3}\left[T_{MN}q_{\alpha}^{\text{ }M}q_{\beta}^{\text{ }N}+q_{\alpha\beta}\left(T_{MN}n^{M}n^{N}-\frac{T}{2}\right)\right]\ ,
	\end{equation}
	where
	\begin{equation}\label{1.3.1-12}
		\Psi_{\alpha\beta}=\text{ }^{\left(5\right)}C_{MNPQ}n^{P}n^{N}q_{\alpha}^{\text{ }M}q_{\beta}^{\text{ }Q}\ ,
	\end{equation}
	is the electric part of the Weyl tensor (Check the appendix \ref{app:1} for more details on this quantity). This follows from the identification \eqref{1.3.1-5}: applying the decomposition of the Riemann tensor in the symmetric, anti-symmetric and trace parts, and identifying such parts with the energy momentum-tensor correspondents via \eqref{1.3.1-8} and \eqref{1.3.1-9}. Comparing the first term of the right hand side in \eqref{1.3.1-10} and the expression \eqref{1.3.1-12}, we see that the Einstein's equations can be rewritten as 
	\begin{align}
		\begin{aligned}\label{1.3.1-13}
			^{\left(4\right)}G_{\alpha\beta}&=\frac{16}{3}\pi G_{5}\left[T_{MN}q_{\alpha}^{\text{ }M}q_{\beta}^{\text{ }N}+q_{\alpha\beta}\left(T_{MN}n^{M}n^{N}-\frac{T}{4}\right)\right]\\&\text{ }\text{ }\text{ }+K_{\alpha\beta}K-K_{\mu\beta}K_{\alpha}^{\text{ }\mu}-\frac{q_{\alpha\beta}}{2}\left(K^{2}-K_{\mu\nu}K^{\mu\nu}\right)-\Psi_{\alpha\beta}\ .
		\end{aligned}	
	\end{align}
	
	\par To make sense of the terms involving extrinsic curvature we have to introduce the so-called \textit{brane tension}, $\sigma$. This quantity is sometimes identified with the vacuum energy in the braneworld \cite{Shiromizu:1999wj}. We introduce it by defining
	\begin{equation}\label{1.3.1-14}
		S_{\alpha \beta}=\eta_{\alpha \beta}-\sigma q_{\alpha \beta}\ ,
	\end{equation}
	where $\eta_{\alpha \beta}$ is an energy-momentum tensor containing additional fields on the brane, rendering $S_{\alpha \beta}$ as an energy-momentum tensor itself, where we have an explicit contribution of the brane tension. We relate $S_{\alpha \beta}$ to the actual energy-momentum tensor (the one appearing on Einstein's equations) via \cite{Shiromizu:1999wj}
	\begin{equation} \label{1.3.1-15}
		T_{MN}=S_{MN}\delta(\xi)-\Lambda_5 g_{MN}\ ,
	\end{equation}
	$\Lambda_5$ is the cosmological constant in the bulk. The delta function is used to localize the contribution of $S_{\alpha \beta}$ in the brane - recall that \eqref{1.3.1-15} is defined for the entire bulk, whereas \eqref{1.3.1-14} only on the brane.
	
	\par Using \eqref{1.3.1-15} we can write terms in \eqref{1.3.1-13} as
	\begin{align}
		\begin{aligned} \label{1.3.1-16}
			T_{MN}q_{\alpha}^{\text{ }M}q_{\beta}^{\text{ }N}&=S_{MN}q_{\alpha}^{\text{ }M}q_{\beta}^{\text{ }N}\delta\left(\xi \right)-\Lambda_{5}g_{\alpha\beta}\ ,  \\
			T_{\mu\nu}n^{\mu}n^{\nu}-\frac{T}{4}&=\left(S_{\mu\nu}n^{\mu}n^{\nu}-\frac{S}{4}\right)\delta\left(\xi \right)+\frac{\Lambda_{5}}{4}\ ,
		\end{aligned}
	\end{align}
	and therefore 
	\begin{equation} \label{1.3.1-17}
		T_{\mu\nu}q_{\alpha}^{\text{ }\mu}q_{\beta}^{\text{ }\nu}+q_{\alpha\beta}\left(T_{\mu\nu}n^{\mu}n^{\nu}-\frac{T}{4}\right)=-\frac{3\Lambda_{5}}{4}q_{\alpha\beta}\ .
	\end{equation}
	The 4D Einstein's equations now read
	\begin{equation} \label{1.3.1-18}
		^{\left(4\right)}G_{\alpha\beta}=-4\pi G_{4}q_{\alpha\beta}+KK_{\alpha\beta}-K_{\mu\beta}K_{\alpha}^{\text{ }\mu}-\frac{q_{\alpha\beta}}{2}\left(K^{2}-K_{\mu\nu}K^{\mu\nu}\right)-\Psi_{\alpha\beta}\ .
	\end{equation}
	
	\par With the definition \eqref{1.3.1-15}, Einstein's equation in 5D is
	\begin{equation}\label{1.3.1-19}
		^{\left(5\right)}G_{MN}=8\pi G_{5}\left(S_{MN}\delta\left(\xi\right)-\Lambda_{5}g_{MN}\right)\ ,
	\end{equation}
	From the definition of the Einstein tensor, we can compute the scalar curvature in terms of the quantities on the right hand-side
	\begin{equation}\label{1.3.1-20}
		^{\left(5\right)}R=-\frac{16}{3}\pi G_{5}\left(S\delta\left(\xi\right)-5\Lambda_{5}\right)\ ,
	\end{equation}
	and therefore the Ricci tensor can be written as
	\begin{align}
		\begin{aligned}\label{1.3.1-21}
			^{\left(5\right)}R_{MN}-\frac{1}{2}g_{MN}\left[-\frac{16}{3}\pi G_{5}\left(S\delta\left(\xi\right)-5\Lambda_{5}\right)\right]&=8\pi G_{5}\left(S_{MN}\delta\left(\xi\right)-\Lambda_{5}g_{MN}\right)\ ,	\\^{\left(5\right)}R_{MN}&=8\pi G_{5}\left[\left(S_{MN}-\frac{S}{3}g_{MN}\right)\right]\ .
		\end{aligned}
	\end{align}
	In Gaussian coordinates, we have the following identities
	\begin{align}
		K_{\alpha\beta}&=\frac{1}{2}\partial_{\xi}q_{\alpha\beta}	\label{1.3.1-22}\ ,	\\ 
		\partial_{\xi}K_{\alpha\beta}&=K_{\beta\mu}K_{\alpha}^{\text{ }\mu}-\tilde{\Psi}_{\alpha\beta} \label{1.3.1-23}\ .
	\end{align}
	Then from Gauss equation \eqref{1.3.1-2}
	\begin{align}
		\begin{aligned} \label{1.3.1-24}
			^{\left(4\right)}R_{\alpha\beta}&=^{\left(5\right)}R_{\mu\nu}q_{\alpha}^{\text{ }\mu}q_{\beta}^{\text{ }\nu}-\tilde{\Psi}_{\alpha\beta}+K_{\alpha\beta}K-K_{\mu\beta}K_{\alpha}^{\text{ }\mu}\ , \\
			& =^{\left(5\right)}R_{\mu\nu}q_{\alpha}^{\text{ }\mu}q_{\beta}^{\text{ }\nu}+\left(\partial_{\xi}K_{\alpha\beta}-K_{\beta\mu}K_{\alpha}^{\text{ }\mu}\right)+K_{\alpha\beta}K-K_{\mu\beta}K_{\alpha}^{\text{ }\mu}\ ,		\\
			^{\left(5\right)}R_{\mu\nu}q_{\alpha}^{\text{ }\mu}q_{\beta}^{\text{ }\nu}&=P_{\alpha\beta}-\partial_{\xi}K_{\alpha\beta}\ ,
		\end{aligned}
	\end{align}
	where $P_{\alpha\beta}={}^{\left(4\right)}R_{\alpha\beta}-K_{\alpha\beta}K+2K_{\mu\beta}K_{\alpha}^{\text{ }\mu}$. We can now compare Eq's \eqref{1.3.1-21} and \eqref{1.3.1-24}
	\begin{align}
		\begin{aligned} \label{1.3.1-25}
			P_{\alpha\beta}-\partial_{\xi}K_{\alpha\beta}&=8\pi G_{5}\left[\left(S_{MN}-\frac{S}{3}g_{MN}\right)\delta\left(\xi\right)+\frac{16\pi}{3}G_{5}\Lambda_{5}g_{MN}\right]q_{\alpha}^{\text{ }M}q_{\beta}^{\text{ }N}\ ,	\\ &=8\pi G_{5}\left[\left(S_{\alpha\beta}-\frac{S}{3}q_{\alpha\beta}\right)\delta\left(\xi\right)+\frac{16\pi}{3}G_{5}\Lambda_{5}q_{\alpha\beta}\right]\ .
		\end{aligned}
	\end{align}
	Finally, we integrate on the brane (in the $\xi$ coordinate). The interval in which the integration is done is $(-\epsilon,+\epsilon)$, and the limit $\epsilon\to 0$ is taken, in order to restrict the quantities to the brane:
	\begin{align}
		\begin{aligned}\label{1.3.1-26}
			\lim_{\epsilon\to0}\int_{-\epsilon}^{\epsilon}\left(P_{\alpha\beta}-\partial_{\xi}K_{\alpha\beta}\right)d\xi&=8\pi G_{5}\lim_{\epsilon\to0}\int_{-\epsilon}^{\epsilon}\left[\left(S_{\alpha\beta}-\frac{S}{3}q_{\alpha\beta}\right)\delta\left(\xi\right)+\frac{16\pi}{3}G_{5}\Lambda_{5}q_{\alpha\beta}\right]d\xi\ ,	\\
			\lim_{\epsilon\to0}\left(P_{\alpha\beta}\left.\xi\right|_{-\epsilon}^{+\epsilon}-\left.K_{\alpha\beta}\right|_{-\epsilon}^{+\epsilon}\right)&=8\pi G_{5}\left[\left(S_{\alpha\beta}-\frac{S}{3}q_{\alpha\beta}\right)\lim_{\epsilon\to0}\int_{-\epsilon}^{\epsilon}\delta\left(\xi\right)d\xi+\frac{16\pi}{3}G_{5}\Lambda_{5}q_{\alpha\beta}\lim_{\epsilon\to0}\left(\left.\xi\right|_{-\epsilon}^{+\epsilon}\right)\right]\ ,		\\
			\lim_{\epsilon\to0}\left(\left.K_{\alpha\beta}\right|_{-\epsilon}^{+\epsilon}\right)&=-8\pi G_{5}\left(S_{\alpha\beta}-\frac{S}{3}q_{\alpha\beta}\right)\ .
		\end{aligned}
	\end{align}
	The quantity $\lim_{\epsilon\to0}\left(\left.K_{\alpha\beta}\right|_{-\epsilon}^{+\epsilon}\right)$ is the limits of extrinsic curvature from the left and from the right. This is where the $\mathbb{Z}^{2}$ symmetry along the $\xi$ direction comes in. Essentially we assume that locally there is no distinction if $\xi$ increases in the positive or negative direction. This means $\xi \mapsto -\xi$ in the vicinity of the brane. Therefore
	\begin{equation}\label{1.3.1-27}
		\lim_{\epsilon\to0}\left(\left.K_{\alpha\beta}\right|_{-\epsilon}^{+\epsilon}\right)=2K_{\alpha\beta}\ .
	\end{equation}
	With this result we are able to relate the extrinsic curvature to the tensor $S_{\alpha\beta}$
	\begin{equation}\label{1.3.1-28}
		K_{\alpha\beta}=-4\pi G_{5}\left(S_{\alpha\beta}-\frac{S}{3}q_{\alpha\beta}\right)\ ,
	\end{equation}
	recall Eq. \eqref{1.3.1-14} for the definition of $S_{\alpha\beta}$ - essentially it contains a generic matter term and the brane tension. From \eqref{1.3.1-28} we can compute all terms involving the extrinsic curvature in the Einstein's equation \eqref{1.3.1-18}:
	\begin{align} 
		K_{\alpha\beta}K&=-\frac{16}{3}\left(\pi G_{5}\right)^{2}\left[\eta\eta_{\alpha\beta}-4\sigma\eta_{\alpha\beta}-\frac{1}{3}\left(\eta^{2}-5\eta\sigma+4\sigma^{2}\right)q_{\alpha\beta}\right]\ , \\
		K_{\beta\mu}K_{\alpha}^{\text{ }\mu}&=16\left(\pi G_{5}\right)^{2}\left(\eta_{\beta\mu}\eta_{\alpha}^{\text{ }\mu}-\frac{2}{3}\left(\eta-\sigma\right)\eta_{\alpha\beta}+\left(\frac{\eta-\sigma}{3}\right)^{2}q_{\alpha\beta}\right)\ ,	\\
		K^{2}&=\frac{16}{9}\left(\pi G_{5}\right)^{2}\left(\eta^{2}-8\eta\sigma+16\sigma^{2}\right)\ ,		\\
		K_{\mu\nu}K^{\mu\nu}&=16\left(\pi G_{5}\right)^{2}\left(\eta_{\beta\mu}\eta^{\beta\mu}-\frac{2}{3}\left(\eta-\sigma\right)\eta+4\left(\frac{\eta-\sigma}{3}\right)^{2}\right)\ .
	\end{align}
	Such that the terms containing the extrinsic curvature can be written as
	\begin{align}
		\begin{aligned} \label{1.3.1-30}
			KK_{\alpha\beta}-K_{\mu\beta}K_{\alpha}^{\text{ }\mu}-\frac{q_{\alpha\beta}}{2}\left(K^{2}-K_{\mu\nu}K^{\mu\nu}\right)&=16\left(\pi G_{5}\right)^{2}\left[-\eta_{\mu\alpha}\eta_{\beta}^{\text{ }\mu}+\frac{1}{2}q_{\alpha\beta}\eta_{\mu\nu}\eta^{\mu\nu}+\frac{\eta}{3}\eta_{\alpha\beta}-\frac{1}{6}\eta^{2}q_{\alpha\beta}\right]\\&+\frac{32}{3}\left(\pi G_{5}\right)^{2}\sigma\eta_{\alpha\beta}-\frac{16}{3}\left(\pi G_{5}\right)^{2}\sigma^{2}q_{\alpha\beta}\ .
		\end{aligned}
	\end{align}
	
	\par Finally, the Einstein's equations on the brane can be written concisely in the form
	\begin{equation}\label{1.3.1-31}
		^{\left(4\right)}G_{\alpha\beta}=-\Lambda_{4}q_{\alpha\beta}+8\pi G_{4}\eta_{\alpha\beta}+\left(8\pi G_{5}\right)^{2}\tau_{\alpha\beta}-\Psi_{\alpha\beta}\ ,
	\end{equation}
	where we have denoted
	\begin{align}
		\Lambda_{4}&=4\pi G_{5}\left(\Lambda_{5}+\frac{4}{3}\pi G_{5}\sigma^{2}\right)	\label{1.3.1-32,1}\ ,	\\
		G_{4}&=\frac{4}{3}\pi\sigma G_{5}^{2}	\label{1.3.1-32,2}\ ,	\\
		\tau_{\alpha\beta}&=\frac{1}{4}\left[-\eta_{\mu\alpha}\eta_{\beta}^{\text{ }\mu}+\frac{1}{2}q_{\alpha\beta}\eta_{\mu\nu}\eta^{\mu\nu}+\frac{\eta}{3}\eta_{\alpha\beta}-\frac{1}{6}\eta^{2}q_{\alpha\beta}\right]	\label{1.3.1-32,3}\ .
	\end{align}
	And recall Eq. \eqref{1.3.1-12} for the definition of $\Psi_{\alpha\beta}$. One can notice a similarity between \eqref{1.3.1-31} and Einstein's equations from General Relativity, in fact, one can recover it in the limiting case when $G_5 \to 0$ while $G_4$ is kept constant. For this reason, it is common to refer to the GR limit of braneworld theory as $\sigma \to \infty$, since they lead to formally the same result. Moreover, the brane tension has to be strictly positive, given \eqref{1.3.1-32,2} we have to keep Newton's constant sign. The extra, or correction terms, $\tau_{\alpha\beta}$ and $\Psi_{\alpha\beta}$ have their own interpretation \cite{daRocha:2012pt,Abdalla:2009pg,HoffdaSilva:2009zza,Bazeia:2014tua,Bazeia:2013bqa}. From Eq. \eqref{1.3.1-32,3} it is immediate to notice that it depends only on squares of the energy-momentum tensor and therefore is deemed as a high energy correction to the equation. 
	
	
	\subsubsection*{Effective Einstein's equations on the brane for general dimension}
	\par Previously we derived the effective Einstein's equations on the brane assuming $d=5$ and bulk for codimension 1. From the procedure, one can see that the generalization for general dimensions should be straightforward. We will present the equations for general dimension for completeness since one of our solutions makes use of a 6-dimensional bulk to obtain a 5-dimensional brane. We follow \cite{Chakraborty:2015bja}, where all these results were obtained similarly to our derivation - except for the fact that the dimension is arbitrary.
	
	\par The Einsteins equations after application of the Gauss equation reads
	\begin{align}
		\begin{aligned} \label{gend-1}
			^{\left(d-1\right)}G_{\alpha\beta}&=\kappa_{n}^{2}\frac{d-3}{d-2}\left[T_{PQ}h_{\alpha}^{\text{ }P}h_{\beta}^{\text{ }Q}+h_{\alpha\beta}\left(T_{PQ}n^{P}n^{Q}-\frac{1}{d-1}T\right)\right]\\&-\Psi_{\alpha\beta}+\left[KK_{\alpha\beta}-K_{\alpha\mu}K_{\beta}^{\text{ }\mu}-\frac{1}{2}h_{\alpha\beta}\left(K^{2}-K_{\mu\nu}K^{\mu\nu}\right)\right]\ .
		\end{aligned}
	\end{align}
	Following similar steps, we define
	\begin{equation}
		T_{MN}=-\Lambda_{d}g_{MN}+S_{\alpha\beta}h_{M}^{\text{ }\alpha}h_{N}^{\text{ }\beta}\delta\left(\xi\right)\ ,
	\end{equation}
	where $S_{\alpha\beta}$ is the energy-momentum tensor of matter on the brane. From the junction conditions for the extrinsic curvature and $\mathbb{Z}_2$ symmetry of the bulk, we have
	\begin{equation}
		K^{+}=-K^{-}=-\kappa_{n}^{2}\frac{S}{2\left(d-2\right)}\ .
	\end{equation}
	Via $S_{\alpha\beta}$ we introduce the brane tension
	\begin{equation}
		S_{\alpha\beta}=\eta_{\alpha\beta}-\sigma h_{\alpha\beta}\ .
	\end{equation}
	From these expressions, the computation of the extrinsic curvature terms in Eq. \eqref{gend-1} is similar to the calculations done at the beginning of this section, and the effective Einstein's equations read
	\begin{equation}
		^{\left(d-1\right)}G_{\alpha\beta}=-\kappa_{d-1}^{2}\Lambda_{d-1}h_{\alpha\beta}+\kappa_{d-1}^{2}\eta_{\alpha\beta}+\kappa_{d}^{4}\tau_{\alpha\beta}-\Psi_{\alpha\beta}\ ,
	\end{equation}
	where
	\begin{align}
		\tau_{\alpha\beta}&=-\frac{1}{4}\eta_{\alpha\mu}\eta_{\beta}^{\text{ }\mu}+\frac{1}{8}h_{\alpha\beta}\eta_{\mu\nu}\eta^{\mu\nu}+\frac{1}{4\left(d-2\right)}\eta\eta_{\alpha\beta}-\frac{1}{8\left(d-2\right)}\eta^{2}h_{\alpha\beta}\ , \\
		\Lambda_{d-1}&=\frac{\left(d-3\right)}{\left(d-1\right)}\frac{\kappa_{d}^{2}}{\kappa_{d-1}^{2}}\left(\Lambda_{d}+\frac{d-1}{8\left(d-2\right)}\kappa_{d}^{2}\sigma\right)\ ,\\
		\kappa_{d-1}^{2}&=8\pi G_{d-1}=\sigma\kappa_{d}^{4}\frac{d-3}{4\left(d-2\right)}\ .	
	\end{align}
	
	\subsubsection*{ADM-like decomposition of bulk equations}
	\par Shortly after the formalism presented in this section was introduced, investigation towards alternative methods to find solutions to the equations started. A quick look at the effective Einstein's equations projected onto the brane, c.f. Eq. \eqref{1.3.1-31}, is enough for one to realize the complexity of the system, given the correction terms. A method to simplify the system of equations was proposed \cite{Casadio:2001jg}, where the authors were able to generate new solutions associated with branes.
	
	\par Since we apply the same method to obtain solutions, we will briefly describe how it is implemented. We shall look at Einstein's equations in the bulk, in the form \eqref{1.3.1-1}. Using Gaussian coordinates, the metric in the bulk assumes the general form
	\begin{equation} \label{admdec-1}
		ds^{2}=dy^{2}+\sum_{i}F_{i}\left(r\right)\left[dx_{M}dx^{M}\right]\ ,
	\end{equation}
	where $y$ is the coordinate along the bulk, whereas $dx^M$ are the coordinates on the brane, and $\sum_{i}F_{i}\left(r\right)\left[dx_{M}dx^{M}\right] = F_{1}\left(r\right)dx_{1}dx^{1}+F_{2}\left(r\right)dx_{2}dx^{2}+\ldots$ simply indicates that the functions can be different. The decomposition consists of projecting the bulk Einstein's equations onto the brane by selecting the component of the Einstein's equations associated to the $y$ coordinate, such that
	\begin{equation}\label{admdec-2}
		G_{y\alpha}=0\ ,
	\end{equation}
	for $\alpha=0,\ldots , D-1$. This equation resembles the momentum constraint of the ADM decomposition \cite{Arnowitt:1959ah}. Projecting the trace of the Einstein's equations onto the brane we have
	\begin{equation}\label{admdec-3}
		R = \Lambda_{D-1}\ ,
	\end{equation}
	which is similar to the Hamiltonian constraint. Equations \eqref{admdec-2} and \eqref{admdec-3} constitute a weaker constraint on the solution when compared to the set of Einstein's equations. The original ADM decomposition is discussed in Appendix \ref{adm-app}, where the Hamiltonian and momentum constraints are derived, c.f. Eqs. \eqref{eq:adm-19}.
	
	\par Notice that this decomposition is only suitable for finding static solutions \cite{Casadio:2001jg}, due to the number of equations and number of variables, i.e. counting the degrees of freedom. Essentially, Eqs. \eqref{admdec-2} and \eqref{admdec-3} constitute a system of $(D-1)$ equations, and one constraint. This method has been consistently used to obtain so-called deformed solutions, where the metric \eqref{admdec-1} is not kept general, but rather all the functions $F_i(r)$, except for one, are set beforehand, and one determines the remaining function. This will become clear in the next section.
	
	\subsection{AdS$_5$-Schwarzschild deformed black branes}  \label{adsschw-def}
	\par In this section we will solve Einstein's equations on a $5D$ brane, so the bulk is a $6$-dimensional space-time. In this model we consider a fluid flow implemented by the electric part of the Weyl tensor defined at the bulk, and therefore
	\begin{equation}
		\Psi_{MN}(\sigma^{-1}) \!=\!-\frac{6}{\sigma}\!\left[ \mathcal{U}\left(u_M u_N + \frac{1}{3}h_{MN}\right)+\mathit{Q}_{(M} u_{N)}+\mathcal{P}_{MN}\right], \label{eq:1.3.2-1}
	\end{equation}
	where $u_M$ is the four velocity, $\sigma$ is the brane tension, the tensor $h_{MN}$ is the projection orthogonal to the four velocity:
	\begin{equation} \label{eq:1.3.2-2}
		h_{MN}=g_{MN}-u_Mu_N\ ,
	\end{equation}
	$\mathcal{U}$ is the effective energy density of the fluid
	\begin{equation} \label{eq:1.3.2-3}
		\mathcal{U}=-\frac{1}{6}\sigma \Psi_{MN}u^Mu^N\ ,
	\end{equation}
	$\mathcal{P}_{MN}$ is the effective non-local anisotropic stress tensor
	\begin{equation} \label{eq:1.3.2-4}
		\mathcal{P}_{MN}=-\frac16\sigma\left(h_{(M}^{\;P}h_{N)}^{\;Q}-\frac13 h^{PQ}h_{MN}\right)\Psi_{PQ}\ ,
	\end{equation}
	and finally, $Q_M$ is the non-local energy flux
	\begin{equation} \label{eq:1.3.2-5}
		Q_M = -\frac16\sigma h^{\;P}_{\mu}\mathcal{E}_{PN}u^M\ .
	\end{equation}
	
	\par Consider $\eta_{MN}=0$, i.e. vacuum. Then for this system, Eqs. \eqref{admdec-2} and \eqref{admdec-3} read
	\begin{align}
		^{(6)}R_{Aw}=0\ , 	\\
		R=\Lambda_{5} \ ,
	\end{align}
	where we considered $\left\{ x^{0},x^{1},x^{2},x^{3},x^{4},x^{5}\right\} =\left\{ t,r,x,y,z,w\right\}$, and $A=x^0,\ldots,x^5$. The first of these equations is similar to the momentum constraint in the ADM formalism, whereas the second looks like the Hamiltonian constraint. The equations closing the system are given by the relation between the Ricci tensor on the brane and the electric part of the Weyl tensor
	\begin{equation}
		R_{MN}=\Psi_{MN}\ .
	\end{equation}
	Recall the ansatz for this type of solution, c.f. Eq. \eqref{eq:18}. We define $u=r_0/r$, and re-write $g\mapsto g^{-1}$ for convenience. Thus, the ansatz reads
	\begin{equation}
		ds^{2}=-\frac{r_{0}^{2}}{u^{2}}f(u)\mathrm{d}t^{2}+\frac{1}{u^{2}g(u)}\mathrm{d}u^{2}+\frac{r_{0}^{2}}{u^{2}}\delta_{ij}\mathrm{d}x^{i}\mathrm{d}x^{j}\ .
	\end{equation}
	The Hamiltonian constraint then reads
	\begin{equation} \label{adssch-hamconstraint}
		\frac{2f''}{f}-\frac{f'^{2}}{f^{2}}+\frac{2g''}{g}+\frac{g'^{2}}{g^{2}}-\frac{f'g'}{fg}+\frac{4}{r}\,\left(\frac{f'}{f}-\frac{g'}{g}\right)-\frac{4g}{r^{2}}=F(r,r_{0},\beta)\ ,
	\end{equation}
	with $f\equiv f(u)$ and $g\equiv g(u)$, and primes denoting differentiation with respect to $u$. The function appearing on the right hand side is written explicitly in appendix \ref{app-C}. The solution to this equation reads
	\begin{eqnarray}
		f(u) &=& 1 - u^4 + \left (\beta - 1 \right ) u^6,\label{eq:Fu}\ ,		\\
		g(u) &=& \left (1 - u^4 \right ) \left ( \frac{2 - 3u^4}{2- \left (4\beta-1\right ) u^4}\right )\ , 
		\label{eq:Gu}
	\end{eqnarray}
	where $\beta$ is called the deformation parameter. It comes as an integration constant within Eq. \eqref{adssch-hamconstraint}, and satisfies the condition that $\beta \to 1$ recovers the AdS-Schwarzschild solution of General Relativity in 5D - c.f. Eq. \eqref{eq:2-14}, Sect. \ref{sec-2.3}, for this solution in 4D. 
	
	
	\subsection{AdS$_4$-Reissner-Nordstr\"om deformed black brane} \label{adsrn-def} 
	\par This solution generalizes the AdS-RN black hole for the brane. In which case we consider the energy-momentum tensor to be given by (c.f. Eq. \eqref{eq:2-6,5})
	\begin{equation}\label{adsrn-2}
		T_{MN}=F_{MP}F_{N}^{\text{ }P}-\frac{1}{5}g_{MN}F^{2}\ ,
	\end{equation}
	with $F^2=F_{MN}F^{MN}$. Upon projecting the bulk equation onto the brane, c.f. \eqref{admdec-2} and \eqref{admdec-3}, using coordinates $\left\{ x^{1},x^{2},x^{3},x^{4},x^{5}\right\} =\left\{ t,r,x,y,w\right\}$, we have
	\begin{align}
		R_{w\alpha}&=0\ ,	\label{adsrn-3}		\\
		R&=\Lambda_{4}\ , \label{adsrn-4}
	\end{align}
	since $T_{\alpha\beta}g^{\alpha\beta}=0$ for \eqref{adsrn-2}. These equations constitute a weaker system than Einstein's equations on the brane, c.f. Eqs. \eqref{admdec-2} and \eqref{admdec-3}, and require another equation to close the system. Just like in the previous section, the equation is given by
	\begin{equation}\label{adsren-5}
		R_{\alpha\beta}=\Psi_{\alpha\beta}\ .
	\end{equation}
	Using the ansatz \eqref{eq:18}, and constraining the time component to
	\begin{equation}\label{adsrn-6}
		f(r)=1-(1+Q^2)\left(\frac{r_0}{r}\right)^3+Q^2\left(\frac{r_0}{r}\right)^4\ ,
	\end{equation}
	Eq. \eqref{adsrn-3} is satisfied immediately, whereas Eq. \eqref{adsrn-4} leads to
	\begin{align}
		\begin{aligned}\label{adsrn-7}
		\left\{ 22L^{4}r^{12}\left(Q^{2}r_{0}^{4}-r^{4}-r^{2}r_{0}^{2}-rr_{0}^{3}+r^{3}r_{0}\right)+4L^{4}r^{12}p_{Q}\left(\left(Q^{2}+1\right)rr_{0}^{3}-3Q^{2}r_{0}^{4}-r^{4}\right)k_{\beta}^{3}\right.&\\+4L^{4}r^{8}r_{0}p_{Q}^{2}k_{\beta}^{2}\left[9\left(Q^{2}+1\right)r^{2}r_{0}^{2}+9Q^{2}r_{0}^{4}-9\left(2Q^{2}+1\right)rr_{0}^{3}-2(\beta-1)r^{4}\right]-L^{4}r^{12}h_{Q}^{2}k_{\beta}^{3}&\\+p_{Q}k_{\beta}^{2}\left[9r^{2}-(\beta+17)rr_{0}+(2\beta+7)r_{0}^{2}\right]+81r_{0}(r-r_{0})^{2}h_{Q}p_{Q}^{4}\left[r^{2}r_{0}^{2}\left(2\beta-27Q^{2}-29\right)\right.&\\+rr_{0}^{3}\left[6\beta+(4\beta+59)Q^{2}+21\right]-4(2\beta+7)Q^{2}r_{0}^{4}\left.\left.+2(\beta-1)(r^{4}+r^{3}r_{0})\right]\right\} &=0,
		\end{aligned}
	\end{align}
	Where
	\begin{align}
		\begin{aligned} \label{adsrn-8}
			p_Q=p_Q(r)&=\left(Q^2+1\right) r r_0^3-Q^2 r_0^4-r^4,\\
			h_Q=h_Q(r)&=p_Q-Q^2 r_0^4, \\
			k_\beta=k_\beta(r) &= 9 r-(2 \beta
			+7) r_0,
		\end{aligned}
	\end{align}
	$\beta$ is a parameter that equals the difference between the unity and a factor that is proportional to the product of suitable negative powers of the brane tension and the 5D cosmological constant.
	
	\par The solution can be written as 
	\begin{equation} \label{adsrn-9}
		g(r)=\frac{1}{f\left(r\right)}\left\{ \frac{1-\frac{r_{0}}{r}}{1-\frac{r_{0}}{r}\left[1+\frac{1}{3}\left(\beta-1\right)\right]}\right\}\ .
	\end{equation}
	The parameter $\beta$ governs the deformation, and the reduces the solution to AdS-RN for $\beta \to 1$. We can now determine the electromagnetic field associated with this solution. Assuming an electrostatic configuration, the gauge potential is $A_{\mu}dx^{\mu}=A(r)dt$, then from Maxwell's equations \eqref{eq:2-7}, the solution is
	\begin{equation} \label{adsrn-10}
		A^{t}=\frac{-2Q\sqrt{r-r_{0}}\sqrt{\beta+2}\sqrt{3r-r_{0}\left(\beta+2\right)}-Qr\left(\beta-1\right)\left[\pi-2\arctan\left(\frac{\sqrt{r-r_{0}}\sqrt{\beta+2}}{\sqrt{3r-r_{0}\left(\beta+2\right)}}\right)\right]}{2\sqrt{3}r_{0}\sqrt{\beta+2}r}\text{ }.
	\end{equation}
	Notice that
	\begin{equation} \label{adsrn-11}
		\lim_{\beta\to1}A^{t}=Q\left(\frac{1}{r}-\frac{1}{r_{0}}\right)\text{ },
	\end{equation}
	recovering Coulomb's law, i.e. the electromagnetic gauge potential producing the AdS-RN solution.
	
	\par We also notice the following, the radius
	\begin{equation} \label{adsrn-12}
		r_{\beta}=r_{0}\left[1+\frac{1}{3}\left(\beta-1\right)\right]\ ,
	\end{equation}
	leads to $g^{rr}\to 0$, a coordinate singularity. Moreover, for $\beta > 1$, Eq. \eqref{adsrn-12} gives $r_{\beta}>r_0$. In order for $r_{\beta}$ to be an event horizon we follow the prescription that the surface it defines must be a Killing horizon \cite{Wald:1984rg}, such that $\xi_{\mu}\xi^{\mu}=0$, for $\xi^{\mu}$ the time-like Killing vector. There are two values of $\beta$ that satisfy this requirement
	\begin{align}
		\beta&=1\ , \label{adsrn-13}								\\
		\color{black}\beta&\color{black}=-3+\frac{2\times2^{1/3}}{\left(-7-27Q^{2}+3\sqrt{3}\sqrt{3+14Q^{2}+27Q^{4}}\right)^{1/3}}+\frac{\left(7+27Q^{2}-3\sqrt{3}\sqrt{3+14Q^{2}+27Q^{4}}\right)^{1/3}}{2^{1/3}} \ . \label{adsrn-14}
	\end{align}
	Solution \eqref{adsrn-13} is expected, since it recovers $r_0 = r_{\beta}$. The second solution, Eq. \eqref{adsrn-14} is interesting because it is dependent on the charge $Q$, constraining the deformation parameter to the charge. Notice that both quantities are constant, but completely arbitrary except if one considers relation \eqref{adsrn-14}. We can plot solution \eqref{adsrn-14} as a function
	\begin{figure} [h!]
		\centering
		\includegraphics[scale=0.7]{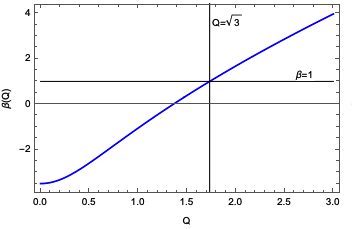}
		\label{fig:1}
		\caption{Plot of Eq. \eqref{adsrn-14}, $\beta(Q) \times Q$. It is  shown that $Q=\sqrt{3}$ yields $\beta=1$.}
	\end{figure}
	

%% file: chapters/AdSCFT_Review.tex
\newpage
\chapter{Review of AdS/CFT correspondence} \label{ch-3}
\par The correspondence between anti-de Sitter space-time and conformal field theory is the main theoretical subject of this work, as it is the mechanism applied to compute quantities of interest, which are transport coefficients. The correspondence is, in fact, a conjecture, meaning it has yet to be proven (either true or false, although the majority of work regarding the subject indicates it is true). This review is intended to introduce the correspondence as a computational device, and when it is convenient the technical aspects of each topic will be only described, and the specialized literature will be referenced.

\par It is important to remark that the AdS/CFT conjecture presented here is not its complete form, meaning that we intentionally will not mention the compactification techniques applied to super-symmetric spaces. Here we will focus on understanding the rationale underlying the duality, and make sense of the various expressions. For a complete review one is referred to the original works of Maldacena \cite{Maldacena:1997re} and Witten \cite{Witten:1998qj}.

\par Understanding the AdS/CFT correspondence requires knowledge of a variety of topics, such as the conformal group, conformal field theory, gauge theory, renormalization group, and general relativity. Besides the latter - which was discussed in detail in a previous chapter -, the other subjects will be  briefly discussed here, but only the key aspects that are relevant to state the conjecture.

\section{Conformal group}\label{adscft-sec1}
\par A conformal transformation is defined as follows
\begin{equation}\label{eq:adscft-1}
	g_{\mu\nu}^{\prime}\left(x^{\prime}\right)=\Omega\left(x\right)g_{\mu\nu}\left(x\right)\text{ },
\end{equation}
i.e. an invertible mapping $x\mapsto x^{\prime}$ which leaves the metric invariant up to a scaling factor. Sometimes conformal transformations are referred to as a change in scale since they only preserve angles, but not lengths. A direct consequence of this fact is that the causal structure of space-time connected by a conformal transformation is completely equal. The set of conformal transformations form a group, usually called $\mathcal{C}\left(1,d-1\right)$, and to work out its Lie algebra one has to look at the infinitesimal transformations, the derivation can be found in detail at \cite{DiFrancesco:1997nk}, the transformations fall in one of the four kinds
\begin{align}
\begin{aligned}\label{eq:adscft-2}
	\text{(T)}\ \ \ \ \ \ x^{\mu}\mapsto x^{\prime\mu}&=x^{\mu}+a^{\mu}\ \ \ \ \ \ \ \ \
	\text{(R)}\ \ \ \ \ \ x^{\mu}\mapsto x^{\prime\mu}=x^{\mu}+\omega_{\text{ }\nu}^{\mu}x^{\nu}\\
	\text{(D)}\ \ \ \ \ \ x^{\mu}\mapsto x^{\prime\mu}&=\left(1+\alpha\right)x^{\mu}\ \ \ \
	\text{(SCT)}\ \ \ \ x^{\mu}\mapsto x^{\prime\mu}=x^{\mu}+2\left(x^{\nu}b_{\nu}\right)x^{\mu}-b^{\mu}x^{2}
\end{aligned}
\end{align}
The labels to the left of each transformation are just a reminder of what each one does: the first one is a translation in space-time, the second is a rotation, including Lorentz boosts, and the last two are exclusive to the conformal transformations. The one labeled with a (D) is called a dilatation, and it stretches or shrinks the axis. The last one is named Special Conformal Transformation (SCT).

\par Notice that (T) and (R) alone form, themselves, a group: the Poincare group. Thus, the conformal group has the Poincarè group as a subgroup. Also, from Eqs. \eqref{eq:adscft-2} one can obtain the generators of the conformal group
\begin{align}
\begin{aligned}\label{eq:adscft-3}
P_{\mu}=-i\partial_{\mu}, \ \ \ \
D=-ix^{\mu}\partial_{\mu},\ \ \ \
L_{\mu\nu}=i\left(x_{\mu}\partial_{\nu}-x_{\nu}\partial_{\mu}\right), \ \ \ \
K_{\mu}=-i\left(2x_{\mu}x^{\nu}\partial_{\nu}-x^{2}\partial_{\mu}\right),
\end{aligned}
\end{align}
which satisfy the algebra
\begin{align}
\begin{aligned}\label{eq:adscft-4}
\left[D,P_{\mu}\right]=iP_{\mu},\ \ \ \
\left[D,K_{\mu}\right]=-iK_{\mu},\ \ \ \
\left[K_{\mu},P_{\mu}\right]=2i&\left(\eta_{\mu\nu}D-L_{\mu\nu}\right),\ \ \ \
\left[K_{\rho,}L_{\mu\nu}\right]=i\left(\eta_{\rho\mu}K_{\nu}-\eta_{\rho\nu}K_{\mu}\right),\\
\left[P_{\rho},L_{\mu\nu}\right]=i\left(\eta_{\rho\mu}P_{\nu}-\eta_{\rho\nu}P_{\mu}\right),\ \ \ \
\left[L_{\mu\nu},L_{\rho\sigma}\right]&=i\left(\eta_{\nu\rho}L_{\mu\sigma}+\eta_{\mu\sigma}L_{\nu\rho}-\eta_{\mu\rho}L_{\nu\sigma}-\eta_{\nu\sigma}L_{\mu\rho}\right).
\end{aligned}
\end{align}
It is possible to write these as a single equation, encoding all the generators in one matrix
\begin{equation}\label{eq:adscft-5}
	J_{MN}=\begin{pmatrix}L_{\mu\nu} & \frac{K_{\mu}-P_{\mu}}{2} & -\frac{K_{\mu}-P_{\mu}}{2}\\
	-\frac{K_{\mu}-P_{\mu}}{2} & 0 & D\\
	\frac{K_{\mu}-P_{\mu}}{2} & -D & 0
	\end{pmatrix}\text{ },
\end{equation}
where $M,N=1,\ldots,d+2$, which is clearly a rotation in $d+2$ dimensions. All the commutation relations, Eqs. \eqref{eq:adscft-4}, can now be cast into a single expression using $J_{MN}$
\begin{equation}\label{eq:adscft-6}
	\left[J_{MN},J_{RS}\right]=i\left(\eta_{NR}J_{MS}+\eta_{MS}J_{NR}-\eta_{MR}J_{NS}-\eta_{NS}J_{MR}\right)\text{ }.
\end{equation}
This is, in fact, the same algebra of the group $SO(2,d)$, and the expression above proves the isomorphism between these groups. Recall that $SO(2,d)$ is precisely the symmetry group of the AdS$_{d}$ described in Sect. \ref{adscft-sec2}, Subsect. \ref{adscft-subsec21}.

\section{Conformal invariance in field theory}\label{adscft-sec2}
\par Having described how conformal symmetry manifests, it is worthwhile to have a look on how fields transform once conformal invariance is imposed. One has to know how fields behave under this kind of transformation since it is one of the ways to verify if the results obtained through the prescription of the (AdS/CFT) correspondence are in accordance with what is expected.

\par At the classical level, a field theory is invariant under the conformal symmetry if it is invariant under the group transformations. This does not mean that the associated quantum theory is conformally invariant, for a quantum field theory needs a regularization prescription, which necessarily introduces a scale - recall that conformal invariance is in close relation with scale invariance -, hence conformal symmetry is broken, except at the so-called fixed points of the renormalization group. 

\subsection{Conformal invariance in classical field theory}\label{adscft-subsec21}
\par Let $T_{g}$ be a matrix representation of a conformal transformation, parametrized by $\omega_{g}$. The field $\Phi\left(x\right)$ transforms as
\begin{equation}\label{eq:adscft-7}
	\Phi^{\prime}\left(x^{\prime}\right)=\left(1-\omega_{g}T_{g}\right)\Phi\left(x\right)\text{ }.
\end{equation}
To obtain the full form of the generators the Hausdorff Formula is used: one considers the action of the generators at the origin, $x=0$, then translates the generator to an arbitrary point in space-time. Consider, for example, the angular momentum, for which the spin matrix $S_{\mu\nu}$ is used to define the action on the field $\Phi\left(0\right)$
\begin{equation}\label{eq:adscft-8}
	L_{\mu\nu}\Phi\left(0\right)=S_{\mu\nu}\Phi\left(0\right)\text{ }.
\end{equation}
Applying the Hausdorff Formula one finds
\begin{equation}\label{eq:adscft-9}
	e^{ix^{\lambda}P_{\lambda}}L_{\mu\nu}e^{-ix^{\sigma}P_{\sigma}}=S_{\mu\nu}-x_{\mu}P_{\nu}+x_{\nu}P_{\nu}\text{ },
\end{equation}
so the action of the generators on the field at an arbitrary point is given by
\begin{align}
	P_{\mu}\Phi\left(x\right)&=-i\partial_{\mu}\Phi\left(x\right)\ , \label{eq:adscft-10}  \\
	L_{\mu\nu}\Phi\left(x\right)&=i\left(x_{\mu}\partial_{\nu}-x_{\nu}\partial_{\mu}\right)\Phi\left(x\right)+S_{\mu\nu}\Phi\left(x\right)\ . \label{eq:adscft-11}
\end{align}
This argument can be extended to the full conformal group by looking at the subgroup that does not translate the origin $x=0$, such a subgroup is generated by rotations, dilatations, and special conformal transformations. Let $S_{\mu\nu}, \tilde{\Delta}$, and $\kappa_{\mu}$ be the eigenvalues of $L_{\mu\nu}, D, K_{\mu}$ respectively, so from Eqs. \eqref{eq:adscft-4} the commutation relations amongst the eigenvalues are
\begin{align}
\begin{aligned}\label{eq:adscft-12}
	\left[\tilde{\Delta},S_{\mu\nu}\right]=0, \ \ \ \
	\left[\tilde{\Delta},K_{\mu}\right]&=-i\kappa_{\mu}, \ \ \ \
	\left[\kappa_{\mu},\kappa_{\nu}\right]=0,\ \ \ \
	\left[\kappa_{\lambda},S_{\mu\nu}\right]=i\left(\eta_{\lambda\mu}\kappa_{\nu}-\eta_{\lambda\nu}\kappa_{\mu}\right)\ ,  \\
	\left[S_{\mu\nu},S_{\lambda\rho}\right]&=i\left(\eta_{\nu\lambda}S_{\mu\rho}+\eta_{\mu\rho}S_{\nu\lambda}-\eta_{\mu\lambda}S_{\nu\rho}-\eta_{\nu\rho}S_{\mu\lambda}\right)\ .
\end{aligned}
\end{align}
Application of the Hausdorff formula, together with these commutation relations, leads to the action of $K_{\mu}$ and $D$ on the field $\Phi\left(x\right)$ at an arbitrary point $x$:
\begin{align}
	D\Phi\left(x\right)&=\left(-ix^{\rho}\partial_{\rho}+\tilde{\Delta}\right)\Phi\left(x\right)\ , \label{eq:adscft-13}  \\
	K_{\mu}\Phi\left(x\right)&=\left(\kappa_{\mu}+2x_{\mu}\tilde{\Delta}-x^{\lambda}S_{\mu\lambda}-2ix_{\mu}x^{\lambda}\partial_{\lambda}+2ix^{2}\partial_{\mu}\right)\Phi\left(x\right)\ . \label{eq:adscft-14}
\end{align}

\par Demanding that $\Phi\left(x\right)$ belongs to an irreducible representation of the Lorentz group leads to the interesting result that any matrix that commutes with all $S_{\mu\nu}$ is a multiple of the identity (Schur's lemma). Consequently, from the first Eq. \eqref{eq:adscft-12}, $\tilde{\Delta}$ must be proportional to the identity matrix and, thus, forcing all $\kappa_{\mu}=0$. Also, the number $-i\Delta$ is the number such that $\tilde{\Delta}=-i\Delta Id$, $\Delta$ being the scaling dimension of the field $\Phi$. For a scalar field $\phi \left(S_{\mu\nu}=0\right)$
\begin{equation}\label{eq:adscft-15}
	\phi^{\prime}\left(x^{'}\right)=\Omega^{\Delta/2}\left(x\right)\phi\left(x\right)\text{ },
\end{equation}
$\Omega$ being the conformal factor in Eq. \eqref{eq:adscft-1}. For an arbitrary (irreducible) representation $R$, with non-zero spin, the transformation depends on the rotation matrix $M_{\text{ }\nu}^{\mu}$ of the representation being considered
\begin{equation}\label{eq:adscft-16}
	\Phi^{\prime}\left(x^{'}\right)=\Omega^{\Delta/2}R[M_{\text{ }\nu}^{\mu}]\left(x\right)\Phi\left(x\right)\text{ }.
\end{equation}
The factor $R\left[M_{\text{ }\nu}^{\mu}\right]$ is obtained through the usual procedure of computing the representations of the Lorentz Group, which can be found in any standard textbook of classical field theory such as \cite{Barut:1980aj}.

\par Finally, to close this section of conformal invariance in classical field theory consider the energy-momentum tensor. It is known that a (classical) field theory that is invariant under rotations, translations, and scale transformations is conformally invariant \cite{Polyakov:1970xd}. To see that, one considers an infinitesimal change in the coordinates $x_{\mu}^{\prime}=x_{\mu}+\epsilon_{\mu}\left(x\right)$, and notices that the action changes as
\begin{equation}\label{eq:adscft-17}
	\delta S=\int d^{n}xT^{\mu\nu}\partial_{\mu}\epsilon_{\nu}\text{ },
\end{equation}
where $T^{\mu\nu}$ is the energy-momentum tensor. This result also means that locality implies the existence of a privileged tensor $T^{\mu\nu}$, conjugate to the metric tensor $g_{\mu\nu}$.

\par Invariance under rotations means that $T_{\mu\nu}=T_{\nu\mu}$, i.e. the energy-momentum tensor is symmetric, while invariance under translations leads to the conservation equations $\partial_{\mu}T^{\mu\nu}=0$. When invariance under scale transformations is imposed the property of tracelessness emerges: $T_{\text{ }\mu}^{\mu}=0$ \cite{DiFrancesco:1997nk}. Thus, an energy-momentum tensor associated with a CFT must have all these properties. However, it is worth noticing that conformal invariance, in general, does not imply that $T^{\mu\nu}$ is traceless unless some more conditions are supplied \cite{DiFrancesco:1997nk}.

\subsection{Conformal invariance in quantum field theory}\label{adscft-subsec22}
\par Throughout this section the fields are assumed to be primary fields, i.e. the fields belong to an irreducible representation of the Lorentz group.

\par Start by considering the two-point function
\begin{equation}\label{eq:adscft-18}
	\left\langle \phi_{1}\left(x_{1}\right)\phi_{2}\left(x_{2}\right)\right\rangle =\frac{1}{Z}\int D\Phi\phi_{1}\left(x_{1}\right)\phi_{2}\left(x_{2}\right)e^{-iS\left[\Phi\right]}\text{ },
\end{equation}
where $\phi_{1,2}$ are scalar fields, $\Phi$ is the set of all functionally independents fields and $S\left[\Phi\right]$ is the conformally invariant action. Assuming that all the quantities in Eq. \eqref{eq:adscft-18} are conformally invariant, then from result in Eq.  \eqref{eq:adscft-15} one may infer directly
\begin{equation}\label{eq:adscft-19}
	\left\langle \phi_{1}\left(x_{1}\right)\phi_{2}\left(x_{2}\right)\right\rangle =\Omega^{\Delta_{1}/2}\Omega^{\Delta_{2}/2}\left\langle \phi_{1}^{\prime}\left(x_{1}^{\prime}\right)\phi_{2}^{\prime}\left(x_{2}^{\prime}\right)\right\rangle \text{ }.
\end{equation}
From the previous it is possible to gain insight about the correlation $\left\langle \phi_{1}\left(x_{1}\right)\phi_{2}\left(x_{2}\right)\right\rangle$. First, notice that if only a scale transformation is considered, $x\mapsto\lambda x$, Eq. \eqref{eq:adscft-19} becomes
\begin{equation}\label{eq:adscft-20}
	\left\langle \phi_{1}\left(x_{1}\right)\phi_{2}\left(x_{2}\right)\right\rangle =\lambda^{\Delta_{1}+\Delta_{2}}\left\langle \phi_{1}\left(\lambda x_{1}\right)\phi_{2}\left(\lambda x_{2}\right)\right\rangle \text{ },
\end{equation}
whereas rotational and translation invariance require that
\begin{equation}\label{eq:adscft-21}
	\left\langle \phi_{1}\left(x_{1}\right)\phi_{2}\left(x_{2}\right)\right\rangle =f\left(\left|x_{1}-x_{2}\right|\right)\text{ },
\end{equation}
where $f\left(x\right)$ is scale invariant as a consequence of Eq. \eqref{eq:adscft-19}. This leads to
\begin{equation}\label{eq:adscft-22}
	\left\langle \phi_{1}\left(x_{1}\right)\phi_{2}\left(x_{2}\right)\right\rangle =\frac{C_{12}}{\left|x_{1}-x_{2}\right|^{\Delta_{1}+\Delta_{2}}}\text{ },
\end{equation}
with $C_{12}$ a constant. Finally, invariance under SCTs imply
\begin{equation}\label{eq:adscft-23}
	\frac{C_{12}}{\left|x_{1}-x_{2}\right|^{\Delta_{1}+\Delta_{2}}}=\frac{C_{12}}{\left|x_{1}-x_{2}\right|^{\Delta_{1}+\Delta_{2}}}\frac{\left(\gamma_{1}\gamma_{2}\right)^{\frac{\Delta_{1}+\Delta_{2}}{2}}}{\gamma_{1}^{\Delta_{1}}\gamma_{2}^{\Delta_{2}}}\text{ },
\end{equation}
where $\gamma_{i}=1-2b_{\mu}x_{i}^{\mu}+b^{2}x_{i}^{2}$. It is clear that constraint \eqref{eq:adscft-23} is satisfied if $\Delta_{1}=\Delta_{2}$, thus the scalar fields are correlated only if they have the same scaling dimension, i.e.
\begin{align}\label{eq:adscft-24}
	\left\langle \phi_{1}\left(x_{1}\right)\phi_{2}\left(x_{2}\right)\right\rangle =\begin{cases}
	\frac{C_{12}}{\left|x_{1}-x_{2}\right|^{2\Delta}}, & \Delta_{1}=\Delta_{2}=\Delta\ ,	\\
	0, & \Delta_{1}\neq\Delta_{2}\ .
	\end{cases}
\end{align}

\par For the three-point function, a similar analysis can be carried and the result is
\begin{equation}\label{eq:adscft-25}
	\left\langle \phi_{1}\left(x_{1}\right)\phi_{2}\left(x_{2}\right)\phi_{3}\left(x_{3}\right)\right\rangle =\frac{C_{123}}{x_{12}^{\Delta_{1}+\Delta_{2}-\Delta_{3}}x_{23}^{\Delta_{2}+\Delta_{3}-\Delta_{1}}x_{13}^{\Delta_{1}+\Delta_{3}-\Delta_{2}}}\text{ },
\end{equation}
with $x_{ij}=\left|x_{i}-x_{j}\right|$ and $C_{123}$ a constant. For $n$-point functions, $n\geq4$, it is not possible to determine the form of the correlators due to the appearance of the so called anharmonic ratios, and the $n-$point functions may depend arbitrarily on these ratios, for more details see \cite{DiFrancesco:1997nk}. Also, recall that these expressions are for a scalar field only, but the same kind of analysis may be carried for higher spin fields \cite{Rychkov:2016iqz, Bzowski:2013sza}.

\section{Conformal boundary of AdS space-time}\label{adscft-sec3}
\par In chapter \ref{ch-2} the AdS space-time was derived as a maximally symmetric solution of Einstein's equations, in that section (c.f. subsection \ref{sec:2.1}) the symmetry properties were discussed and various coordinate systems were introduced, but the discussion of the so-called conformal boundary of AdS space-time was left out on purpose. The correspondence between AdS space-times and Conformal Field Theory is, roughly speaking, a correspondence between quantities defined on the bulk (the AdS space-time) and on the boundary (the CFT), the question addressed now is: what is this boundary? And in what sense it can be called a boundary?

\par To gain insight of what is this boundary of AdS space-time, a look at the geodesic motion of massless particles is useful. Recall the geodesic equation
\begin{equation}\label{eq:adscft-26}
	\frac{d^2x^{\mu}}{d\tau^2}-\Gamma_{\nu\lambda}^{\mu}\frac{dx^{\nu}}{d\tau}\frac{dx^{\lambda}}{d\tau}=0\text{ },
\end{equation}
where $x^{\mu}\equiv x^{\mu}\left(\tau\right)$ describes the trajectory of the particle as the affine parameter varies, the parameter used here is the proper time, hence the choice of letter $\tau$.

\par Considering the radial motion is enough for the discussion intended here (a more general analysis can be done, but it is of no interest here). In such case the velocity, $u^{\mu}\equiv dx^{\mu}/d\tau$, assumes the form
\begin{equation}\label{eq:adscft-27}
	u^{\mu}=\left(\dot{t}\left(\tau\right),\dot{r}\left(\tau\right),0,0\right)\text{ },
\end{equation}
where the dot denotes differentiation with respect to $\tau$. In this context, the geodesic equation assumes the simplest form when using the global coordinates defined by Eq. \eqref{eq:39} in section \ref{sec:2.1}]\footnote{A slightly different notation is used here so the transformation of coordinates is displayed in the $\left\{ t,r\right\}$ variables.}
\begin{equation}\label{eq:adscft-28}
	ds^{2}=L^{2}\left(-\cosh^{2}\left(\rho\right)d\tilde{t}^{2}+d\rho^{2}+\sinh^{2}\left(\rho\right)d\Omega_{d-2}^{2}\right)\text{ }.
\end{equation}
Transforming $r=L\sinh\left(\rho\right)$ and $t=L\tilde{t}$, Eq. \eqref{eq:adscft-28} reads
\begin{equation}\label{eq:adscft-29}
	ds^{2}=-\left(1+\frac{r^{2}}{L^{2}}\right)dt^{2}+\frac{dr^{2}}{\left(1+\frac{r^{2}}{L^{2}}\right)}+r^{2}d\Omega_{d-2}^{2}\text{ }.
\end{equation}
It is now a simple matter of computation to verify that Eqs. \eqref{eq:adscft-26} are
\begin{align}
\begin{aligned}\label{eq:adscft-30}
\ddot{t}+2\dot{t}\dot{r}\frac{r}{r^{2}+L^{2}} =0, \ \ \ \ \
\ddot{r}+\dot{t}^{2}\frac{r}{L^{2}}\left(1+\frac{r^{2}}{L^{2}}\right)-\dot{r}^{2}\frac{r}{r^{2}+L^{2}}=0\ .
\end{aligned}
\end{align}

\par For massless particles the constraint $u_{\mu}u^{\mu}=0$ must be satisfied, which leads to a great simplification in Eqs. \eqref{eq:adscft-30}, in particular the equation for $r\left(\tau\right)$, second one in Eqs. \eqref{eq:adscft-30}, becomes
\begin{equation}\label{eq:adscft-31}
	\ddot{r}=0\text{ }.
\end{equation}
It is solved by
\begin{equation}\label{eq:adscft-32}
	r\left(\tau\right)=r_{0}\left(\tau-\tau_{0}\right)\text{ },
\end{equation}
with $r_{0}$ and $\tau_{0}$ constants to be determined by initial conditions. The first among Eqs. \eqref{eq:adscft-30} is
\begin{equation}\label{eq:adscft-33}
	\ddot{t}+\frac{2r^{2}}{L^{2}}\dot{t}^{2}=0\text{ },
\end{equation}
and its solution is
\begin{equation}\label{eq:adscft-34}
	t\left(\tau\right)=t_{0}+L\arctan\left[\frac{r\left(\tau\right)}{L}\right]\text{ }.
\end{equation}

From these solutions for $r\left(\tau\right)$ and $t\left(\tau\right)$ it is seen that $r\to\infty$ as $\tau\to\infty$, on the other hand, if one writes $r\equiv r\left(t\right)$ it can be checked that $r\to\infty$ as $t\to\frac{\pi L}{2}$. Put in another way, a massless particle, for example a light ray, will reach the boundary of AdS in a finite coordinate time interval. It is in this sense that AdS space-time has a boundary, it is such that a light ray takes a finite (coordinate) time interval do reach (null) infinity.

\par
\begin{figure}[h!]
	\centering
	\includegraphics[scale=0.7]{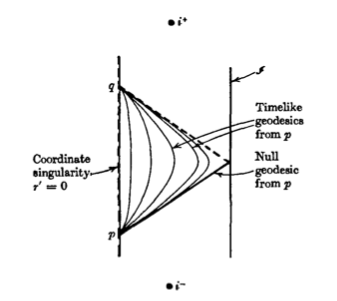}
	\caption{AdS space-time Penrose diagram: vertical direction represents time, $t$; horizontal direction represents the coordinate $\theta$. $i^+$ and $i^-$ represent future and past infinity, respectively. Source: \cite{Hawking:1973uf}.}
	\label{fig:adsdiagram}
\end{figure}

 In fact, the surface $r=\infty$ is a conformal boundary, to understand exactly what this means it is convenient to make a change of coordinates and draw a Carter-Penrose diagram, c.f. Fig. \ref{fig:adsdiagram}. The whole idea of such a diagram is to represent a non-compact space-time as a compact one. In order to do so, a transformation of coordinates is necessary, so that finite values in the new coordinate system correspond to infinity in the previous one. Also, the transformation should relate both metrics by a Weyl rescaling\footnote{Notice that this transformation takes different names depending on the context. A Weyl rescaling is also called a conformal transformation, i.e. relation between two metrics like in Eq. \eqref{eq:adscft-1}. In General Relativity it is more common to find the term Weyl rescaling because this kind of transformation is such that the Weyl tensor (or conformal tensor) is unchanged.}, so that the causal structure of the space-time is kept the same.

\par The transformation $r\mapsto\tan\theta$, with $\theta\in[0,\pi/2)$ maps Eq. \eqref{eq:adscft-29} to
\begin{equation}\label{eq:adscft-35}
	ds^{2}=\frac{1}{\cos^{2}\theta}\left(-dt^{2}+d\theta^{2}+\sin^{2}\theta d\Omega_{d-2}^{2}\right)\text{ },
\end{equation}
so, a conformal transformation, with conformal factor $\Omega\left(\theta\right)=\cos^{2}\theta$, means that Eq. \eqref{eq:adscft-35} has the equivalent causal structure of the so called Einstein static cylinder - actually it is half of that space-time, due to the interval where $\theta$ is defined
\begin{equation}\label{eq:adscft-36}
	ds_{ESC}^{2}=-dt^{2}+d\theta^{2}+\sin^{2}\theta d\Omega_{d-2}^{2}\text{ }.
\end{equation}
The surface $\theta=\pi/2$ is the conformal boundary of AdS. As it is not possible to map the proper time to a finite interval, the future and past infinity are disjoint points in the diagram, technically this is because the hypersurface $\theta=\pi/2$ is time-like, thus the space-time is not globally hyperbolic \cite{Hawking:1973uf}. 

\section{Large N limit of gauge theories}\label{adscft-sec4}

\par In physics, all theories, besides gravity, are formulated as gauge theories. Roughly speaking, these theories are invariant under the group $SU\left(N\right)$, and the $N\to\infty$ of these theories have special properties that are fundamental to the AdS/CFT correspondence. Notice that the CFT under investigation is nonetheless a gauge theory which, in addition, is invariant under the conformal group transformations. The large $N$ limit here is discussed only in the context that matters for AdS/CFT correspondence, but bear in mind that this limit has a wide range of applications, AdS/CFT being only one of them, for instance, some other applications can be found at \cite{Migdal:1984gj, Yaffe:1981vf, tHooft:1973alw}

\par One feature of the limit is that non-planar Feynman diagrams are not important, and the correlation functions satisfy some rules, which we will now describe. Consider a $U\left(N\right)$ gauge theory\footnote{In the $N\to\infty$ limit $SU\left(N\right)\sim U\left(N\right)$.} with fields valued in the adjoint representation, the action can be written as
\begin{equation} \label{eq:adscft-37}
	S=\frac{N}{\lambda}\int dx\text{Tr}\left[\left(D\Phi\right)^{2}+c_{3}\Phi^{3}+c_{4}\Phi^{4}+\ldots\right]\text{ },
\end{equation}
where notation of Sect. \ref{adscft-sec2} is used, and $\Phi$ belongs to an irreducible representation of Lorentz group that depends on $S_{\mu\nu}$. The $c_{i}$'s are coupling constants independent of $N$, and $\lambda=g_{YM}^{2}N$ is the so called 't Hooft coupling (YM stands for Yang-Mills). The coupling constant $\lambda$ is named after 't Hooft because he was the first to propose this idea of large $N$ \cite{tHooft:1973alw}, which is basically take $N\to\infty$ while keeping $\lambda$ constant, hence $g_{YM}\to0$, otherwise it makes no sense. The propagator, from Eq. \eqref{eq:adscft-37}, obeys
\begin{equation}\label{eq:adscft-38}
	\left\langle \Phi_{j}^{i}\Phi_{l}^{k}\right\rangle \propto\frac{\lambda}{N}\delta_{l}^{i}\delta_{j}^{k}\text{ }.
\end{equation}
This expression suggests that the propagator can be represented by a double line, each line denoting the flow of a fundamental index. A (vacuum) diagram with $V$ vertices, $E$ propagators (or edges) and $F$ lines (or faces) scales as
\begin{equation}\label{eq:adscft-39}
	\left(\frac{N}{\lambda}\right)^{V}\left(\frac{\lambda}{N}\right)^{E}N^{F}=\left(\frac{N}{\lambda}\right)^{\chi}\lambda^{F}\text{ },
\end{equation}
where $\chi=V+F-E=2-2g$ is the minimal Euler characteristic of the two-dimensional surface where the diagram is embedded. $g$ is called the genus, which is the (genus) genera of the surface.\footnote{For instance, a sphere has g=0 while a torus has g=1.} On the large $N$ limit, diagrams with $g=0$ dominate the expansion, such diagrams are called planar. This is because one can draw these diagrams in a 2$D$ surface in a way that no lines intersect each other, when $g\geq1$ this is just impossible. Hence, the genus, $g$, defines a topology, therefore the expansion in diagrams is, sometimes, called topological expansion. Notice that for a given topology there is an infinite number of diagrams contributing with the increasing power of the coupling $\lambda$, which in turn corresponds to increasing the number of faces in the 2$D$ figure.
\begin{figure}[h!]
	\centering
	\includegraphics[scale=0.5]{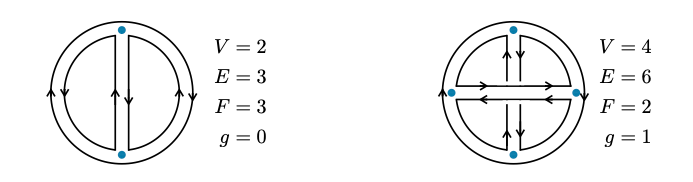}
	\caption{Planar versus toroidal diagram. Interaction vertices are marked with the dot. Notice how one has to draw lines over each other for $g=1$, if $g>1$ the figures become even more complicated. Source: \cite{Penedones:2016voo}.}
	\label{fig:planardiagram}
\end{figure}
\par The large N expansion has implications in the correlation functions of the theory, which is the main interest in this work. Consider single-trace \footnote{In holography single-trace operator means that the field in the CFT is dual to a single field in the bulk, i.e. one to one correspondence of sources.} local operators of the form $c_{i}\text{Tr}\left(\Phi^{i}\right)$, where $c_{i}$ is a constant independent of $N$. From the arguments above concerning the large $N$ expansion, one can derive \cite{McGreevy:2009xe} the expression for the correlator in this regime
\begin{equation}\label{eq:adscft-40}
	\left\langle O_{1}\ldots O_{n}\right\rangle =\sum_{g}N^{2-n-2g}f_{g}\left(\lambda\right)\text{ },
\end{equation}
where the sum is over the topologies, ranging from $g=0$ (planar diagram) to infinity. Expansion \eqref{eq:adscft-40} is dominated by planar diagrams, but maybe more interesting is that two-point functions, i.e. $n=2$ in Eq. \eqref{eq:adscft-40}, are independent of $N$. Meanwhile, for $n\geq2$ the correlators are suppressed by powers of $N$, this is called the large $N$ factorization. A consequence is that the two-point function of multi-trace operators, $\tilde{O}\left(x\right)$, is given by the product of the two-point functions of its single-trace constituents
\begin{equation}\label{eq:adscft-41}
	\left\langle \tilde{O}\left(x\right)\tilde{O}\left(y\right)\right\rangle \approx\prod_{i}\left\langle O_{i}\left(x\right)O_{i}\left(y\right)\right\rangle =\frac{1}{\left(x-y\right)^{2\sum_{i}\Delta_{i}}}\text{ },
\end{equation}
where suitable normalization is assumed. The scaling dimension of multi-trace operators is given by $\sum_{i}\Delta_{i}+\mathcal{O}\left(1/N^{2}\right)$, this indicates that on the large $N$ expansion the space of local operators in the CFT has a Fock structure, where single-trace operators play the role of single particle states of a weakly couple theory. This concludes the relevant features of the large $N$ expansion for the AdS/CFT correspondence, for a more detailed review on the topic covering other aspects of field theory, gauge theory as well as other applications of the expansion the reader is referred to \cite{tHooft:1973alw, McGreevy:2009xe}. 

\section{AdS/CFT duality}\label{adscft-sec5}
\par Having introduced all these seemingly unrelated subjects, CFTs, AdS space-time, and some features of the large $N$ expansion of gauge theory, it is now possible to formulate the AdS/CFT conjecture. Notice, however, that it is introduced here as something ``taken out of a hat”, and later justified by reasonable arguments regarding the statement. There are ways to derive this conjecture that require String Theory, which of course is not dealt with in this work. In fact, the celebrated work of Maldacena \cite{Maldacena:1997re} does exactly that, by using compactification techniques on superstring theory he was able to obtain the duality from fundamental principles. In the literature, this process is usually called a top-down formulation, when one starts from fundamental principles and builds up the conjecture. Here a different approach is taken, which is usually called bottom-up, since the main interests are the results obtained from the duality, this is seen as an argument to justify it as a reasonable statement.

\par Considering any of both approaches, it must be said that the AdS/CFT correspondence has the status of a conjecture, i.e. a statement that lacks proof but has not been proved wrong as well. To this date, there is no formal proof (in the form of a theorem) that the statement is true in all situations, regardless of it being based on a fundamental theory (string theory) or on experimental evidence (top-down approach). Nonetheless, it is the only method that enables one to obtain a large number of results and has proven, throughout the last years, to be worthwhile understanding.

\subsection{Statement of the correspondence}\label{adscft-subsec51}
\par The intuitive way to state the correspondence in one sentence is as follows:
\begin{quote}
	$d+1$-dimensional classical gravity theories on AdS$_{d+1}$ vacuum are equivalent to the large $N$ (degrees of freedom per site) limit of strongly coupled $d$-dimensional CFTs in flat space.
\end{quote}

\par There is some explanation needed in order to make sense of this statement. First, the relation between AdS space-time and CFTs, and not any other maximally symmetric space-time (or any solution to Einstein's equations) and CFTs (and not any kind of field theory), is because these theories have the same group of symmetries, namely $SO\left(2,d\right)$. This is the only reason motivating the identification of both theories.

\par Second, there is a difference in the dimensions of the gravity theory and CFT. To gain insight on that, recall the discussion on black hole thermodynamics c.f. section \ref{sec:3}, in particular the area law for the entropy of a black hole. It is possible to show \cite{Nastase:2017cxp} that for any number of dimensions in the space time the entropy of a black hole reads\footnote{This expression is valid as long as the black hole solution is obtained from Einstein's equations, i.e. equations of motion from Einstein-Hilbert action.}
\begin{equation}\label{eq:adscft-42}
	S_{BH}=\frac{A}{4G_{d+1}}\text{ },
\end{equation}
where $G_{d+1}$ is Newton's gravitation constant in $d+1$-dimensions. Turning to the fundamental definition of entropy, which is the number of micro-states on a given system, one can see that for a black hole covering a volume $V$, the entropy is encoded in the boundary of that volume, $A$. In fact, there is a name for this fact: holographic principle \cite{Susskind:1994vu}, it states that a theory of gravity in $d+1$-dimensions, in a local region of space has a number of degrees of freedom that scales like those of a quantum field theory on the boundary of that region.

\par As the system being considered is a black hole, it has to have the maximal entropy of anything that is encoded by its volume, thus every region of space has a maximum entropy scaling with the area of its boundary, and not the volume. On the other hand, the entropy of a (local) QFT behaves like $S\sim\log N$, with $N\sim e^{V}$ micro-states, which is, of course, much higher than the entropy of the boundary alone. This is why one has to consider the number of degrees of freedom of the QFT in a space of fewer dimensions, call it $N_{d}$, so
\begin{equation}\label{eq:adscft-43}
	S=\frac{A}{4G_{d+1}}=N_{d}\text{ }.
\end{equation}
The area of AdS space can be evaluated using the metric in the form given by Eq. \eqref{eq:43}
\begin{equation}\label{eq:adscft-44}
	A=\lim_{z\to0}\int d^{d-1}x\sqrt{g_{d-1}}\text{ },
\end{equation}
where the limit of $z\to0$ is taken because that is the value corresponding to the boundary in this coordinate system, also the volume element and determinant of the metric are already considering this fact, hence the $d-1$. This integral will be indeterminate if one tries to evaluate it right away, for $\sqrt{g_{d-1}}=\left(L/z\right)^{d-1}$ which goes to infinity in the limit $z\to0$, as well as $\int d^{d-1}x\to\infty$ upon integration. To solve this issue one has to consider the limit as a small (but finite) number, $R$, as well as enclose the by a volume $V_{d-1}$ such that
\begin{equation}\label{eq:adscft-45}
	A=\frac{L^{d-1}}{R^{d-1}}V_{d-1}\text{ },
\end{equation}
so the entropy \eqref{eq:adscft-43}, with area given by Eq. \eqref{eq:adscft-44} while imposing previous considerations reads
\begin{equation}\label{eq:adscft-46}
	S=\frac{L^{d-1}}{4G_{d+1}R^{d-1}}V_{d-1}\text{ }.
\end{equation}

\par From the QFT side, one can expect divergences on both, IR and UV, which need to be regularized. A usual procedure is to introduce a box of volume $V_{d-1}$ surrounding the theory (for the IR regime), and a short distance cut-off $\alpha$ (for the UV regime). This cut-off $\alpha$ is arbitrarily small, so to connect it to the gravity side we need to build a quantity that can be arbitrarily small as well, using parameters that are meaningful from the gravity theory point of view. Noticing that $R$ was introduced as an arbitrarily small number to avoid divergence in Eq. \eqref{eq:adscft-44}, it is a natural candidate in this context. A look at Eq. \eqref{eq:adscft-45} reveals that $R$ has units of energy (because of the natural units), so $\alpha\sim RL^{2}$ has the dimension of length and is written in terms of quantities defined in the gravitational theory. The number of degrees of freedom in a quantum field theory can be counted by multiplying the number of spatial cells (lattice approximation) by the number of degrees of freedom per site. The amount of cells in the lattice is simply the total volume $V_{d-1}$ divided by the size $\alpha$, in $d-1$ dimensions
\begin{equation}\label{eq:adscft-47}
	\frac{V_{d-1}}{\alpha^{d-1}}\sim\frac{V_{d-1}}{R^{d-1}L^{2\left(d-1\right)}}\text{ }.
\end{equation}
The number of degrees of freedom per lattice depends on the kind of field considered in the QFT, for instance, if one considers matrix fields in the adjoint representation of the symmetry group $U\left(N\right)$ there are $N^{2}$ degrees of freedom per site. Hence
\begin{equation}\label{eq:adscft-48}
	N_{d}\sim\frac{V_{d-1}N^{2}}{R^{d-1}L^{2\left(d-1\right)}}\text{ }.
\end{equation}

\par The estimations of degrees of freedom above give us insight on a third point of the statement of the correspondence, which is the large $N$. The fact that the theory of gravity is classical demands that the excitation length of the theory be much larger than the Planck length, $l_{p}$. Considering AdS, the typical length scale is given by the AdS radius, $L$, then recalling that $G_{d+1}\sim l_{p}^{d-1}$, and using Eqs. \eqref{eq:adscft-45} and \eqref{eq:adscft-48} one obtains
\begin{equation}\label{eq:adscft-49}
	\frac{L^{d-1}}{G_{d+1}}\sim\frac{L^{d-1}}{l_{p}^{d-1}}\sim N^{2}\gg1\text{ },
\end{equation}
considering that $l_{p}$ is extremely small, regardless of the units being used - for instance, $l_{p}\sim10^{-36}\text{m}$ in natural units.

\par Last, the fourth part of the statement is regarding that the CFT is strongly coupled, which, as one may expect, can be solved by elaborating more on the parameter $R$, introduced as a cut-off of the radial coordinate in AdS space-time and, later, identified with the UV cut-off of the field theory. For this identification to be correct, the extra dimension of the AdS space-time, with respect to the field theory, which is the radial coordinate, $z$, must be related to the renormalization group flow. One way of describing a QFT is organizing it in length or energy scales, then taking the appropriate limits of $r$ with respect to $\alpha$ according to the scale of interest. Say one is interested in the regime of large lengths of a theory, so one would consider $r\gg\alpha$ and integrate out the short distance degrees of freedom to obtain an effective theory at the scale $r$. In this way, one defines a renormalization group flow and generates a continuous family of effective theories in $d$-dimensional Minkowski space-time, labeled by $r$. The interesting fact is that, when the change of variables $u=1/r$ is done, the equations of the renormalization group become local\footnote{Notice that, as $r$ has the dimension of length, $u$ has the dimension of energy.} \cite{Kikuchi:2015hcm}. This means that it is not necessary to know the behavior of the theory in the UV, nor in the IR, to understand how physics changes with $u$.

\par To identify the radial AdS coordinate with the extra scale dimension of the CFT, recall the metric AdS metric as in equation \eqref{eq:43}, i.e.
\begin{equation}\label{eq:adscft-50}
	ds^{2}=\frac{r^{2}}{L^{2}}\eta_{\mu\nu}dx^{\mu\nu}+\frac{L^{2}}{r^{2}}dr^{2}\ .
\end{equation}
where $\eta_{\mu\nu}$ is the Minkowski space-time metric, the (conformal) boundary is located at $r\to \infty$. Using these coordinates one can see that for each $r=\text{const.}$ the AdS space is, in fact, Minkowski space-time, and the $r$ coordinate just shrinks or expands it. As the CFT is located on the boundary, it corresponds to the short UV distance physics, whereas this regime is of large scale in gravitational physics.
\begin{figure}[h!]
	\centering
	\includegraphics[scale=0.6]{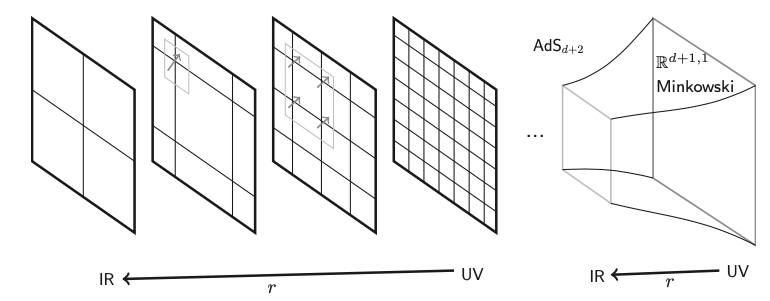}
	\caption{Pictorial representation of AdS/CFT correspondence, where the UV regime is at the boundary, while IR in the bulk. Source: \cite{McGreevy:2009xe}}
	\label{fig:buld_boundary_AdSCFT}
\end{figure}

\subsection{GKPW relation}\label{adscft-subsec52}
\par So far the discussion of AdS/CFT correspondence is only on the conceptual level, the conceptual discussion of the previous section enables one to understand and make sense of the statement, but is of no practical use when it comes to computation of correlation functions or transport coefficients. The GKPW is a powerful tool when it comes to computation, it relates partition functions and gives a prescription to extract quantities of the CFT from knowledge of classical gravitation theory.

\par Consider the Lagrangian of a conformal field theory, $\mathcal{L}_{CFT}$. One can perturb it by adding arbitrary functions
\begin{equation}\label{eq:adscft-51}
	\mathcal{L}_{CFT}\mapsto\mathcal{L}_{CFT}+\sum_{a}h^{a}\left(x\right)O_{a}\left(x\right)\text{ },
\end{equation}
where $h^{a}\left(x\right)$ is a source and $O\left(x\right)$ is the local operator associated to the source. Since this is a perturbation on the bare Lagrangian by local operators this perturbation is in the UV regime. From the discussion in Sect. \ref{adscft-subsec51} this kind of perturbation must be related to $r\to\infty$ limit in AdS space-time, i.e. adding a source as in Eq. \eqref{eq:adscft-51} corresponds to adding a boundary condition on the bulk fields.\footnote{The terms bulk and boundary will be used very often from now on. The bulk is the whole, $d+1$ space-time, where the gravitation theory is defined, whereas the boundary is the conformal boundary. One often refers to bulk or boundary fields meaning that such fields are defined by the gravitation theory or on the field theory.}

\par To find what $h^{a}\left(x\right)$ looks like in the bulk one makes the extension $h^{a}\left(x\right)\mapsto h^{a}\left(x_{\mu},r\right)$, where the extra coordinate $r$ is, of course, the radial coordinate of AdS space-time. One then demands that $h^{a}\left(x_{\mu},r\right)$ is a solution of the equations of motion in the bulk such that
\begin{equation}\label{eq:adscft-52}
	\lim_{r\to\infty}h\left(x_{\mu},r\right)=h\left(x\right)\text{ },
\end{equation}
is satisfied. To determine $h\left(x_{\mu},r\right)$ uniquely another condition is needed, namely, its value at the horizon (i.e. $r\to0$ limit) must be fixed. This is so because AdS space is not globally hyperbolic (for more on that see \cite{Hawking:1973uf}), which basically means that one boundary condition is not enough to determine the field in the bulk completely. Once both conditions are established one has a one-to-one map from bulk to boundary: the source $h\left(x_{\mu},r\right)$ \cite{Witten:1998qj, Gubser:1998bc}. For each observable $O\left(x\right)$ in the CFT there will be a source, corresponding to a field defined in the bulk, to determine which field on the bulk corresponds to each operator in the boundary on has to make use of symmetries. The internal symmetries of the fields in the gravitational sector are preserved in the dual field theory, for instance, one of such symmetries is the spin, thus fields in the bulk and boundary will have the same spin.

\par The proposal of Gubser, Klebanov, Polyakov and Witten (hence GKPW) \cite{Witten:1998qj, Gubser:1998bc} says that the partition function of the CFT, $Z_{CFT}\left[\left\{ h\left(x\right)\right\} \right]$ is equal to the partition function of the gravitational theory $Z_{AdS}\left[\left\{ h\left(x_{\mu},r\right)\right\} \right]$:
\begin{equation}\label{eq:adscft-53}
	Z_{CFT}\left[\left\{ h\left(x\right)\right\} \right]=Z_{AdS}\left[\left\{ h\left(x_{\mu},r\right)\right\} \right]\text{ },
\end{equation}
where the arguments of the partition functions are the fields discussed in the previous paragraphs. This equality by itself does not mean a lot, but in the large \textit{N} limit the gravitational theory becomes classical, and the right hand side can be solved by using a saddle point approximation, only in this limit one is allowed to write Eq. \eqref{eq:adscft-53} as
\begin{equation}\label{eq:adscft-54}
	Z_{CFT}\left[\left\{ h\left(x\right)\right\} \right]=e^{i\tilde{S}\left[\left\{ h_{\mu},r\right\} \right]}\text{ },
\end{equation}
where $\tilde{S}$ is the on-shell action, i.e. action integral evaluated with fields assuming the values that solve the equations of motion.

\par It is common in the literature to make a dictionary of holography, this comes in the form of a set of rules which have to be followed in order to accomplish the results provided by the computational devices of AdS/CFT, the first rule is the following
\begin{quote}
	\textbf{Rule 1:}\\
	The duality between gauge and gravity theories is a duality between the partition function of a CFT in $d$-dimensions and the on-shell action of the gravitational theory in AdS$_{d+1}$, as in Eq. \eqref{eq:adscft-54}.
	
	Operators in the CFT are related to fields in the bulk in accordance with the boundary condition given by Eq. \eqref{eq:adscft-52}, and are such that the spin is the same in the bulk and the boundary.
\end{quote}

\par Typically, the on-shell action, as well as the partition function on the left-hand side of Eq. \eqref{eq:adscft-54}, will be divergent in some limits. It is obvious that integration over all the volume of AdS space-time on the on-shell action will lead to infinity. To solve this issue, a method similar to the one used to compute the area, Eq. \eqref{eq:adscft-44} on section \ref{adscft-subsec51} is used, in order to identify divergent terms. Then, by adding suitable counter terms one is able to ``renormalize” the on-shell action and extract a finite value. This procedure has the name Holographic Renormalization, and its outlined in the next paragraphs, for the technical aspects one is referred to \cite{deHaro:2000vlm}.

\par With appropriate boundary conditions set (this shall be clear in the examples to come), the on-shell action of the AdS space-time must be evaluated but, instead of setting the $r$ limit to infinity, one puts a cut-off at $r=R$, and splits the on shell action in two parts:
\begin{equation}\label{eq:adscft-55}
	\tilde{S}\left(\left\{ \Phi\right\} ,R\right)=\tilde{S}^{\text{(reg)}}\left(\left\{ \Phi\right\} ,R\right)+\tilde{S}^{\text{(div)}}\left(\left\{ \Phi\right\} ,R\right)\text{ },
\end{equation}
where reg stands for regular while div for divergent parts. It is clear by the label that on the limit $R\to\infty$, the first term on the right-hand side of Eq. \eqref{eq:adscft-55} is finite and the second diverges. Having recognized the second term, one adds a counter term which cancels out the second term at $r=R$, call it $\tilde{S}^{\text{(c.t.)}}$. This counter term must be a local function of the embedded metric $\left.g_{\mu\nu}\right|_{r=R}\equiv\gamma_{ij}$ and cannot change the boundary conditions determined previously.

\par After following the previous steps, outlined in the last paragraph, define the renormalized on-shell action
\begin{equation}\label{eq:adscft-56}
	\tilde{S}^{\text{(ren)}}=\lim_{R\to\infty}\left(\tilde{S}\left(\left\{ \Phi\right\} ,R\right)+\tilde{S}^{\text{(c.t.)}}\left(\left\{ \Phi\right\} ,R\right)\right)\ .
\end{equation}
The correlation functions of the CFT can be obtained from the renormalized on-shell gravitational action, this procedure is so important in the correspondence that it has the status of rule, according to our dictionary of AdS/CFT
\begin{quote}
	\textbf{Rule 2:}
	Having obtained the renormalized gravitational action as a function of boundary values of fields in the gravity side, the correlators of the boundary CFT are given by
	\begin{equation}\label{eq:adscft-57}
		\left\langle O_{i}\left(x_{1}\right)\ldots O_{j}\left(x_{l}\right)\ldots O_{k}\left(x_{n}\right)\right\rangle =\frac{\delta\tilde{S}^{\text{(ren)}}\left(h_{i}\left(x\right)\right)}{\delta\Phi_{i}\left(x_{1}\right)\ldots\delta\Phi_{j}\left(x_{l}\right)\ldots\delta\Phi_{k}\left(x_{n}\right)}\ .
	\end{equation}
\end{quote}
\subsection{Example: scalar field} \label{adscft-subsec53}

\par The simplest case is chosen in order to avoid technical details that would not bring new physical insights. The idea is to apply the GKPW relation to compute the two-point function associated with a scalar field in the AdS background. Also, as will be shown, a relation between the mass of the scalar field and the conformal dimension of the source of the operator in the boundary is found as part of the computation.

\par Starting from the action for the scalar field\footnote{The backreaction effects on the space-time are not considered - i.e. the fact that $\phi$ has an energy-momentum tensor associated with it and would affect Einstein's equations, thus modifying the metric, is disregarded.}
\begin{equation} \label{eq:adscft-58}
	S=-\frac{1}{2\kappa_{d+1}^{2}}\int d^{d+1}x\sqrt{\left|g\right|}\left(g^{AB}\partial_{A}\phi\partial_{B}\phi+m^{2}\phi^{2}\right)\text{ },
\end{equation}
with the AdS metric in the Poincar\`e coordinates $\{t,z,x_1,\ldots,x_d\}$ (c.f. \eqref{eq:43})
\begin{equation} \label{eq:adscft-59}
	ds^{2}=g_{AB}dx^{A}dx^{B}=\frac{L^{2}}{z^{2}}\left(dz^{2}+\eta_{\mu\nu}dx^{\mu}dx^{\nu}\right)\text{ }.
\end{equation}
In these coordinates $z\to0$ is the conformal boundary and $\eta_{\mu\nu}$ is the Minkowski $d$-dimensional space-time metric. The equations of motion from Eq. \eqref{eq:adscft-58} are
\begin{equation}\label{eq:adscft-60}
	\frac{1}{\sqrt{\left|g\right|}}\partial_{A}\left(\sqrt{\left|g\right|}g^{AB}\partial_{B}\phi\right)-m^{2}=z^{d+1}\partial_{z}\left(z^{d-1}\partial_{z}\phi\right)+z^{2}\partial^{\mu}\partial_{\mu}\phi-m^{2}L^{2}=0\ ,
\end{equation}
where the indices $A,B$ run over all coordinates (including $z$), while $\mu,\nu$ run over $\{t,x_1,\ldots,x_d\}$.

\par Eq. \eqref{eq:adscft-60} is of the Fuchsian type, and $z=0$ is a singular point, so around this point, one can apply the Frobenius method with ansatz $\phi \sim z^{a}$, the indicial equation is
\begin{equation}\label{eq:adscft-61}
	a\left(a-d\right)=m^{2}L^{2}\text{ },
\end{equation}
the solution to Eq. \eqref{eq:adscft-61} is found immediately
\begin{equation} \label{eq:adscft-62}
	a_{\pm}=\frac{d}{2}\pm\sqrt{\frac{d}{4}+m^{2}L^{2}}\text{ },
\end{equation}
so the form of the field in the asymptotic regime is
\begin{equation}\label{eq:adscft-63}
	\phi=\phi_{0}\left(x_{\mu}\right)z^{a_{-}}+\phi_{+}\left(x_{\mu}\right)z^{a_{+}}\text{ }.
\end{equation}
From \cite{Wald:1992fr} the inner product
\begin{equation}\label{eq:adscft-64}
	\left(\phi_{1},\phi_{2}\right)=i\int\left(\phi_{2}^{\star}\overleftrightarrow{\partial}_{\mu}\phi_{1}\right)n^{\mu}d\Sigma\ ,
\end{equation} 
with $d\Sigma$ a space-like volume element on the hyper surface and $n_{\mu}$ a time-like unit vector, normal to the surface, ensures that the theory can be regularly quantized. \cite{Wald:1992fr} proves that the inner product is independent of the choice of $n_{\mu}$, given that it satisfies the property of being time-like orthogonal to the (hyper) surface $\Sigma$. For instance, choosing $n_{\mu}\propto\partial_t$ the inner product becomes
\begin{equation}\label{eq:adscft-65}
	\left(\phi_{1},\phi_{2}\right)=-i\int_{\Sigma}dzd^{d-1}x\sqrt{\left|g\right|}g^{tt}\left(\phi_{1}\partial_{t}\phi_{2}-\phi_{2}\partial_{t}\phi_{1}\right)\text{ },
\end{equation}
evaluating Eq. \eqref{eq:adscft-65} for a solution of the form $z^a$ one finds that the integral goes as $z^{2a-d+1}$, meaning it converges for
\begin{equation}\label{eq:adscft-66}
	2a-d+1>-1\implies a>\frac{d}{2}-1\text{ }.
\end{equation}
Notice from Eq. \eqref{eq:adscft-62} that $a_+$ always satisfies the inequality in Eq. \eqref{eq:adscft-66}, while for $a_-$ there is an interval
\begin{equation}\label{eq:adscft-67}
	0\leq\sqrt{\frac{d}{4}+m^{2}L^{2}}<1\text{ }.
\end{equation}
The modes associated with $a_{\pm}$ receive different names, $\phi_0$ is usually called the leading term, or non-normalizable, whereas $\phi_+$ is the sub-leading term, or normalizable.

\par Computing the on-shell action, Eq. \eqref{eq:adscft-58}, for the asymptotic solution \eqref{eq:adscft-63} one finds, after some manipulation, that the boundary term is
\begin{equation}\label{eq:adscft-68}
	\tilde{S}_{\partial}=\int d^{d}x\sqrt{\left|\gamma\right|}\phi_{0}\left(x_{\mu}\right)\phi_{+}\left(x_{\mu}\right)\ ,
\end{equation}
where the subscript $\partial$ indicates that the boundary term is left - the other term vanishes identically since it is the equation of motion. Now, identify
\begin{equation}\label{eq:adscft-69}
	\phi_{0}\left(x_{\mu}\right)\equiv\lim_{z\to0}z^{-a_{-}}\phi\left(z,x_{\mu}\right)\text{ },
\end{equation}
i.e. the non-normalizable term is the source of the operator. Notice that this identification follows the prescription given by Eq. \eqref{eq:adscft-52}. The sub-leading term is identified with the expectation value of the operator associated to the source \eqref{eq:adscft-69}, action \eqref{eq:adscft-68} can then be written as
\begin{equation}\label{eq:adscft-70}
	\int d^{d}x\phi_{0}\left(x_{\mu}\right)O\left(x_{\mu}\right)=\int d^{d}x\lim_{z\to0}z^{-a_{-}}\phi\left(z,x_{\mu}\right)O\left(x_{\mu}\right)\text{ }.
\end{equation}
From results in Sects. \ref{adscft-sec1} and \ref{adscft-sec2}, the conformal transformation of the coordinates and scalar field operator are
\begin{align}
	x^{\mu}\mapsto x^{\prime\mu}&=\lambda x^{\mu},	\ \ \ \	\ 
	z\mapsto z^{\prime}=\lambda^{-1}z,	\label{eq:adscft-72}\\
	O\left(x\right)&\mapsto O^{\prime}\left(x^{\prime}\right)=\lambda^{-\Delta}O\left(x\right)\ .	\label{eq:adscft-73}
\end{align}
Eq. \eqref{eq:adscft-72} follows from the fact that $z\propto r^{-1}$. Another result, from Sect. \ref{adscft-sec2}, is that the action is invariant under conformal transformations, i.e.
\begin{equation}\label{eq:adscft-74}
	\int d^{d}x\phi_{0}\left(x_{\mu}\right)O\left(x_{\mu}\right)=\int d^{d}x^{\prime}\phi_{0}\left(x_{\mu}^{\prime}\right)O^{\prime}\left(x_{\mu}^{\prime}\right)\text{ }.
\end{equation}
Putting these results altogether, plus the result $a_{+}+a_{-}=d$, one can arrive at the following
\begin{align}
	\begin{aligned}\label{eq:adscft-75}
	\int d^{d}x^{\prime}\phi_{0}\left(x_{\mu}^{\prime}\right)O^{\prime}\left(x_{\mu}^{\prime}\right)&=\int d^{d}x^{\prime}\lim_{z^{\prime}\to0}z^{\prime a_{-}}\phi\left(z^{\prime},x_{\mu}^{\prime}\right)O^{\prime}\left(x_{\mu}^{\prime}\right)\ ,		\\
	&=\int d^{d}\left(\lambda x\right)\lim_{z\to0}\left(\lambda^{-1}z\right)^{a_{-}}\phi\left(z,x_{\mu}\right)O\left(x_{\mu}^{\prime}\right)\ ,		\\
	&=\lambda^{a_{+}}\int d^{d}x\phi_{0}\left(x_{\mu}\right)\lambda^{-\Delta}O\left(x_{\mu}\right)\ .
	\end{aligned}
\end{align}
So, the conformal dimension of the scalar operator is given by
\begin{equation}\label{eq:adscft-76}
	\Delta=a_{+}\text{ },
\end{equation}
or, more specifically, it is related to the mass of the scalar field. One can proceed with a more general calculation in the same fashion (if one is interested, the result is proven in \cite{DiFrancesco:1997nk}), and obtain the following result:
\begin{quote}
	\textbf{Rule 3:}\\
	The conformal dimension of the operator in the CFT is related to the dimension, the mass of the field, and $p-$form originating that field in the bulk by the following equation
	\begin{equation}\label{eq:adscft-77}
		\Delta=\frac{1}{2}\left(d+\sqrt{d^{2}-4dp+4L^{2}m^{2}+4p^{2}}\right)\ .
	\end{equation}
	Notice that for a massless $1-$form this value is $d-1$, the usual dimension for a conserved current.
\end{quote}

\subsubsection*{Correlator of the scalar field}
\par To compute the correlator, rule number 2 will be applied. In order to apply the rule one has to follow the procedure described in the paragraphs prior to the statement of the rule, i.e. compute the on-shell action, identify the divergences and write the renormalized action by adding suitable counter-terms.

\par Expansion of the action to first-order gives\footnote{The constants are set to unity here for simplicity.}
\begin{equation}\label{eq:adscft-78}
S=\int d^{d+1}x\sqrt{\left|g\right|}\left(\Box-m^{2}\right)\phi\delta\phi-\int d^{d}x\sqrt{\left|\gamma\right|}n^{\mu}\partial_{\mu}\phi\delta\phi\ ,	
\end{equation}
where $\Box \equiv \nabla^{A}\nabla_{A}$, the second term is a boundary term, and $n^{\mu}$ is a vector normal to the surfaced defined by the induced metric $\gamma$. Evaluating Eq. \eqref{eq:adscft-78} on the classical solution makes the first term vanish, and one is left with the boundary term. The induced metric is $\gamma = \frac{1}{z^2}\eta_{\mu\nu}$, while $n^{\mu}=\sqrt{g^{zz}}\delta^{\mu}_{z}$, so the on-shell action \eqref{eq:adscft-78} is
\begin{equation}\label{eq:adscft-79}
	\tilde{S}=-\left.\int d^{d}x\left(\frac{1}{z}\right)^{d}\phi\partial_{z}\phi\right|_{z=0}^{z=\infty}\ ,
\end{equation}
where the boundaries are indicated.

\par In order to have a regular solution at infinity set $\lim_{z\to\infty}\phi(z,x_{\mu})=0$, on the other hand from Eq. \eqref{eq:adscft-63} one of the terms diverges as $z\to 0$. This is no surprise since in the $z\to 0$ limit one approaches the UV regime, and it is well known from QFT that such divergences occur. To solve this issue set a cutoff $z=\epsilon$ such that $\epsilon$ is at an infinitesimal distance from the boundary $z=0$. Then identify the divergent terms and add a counter term that exactly cancels the divergent one, and then take the limit $\epsilon\to 0$. This procedure, and the actual computation of the correlator, are done easier when working on momentum space, hence, consider the Fourier transformation of Eq. \eqref{eq:adscft-79}:
\begin{equation}\label{eq:adscft-80}
	\tilde{S}=-\left.\int\frac{d^{d}k}{\left(2\pi\right)^{d}}\left(\frac{1}{z}\right)^{d-1}\phi_{k}\partial_{z}\phi_{-k}\right|_{z=0}^{z=\infty}\ ,
\end{equation}
the transformation is, of course, in the $\{t,x_i\}$ variables. Applying the cutoff and considering the asymptotic solution, c.f. Eq.  \eqref{eq:adscft-63}, the on-shell action \eqref{eq:adscft-80} reads
\begin{align}
\begin{aligned}\label{eq:adscft-81}
\tilde{S}&=-\frac{1}{2}\lim_{\epsilon\to0}\left.\int\frac{d^{d}k}{\left(2\pi\right)^{d}}\left(\frac{1}{z}\right)^{d-1}\phi_{k}\partial_{z}\phi_{-k}\right|_{z=\epsilon}\ ,	\\	&=-\frac{1}{2}\lim_{\epsilon\to0}\int\frac{d^{d}k}{\left(2\pi\right)^{d}}\left(\frac{1}{z}\right)^{d-1}\left[\left(\phi_{0}\left(k\right)z^{a_{-}}+\phi_{+}\left(k\right)z^{a_{+}}\right)\partial_{z}\left(\phi_{0}\left(-k\right)z^{a_{-}}\right)\right.\\&\hspace{8cm}\left.\left.+\phi_{+}\left(-k\right)z^{a_{+}}\right]\right|_{z=\epsilon}\ ,	\\	&=-\frac{1}{2}\lim_{\epsilon\to0}\int\frac{d^{d}k}{\left(2\pi\right)^{d}}\left[a_{-}\phi_{0}\left(k\right)\phi_{0}\left(-k\right)z^{a_{-}-a_{+}}+a_{+}\phi_{0}\left(k\right)\phi_{+}\left(-k\right)\right.\\&\hspace{3cm}\left.\left.+a_{-}\phi_{+}\left(k\right)\phi_{0}\left(-k\right)+a_{+}\phi_{+}\left(k\right)\phi_{+}\left(-k\right)z^{a_{+}-a_{-}}\right]\right|_{z=\epsilon}\ ,
\end{aligned}
\end{align}
where $a_{+}+a_{-}=d$ has been used. Now, for shortness write $a_{\pm}=\frac{d}{2}\pm\nu$, where the definition of $\nu$ is obvious from Eq. \eqref{eq:adscft-62}. Eq. \eqref{eq:adscft-81} reads
\begin{align}
\begin{aligned}\label{eq:adscft-82}
	\tilde{S}&=-\frac{1}{2}\lim_{\epsilon\to0}\int\frac{d^{d}k}{\left(2\pi\right)^{d}}\left[a_{-}\phi_{0}\left(k\right)\phi_{0}\left(-k\right)z^{-2\nu}+a_{+}\phi_{0}\left(k\right)\phi_{+}\left(-k\right)+a_{-}\phi_{+}\left(k\right)\phi_{0}\left(-k\right)\right.\\&\hspace{8cm}\left.\left.+a_{+}\phi_{+}\left(k\right)\phi_{+}\left(-k\right)z^{2\nu}\right]\right|_{z=\epsilon}\ .
\end{aligned}
\end{align}
The divergent term is now clear, it is the term $\propto z^{-2\nu}$. At this point there is not a formal procedure to find a suitable counter-term, one literally has to guess. In the present case, it is not very hard to figure out that the suitable term is
\begin{align}
\begin{aligned}\label{eq:adscft-83}
	\tilde{S}^{\mbox{(c.t.)}}&=-\frac{1}{2}\lim_{\epsilon\to0}\left.\int\frac{d^{d}k}{\left(2\pi\right)^{d}}\left[a_{-}\sqrt{\left|\gamma\right|}\phi\left(k\right)\phi\left(-k\right)\right]\right|_{z=\epsilon}\ ,	\\	&=-\frac{1}{2}\lim_{\epsilon\to0}\left.\int\frac{d^{d}k}{\left(2\pi\right)^{d}}\left[a_{-}\phi_{0}\left(k\right)\phi_{0}\left(-k\right)z^{-2\nu}+2a_{-}\phi\left(k\right)\phi_{+}\left(-k\right)+\mathcal{O}\left(z^{2\nu}\right)\right]\right|_{z=\epsilon}\ .
\end{aligned}
\end{align}
The renormalized on-shell action is obtained according to Eq. \eqref{eq:adscft-56}\footnote{Notice that $z\to0$ corresponds to $R\to\infty$ in that equation.}
\begin{equation}\label{eq:adscft-84}
	\tilde{S}^{\text{(ren)}}=\frac{1}{2}\int\frac{d^{d}k}{\left(2\pi\right)^{d}}\left[2\nu\phi_{0}\left(k\right)\phi_{+}\left(-k\right)\right]\ ,
\end{equation}
where terms of $\mathcal{O}(z^{2\nu})$ are neglected since they go to zero faster.

\par The expectation value of the operator associated to $\phi_0(k)$ can now be computed according to rule 2, c.f. Eq. \eqref{eq:adscft-57}
\begin{equation}\label{eq:adscft-85}
	\left\langle O\left(k\right)\right\rangle =\left(2\pi\right)^{d}\frac{\delta\tilde{S}^{\text{(ren)}}\left(\phi\right)}{\delta\phi_{0}\left(-k\right)}=2\nu\phi_{+}\left(k\right)\ .
\end{equation}
The two-point function is easily obtained from the linear response of the one-point function to the source (more on this later)
\begin{equation}\label{eq:adscft-86}
	\left\langle O\left(k\right)\right\rangle =G\left(k\right)\phi_{0}\left(k\right)\ ,
\end{equation}
$G(k)=\left\langle O\left(k\right)O\left(-k\right)\right\rangle $ is called the Green function. Putting Eqs. \eqref{eq:adscft-85} and \eqref{eq:adscft-86} together
\begin{equation}\label{eq:adscft-87}
	G\left(k\right)=\left\langle O\left(k\right)O\left(-k\right)\right\rangle =2\nu\frac{\phi_{+}\left(k\right)}{\phi_{0}\left(k\right)}\ .
\end{equation}
Now, all that is left is to compute the explicit form of the coefficients $\phi_{+}$ and $\phi_{0}$. In order to do so, one must solve Eq. \eqref{eq:adscft-60} - recall that the solution \eqref{eq:adscft-63} is valid only in the regime $z\to 0$. In general, the equation of motion cannot be solved by analytical methods and one has to rely on numerical methods, however, in this simple example, the analytical solution is available.

\par Fourier transforming Eq. \eqref{eq:adscft-60} and redefining the field as $\phi(z,k)\mapsto z^{d/2}\phi(z,k)$ leads to the equation
\begin{equation} \label{eq:adscft-88}
	z^{2}\partial_{z}^{2}\phi\left(z,k\right)+z\partial_{z}\phi\left(z,k\right)-\left(\nu^{2}+k^{2}z^{2}\right)\phi\left(z,k\right)=0\ ,
\end{equation}  
which is a Bessel equation. The solution to this particular kind of equation is given in terms of modified Bessel functions
\begin{equation}\label{eq:adscft-89}
	\phi\left(z,k\right)=c_{1}z^{\frac{d}{2}}K_{\nu}\left(z,k\right)+c_{2}z^{\frac{d}{2}}I_{\nu}\left(z,k\right)\ ,
\end{equation}
where the redefinition done prior to Eq. \eqref{eq:adscft-88} has been considered. The behaviour of modified Bessel functions at infinity is
\begin{align}
\begin{aligned}\label{eq:adscft-90}
	\lim_{x\to \infty}K_{\nu}\left(x\right)\sim\frac{e^{-x}}{\sqrt{x}}, \ \ \ \
	\lim_{x\to \infty}I_{\nu}\left(x\right)\sim\frac{e^{x}}{\sqrt{x}}, 
\end{aligned}
\end{align}
i.e. $I_{\nu}$ diverges, hence $c_2$ is set to zero in order to have a regular solution. Next, consider the expansion of $K_{\nu}$ near the origin
\begin{equation}\label{eq:adscft-91}
		\lim_{x\to 0}K_{\nu}\left(x\right)=\frac{\Gamma\left(\nu\right)}{2}\left(\frac{x}{2}\right)^{-\nu}+\frac{\Gamma\left(-\nu\right)}{2}\left(\frac{x}{2}\right)^{\nu}\ ,
\end{equation}
up to first order in the expansion. From these expressions the solution is obtained
\begin{equation}\label{eq:adscft-92}
	\phi\left(z,k\right)=c_{1}\left[\underbrace{\frac{\Gamma\left(\nu\right)}{2}\left(\frac{k}{2}\right)^{-\nu}}_{\phi_{0}}z^{a_{-}}+\underbrace{\frac{\Gamma\left(-\nu\right)}{2}\left(\frac{k}{2}\right)^{\nu}}_{\phi_{+}}z^{a_{+}}\right]\ ,
\end{equation}
the identification with $\phi_0$ and $\phi_+$ is obtained through comparison with Eq. \eqref{eq:adscft-63}, of the solution in the asymptotic regime. Substituting in Eq. \eqref{eq:adscft-87} one obtains the two-point function
\begin{equation}\label{eq:adscft-93}
	\left\langle O\left(k\right)O\left(-k\right)\right\rangle =2\nu\frac{\Gamma\left(-\nu\right)}{\Gamma\left(\nu\right)}\left(\frac{k}{2}\right)^{2\nu}\ .
\end{equation}
The inverse Fourier transformation can be easily done by applying the formula
\begin{equation}\label{eq:adscft-94}
	\int\frac{d^{d}k}{\left(2\pi\right)^{d}}e^{ik^{\mu}x_{\nu}}k^{n}=\frac{2^{n}}{\pi^{\frac{d}{2}}}\frac{\Gamma\left(\frac{d+n}{2}\right)}{\Gamma\left(-\frac{n}{2}\right)}\frac{1}{\left|x\right|^{d+n}}\ ,
\end{equation}
such that, for $n=2\nu$, the form of Eq. \eqref{eq:adscft-93} in position space is
\begin{equation}\label{eq:adscft-95}
	\left\langle O\left(x\right)O\left(0\right)\right\rangle =\frac{2\nu}{\pi^{\frac{d}{2}}}\frac{\Gamma\left(\frac{d}{2}+\nu\right)}{\Gamma\left(-\nu\right)}\frac{1}{\left|x\right|^{2\Delta}}\ .
\end{equation}
Notice how the conformal dimension shows up, as well as the expression derived in Sect. \ref{adscft-sec4}, c.f. Eq. \eqref{eq:adscft-40}, as well as Eq. \eqref{eq:adscft-24} - result from CFT only -, is obtained.

\par In fact, there is room for more discussion, but for time's sake some details will be mentioned and references to such details are given. First, it is important to notice that throughout the example the determinant of the metric was always written as $\sqrt{|g|}$, instead of $\sqrt{-g}$ as usual. This is because in these problems there is an ambiguity in the signature of the metric which leads to some technical complications to define the boundary conditions, precisely, the condition for $z\to\infty$, which corresponds to $r\to 0$. In Eq. \eqref{eq:adscft-88} the Euclidean time signature was chosen, i.e. $\omega \to i\omega_E$. One could have chosen to keep the Minkowski signature and proceed with the calculation, the results in Eqs. \eqref{eq:adscft-93} and \eqref{eq:adscft-95} would be the same except for a a complex exponential containing a $\mbox{sgn}(\omega)$ function in its argument. This accounts for the choice of ingoing/outgoing conditions that would show up in the equation of motion for the scalar field -  Eq. \eqref{eq:adscft-88} is a Henkel equation in this scenario, and its solution are Henkel functions, complex combinations of modified Bessel functions.

\par Second, there are 3 possible cases when it comes to the value of $\nu$, in one of such cases the modes $\phi_0$ and $\phi_+$ interchange their roles, i.e. the leading term becomes sub-leading and vice-versa. This is because Eq. \eqref{eq:adscft-67}, this leads to the Breitenlohner-Freedman bound on the mass of the scalar field. Both topics, in this and the previous paragraph, are discussed in great detail in \cite{Minces:1999eg}. The aim of this section was to give an example of how to work with the correspondence, and not to go in deep into all the technical aspects of the calculation.

\section{AdS/CFT at finite temperature} \label{adscft-sec6}

\par Everything presented to this point in this chapter is considered to be at zero temperature, in fact, the temperature is not even mentioned anywhere else. Now it is time to see how one adds temperature and, consequently, thermodynamics to AdS/CFT. This notion has to come from the gravity side, and then be identified on the field theory side, as usual. The discussion in Sect. \ref{sec:3} of Chapter \ref{ch-2}, black hole thermodynamics, gives a hint on how one introduces temperature from the gravity side: one considers the space-time of a black hole as the bulk geometry. Notice, however, that introducing a temperature to the system adds a scale to the problem, whereas scale invariance is at the core of the correspondence. At high energies scale invariance is recovered so, in order to keep the theory located at the boundary scale-invariant, the black hole space-time must be asymptotically AdS. On the field theory side, one has to apply statistical mechanics and find a link to the black hole space-time in the bulk.

\subsection{Thermal field theory} \label{adscft-subsec61}

\par The expectation value of an operator $A$ at zero temperature is
\begin{equation}\label{eq:adscft-96}
	\left\langle A\right\rangle =\sum_{n}\left\langle n\right|A\left|n\right\rangle \ ,
\end{equation}
for a set of orthonormal states $\left|n\right\rangle$. When this system is at a temperature $T$, one introduces the density operator
\begin{equation}\label{eq:adscft-97}
	\rho=e^{-\beta H}\ ,
\end{equation}
with $\beta=1/k_BT$ the inverse temperature multiplied by Boltzmann's constant, and $H$ the Hamiltonian operator. Then, Eq. \eqref{eq:adscft-96} becomes an ensemble average:
\begin{equation}\label{eq:adscft-98}
	\left\langle A\right\rangle _{\beta}=\frac{1}{\text{Tr}\rho}\sum_{n}\left\langle n\right|A\left|n\right\rangle e^{-\beta H}=\frac{\text{Tr}\left(A\rho\right)}{\text{Tr}\rho}\ .
\end{equation}
Now, consider the following manipulation in the two-point correlation function of an operator $\Phi$ at temperature $T$
\begin{align}
\begin{aligned}\label{eq:adscft-99}
	\left\langle \Phi\left(t,x\right)\Phi\left(0,y\right)\right\rangle _{\beta}&=\frac{1}{\text{Tr}\rho}\text{Tr}\left(\Phi\left(t,x\right)\Phi\left(0,y\right)e^{-\beta H}\right)\ ,	\\	&=\frac{1}{\text{Tr}\rho}\text{Tr}\left(\Phi\left(t,x\right)e^{-\beta H}e^{\beta H}\Phi\left(0,y\right)e^{-\beta H}\right)\ ,	\\	&=\frac{1}{\text{Tr}\rho}\text{Tr}\left(\Phi\left(t,x\right)e^{-\beta H}e^{i\left(-i\beta H\right)}\Phi\left(0,y\right)e^{-i\left(-i\beta H\right)}\right)\ ,	\\&=\frac{1}{\text{Tr}\rho}\text{Tr}\left(\Phi\left(t,x\right)\Phi\left(-i\beta,y\right)e^{-\beta H}\right)\ ,	\\&=\left\langle \Phi\left(t,x\right)\Phi\left(-i\beta,y\right)\right\rangle _{\beta}\ .
\end{aligned}
\end{align}
Where the identity $\mbox{Tr}(AB)=\mbox{Tr}(BA)$ for any matrices $A,B$ was used, and the fact that $A\left(t,x\right)=e^{iH}A\left(0,x\right)e^{-iH}$ is the time evolution of an operator from time $t=0\mapsto t$. The conclusion is that considering imaginary time $t\mapsto i\tau$ is the same as considering that the theory is at temperature $\beta$, for one can clearly see that
\begin{equation}\label{eq:adscft-100}
	\left\langle \Phi\left(t,x\right)\Phi\left(-i\beta,y\right)\right\rangle _{\beta}\stackrel{t\mapsto i\tau}{\implies}\left\langle \Phi\left(\tau,x\right)\Phi\left(\beta,y\right)\right\rangle _{\beta} \ .
\end{equation}
This equation is known as the Kubo-Martin-Schwinger relation. It is important to mention that one needs to impose a condition of periodicity for the imaginary time, i.e.
\begin{equation}\label{eq:adscft-101}
	\Phi\left(0,x\right)=\pm\Phi\left(\beta,x\right)\ ,
\end{equation}
where the plus or minus sign is chosen whether one is considering bosons or fermions, respectively.

\par As the GKPW formula relates partition functions it is convenient to write down the same statement using this language. Following the same steps and using the path integral notation one can write the partition function for a field theory at temperature $T$ defined by Lagrangian density $\mathcal{L}$ as
\begin{equation}\label{eq:adscft-102}
	Z=\int\mathcal{D}\Phi e^{-\int_{0}^{\beta}d\tau\mathcal{L}\left(\tau,\Phi\right)}\ ,
\end{equation}
where the periodic condition must be applied to bosons or fermions accordingly to the statement in the previous paragraph.

\subsection{Temperature in Holography} \label{adscft-subsec62}
\par In Ch. \ref{ch-2} it was shown how to compute the temperature associated with a black hole in two ways: computing its surface gravity (c.f. Eq. \eqref{eq:64}), or applying the first law of black hole mechanics (c.f. Eq. \eqref{eq:76} and discussion prior to it), i.e. the law analogous to the first law of thermodynamics for black holes. In fact, there is another way to compute the temperature of a black hole, which is writing the metric using the imaginary time and transforming the coordinates in a specific way, such that a periodic condition is required. This leads to the temperature by means of the AdS/CFT correspondence, since the CFT on the boundary inherits the time from the bulk theory and, it was shown in the previous section that using the imaginary time in the field theory case is equivalent to considering the theory in a heat bath of temperature $T\propto\beta^{-1}$.

\par Consider a general static metric\footnote{Stationary or even more general metrics will not be considered because in this work such space-times will not be relevant.}
\begin{equation}\label{eq:adscft-103}
	ds^{2}=-f\left(r\right)dt^{2}+\frac{dr^{2}}{g\left(r\right)}+r^2d\vec{x}^{2}\ ,
\end{equation}
and suppose it has a horizon at $r=r_h$ such that $f(r_h)=g(r_h)=0$. The analytic continuation of metric \eqref{eq:adscft-103} to imaginary time, $t\mapsto i\tau$, only changes the signature to the Euclidean one:
\begin{equation}\label{eq:adscft-104}
ds^{2}=f\left(r\right)d\tau^{2}+\frac{dr^{2}}{g\left(r\right)}+r^2d\vec{x}^{2}\ .
\end{equation}
Now expand Eq. \eqref{eq:adscft-104} near the horizon, that is, the functions $f(r)$ and $g(r)$ become
\begin{align}
	\begin{aligned}\label{eq:adscft-105}
	\left.f\left(r\right)\right|_{r\to r_{h}}&\approx f\left(r_{h}\right)+f^{\prime}\left(r_{h}\right)\left(r-r_{h}\right)=f^{\prime}\left(r_{h}\right)\left(r-r_{h}\right)\ ,	\\
	\left.g\left(r\right)\right|_{r\to r_{h}}&\approx g\left(r_{h}\right)+g^{\prime}\left(r_{h}\right)\left(r-r_{h}\right)=g^{\prime}\left(r_{h}\right)\left(r-r_{h}\right)\ .
	\end{aligned}
\end{align}
such that
\begin{equation}\label{eq:adscft-106}
	\left.ds^{2}\right|_{r\to r_{h}}=f^{\prime}\left(r_{h}\right)\left(r-r_{h}\right)d\tau^{2}+\frac{dr^{2}}{g^{\prime}\left(r_{h}\right)\left(r-r_{h}\right)}+d\vec{x}^{2}\ .
\end{equation}
Defining the new coordinate $R$ as
\begin{equation}\label{eq:adscft-107}
	R=2\sqrt{\frac{r-r_{h}}{g^{\prime}\left(r_{h}\right)}}\ ,
\end{equation}
allows one to rewrite Eq. \eqref{eq:adscft-106} as
\begin{equation}\label{eq:adscft-108}
	\left.ds^{2}\right|_{r\to r_{h}}=\frac{f^{\prime}\left(r_{h}\right)g^{\prime}\left(r_{h}\right)}{4}R^{2}d\tau^{2}+dR^{2}+d\vec{x}^{2}\ ,
\end{equation}
where the expression for $r-r_h$ was obtained from Eq. \eqref{eq:adscft-107}. Finally, redefining $\tau$ as
\begin{equation}\label{eq:adscft-109}
	\tau=\frac{\sqrt{f^{\prime}\left(r_{h}\right)g^{\prime}\left(r_{h}\right)}}{2}\theta\ ,
\end{equation}
makes the $\tau-R$ plane of the metric near the horizon look very familiar
\begin{equation}\label{eq:adscft-110}
	\left.ds^{2}\right|_{r\to r_{h}}=R^{2}d\theta^{2}+dR^{2}\ .
\end{equation}
This is, in fact, the plane written in polar coordinates. In order to make $R=0$ the center of the plane one needs to impose the condition of periodicity on $\theta$, with period $2\pi$. This condition implies directly on the imaginary time is a periodic coordinate which, from Eq. \eqref{eq:adscft-109}, has period
\begin{equation}\label{eq:adscft-111}
	\beta=\left(\frac{2}{\sqrt{f^{\prime}\left(r_{h}\right)g^{\prime}\left(r_{h}\right)}}\right)2\pi=\frac{4\pi}{\sqrt{f^{\prime}\left(r_{h}\right)g^{\prime}\left(r_{h}\right)}}\ .
\end{equation}
The nomenclature $\beta$ is not occasional. In fact, it is given conveniently because this period is identified with the period of the imaginary time of the previous section, c.f. Subsect. \ref{adscft-subsec61}, Eq. \eqref{eq:adscft-101}, so
\begin{equation}\label{eq:adscft-112}
	T=\frac{\sqrt{f^{\prime}\left(r_{h}\right)g^{\prime}\left(r_{h}\right)}}{4k_{B}\pi}\ .
\end{equation}
One can check that this expression gives the exact same result as the formulas presented in the previous chapter (mentioned in the beginning of this section). Hence, we state another rule of AdS/CFT:
\begin{quote}
	\textbf{Rule 4:}\\
	A strongly coupled CFT at finite temperature is dual to a black hole, moreover, the temperature of the CFT is exactly the temperature of the black hole.
\end{quote}

\subsection{Thermodynamics} \label{adscft-subsec63}

\par The next natural step is to obtain thermodynamic quantities such as pressure, internal energy, entropy, etc. In this subsection the focus will be on the computation of these quantities, associated to the CFT, from knowledge of the bulk theory, this is done easier by an example. 

\par Consider the AdS-Schwarzschild black hole, i.e. Eq. \eqref{eq:18} with $k=0$. In this case $h_{ij}=\delta_{ij}$, so:
\begin{align}
\begin{aligned}\label{eq:adscft-116}
	ds^{2}&=\frac{L^{2}}{z^{2}}\left(-f\left(z\right)dt^{2}+\frac{dz^{2}}{f\left(z\right)}+dx_{i}dx^{i}\right)\ ,\\
	f\left(z\right)&=1-\left(\frac{z_{h}}{z}\right)^{d}\ ,
\end{aligned}
\end{align}
where $z=r^{-1}$ and $z_h$ is the surface that defines the horizon, c.f. last paragraph of the previous section. The temperature is easily computed by direct application of Eq. \eqref{eq:adscft-112}
\begin{equation}\label{eq:adscft-117}
	T=\frac{4\pi z_h}{d}\ .
\end{equation}
Notice that this allows one to switch the horizon $z_h$ for the temperature, i.e. write the equations in the bulk as a function of temperature.

\par The on-shell action is given by
\begin{equation}\label{eq:adscft-118}
	\tilde{S}_{E}=-\frac{V\left(4\pi\right)^{d}L^{d-1}}{2\kappa_{d+1}^{2}d^{d}}T^{d-1}\ ,
\end{equation}
where the subscript $E$ indicates that before evaluating the on-shell action on Eq. \eqref{eq:adscft-116} a Wick rotation is performed on the time coordinate, therefore, this is the Euclidean on-shell action. $V$ is a finite volume that encloses the system, it is needed in order to avoid a divergence because technically the volume of the space-time is infinite -- the same thing was done in section\ref{adscft-subsec51}. The partition function is then
\begin{equation}\label{eq:adscft-119}
	Z=e^{-\tilde{S}_E}\ ,
\end{equation}
and the thermodynamic quantities are computed according to the usual scheme of statistical mechanics. Pressure, $p$, is
\begin{equation}\label{eq:adscft-120}
	p=T\frac{\partial\log Z}{\partial V}=\frac{\left(4\pi\right)^{d}L^{d-1}}{2\kappa_{d+1}^{2}d^{d}}T^{d}=-F\ ,
\end{equation}
which equals minus the free energy, $F$. Entropy is
\begin{equation}\label{eq:adscft-121}
	s=T\frac{\partial\log Z}{\partial T}=\frac{V\left(4\pi\right)^{d}L^{d-1}\left(d-1\right)}{2\kappa_{d+1}^{2}d^{d}}T^{d-1}\ ,
\end{equation}
notice that this provides another method of evaluating the temperature. One can check that this expression matches the computation using the area law, c.f. Eq. \eqref{eq:adscft-43}. The energy of the system can then be computed according to the first law of thermodynamics
\begin{equation}\label{eq:adscft-122}
	\varepsilon=Ts-p=\frac{\left(4\pi\right)^{d}L^{d-1}}{2\kappa_{d+1}^{2}d^{d}}T^{d}\ .
\end{equation}

\par This simple example shows how to compute the thermodynamic quantities from knowledge of the gravitational theory. The CFT at the boundary simply inherits all thermodynamic properties defined by the black hole in the bulk.

\subsection{Finite charge density in holography}\label{adscft-subsec64}

\par Condensed matter problems are in close relation to the existence of conserved currents, so in order to attack a problem of condensed matter physics within the holography setting it is fundamental to describe how one introduces a conserved current using the AdS/CFT prescription. The most common case of a conserved current is the $U(1)$ symmetry, and this is the case of interest in the present context. In field theory these symmetries, gauge symmetries, are local, however, in the holographic framework, gauge symmetries are global \cite{Witten:1998qj,Horowitz:2006ct}.

\par From the discussion of the conformal group, c.f. Sect. \ref{adscft-sec2}, one realizes that the diffeomorphism invariance under $SO(d-1)$ of the AdS bulk implies that the field theory in the boundary is globally invariant under Lorentz transformation. From this intuition, the next rule is stated as
\begin{quote}
	\textbf{Rule 5:}\\
	Gauge symmetries in the gravitational theory are dual to global symmetries in the field theory.
\end{quote}
Hence, if one desires to describe a field theory at the boundary with a conserved $U(1)$ current, the Einstein--Hilbert action plus the Maxwell term is the simplest way to do so, this leads to the so-called AdS-Reissner-Nordstr\"om (AdS-RN) black hole and a nice interpretation to the gauge potential, as it will be seen.

\par The action originating AdS-RN space-time is given by Eq. \eqref{eq:2-13} plus the term associated to the gauge field, this action is often called Einstein--Maxwell:
\begin{align}
\begin{aligned}\label{eq:adscft-123}
	S&=\frac{1}{2\kappa_{d+1}^{2}}\int d^{d+1}x\sqrt{-g}\left[R+\frac{d\left(d-1\right)}{L^{2}}\right]-\frac{1}{4b^{2}}\int d^{d+1}x\sqrt{-g}F_{\mu\nu}F^{\mu\nu}\\&\hspace{4cm}+\frac{1}{2\kappa_{d+1}^{2}}\int d^{d}x\sqrt{-\gamma}\left[-2K+\frac{2\left(d-1\right)}{L^{2}}\right]\ .
\end{aligned}
\end{align}
the coupling constant $b$ takes into account the backreaction of the gauge field on the gravity sector, and $F_{\mu\nu}=2\partial_{\mu}A_{\nu}$ is the field strength of the gauge field $A_{\mu}$. The equations of motion for this action are
\begin{align}
	R_{\mu\nu}-\frac{1}{2}Rg_{\mu\nu}-\frac{d\left(d-1\right)}{2L^{2}}g_{\mu\nu}&=\frac{\kappa_{d+1}^{2}}{2b^{2}}T_{\mu\nu}\ ,	\label{eq:adscft-124}	\\
	\nabla_{\mu}F^{\mu\nu}&=0\ , \label{eq:adscft-125}
\end{align}
i.e. Eq. \eqref{eq:adscft-124} are Einstein's equations and Eq. \eqref{eq:adscft-125} are Maxwell's equations. The energy momentum tensor was obtained in Chapter \ref{ch-2}, c.f. Sect. \ref{rnBH}, and is given by
\begin{equation}\label{eq:adscft-126}
	T_{\mu\nu}=F_{\mu}^{\text{ }\lambda}F_{\lambda\nu}-\frac{1}{d}g_{\mu\nu}F_{\lambda\rho}F^{\lambda\rho}\ .
\end{equation}

\par Before proceeding to the solution of this system, some considerations are in order, to understand how the charge density arises in the holographic setting. As it was seen is Sect. \ref{adscft-sec5}, the field on the gravity side acts as a source for the operator in the boundary, for a finite density, $\rho$, the gauge field must be such that its time component has a non-zero expectation value, i.e. $\left\langle J_{t}\right\rangle =\rho$. At the dual level, the source of charge density, $\rho$, is the chemical potential, $\mu$, so, putting this all together one obtains the condition
\begin{equation}\label{eq:adscft-127}
	\lim_{r\to\infty}A_{t}=\mu\ .
\end{equation}
Another condition imposed is that scale invariance is recovered at energy scales much higher than $\mu$, this is actually imposed on the space-times, and requires that the space-time be asymptotically AdS. Moreover, it was mentioned in the previous section that the gravity dual of a CFT at finite temperature in the boundary is a black hole space-time, so these are all the ingredients needed on the gravity side such that the theory on the boundary is at finite temperature with non-zero charge density.

\par The procedure of solving the equations of motion is very similar to the previous ones, in fact, the ansatz for the metric is exactly the same as the one used to solve the AdS-Schwarzschild, c.f. Eq. \eqref{eq:2-14} -- the first equation --, the function $f(r)$ will be different, of course. For the gauge field, the ansatz is
\begin{equation}\label{eq:adscft-128}
	A=A_{t}\left(r\right)dt\ .
\end{equation}
It is a computational matter to solve these equations, here only the solution is presented:
\begin{align}
	ds^{2}&=\frac{r^{2}}{L^{2}}\left(-f\left(r\right)dt^{2}+dx_{i}dx^{i}\right)+\frac{L^{2}}{r^{2}f\left(r\right)}dr^{2}\ , \label{eq:adscft-129}	\\
	f\left(r\right)&=1-\frac{M}{r^{d}}+\frac{Q^{2}}{r^{2d-2}}\ , \label{eq:adscft-130}	\\
	A_{\mu}dx^{\mu}&=\mu\left[1-\left(\frac{r_{h}}{r}\right)^{d-2}\right]dt\ . \label{eq:adscft-131}
\end{align}
Notice that this solution, like AdS-Schwarzschild, is obtained by setting $k=0$ in Eq. \eqref{eq:2-15}, therefore its horizon is planar. Also, the surface defined by $r=r_h$ is the outer horizon -- remember that the Reissner--Nordstr\"om black hole has two horizons --, the same is true for the AdS version of this space-time and, in holography, one is usually concerned with the outer horizon. The thermodynamic quantities of this black hole can be easily computed following the same steps of Sect. \ref{adscft-subsec64}, the difference is in the ensemble. While for the AdS-Schwarzschild the canonical ensemble is convenient, here the grand canonical ensemble proves to be more useful, this is because the chemical potential is present. Besides that, the computations are the same and the results will be shown in the next chapter when the AdS-Reissner--Nordstr\"om solution is studied in greater detail.

\par For future reference, notice that on the surface $r=r_h$, defined by $f(r_h)=0$, one can obtain the following relation from Eq. \eqref{eq:adscft-130}:
\begin{equation}\label{eq:adscft-132}
	M=r_h^d+\frac{Q^2}{r_h^{d-2}}\ .
\end{equation}
Eq. \eqref{eq:adscft-132} is particularly useful to eliminate the mass from the expressions in favor of the horizon radius and charge, which are useful when one considers the extremal limit of the black hole, i.e. when its energy content is entirely due to charge. Also, using Eq. \eqref{eq:adscft-132}, one can rewrite the function \eqref{eq:adscft-130} as
\begin{equation}\label{eq:adscft-133}
	f\left(r\right)=1-\left(1-q^{2}\right)\left(\frac{r_{0}}{r}\right)^{d}+q^{2}\left(\frac{r_{0}}{r}\right)^{2d-2}\ ,
\end{equation}
where $q^2=\frac{Q^2}{r_0^{2d-2}}$. If now one changes the radial coordinate $z=L^2r^{-1}$ the metric becomes
\begin{align}
	ds^{2}&=\frac{L^{2}}{z^{2}}\left(-f\left(z\right)dt^{2}+dx_{i}dx^{i}+\frac{dz^{2}}{f\left(z\right)}\right)\ , \label{eq:adscft-134}	\\
	f\left(z\right)&=1-\left(1-q^{2}\right)\left(\frac{z}{z_{h}}\right)^{d}+q^{2}\left(\frac{z}{z_{h}}\right)^{2d-2}\ . \label{eq:adscft-135}
\end{align}

\par The AdS-Reissner--Nordstr\"om solution, given by Eqs. \eqref{eq:adscft-129}-\eqref{eq:adscft-131}, is the cornerstone when it comes to the power of AdS/CFT correspondence applied to condensed matter systems. From this theory, including other matter fields, a holographic superconductor can be described \cite{Hartnoll:2008kx}, and following this work many developments have been done, for example, which  \cite{Hartnoll:2016apf,Seo:2016vks} are particularly interesting to the authors. In order to understand properly these results, it is necessary to introduce the so-called Kubo formula and, in a more general context, the fluid/gravity correspondence, which is the low energy limit of the AdS/CFT correspondence, suited to describe condensed matter phenomena \cite{Rangamani:2009xk}.


%% file: chapters/Transport_coefficients.tex
\newpage
\chapter{Transport coefficients in AdS/CFT context}\label{part:4}

\par Having discussed some features of gravitational theories as well as the basics of AdS/CFT correspondence, including how to extract some quantities from the examples seen in the previous section, it is now possible to introduce the method for computing transport coefficients. In the condensed matter context, the measurement of how rapidly a perturbed system returns to its equilibrium state is given by the so-called transport coefficients. It is natural to expect a variety of such coefficients since one can produce perturbations in different quantities. In this work, the attention will be restricted to a few of these coefficients, namely: electric and thermal conductivity, and shear viscosity. The procedure, however, is very general, and other quantities could be included by considering a similar process \cite{Kubo:1957mj}.

\par Since the aim is to describe a particular kind of system within the AdS/CFT correspondence, the dimensions will be restricted to AdS$_4$/CFT$_3$, hence, the equations appearing in this and remaining chapters are no longer in arbitrary dimensions - when formulas already derived are used, the substitution of $d=4$ should be intended.

\section{Linear response theory and Kubo formula} \label{sec:kubo}
\par When one adds an external source to a given system there is always a response, which usually appears in the form of a change in the expectation value associated with the operator corresponding to the source introduced. Examples include:

\begin{table}[ht]
	\centering
	\begin{tabular}{c| c| c}

		System & Source & Response \\ [0.5ex] 
		\hline
		Charged & Gauge potential $(\mu)$ & Charge density $(\rho)$ \\
		Conductor & Vector potential $(A^{i})$ & Current $(J^{i})$\\
		Fluid & Space-time fluctuation $(h_{\mu\nu})$ & Energy-momentum tensor $(T_{\mu\nu})$ \\	[1ex]

	\end{tabular}
	\label{table:nonlin}
\end{table}
In the end of the previous chapter, it was seen how the introduction of a gauge potential produced a charged density. It is in this fashion that the linear response theory fits into holography.

\par In fact, most of the machinery necessary to carry on computations has already been introduced in  the text, here this will be put together and formalized in a clear way. Consider a system described by a Hamiltonian, where a time-dependent perturbation is added $\delta H(t)$:
\begin{equation}\label{eq:ct-1}
	H=H_0+\delta H(t)\ .
\end{equation}
This perturbation will change the action of the system by
\begin{equation}\label{eq:ct-2}
	\delta S=\int d^{4}x\phi^{\left(0\right)}\left(t,\vec{x}\right)O\left(\vec{x}\right)\ ,
\end{equation}
where $\phi^{\left(0\right)}\left(t,\vec{x}\right)$ is the external source. Assume that at $t=t_0\to -\infty$ the system is in equilibrium, and at $t=t_0$ the source is ``turned on". The ensemble average of the operator $O$ is 
\begin{equation}\label{eq:ct-3}
	\left\langle O\left(t,\vec{x}\right)\right\rangle _{\phi^{\left(0\right)}}=\text{Tr}\left[\rho\left(t\right)O\left(\vec{x}\right)\right]\ ,
\end{equation}
for a density matrix $\rho$ such that $\rho(t_0)$ is an equilibrium state. The subscript $\phi^{(0)}$ reminds that this average is considered when the source is turned on. After some computations involving the time evolution of the density matrix, together with the fact that $\rho(t_0)$ corresponds to an equilibrium state, one finds that\footnote{The details can be found in \cite{Natsuume:2014sfa}.}
\begin{equation}\label{eq:ct-4}
	\delta\left\langle O\left(t,\vec{x}\right)\right\rangle =-i\int_{-\infty}^{\infty}d^{4}x^{\prime}\theta\left(t-t^{\prime}\right)\left\langle \left[O\left(t,\vec{x}\right),O\left(t^{\prime},\vec{x}^{\prime}\right)\right]\right\rangle \phi^{\left(0\right)}\left(t^{\prime},\vec{x}^{\prime}\right)\ .
\end{equation}
Defining the retarded Green's function as
\begin{equation}\label{eq:ct-5}
	G_{R}^{OO}\left(t-t^{\prime},\vec{x}-\vec{x}^{\prime}\right)=-i\theta\left(t-t^{\prime}\right)\left\langle \left[O\left(t,\vec{x}\right),O\left(t^{\prime},\vec{x}^{\prime}\right)\right]\right\rangle \ ,
\end{equation}
Eq. \eqref{eq:ct-4} reads simply
\begin{equation}\label{eq:ct-6}
	\delta\left\langle O\left(t,\vec{x}\right)\right\rangle =-\int_{-\infty}^{\infty}d^{4}x^{\prime}G_{R}^{OO}\left(t-t^{\prime},\vec{x}-\vec{x}^{\prime}\right)\phi^{\left(0\right)}\left(t^{\prime},\vec{x}^{\prime}\right)\ .
\end{equation}
The Fourier transformation of Eq. \eqref{eq:ct-6} is

\begin{equation}\label{eq:ct-7}
	\delta\left\langle O\left(k\right)\right\rangle =-G_{R}^{OO}\left(k\right)\phi^{\left(0\right)}\left(k\right)\ .
\end{equation}
Compare this equation to Eq. \eqref{eq:adscft-87}, it was used in the computation and, as it was claimed, it comes within the framework of linear response theory, which provides an alternative way of computing correlation functions, as it was explained in that section, c.f. Sect. \ref{adscft-subsec53}. As 4 dimensions are being considered, $k_{\mu}\equiv (\omega,\vec{k})$ is the momentum four-vector, and the retarded Green's function in momentum space reads
\begin{equation}\label{eq:ct-8}
	G_{R}^{OO}\left(k\right)=-\int_{-\infty}^{\infty}d^{4}xe^{-ik^{\mu}x_{\mu}}\theta\left(t\right)\left\langle \left[O\left(t,\vec{x}\right),O\left(0,\vec{0}\right)\right]\right\rangle \ .
\end{equation}
It was already seen, c.f. Subsect. \ref{adscft-subsec53}, that the AdS/CFT correspondence can be used to determine the Green's functions. Thus, combining Linear Response Theory with AdS/CFT correspondence provides a useful (in fact unique for strongly coupled theories) way of computing transport coefficients.

\par Looking back at the examples given at the beginning of this chapter
\begin{itemize}
	\item Charged system: Let the perturbation in the Lagrangian be of the form $\delta \mathcal{L}=A_t^{(0)}(t)J^t(x)$, such that
	\begin{align}
	\begin{aligned}\label{eq:ct-9}
	\delta\left\langle \rho\right\rangle &=-G_{R}^{\rho\rho}\mu\ ,		\\
	G_{R}^{\rho\rho}&=-i\int_{-\infty}^{\infty}d^{4}xe^{-ik^{\mu}x_{\mu}}\theta\left(t\right)\left\langle \left[\rho\left(t,\vec{x}\right),\rho\left(0,\vec{0}\right)\right]\right\rangle \ ,
	\end{aligned}
	\end{align}
	where $\rho=J^{t}$ is the charge density, and $\mu=A_t^{(0)}$ is the gauge potential conjugate to $\rho$ - this identification was made in the end of the last chapter.
	\item Conductor: For a conductor the perturbation comes in the form of a vector potential, consider for instance $\delta \mathcal{L}=A_x^{(0)}(t)J^x(x)$. The equations are similar to Eq. \eqref{eq:ct-9}:
	\begin{align}
	\begin{aligned}\label{eq:ct-10}
	\delta\left\langle J^{x}\right\rangle &=-G_{R}^{xx}A_{x}^{\left(0\right)}\ ,	\\
	G_{R}^{xx}&=-i\int_{-\infty}^{\infty}d^{4}xe^{-ik^{\mu}x_{\mu}}\theta\left(t\right)\left\langle \left[J^{x}\left(t,\vec{x}\right),J^{x}\left(0,\vec{0}\right)\right]\right\rangle \ .
	\end{aligned}
	\end{align}
	\item Fluid: In this case the perturbation is on the energy-momentum tensor, i.e. $\delta\mathcal{L}=h_{xy}\left(t\right)T^{xy}\left(x\right)$, so
	\begin{align}
	\begin{aligned}\label{eq:ct-11}
	\delta\left\langle T^{xy}\right\rangle &=-G_{R}^{xy,xy}h_{xy}^{\left(0\right)}\ ,	\\
	G_{R}^{xy,xy}&=-i\int_{-\infty}^{\infty}d^{4}xe^{-ik^{\mu}x_{\mu}}\theta\left(t\right)\left\langle \left[T^{xy}\left(t,\vec{x}\right),T^{xy}\left(0,\vec{0}\right)\right]\right\rangle \ .
	\end{aligned}
	\end{align}
	For a space-time perturbation, there are three kinds of perturbation, corresponding to scalar, vector, and tensor modes, the perturbation above turns out to be on the scalar sector \cite{Natsuume:2014sfa}. 
\end{itemize}
Notice that the source of the perturbation is always time-dependent, this is because of the previous considerations regarding the time when the source is turned on and, also, because deeply this formalism is nothing more than time-dependent perturbation theory from quantum mechanics.

\subsubsection*{Relating Green's functions to transport coefficients}
\par As a simple example, consider the conductivity, usually denoted by $\sigma$. By writing Ohm's law as $\delta\left\langle J^{x}\right\rangle =\sigma E_{x}^{\left(0\right)}$, one is able to relate the conductivity to the Green's function. Recall that $E_i=F_{ti}$, where $F_{\mu\nu}$ is the field strength tensor, and $E_i$ is the electric field. As $F_{\mu\nu}=2\partial_{[\mu}A_{\nu]}$, a convenient choice of gauge, i.e. $A_t=0$, allows one to write
\begin{equation}\label{eq:ct-12}
E_{x}^{\left(0\right)}=-\partial_{t}A_{x}^{\left(0\right)}=i\omega A_{x}^{\left(0\right)}\ ,
\end{equation}
where in the last step the Fourier transform is made. Comparing  Eqs.\eqref{eq:ct-10} and \eqref{eq:ct-12} one can read off
\begin{equation} \label{eq:ct-13}
\sigma\left(\omega\right)=-\frac{1}{i\omega}G_{R}^{xx}\left(\omega,\vec{k}\to0\right)\ .
\end{equation}
This is the AC conductivity or frequency-dependent conductivity. If the limit $\omega\to 0$ is taken then one finds the so-called DC conductivity. A relation such as Eq. \eqref{eq:ct-13} is called a Kubo formula, in general, a Kubo formula relates a transport coefficient to a Green function.

\par Another quantity that will be important is the shear viscosity. For a review of hydrodynamics see \cite{Natsuume:2014sfa,2013rehy.book.....R}, the details will be skipped here for time sake. A viscous fluid is described by the energy-momentum tensor:
\begin{align}
\begin{aligned}\label{eq:ct-14}
T^{\mu\nu}&=\left(\epsilon+P\right)u^{\mu}u^{\nu}+Pg^{\left(0\right)\mu\nu}+t^{\mu\nu}\ ,	\\
t^{\mu\nu}&=-P^{\mu\alpha}P^{\nu\beta}\left[\eta\left(2\nabla_{(\alpha}u_{\beta)}-\frac{2}{3}g_{\alpha\beta}^{\left(0\right)}\nabla_{\lambda}u^{\lambda}\right)+\zeta g_{\alpha\beta}^{\left(0\right)}\nabla_{\lambda}u^{\lambda}\right]\ ,
\end{aligned}
\end{align}
where $\epsilon$ is the energy density, $P$ is the pressure, $\eta$ the shear viscosity and $\zeta$ the bulk viscosity of the fluid. The rest frame is considered, so $u^{\mu}=(1,0,0,0)$ and $P^{\mu\nu}=\text{diag}\left(0,1,1,1\right)$ is sometimes called the projection tensor along spatial directions.\footnote{The general expression for the projection tensor is $P^{\mu\nu}=g^{\left(0\right)\mu\nu}+u^{\mu}u^{\nu}$, thus, one can easily check that, on the rest frame, it assumes the expression showed in the text.} The metric denoted $g^{(0)\mu\nu}$ is the (linear) perturbation in the background metric (assumed to be Minkowski)
\begin{equation} \label{eq:ct-15}
g^{\left(0\right)\mu\nu}=\begin{pmatrix}-1 & 0 & 0 & 0\\
0 & 1 & h_{xy}^{\left(0\right)}(t) & 0\\
0 & h_{xy}^{\left(0\right)}(t) & 1 & 0\\
0 & 0 & 0 & 1
\end{pmatrix}\ ,
\end{equation}
and the covariant derivatives, $\nabla_{\mu}$, are taken with respect to this metric. The non-zero contribution in the covariant derivatives in Eq. \eqref{eq:ct-14} comes from Christoffel symbols only and read
\begin{equation}\label{eq:ct-16}
\nabla_{x}u_{y}=\Gamma_{xy}^{0}\ ,
\end{equation}
the component $\nabla_{y}u_{x}$ is the same by the symmetry of Christoffel symbols. One can check that the other components vanish identically and that the divergence of $u^{\mu}$ is of higher-order in the perturbation, thus
\begin{equation}\label{eq:ct-17}
\delta\left\langle t^{xy}\right\rangle \sim-2\eta\nabla_{(x}u_{y)}=-\eta\partial_{t}h_{xy}^{\left(0\right)}\left(t\right)\stackrel{\text{F.T}}{=}i\omega\eta h_{xy}^{\left(0\right)}\left(\omega\right)\ ,
\end{equation}
where the Christoffel symbol was evaluated according to Eq. \eqref{eq:6}, and equals to $\Gamma=\frac{1}{2}\partial_th_{xy}^{(0)}$. In the last equality the Fourier transformation was taken to get rid of the time derivative. Comparison of Eqs. \eqref{eq:ct-17} and \eqref{eq:ct-11} gives
\begin{equation}\label{eq:ct-18}
\eta=-\lim_{\omega\to0}\frac{1}{\omega}\Im G_{R}^{xy,xy}\left(\omega,\vec{k}\to0\right)\ ,
\end{equation}
which is the Kubo formula for the shear viscosity.

\par Some comments are in order before this section is closed. Many technical details have been skipped in order to keep this review as short as possible, one may notice that Eqs. \eqref{eq:ct-13} and \eqref{eq:ct-18} involve limits that were not mentioned previously. This is because, when computing the general expression for Green's functions in each of those formulas there is a need of introducing a regulator in the integrals and, in order to obtain the desired result, the limits have to be taken. From a physical perspective, these limits suggest that these relations are valid only on the low energy/long-wavelength regime, which is characteristic of hydrodynamics\footnote{By hydrodynamics here it is meant the effective field theory approach and not the theory of fluid flow. Although similar, the first describes a wider variety of phenomena, including condensed matter systems.}, and constitutes the study of phenomena that are manifested regardless of the microscopic behavior of the system. This makes sense, considering that linear response theory is being used to evaluate transport coefficients, which are properties that only make sense on large scales. For this reason, the limit of low energy/long-wavelength ($\omega\to 0$, $\vec{k}\to 0$, respectively) is called fluid/gravity correspondence, given that one is technically computing hydrodynamic quantities from classical gravity  \cite{Bhattacharyya:2008jc,Hubeny:2011hd}.

\section{AdS-Reissner-Nordstr\"om black hole and its transport coefficients}

\par In Sect. \ref{adscft-subsec64}, of Chapter \ref{ch-3}, the charged black hole with a planar horizon was presented. Now the goal is to describe its thermodynamic properties and compute the electric conductivity as well as the shear viscosity associated with this black hole. From these one can gain insight into the connection between holography and condensed matter because $i)$ a minor modification in action \eqref{eq:adscft-123}, from last chapter, describes a model of superconductivity \cite{Hartnoll:2008kx}. $ii)$ The ratio of shear viscosity-to-entropy density is a quantity that is bounded to a minimum value \cite{Son:2009zzc,Cremonini:2011iq}, and is usually a good way to verify if the theory is in agreement with experiments. This does not mean that the value always agrees with this lower bound, a known case where the bound is violated is the Gauss-Bonnet gravity \cite{Brigante:2007nu,Brigante:2008gz}.

\par As in this section the dimension is fixed to $d=3$, so the gravity theory is $4$-dimensional, the solution given by Eqs. \eqref{eq:adscft-129}-\eqref{eq:adscft-131}, is
\begin{align}
	ds^{2}&=\frac{1}{z^{2}}\left[-f\left(z\right)dt^{2}+\frac{dz^{2}}{f\left(z\right)}+dx_{i}dx^{i}\right]\ , \label{eq:ct-19}	\\
	f\left(z\right)&=1-\left(1+\frac{1}{2}z_{h}^{2}a^{2}\mu^{2}\right)\left(\frac{z}{z_{h}}\right)^{3}+\frac{1}{2}z_{h}^{2}a^{2}\mu^{2}\left(\frac{z}{z_{h}}\right)^{4}\ , \label{eq:ct-20}	\\
	A_{t}\left(z\right)&=\mu\left[1-\left(\frac{z}{z_{h}}\right)\right]\ . \label{eq:ct-21}	
\end{align}
In Eq. \eqref{eq:ct-20}, $a=\frac{\kappa_4}{b}$ is the ratio between gravitational and electrodynamic coupling constants. $\mu$ is the chemical potential, set by the condition at the boundary
\begin{equation}\label{eq:ct-22}
	\lim_{z\to0}A_t(z)=\mu\ .
\end{equation}
Finally, notice that $L=1$, which means that distances are measured in units of $L$.

\subsection{Thermodynamics of AdS-RN}
\par As usual, starting with the temperature, according to equation Eq. \eqref{eq:adscft-112}  (where $k_B=1$)
\begin{equation}\label{eq:ct-23}
	T=-\left.\frac{f^{\prime}\left(z\right)}{4\pi}\right|_{z=z_{h}}=\frac{1}{4\pi}\left[\frac{3}{z_{h}}-\frac{1}{2}z_{h}a^{2}\mu^{2}\right]\ ,
\end{equation}
the minus sign appears in this definition of temperature because of the derivative with respect to $z$. In order to write the other thermodynamic functions as a function of temperature one has to solve Eq. \eqref{eq:ct-23} for $z_h$, this is a simple task:
\begin{equation}\label{eq:ct-24}
	z_{h}=\frac{-4\pi T+\sqrt{16\pi^{2}T^{2}+6a^{2}\mu^{2}}}{a^{2}\mu^{2}}\ ,
\end{equation}
the plus so that the horizon radius is positive.

\par Other thermodynamic quantities are obtained from the partition function. Since the chemical potential is fixed the grand canonical ensemble is the best choice. The Landau potential is given by
\begin{equation}\label{eq:ct-25}
	\Omega = -T\log Z=T\tilde{S}^E\ ,
\end{equation}
where $Z=\exp(-\tilde{S}^E)$ is the partition function and $\tilde{S}^E$ is the Euclidean on-shell action. Plugging Eq. \eqref{eq:ct-19} in Eq. \eqref{eq:adscft-123} one finds, after a wick rotation, the Landau potential
\begin{equation}\label{eq:ct-26}
	\Omega=-\frac{V}{2\kappa_{4}^{2}z_{h}^{3}}\left(1+\frac{1}{2}z_{h}^{2}\mu^{2}a^{2}\right)=F\left(\frac{T}{\mu}\right)VT^{3}\ ,
\end{equation}
where the function $F(\frac{T}{\mu})$ is
\begin{equation} \label{eq:ct-27}
	F\left(\frac{T}{\mu}\right)=\frac{2a^{4}}{\kappa_{4}^{2}}\left(\frac{\mu}{T}\right)^{4}\left[\frac{4\pi^{2}+a^{2}\left(\frac{\mu}{T}\right)^{2}-\pi\sqrt{16\pi^{2}+6a^{2}\left(\frac{\mu}{T}\right)^{2}}}{\left(-4\pi+\sqrt{16\pi^{2}+6a^{2}\left(\frac{\mu}{T}\right)^{2}}\right)^{3}}\right]\ .
\end{equation}
From Eq. \eqref{eq:ct-26} and knowledge of statistical mechanics one can then compute the relevant thermodynamic quantities:
\begin{itemize}
	\item Entropy density
	\begin{equation}\label{eq:ct-28}
		\frac{S}{V}\equiv s=-\frac{1}{V}\left(\frac{\partial\Omega}{\partial T}\right)_{V,\mu}=\frac{2\pi}{\kappa_{4}^{2}z_{h}^{2}}=\frac{2a^{4}\pi\mu^{4}}{\kappa_{4}^{2}\left(-4\pi T+\sqrt{16\pi^{2}T^{2}+6a^{2}\mu^{2}}\right)^{2}}\ .
	\end{equation}
	\item Pressure
	\begin{equation}\label{eq:ct-29}
		P=-\left(\frac{\partial\Omega}{\partial V}\right)_{T,\mu}=-\frac{\Omega}{V}=\frac{1}{2z_{h}^{3}\kappa_{4}^{2}}\left(1+\frac{1}{2}z_{h}^{2}a^{2}\mu^{2}\right)\ .
	\end{equation}
	\item Energy density
	\begin{equation}\label{eq:ct-30}
		\varepsilon=-P+Ts=2P=\frac{1}{z_{h}^{3}\kappa_{4}^{2}}\left(1+\frac{1}{2}z_{h}^{2}a^{2}\mu^{2}\right)\ .
	\end{equation}
\end{itemize}

\par Looking at Eq. \eqref{eq:ct-23}, or \eqref{eq:ct-24}, one can see that the value $z_h^{\star}=\frac{\sqrt{6}}{a\mu}$ sets $T=0$. This is the well-known case of an extremal black hole, where both horizons coincide and all the energy associated with the black hole is due to its charge. As it was mentioned previously, in this limit the entropy of a regular (non AdS) Reissner-Nordstr\"om black hole was non-zero, and the same can be checked here by inspecting Eq. \eqref{eq:ct-28} in the $T\to0$ limit, in fact
\begin{equation}\label{eq:ct-31}
	\left.s\right|_{T=0}=\frac{\pi a^{2}\mu^{2}}{3\kappa_{4}^{2}}\ .
\end{equation}
For very low temperatures one can expand the entropy \eqref{eq:ct-28} as a power series
\begin{equation}\label{eq:ct-32}
	\left.s\right|_{T\ll1}\simeq\frac{\pi a^{2}\mu^{2}}{3\kappa_{4}^{2}}+\frac{4\pi^{2}a\mu}{3\kappa_{4}^{2}}\sqrt{\frac{2}{3}}T+\ldots\ 
\end{equation}
this means that the specific heat $C=T\frac{\partial s}{\partial T}$ is linear in this regime. In the opposite case, $T\to \infty$, entropy behaves as
\begin{equation}\label{eq:ct-33}
	\left.s\right|_{T\to\infty}\simeq\frac{32\pi^{3}}{9\kappa_{4}^{2}}T^{2}\ .
\end{equation}
These results match what is expected from a CFT at finite temperature in 3 dimensions \cite{Amoretti:2017tbk}.

\par One remarkable property of the AdS-RN black hole can be seen by looking at its near-horizon geometry. To do so, set the temperature to zero, expand metric \eqref{eq:ct-19} near $z=z_h^{\star}$ and define the new coordinate
\begin{equation}\label{eq:ct-34}
	\xi=\frac{\sqrt{6}}{z_{h}^{\star}}\left(1-\frac{z}{z_{h}^{\star}}\right)\ ,
\end{equation}
then the metric reads
\begin{equation}\label{eq:ct-35}
	ds^{2}=\frac{1}{\xi^{2}}\left(-dt^{2}+d\xi^{2}\right)+\frac{1}{\left(z_{h}^{\star}\right)^{2}}\left(dx^{2}+dy^{2}\right)\ .
\end{equation}
Notice that the scale isometry of Eq. \eqref{eq:ct-35} is
\begin{align}
	\begin{aligned}\label{eq:ct-36}
	t&\mapsto\lambda t\ ,\\\xi&\mapsto\lambda\xi\ ,\\x^{i}&\mapsto x^{i}\ .
	\end{aligned}
\end{align}
While for pure AdS such an isometry reads
\begin{align}
\begin{aligned}\label{eq:ct-37}
t&\mapsto\lambda t\ ,\\\xi&\mapsto\lambda\xi\ ,\\x^{i}&\mapsto\lambda x^{i}\ .
\end{aligned}
\end{align}
This indicates a quasi-local quantum critical state in the boundary field theory. The absence of scale invariance in spatial directions while the dynamics of the fields are scale-invariant in pure temporal regards. Such a phenomenon is ubiquitous to strange metals \cite{Amoretti:2017tbk}. In fact, both properties presented here, non-vanishing entropy at zero temperature and local quantum criticality are unique to the so-called strange metals, which find a suitable description in AdS$_4$/CFT$_3$.

\subsection{Conductivities} \label{conductivities}
\par The conductivities computed here appeared first in \cite{Hartnoll:2009sz} and following this work, every other point to this one when it comes to the calculation of electrical, thermal, and thermoelectric conductivities on the AdS-RN background. Following the prescription described in Sect. \ref{sec:kubo}, the perturbation that must be added, which has the electric conductivity as a response, is in the gauge field. In fact, due to the charge density, a dissipative effect, in the form of heat, appears \cite{Hartnoll:2009sz}. This requires that the Ohm's law have to be generalized:
\begin{equation}\label{eq:ct-38}
\begin{pmatrix}\left\langle J_{x}\right\rangle \\
\left\langle Q_{x}\right\rangle 
\end{pmatrix}=\begin{pmatrix}\sigma & \alpha T\\
\alpha T & \overline{\kappa}T
\end{pmatrix}\begin{pmatrix}E_{x}\\
-\left(\nabla_{x}T\right)/T
\end{pmatrix}\ ,
\end{equation}
instead of $\delta\left\langle J_{x}\right\rangle =\sigma E_{x}$, as stated prior to Eq. \eqref{eq:ct-12}. The heat current is given by $Q_{x}=T_{tx}-\mu J_{x}$, and the conductivities are the electrical ($\sigma$), the thermal ($\alpha$), and thermoelectric ($\bar{\kappa}$) ones. In the absence of fields that break time-reversal symmetry (such as a magnetic field) the Green's functions are symmetric, meaning that $G_{J_{i}J_{j}}=G_{J_{j}J_{i}}$\footnote{Here a slightly different notation is introduced for Green's functions: notice that the $R$, which indicated it is a retarded Green's function is omitted, this is because all the Green's functions being considered are of this kind. Second, instead of denoting the ``indices" of the Green's functions by the components of the source on the boundary, the source itself is written, in this case, the current $J_i$.} The perturbations considered here are isotropic, i.e. $k=0$, implying that other conductivities, such as the Hall conductivity, are identically zero. This sort of perturbation, as well as the breaking of translation symmetry, are treated by \cite{Amoretti:2017tbk}.

\par The sources $E_x$ (electric field in the $x$-direction) and $\nabla_xT$ (thermal gradient in the $x$-direction) are related to background perturbations $\delta A_{x}^{\left(0\right)}$ and $\delta g_{ts}^{\left(0\right)}$, respectively. The relation between the former was already seen in Sect. \ref{sec:kubo}:
\begin{equation}\label{eq:ct-39}
E_{j}=i\omega\delta A_{j}^{\left(0\right)}\text{ }.
\end{equation}
The thermal gradient reads 
\begin{align}\label{eq:ct-40}
i\omega\delta g_{tj}^{\left(0\right)}=-\frac{\nabla_{j}T}{T},&\ \ \ \ i\omega\delta A_{j}^{\left(0\right)}=\mu\frac{\nabla_{j}T}{T}\text{ }.
\end{align}
This can be seen from the following explanation, taken from \cite{Hartnoll:2009sz}. Recall that the period of the Euclidean time coordinate is $T^{-1}$, so a redefinition of the time coordinate is useful to keep track of the temperature dependence that appears: $t=\overline{t}/T$. This dimensionless time coordinate leads to $g_{\overline{t}\overline{t}}^{\left(0\right)}=-T^{-2}$. For the sake of argument, the metric being perturbed is Minkowski space-time \footnote{Since the main interest here is to analyze the effects from the perturbation, the background metric is irrelevant, therefore the simplest one is chosen.}, then for a small, constant, thermal gradient $T\mapsto T+x\nabla_{x}T$, the perturbation reads
\begin{equation}\label{eq:ct-41}
\delta g_{\overline{t}\overline{t}}^{\left(0\right)}=-\frac{2x\nabla_{x}T}{T^{3}}\text{ }.
\end{equation}
All the quantities can be endowed with time dependence by including the factor $e^{-i\overline{\omega}\overline{t}}$. Moreover, diffeomorphisms acting on the background fields give the fluctuations
\begin{align}\label{eq:ct-42}
\delta g_{\mu\nu}^{\left(0\right)}=2\partial_{(\mu}\xi_{\nu)},\ \ \ \ &\text{ }\delta A_{\mu}^{\left(0\right)}=A_{\nu}^{\left(0\right)}\partial_{\mu}\xi^{\nu}+\xi^{\nu}\partial_{\nu}A_{\mu}^{\left(0\right)}\text{ }.
\end{align}
These can be viewed as gauge transformations, thus, terms like \eqref{eq:ct-42} can be added to the perturbations. Taking
\begin{align}\label{eq:ct-43}
\xi_{\overline{t}}=ix\frac{\nabla_{x}T}{\omega T^{3}},\ \ \ \ &\xi_{x}=0\ ,
\end{align}
change the perturbations to
\begin{align}\label{eq:ct-44}
\delta g_{\overline{t}\overline{t}}^{\left(0\right)}=0,\ \ \ \ &\delta g_{x\overline{t}}^{\left(0\right)}=i\frac{\nabla_{x}T}{\overline{\omega} T^{3}},\ \ \ \ \delta A_{x}^{\left(0\right)}=-i\mu\frac{\nabla_{x}T}{\overline{\omega} T^{3}}\ .
\end{align}
Transforming back to $\overline{t}=tT$ gives expressions \eqref{eq:ct-40}.

\par Combining these terms, the perturbed action becomes\footnote{Notice that this action has the form of Eq. \eqref{eq:ct-2}, but with two terms.}
\begin{align}
\begin{aligned}\label{eq:ct-45}
\delta S&=\int d^{2}xdt\sqrt{-g^{\left(0\right)}}\left(T^{tx}\delta g_{tx}^{\left(0\right)}+J^{x}\delta A_{x}^{\left(0\right)}\right)\ ,\\
&=\int d^{2}xdt\sqrt{-g^{\left(0\right)}}\left[\left(T^{tx}-\mu J^{x}\right)\left(-\frac{\nabla_{x}T}{i\omega T}\right)+J^{x}\frac{E_{x}}{i\omega}\right]\ .
\end{aligned}
\end{align}
So the current associated with the thermal gradient is, in fact, $Q_{x}=T_{tx}-\mu J_{x}$, and Ohm's law can be written as
\begin{equation}\label{eq:ct-46}
\begin{pmatrix}\left\langle J_{x}\right\rangle \\
\left\langle Q_{x}\right\rangle 
\end{pmatrix}=\begin{pmatrix}\sigma & \alpha T\\
\alpha T & \overline{\kappa}T
\end{pmatrix}\begin{pmatrix}i\omega\left(\delta A_{x}^{\left(0\right)}+\mu\delta g_{tx}^{\left(0\right)}\right)\\
i\omega\delta g_{tx}^{\left(0\right)}
\end{pmatrix}\ .
\end{equation}
Applying the Kubo formula, c.f. Eq. \eqref{eq:ct-13}, the transport coefficients are
\begin{align}\label{eq:ct-47}
\sigma\left(\omega\right)=-i\frac{G_{J_{x}J_{x}}\left(\omega\right)}{\omega},\ \ \ \
&T\alpha\left(\omega\right)=-i\frac{G_{Q_{x}J_{x}}\left(\omega\right)}{\omega},\ \ \ \
T\overline{\kappa}\left(\omega\right)=-i\frac{G_{Q_{x}Q_{x}}\left(\omega\right)}{\omega}.
\end{align}

\par To calculate the Green's functions one has to actually solve the equations of motion for the perturbation in the bulk. Consider the perturbations in Eqs. \eqref{eq:ct-19} and \eqref{eq:ct-21}, keeping terms up to order one in the perturbation one finds
\begin{align}
\delta g_{tx}^{\prime}+\frac{2}{z}\delta g_{tx}+4a^{2}A_{t}^{\prime}\delta A_{x}&=0\ ,	\label{eq:ct-48}	\\
\left(f\delta A_{x}^{\prime}\right)^{\prime}+\frac{\omega^{2}}{f}\delta A_{x}+z^{2}A_{t}^{\prime}\left(\delta g_{tx}^{\prime}+\frac{2}{z}\delta g_{tx}\right)&=0\ .	\label{eq:ct-49}
\end{align}
the first equation is Einstein's equation, while the second is the (non-vanishing) Maxwell equation. Primes denote differentiation with respect to $z$, $f\equiv f(z)$, c.f. Eq. \eqref{eq:ct-20}, and $A_t$ is given by Eq. \eqref{eq:ct-21}. One can easily see that the metric perturbation can be eliminated from Eqs. \eqref{eq:ct-48} and \eqref{eq:ct-49}, and an ordinary differential equation can be written for $\delta A_{x}$:
\begin{equation}\label{eq:ct-50}
\left(f\delta A_{x}^{\prime}\right)^{\prime}+\left(\frac{\omega^{2}}{f}-\frac{4\mu^{2}a^{2}z^{2}}{z_{h}^{2}}\right)\delta A_{x}=0\ ,
\end{equation}
notice that $A_{t}^{\prime}=\mu z_{h}^{-1}$ was used. Near the boundary one can verify that the equation above is solved by
\begin{equation}\label{eq:ct-51}
\delta A_{x}=\delta A_{x}^{\left(0\right)}+z\delta A_{x}^{\left(1\right)}+\ldots\text{ }\text{as}\text{ }z\to0\ .
\end{equation}
As Eq. \eqref{eq:ct-50} is a linear equation, the coefficient $\delta A_{x}^{\left(1\right)}$ will depend linearly on $\delta A_{x}^{\left(0\right)}$ \cite{Hartnoll:2009sz}, however the coefficient of proportionality must be found numerically. Also, knowing that the equation for $\delta g_{tx}$ is of first-order allows one to obtain two out of the three transport coefficients without solving any differential equation. As it will be seen shortly, this is due to the fact that these two coefficients ($\alpha,\bar{\kappa}$) depend on the third ($\sigma$).

\par From Eqs. \eqref{eq:adscft-85}-\eqref{eq:adscft-87}, one can write the Green's functions as
\begin{equation}\label{eq:ct-52}
G_{O_{A}O_{B}}=\left.\frac{\delta\left\langle O_{A}\right\rangle }{\delta\Phi_{B}^{\left(0\right)}}\right|_{\delta\Phi=0}\equiv\lim_{z\to0}\left.\frac{\delta\Pi_{A}}{\delta\Phi_{B}^{\left(0\right)}}\right|_{\delta\Phi=0}\ ,
\end{equation}
where the momentum conjugate was defined in the third equality. Varying the full action, c.f. Eq. \eqref{eq:adscft-123},  the momenta are
\begin{align}
\Pi_{g_{tx}}&=\frac{\delta S}{\delta g_{tx}^{\left(0\right)}}=-\rho\delta A_{x}^{\left(0\right)}+\frac{2}{\kappa^{2}z^{3}}\left(1-\frac{1}{\sqrt{f\left(r\right)}}\right)\delta g_{tx}^{\left(0\right)}\ ,	\label{eq:ct-53}	\\
\Pi_{A_{x}}&=\frac{\delta S}{\delta A_{x}^{\left(0\right)}}=\frac{f\left(r\right)\left(A_{x}^{\left(0\right)}\right)^{\prime}}{b^{2}}-\rho\delta g_{tx}^{\left(0\right)}\ . \label{eq:ct-54}
\end{align}
The charge density, $\rho$, was introduced. It is defined by $\rho=-V_{2}^{-1}\frac{\partial\Omega}{\partial\mu}$, where $\Omega$ is the Landau potential, c.f. Eq. \eqref{eq:ct-26}. Some simplifications were made in order to obtain the expressions above: terms with a second derivative are eliminated using integration by parts, and the resulting terms on the boundary are canceled by the Gibbons-Hawking term. Also, the equations of motion were used to simplify the expressions, and finally, at finite $z$ the quantities $g_{tx}^{\left(0\right)}=g^{xx}g_{tx}$ and $A_{x}^{\left(0\right)}=A_{x}$ are defined.

\par In the limit, $z\to 0$ one finds that Ohm's law, Eq. \eqref{eq:ct-38}, becomes
\begin{equation}\label{eq:ct-55}
\begin{pmatrix}\left\langle J_{x}\right\rangle \\
\left\langle T_{tx}\right\rangle 
\end{pmatrix}=\begin{pmatrix}\frac{1}{b^{2}}\frac{\delta A_{x}^{\left(1\right)}}{\delta A_{x}^{\left(0\right)}} & -\rho\\
-\rho & -\epsilon
\end{pmatrix}\begin{pmatrix}\delta A_{x}^{\left(0\right)}\\
\delta g_{tx}^{\left(0\right)}
\end{pmatrix}\ .
\end{equation}
This identification comes from equation \eqref{eq:ct-51} and what is written below. The energy density was derived previously, c.f. Eq. \eqref{eq:ct-30}. Comparing \eqref{eq:ct-47} and \eqref{eq:ct-55} one finds
\begin{align}\label{eq:ct-56}
\sigma\left(\omega\right)=-\frac{i}{b^{2}\omega}\frac{\delta A_{x}^{\left(1\right)}}{\delta A_{x}^{\left(0\right)}},\ \ \ \
&T\alpha\left(\omega\right)=\frac{i\rho}{\omega}-\mu\sigma\left(\omega\right),\ \ \ \
T\overline{\kappa}\left(\omega\right)=\frac{i\left(\epsilon+P-2\mu\rho\right)}{\omega}+\mu^{2}\sigma\left(\omega\right).
\end{align}
The extra term, $P$, in the thermoelectric conductivity comes from a contact term that is present due to translational invariance \cite{Hartnoll:2007ip}. Finally, it remains to solve the differential equation for the perturbation $\delta A_{x}$ in the bulk. It can only be done numerically, so the results will be only shown here and commented on it according to the paper where these were obtained.
\subsubsection*{Numerical Results}
\par
\begin{figure}[h!]
	\centering
	\includegraphics[scale=0.35]{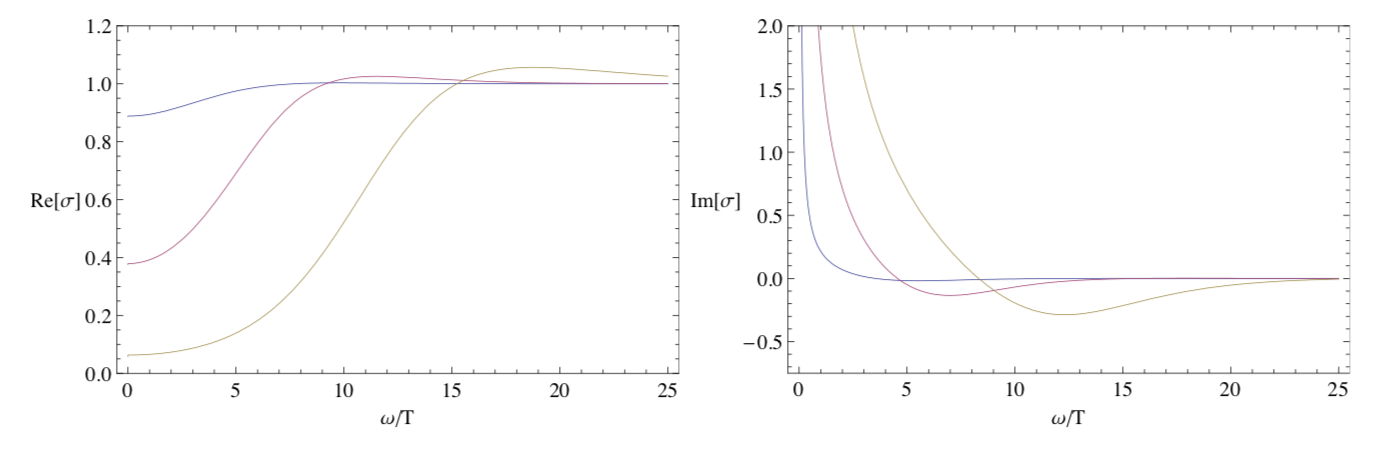}
	\caption{Real (left) and imaginary (right) parts of $\sigma(\omega)$. Different curves represent different values of chemical potential for fixed temperature. $b=1$ in all plots. Source \cite{Hartnoll:2009sz}.	}
	\label{fig:sigma_numerical}
\end{figure}

 The numerical results are shown here, Fig. \ref{fig:sigma_numerical}, are taken from \cite{Hartnoll:2009sz}, there are some technical details regarding the numerical stability of the solution which will not be discussed here. Solving equation \eqref{eq:ct-50} subject to \eqref{eq:ct-51} and to the so-called ingoing boundary conditions allows one to compute the conductivity $\sigma$ as a function of frequency, the results are shown in Figure \ref{fig:sigma_numerical}. In \cite{Hartnoll:2009sz} a comparison with conductivity in graphene at low energies is made, and the agreement with experimental results is remarkable.

\section{Shear viscosity} \label{eta-calc}
\par To evaluate the shear viscosity we apply the calculations for the
scalar field since the perturbation behaves the same way  \cite{Son:2009zzc,Kovtun:2004de}. Consider
a massless scalar field
\begin{equation} \label{etas-1}
	S=-\frac{1}{2}\int d^{5}x\sqrt{-g}g_{MN}\nabla^{M}\phi\nabla^{N}\phi \ ,
\end{equation}
We are interested in the on-shell action at the boundary. Since we
only consider space-times that are asymptotically AdS it is safe to
assume that
\begin{equation}\label{etas-2}
	\sqrt{-g}\sim u^{-5}\ ,
\end{equation}
and
\begin{equation}\label{etas-3}
	g^{uu}\sim u^{2}\ .
\end{equation}
(In fact, this can be immediately checked for all the solutions within this work). This readily gives
\begin{align}
\begin{aligned}\label{etas-4}
		S & \sim\int d^{4}xdu\left[-\frac{1}{2u^{3}}\left(\phi^{\prime}\right)^{2}\right]\ ,	\\
	& =\int d^{4}x\int_{0}^{1}du\left[\left(-\frac{1}{2u^{3}}\phi\phi^{\prime}\right)^{\prime}+\left(\frac{1}{2u^{3}}\phi^{\prime}\right)^{\prime}\phi\right]\ ,	\\
	& =\left[\int d^{4}x\frac{1}{2u^{3}}\phi\phi^{\prime}\right]_{u=0}+\int d^{4}xdu\left(\frac{1}{2u^{3}}\phi^{\prime}\right)^{\prime}\phi	\ .
\end{aligned}
\end{align}
The second term is just the equation of motion
\begin{equation}\label{etas-5}
	\left(\frac{1}{2u^{3}}\phi^{\prime}\right)^{\prime}\sim0\ .
\end{equation}
The asymptotic form of the solutions are
\begin{equation}\label{etas-6}
	\phi\sim\phi^{\left(0\right)}\left(1+\phi^{\left(1\right)}u^{4}\right)\ .
\end{equation}

\par Using the asymptotic solution, the action becomes the boundary term
\begin{equation}\label{etas-7}
	S\sim\left[\int d^{4}x\frac{1}{2u^{3}}\phi\phi^{\prime}\right]_{u=0}=\int d^{4}x2\left(\phi^{\left(0\right)}\right)^{2}\phi^{\left(1\right)}\ ,
\end{equation}
therefore the one-point function is, from the GKPW relation, c.f. \eqref{eq:adscft-57}:
\begin{equation}\label{etas-8}
	\left\langle O\right\rangle =\frac{\delta S}{\delta\phi^{\left(0\right)}}=4\phi^{\left(1\right)}\phi^{\left(0\right)}\ ,
\end{equation}
From the linear response relation
\begin{equation}\label{etas-9}
	\delta\left\langle O\left(t,\vec{x}\right)\right\rangle =-\int_{\infty}^{\infty}d^{4}x^{\prime}G_{R}^{OO}\left(t-t^{\prime},\vec{x}-\vec{x}^{\prime}\right)\phi^{\left(0\right)}\left(t^{\prime},\vec{x}^{\prime}\right)\ ,
\end{equation}
we can determine the Green's function $G_{R}$
\begin{equation}\label{etas-10}
	G_{R}^{OO}\left(k=0\right)=-4\phi^{\left(1\right)}\ .
\end{equation}

\par This is just the general procedure. Once one has identified the solution
for the equation of motion in the asymptotic regime, the boundary
conditions have to be implemented. We will do this while we obtain
the value for the shear viscosity associated with the Schwarzschild
black hole in AdS.

\par For the shear viscosity the associated perturbation is on the off-diagonal
in the metric,c.f. \eqref{eq:ct-15} and discussion following this equation,
\begin{equation}\label{etas-11}
	\delta\left\langle \tau^{xy}\right\rangle =i\omega\eta h_{xy}^{\left(0\right)}\ ,
\end{equation}
where $h_{xy}^{\left(0\right)}$ is associated to $\phi^{\left(0\right)}$
above. The bulk perturbation is
\begin{equation}\label{etas-12}
	ds^{2}=ds_{0}^{2}+2h_{xy}dxdy\ ,
\end{equation}
where $ds_{0}^{2}$ is the bulk metric. One can show that the contraction
$g^{xx}h_{xy}$ obeys the equation of motion for a massless scalar
field \cite{Kovtun:2004de}, and therefore the development above applies
for this perturbation. 

\par Consider for instance the AdS-Schwarzschild metric in 5 dimensions, Eq. \eqref{eq:adscft-120},
in the following coordinate system 
\begin{equation}\label{etas-13}
	ds_{0}^{2}=\left(\frac{r_{0}}{L}\right)^{2}\frac{1}{u^{2}}\left(-fdt^{2}+dx_{i}dx^{i}\right)+\left(\frac{L}{u}\right)^{2}\frac{du^{2}}{f}\ ,
\end{equation}
where
\begin{equation}\label{etas-14}
	f\equiv f\left(u\right)=1-u^{4}\ ,
\end{equation}
and $u=r_{0}/r$. By computing the on-shell action one can show that \cite{Natsuume:2014sfa}
\begin{equation}\label{etas-15}
	\delta\left\langle \tau^{xy}\right\rangle =\frac{1}{16\pi G_{5}}\frac{r_{0}^{4}}{L^{5}}4h_{xy}^{\left(1\right)}h_{xy}^{\left(0\right)}\ .
\end{equation}
Comparing Eqs. \eqref{etas-11} and \eqref{etas-15}
\begin{equation}\label{etas-16}
	i\omega\eta h_{xy}^{\left(0\right)}=\frac{1}{16\pi G_{5}}\frac{r_{0}^{4}}{L^{5}}4h_{xy}^{\left(1\right)}h_{xy}^{\left(0\right)}\implies i\omega\eta=\frac{1}{16\pi G_{5}}\frac{r_{0}^{4}}{L^{5}}4h_{xy}^{\left(1\right)}\ .
\end{equation}
Or, by using units where $4\pi G_{5}=L=r_{0}=1$, we can write it
in a more elegant form
\begin{equation}\label{etas-17}
	i\omega\eta=h_{xy}^{\left(1\right)}\ .
\end{equation}
Now we have to solve the differential equation and find $h_{xy}^{\left(1\right)}$.

\subsubsection*{Solving the differential equation}

\par Let $\phi=g^{xx}h_{xy}$, then as we mentioned, the equation of motion
is given by \cite{Kovtun:2004de}
\begin{equation}\label{etas-18}
	\partial_{\mu}\left(\sqrt{-g}g^{\mu\nu}\partial_{\nu}\phi\right)=0\ .
\end{equation}
Consider $\phi=\phi_{k}\left(u\right)e^{-i\omega t}$, the equation
becomes
\begin{equation}\label{etas-19}
	\frac{u^{3}}{f}\left(\frac{f}{u^{3}}\phi_{k}^{\prime}\right)^{\prime}+\frac{\mathfrak{w}^{2}}{\pi^{2}f^{2}}\phi_{k}=0\ ,
\end{equation}
where $\mathfrak{w}=\omega/T$, $T$ is the temperature associated
to the black hole. Next we recall that we are interested in the solution
in the hydrodynamic limit, i.e. $\omega\to0$ - this is the condition
that allows us to write an analytic solution to this problem. We will
implement 2 boundary conditions
\begin{itemize}
	\item At the horizon $\left(u\to1\right)$: The incoming-wave condition.
	This is natural since we are dealing with a black hole, and only ingoing waves are expected.
	\item At the boundary $\left(u\to0\right)$: Dirichlet boundary condition,
	basically we demand that the solution is constant at infinity, i.e.
	$\lim_{u\to0}\phi_{k}=\phi_{k}^{\left(0\right)}$.
\end{itemize}
To implement the condition at the horizon we notice that
\begin{equation} \label{etas-20}
	f=1-u^{4}\stackrel{u\to1}{\approx}4\left(1-u\right)\ ,
\end{equation}
and the equation reads
\begin{equation}\label{etas-21}
	\phi_{k}^{\prime\prime}-\frac{1}{1-u}\phi_{k}^{\prime}+\left(\frac{\mathfrak{w}}{4\pi}\right)^{2}\frac{1}{\left(1-u\right)^{2}}\phi_{k}=0\ .
\end{equation}
Now set $\phi_{k}=\left(1-u\right)^{\lambda}$, obtain the characteristic
equation
\begin{equation}\label{etas-22}
	\lambda\left(\lambda-1\right)+\lambda+\left(\frac{\mathfrak{w}}{4\pi}\right)^{2}=0\ ,
\end{equation}
leading to the solution for $u\to1$
\begin{equation}\label{etas-23}
	\phi_{k}=\left(1-u\right)^{\pm\frac{i\mathfrak{w}}{4\pi}}\ .
\end{equation}

\par These solutions are fairly simple to interpret in the so-called tortoise
coordinates. We define them such that the Schwarzschild metric becomes
conformally flat in the $\left\{ t,r_{\star}\right\} $ plane, that
is
\begin{equation}\label{etas-24}
	ds^{2}=-f\left(r\right)dt^{2}+\frac{dr^{2}}{g\left(r\right)}=\overline{f}\left(r\right)\left(-dt^{2}+dr_{\star}^{2}\right)\ .
\end{equation}
Considering the components of metric \eqref{etas-13}, we have
\begin{equation}\label{etas-25}
	ds^{2}=\left(\frac{r_{0}}{L}\right)^{2}\frac{f}{u^{2}}\left\{ -dt^{2}+\frac{L^{4}}{r_{0}^{2}}\frac{du^{2}}{f^{2}}\right\} +\ldots\ ,
\end{equation}
such that the tortoise coordinates are
\begin{align}
	\begin{aligned}\label{etas-26}
		u_{\star} & =-\frac{L^{2}}{r_{0}}\int_{0}^{u}\frac{d\overline{u}}{f}\ ,	\\
	& \sim\frac{1}{4\pi T}\ln\left(1-u\right)\ ,
	\end{aligned}
\end{align}
where $\sim$ indicates that the expression is valid near the horizon,
i.e. $u\to1$. Notice that in tortoise coordinates the horizon
is located at $u_{\star}\to-\infty$.

\par Using the coordinate $u_{\star}$, combined with the time dependence,
the solution near the horizon reads
\begin{equation}\label{etas-27}
	\phi\sim e^{-i\omega t}\left(1-u\right)^{\pm i\mathfrak{w}/4\pi}\approx e^{-i\omega\left(t\mp u_{\star}\right)}\ .
\end{equation}
These are plane waves, and they represent either incoming or
outgoing waves from the horizon. As we expect waves to only go in through the event horizon, we impose the incoming boundary
condition, which leads to the solution for $u\to1$
\begin{equation}\label{etas-28}
	\phi_{k}\propto\left(1-u\right)^{-i\mathfrak{w}/4\pi}\approx\left(1-u^{4}\right)^{-i\mathfrak{w}/4\pi}\ .
\end{equation}
We recall the expansion
\begin{equation}\label{etas-29}
	a^{x}\sim1+x\ln a+\frac{1}{2}x^{2}\left(\ln a\right)^{2}\ldots\ ,
\end{equation}
allowing us to write the solution as
\begin{equation}\label{etas-30}
	\phi_{k}\propto1-\frac{i\mathfrak{w}}{4\pi}\ln\left(1-u^{4}\right)\ldots\ .
\end{equation}
The solution \eqref{etas-28} is valid near the horizon $1-u\ll1$,
or $\ln\left(1-u\right)\gg1$. Whereas the expression above, the $\mathfrak{w}$
expansion, is valid when $\mathfrak{w}\ln\left(1-u\right)\ll1$. Both
conditions are satisfied for small enough values of $\mathfrak{w}$.

\par To solve the equation for all $u$ we make the expansion in a power
series for $\mathfrak{w}$. Suppose the solution has the form
\begin{equation}\label{etas-31}
	\phi_{k}=F_{0}\left(u\right)+\mathfrak{w}F_{1}\left(u\right)\ldots\ .
\end{equation}
Then the equation of motion, Eq. \eqref{etas-5}, reads
\begin{equation}\label{etas-32}
	\left(\frac{f}{u^{3}}F_{i}^{\prime}\right)^{\prime}=0\ ,
\end{equation}
for $i=0,1$. The solution is
\begin{equation}\label{etas-33}
	F_{i}=A_{i}+B_{i}\ln\left(1-u^{4}\right)\ ,
\end{equation}
where $A_{i},B_{i}$ are integration constants. Writing it explicitly
\begin{equation}\label{etas-34}
	\phi_{k}=\left(A_{0}+\mathfrak{w}A_{1}\right)+\left(B_{0}+\mathfrak{w}B_{1}\right)\ln\left(1-u^{4}\right)+\ldots\ .
\end{equation}
The boundary condition for $u\to0$ is $\phi_{k}\left(u\to0\right)=\phi_{k}^{\left(0\right)}$,
such that
\begin{equation}\label{etas-35}
	\left(A_{0}+\mathfrak{w}A_{1}\right)=\phi_{k}^{\left(0\right)}\ .
\end{equation}
At the horizon, we have, from the previous discussion, c.f. Eq. \eqref{etas-30} . Which gives immediately
\begin{equation}\label{etas-36}
	\left(B_{0}+\mathfrak{w}B_{1}\right)=-\frac{i\mathfrak{w}}{4\pi}\phi_{k}^{\left(0\right)}\ ,
\end{equation}
therefore
\begin{align}
	\begin{aligned}\label{etas-37}
		\phi_{k} & =\phi_{k}^{\left(0\right)}\left\{ 1-\frac{i\mathfrak{w}}{4\pi}\ln\left(1-u^{4}\right)+\mathcal{O}\left(\mathfrak{w}^{2}\right)\right\}\ , \\
	& \sim\phi_{k}^{\left(0\right)}\left\{ 1+\frac{1}{4\pi}i\mathfrak{w}u^{4}+\ldots\right\} \ ,
	\end{aligned}
\end{align}
where the first line of this equation is the solution for $0\leq u\leq1$,
whereas the second line is the asymptotic solution for $u\to0$. 

The factor multiplying $u^{4}$ in Eq. \eqref{etas-37} is precisely
$h_{xy}^{\left(1\right)}$, from our previous discussion. Therefore
we have determined the value for $\eta$, c.f. Eq. \eqref{etas-16}. Combining
this with the entropy
\begin{equation}\label{etas-38}
	s=\frac{1}{4G_{5}}\left(\frac{r_{0}}{L}\right)^{3}\ .
\end{equation}
Obtained in \eqref{eq:adscft-121} (with a slightly different notation), we arrive at the celebrated KSS relation
for the shear viscosity-to-entropy density ratio
\begin{equation}\label{etas-39}
	\frac{\eta}{s}=\frac{1}{4\pi}\ .
\end{equation}
Notice that this expression is written with constants $G=L=r_{0}=1$. 

We present the calculation in full detail in this section to highlight
the steps and assumptions used during the calculation, since a similar
method is used for calculations using the Black branes presented in
Ch. \ref{ch-2}, and therefore we can see exactly where the differences
come from and how the extra factors come about in the calculations.

%% file: chapters/thesis_result.tex
\newpage
\chapter{Results}
\par In this chapter, we summarize the main results of this thesis. These are published in \cite{Meert:2018qzk,Ferreira-Martins:2019wym,Ferreira-Martins:2019svk,Meert:2021khi}, and we dedicate one section to each of these works. Since we have already introduced the theoretical background, we only present the developments leading to the results and discuss them in detail.

\section{AdS$_4$-RN deformed black brane}
\par On Sect. \ref{adsrn-def} we have obtained a family of extensions to the AdS-Reissner-Nordstr\"om black brane. Recall the metric
\begin{equation} \label{r:1}
	ds^{2}=-r^{2}fdt^{2}+\frac{1}{r^{2}}gdr^{2}+r^{2}\left(dx^{2}+dy^{2}\right)\ ,
\end{equation}
where
\begin{align}
	\begin{aligned} \
	f\equiv f\left(r\right)&=1-\left(1+Q^{2}\right)\left(\frac{r_{0}}{r}\right)^{3}+Q^{2}\left(\frac{r_{0}}{r}\right)^{4}\ ,		\\
	g\equiv g\left(r\right)&=\frac{1}{f}\left\{ \frac{1-\frac{r_{0}}{r}}{1-\frac{r_{0}}{r}\left[1+\frac{1}{3}\left(\beta-1\right)\right]}\right\} \ .
	\end{aligned}
\end{align}
We investigate the shear viscosity-to-entropy density ratio associated to the dual field theory to this metric, and how the deformation parameter $\beta$ influences it, as well as use the KSS bound, i.e. $\eta/s \leq 1/4\pi$ \cite{Kovtun:2004de},  to impose a constraint on the value of the deformation parameter, thus restricting the family of deformed black branes.

\par Following the prescription introduced in Sect. \ref{eta-calc}. We introduce a perturbation in metric \eqref{r:1} as
\begin{equation}\label{r:2}
	g_{\mu \nu}^{(0)}\mathrm{d}x^\mu \mathrm{d}x^\nu = g_{\mu \nu} \mathrm{d}x^\mu \mathrm{d}x^\nu + 2h_{xy}^{(0)}(t) \mathrm{d}x \mathrm{d}y\ ,
\end{equation}
and find the equation of a massless scalar field. In this context the equation is slightly different, since $g_{tt}\neq g_{rr}$. One now must find $h_{xy}^{(1)}$, by solving the equation of motion for the perturbation $g^{xx}h_{xy} \equiv \varphi$, which is that of a massless scalar in a 4D background, 
\begin{equation} \label{r:13}
	\partial_\mu \left ( \sqrt{-g} g^{\mu \nu} \partial_\nu \varphi \right ) = 0\ .
\end{equation}
Taking $\varphi = \phi(u) e^{-i\omega t}$, the perturbation equation reduces to
\begin{equation} \label{r:14}
	\frac{u^2}{\sqrt{fn}} \left ( \frac{\sqrt{fn}}{u^2}\phi^{\prime}\right )^{\prime}  + \frac{1}{fn} \frac{\omega^2}{r_0^2} \phi  = 0 \ .
\end{equation}
Now, by imposing the incoming wave boundary condition near the horizon and the Dirichlet boundary condition at the AdS boundary, and proceeding as outlined in Sect. \ref{eta-calc} to incorporate these conditions\footnote{Namely: solving Eq. \eqref{r:14} in the limit $u \rightarrow 1$ and afterwards in a power series of $\omega$ up to $\mathcal{O}(\omega)$ in the hydrodynamic limit.} one arrives at the full solution 
\begin{equation} \label{r:15}
	\phi = \phi^{(0)} \left ( 1 - i \frac{\omega}{r_0} \frac{Q^2-3}{|Q^2-3|} \int \frac{u^2}{\sqrt{fn}}\, \mathrm{d} u \right ) \ .
\end{equation}

\par Accordingly, the full time-dependent perturbation is asymptotically given by
\begin{equation} \label{r:16}
	g^{xx}h_{xy} \sim e^{-i\omega t} \phi^{(0)}  \left ( 1 - i \frac{\omega}{r_0} \frac{Q^2-3}{|Q^2-3|}\frac{u^3}{3} \right )  \ .
\end{equation}
Comparing now Eq. \eqref{r:16} to Eq. \eqref{etas-37}, and identifying $h_{xy}^{(0)} =  \phi^{(0)} e^{-i \omega t}$, one promptly obtains 
\begin{equation} \label{r:17}
	h_{xy}^{(1)} = \frac{-i\omega}{3r_0}  \frac{Q^2-3}{|Q^2-3|} \ .
\end{equation}
The shear viscosity can be computed directly from \eqref{etas-16}, yielding
\begin{equation} \label{adsrndef-eta}
	i\omega\eta=\frac{3r_{0}^{3}}{16\pi G_{4}}h_{xy}^{\left(1\right)}\mapsto\eta=\frac{r_{0}^{2}}{16\pi G_{4}}\frac{3-Q^{2}}{\left|3-Q^{2}\right|}\ .
\end{equation}

\par The computation of entropy density follows from the area law. In Sect. \ref{adsrn-def} we have shown that
\begin{equation} \label{r:18}
	r_{\beta}=r_{0}\left[1+\frac{1}{3}\left(\beta-1\right)\right]
\end{equation}
is a horizon\footnote{ For the definition of Kiling vector field see footnote \ref{footnote}.
	A Killing horizon is defined as a surface $\mathcal{H}_{X}$, associated with the Killing vector field $X_{\mu}$ if and and only if $X_{\mu}$ is null on $\mathcal{H}_{X}$, nowhere zero on $\mathcal{H}_{X}$, and tangent to $\mathcal{H}_{X}$. By Hawking's rigidity theorem the event horizon of a stationary hole is a Killing horizon. }

 provided $\beta=1$, which is no more than the well known AdS-RN solution, or
\begin{equation} \label{r:19}
	\beta= -3+\frac{2\times2^{1/3}}{\left(-7-27Q^{2}+3\sqrt{3}\sqrt{3+14Q^{2}+27Q^{4}}\right)^{1/3}}+\frac{\left(7+27Q^{2}-3\sqrt{3}\sqrt{3+14Q^{2}+27Q^{4}}\right)^{1/3}}{2^{1/3}} \ .
\end{equation}
From demanding that the time-like Killing vector vanishes on the surface span by $r_{\beta}$. Therefore the entropy reads
\begin{equation} \label{r:20}
	s=\frac{r_{\beta}^{2}}{4}=\frac{r_{0}^{2}}{4}\left[1+\frac{1}{3}\left(\beta-1\right)\right]^{2}\ ,
\end{equation}
where the horizon radius is deformed by the $\beta$ parameter.

\par Finally, the ratio of shear viscosity-to-entropy density can be readily obtained from \eqref{adsrndef-eta} and \eqref{r:20}, it reads
\begin{equation} \label{eta-sRatio}
	\frac{\eta}{s}=\color{black}\frac{9}{\left(1-\frac{2\times2^{1/3}}{\chi_{Q}}+\frac{\chi_{Q}}{2^{1/3}}\right)^{2}} \left(\frac{3-Q^{2}}{\left|3-Q^{2}\right|}\right)\ ,
\end{equation}
for $\color{black}\chi_Q=\left(-7-27Q^{2}+3\sqrt{9+42Q^{2}+81Q^{4}}\right)^{1/3}$, where the $\beta$ parameter is written as given by Eq. \eqref{r:19}.

\par
\begin{figure} [h]
	\centering
	\includegraphics[scale=0.7]{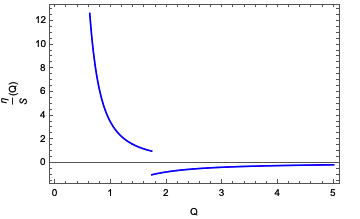}
	\caption[short text]{Plot of Eq. \eqref{eta-sRatio} normalized such that unity corresponds to $\frac{1}{4\pi}$. We see that there is a change of sign in the $\eta/s$ ratio for $Q=\sqrt{3}$.}	\label{eta-s-plot1}
\end{figure}

 The ratio $\eta/s$ is a positive quantity since both the shear viscosity and the entropy density are positive. From the plot in Fig.  \ref{eta-s-plot1}, representing Eq. \eqref{eta-sRatio}, one sees where this condition is met. 
 \vspace{-1,8cm}
Therefore, the following bound for $Q$ can be obtained
\begin{equation}
\begin{gathered}
0 < Q < \sqrt{3} \ , 
\end{gathered}
\end{equation}
for the $\eta/s$ ratio to assume the saturated value $\eta/s = 1/4\pi$ \cite{Kovtun:2004de}, as the initial action is Einstein--Hilbert. Notice that, as defined, the tidal charge $Q$ must be a positive quantity, so that the $-\sqrt{3} < Q < 0$ interval, which also satisfies the $3-Q^2 > 0$ bound, was not considered in this result, which is also very interesting\textcolor{black}{, and worthy of further investigation}.

\par It is worth emphasizing that the value of the $\beta$ deformation parameter, in Eq.\eqref{adsrn-13}, does correspond to the standard AdS$_4$--RN black brane, as expected. In fact by exploring some features of the deformation exposed in Sect. \ref{adsrn-def}, we found that the deformation parameter is restricted to the precise value $\beta=1$, taking us back to the conventional AdS$_4$-RN spacetime, which we see as an argument in favor of the unicity of this solution, whenever the ADM formalism is utilized.


\section{AdS$_5$-Schwarzschild}
\subsection{Thermodynamics and state functions}
Recall the metric obtained in Sect. \ref{adsschw-def}, Eqs. \eqref{eq:Fu},\eqref{eq:Gu}. Using the coordinate $u=r_0/r$
\begin{equation} \label{r52-1}
ds^2 = -\frac{r_0^2}{u^2} N(u) \mathrm{d}t^2 + \frac{1}{u^2 A(u)} \mathrm{d}u^2 + \frac{r_0^2}{u^2} \delta_{ij} \mathrm{d}x^i \mathrm{d}x^j,
\end{equation}
where
\begin{eqnarray}
N(u) &=& 1 - u^4 + \left (\beta - 1 \right ) u^6,\label{eq:Nu}\\
A(u) &=& \left (1 - u^4 \right ) \left ( \frac{2 - 3u^4}{2- \left (4\beta-1\right ) u^4}\right ). \label{eq:Au}
\end{eqnarray}
The constant $\beta$ parameter is referred to as a deformation parameter. We apply the formalism described in Sect. \ref{adscft-sec6} to calculate the thermodynamic quantities associated to the metric. Basically,  the following action must be evaluated
\begin{align}
	\begin{aligned}	\label{Sonshell1}
		S_{E}=-\frac{1}{16\pi G}\overbrace{\int \!d^{5}x\sqrt{g}\left(R\!-2\!\Lambda_5\right)}^{I_{\rm bulk}} - \frac{1}{8\pi G}\overbrace{\lim_{u\to 0}\int \!d^{4}x\sqrt{h}K}^{I_{GH}}\!+\!I_{\text{c.t}}\ ,	
	\end{aligned}
\end{align}				
where the first term is the Einstein--Hilbert action with the cosmological constant, the second term is the Gibbons--Hawking term, and the last is the counter term, which is introduced to ensure that the result is finite. In this case one uses the Euclidean signature, obtained by performing a Wick rotation in the time coordinate $t\mapsto i\tau$. This implies that $\tau$ is a periodic coordinate with period $2\pi$ \cite{Wald:1995yp}.

Each term will be individually computed, starting with the Einstein--Hilbert term. The cosmological constant is $-2\Lambda_5=12$, and the expansion on $u$ of the scalar curvature reads 
\begin{eqnarray} \label{scalCurv}
R=-20-8\left(\beta-1\right)u^{4}+\ldots\ ,
\end{eqnarray}
since the variable $u$ is defined from $0$ to $1$. For the metric determinant, the expansion on $u$ is given by 
\begin{equation}
	\sqrt{g}\approx\frac{r_0^4}{u^5}-\frac{\left(\beta-1\right)r_{0}^{4}}{u}+\frac{1}{2}(\beta-1)r_{0}^{4}u+\frac{\left(1-\beta\right)}{4}\left[6-\left(1-\beta\right)\right]r_{0}^{4}u^{3}.\label{281}
\end{equation}
Hence, the Einstein--Hilbert term becomes
\begin{equation}\label{Ibulk}
	I_{\text{bulk}}=\left[\left(\frac{1}{\epsilon^{4}}-1\right)-2(\beta-1)+\frac{1}{2}\left(\beta^{2}+2(\beta-1)^{2}+\beta-2\right)\right],
\end{equation}
where $\epsilon\to 0$ is used to keep track of divergent terms, which will be cancelled with the counter term.

\par \textcolor{black}{The Gibbons--Hawking term is a surface term. 
By considering the normal vector $n_{\alpha}=g_{uu}^{-1/2}\delta_{\alpha}^{u}$, the induced metric for a hypersurface at constant $u$ is given by $h_{\mu\nu}=g_{\mu\nu}-n_{\mu}n_{\nu}$, using $g_{\mu\nu}$ from \eqref{r52-1} we have
\begin{eqnarray}  \label{inducmetric}
	ds_{\text{HS}}^{2}=-\frac{r_{0}^{2}}{u^{2}}N(u)dt^{2}+\frac{r_{0}^{2}}{u^{2}}\delta_{ij}dx^{i}dx^{j}.
\end{eqnarray}
The computation of $K$ is straightforward, being its expansion near the boundary given by
\begin{eqnarray} 
	K=-4\left[1+\left(\beta-1\right)u^{4}+\ldots\right]\ ,
\end{eqnarray}
as well as for the metric determinant
\begin{eqnarray} 
	\sqrt{h}=r_{0}^{4}\left[\frac{1}{u^{4}}-\frac{1}{2}+\frac{u^{2}}{2}\left(\beta-1\right)-\frac{u^{4}}{8}+\cdots\right]\ . \label{hdet}
\end{eqnarray}
Then, it is just a matter of manipulating terms to find
\begin{eqnarray}
		I_{GH}=-4\left[\frac{1}{\epsilon^{4}}-\frac{1}{2}\left(3-2\beta\right)\right]\ ,
\end{eqnarray}}
where again, the divergent term is left explicit.

\par In dimension $d$, the counter term has a standard form and depends only on the geometry of the boundary theory, explicitly given by  \cite{Emparan:1999pm}
\begin{align}
	\begin{aligned}
		\!\!\!\!I_{\text{c.t}}=\frac{1}{8\pi G}\!\lim_{u\to 0}\int& d^{d}x\sqrt{h}\left\{ \left(d-1\right)+\frac{\mathfrak{R}}{2\left(d-2\right)}\right.\\ 
		&\left.\!\!\!\!\!\!+\frac{1}{2\left(d\!-\!4\right)\!\left(d\!-\!2\right)^{2}}\!\left[\mathfrak{R}_{\mu\nu}\mathfrak{R}^{\mu\nu}\!-\!\frac{d\,\mathfrak{R}^{2}}{4\!\left(d\!-\!1\right)}\right]+\ldots\right\} \ ,
	\end{aligned}
\end{align}
where $\mathfrak{R}$ and $\mathfrak{R}_{\mu\nu}$, respectively, refer to the scalar curvature and Ricci tensor of the induced metric \eqref{inducmetric}, (remembering that $\mu, \nu = 0,1,2,3$), and one can quickly check that these vanish. In dimension $d=4$, remembering that it is a surface term, it leads to the following, 
\begin{eqnarray} 
	I_{\text{c.t}}=\frac{3}{8\pi G}\lim_{u\to 0}\int d^{4}x\sqrt{h}\ .
\end{eqnarray} 
Eq. \eqref{hdet} yields 
\begin{eqnarray}
	I_{\text{c.t}}=\frac{3r_{0}^{4} V b}{{8\pi G}}\left[\frac{1}{\epsilon^{4}}-\frac{1}{2}\right],
\end{eqnarray}
where $V=\int dxdydz$ and $b=\int d\tau$. (Usually this is called $\beta$ in the literature, but to avoid confusion with the deformation parameter, we called it $b$.) Combining the integrals and restoring the constant factors yields 
\begin{equation} \label{partfnc}
	\!\!\!\!S_{E}=\frac{Vbr_{0}^{4}}{8\pi G}\left( \frac{11-15\beta+3\beta^{2}}{2}\right).
\end{equation}
Eq. \eqref{partfnc} is the partition function of the dual theory at the boundary, according to the GKPW relation. Now, from statistical mechanics, one knows that $Z=bF$, where $F$ is the free energy. Therefore we can calculate thermodynamic functions, by taking derivatives of $F$.

\par
\begin{figure}[hb!]
	\centering\includegraphics[width=9.6cm]{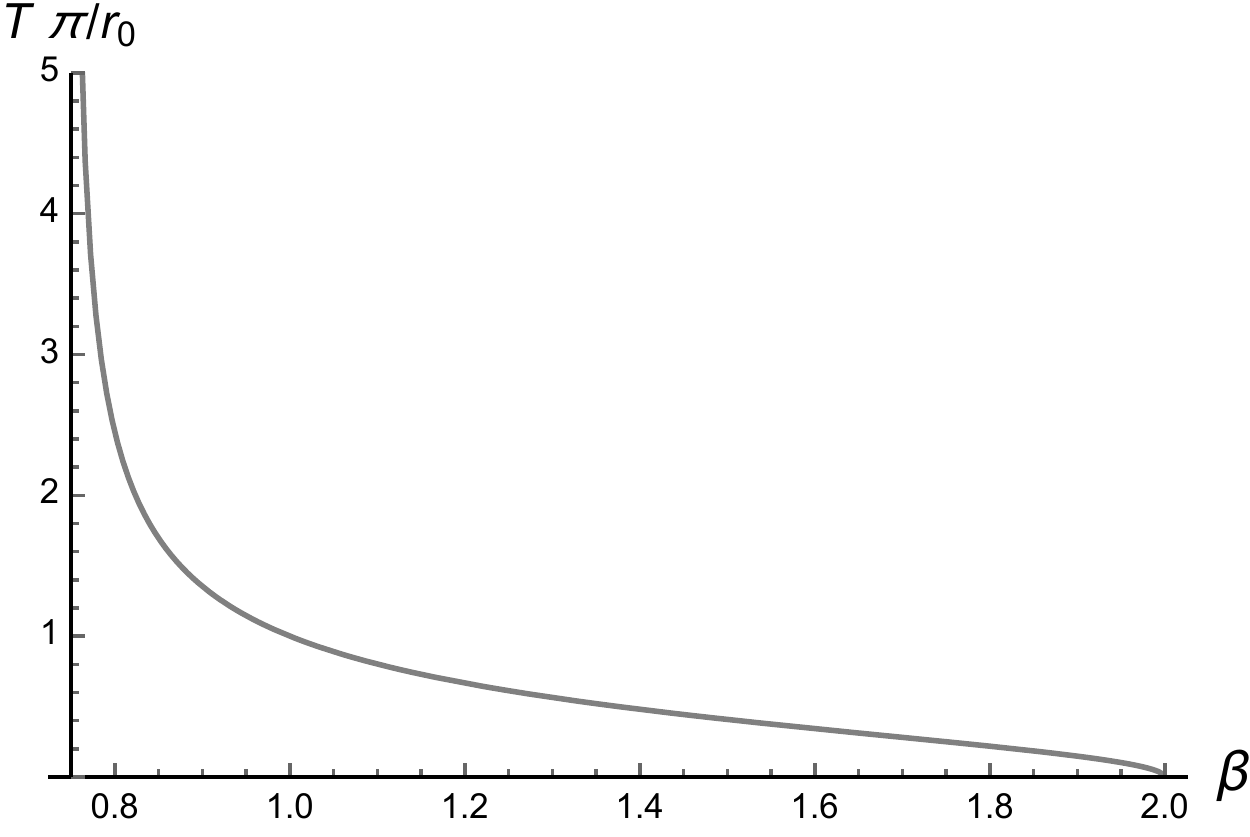}
	\caption{Temperature of the deformed black brane, as a function of $\beta$.}
	\label{fig:1}
\end{figure}

 Since we are going to compute thermodynamic functions, it is convenient to know the temperature. In the AdS/CFT context, the temperature is associated with the Hawking temperature at the horizon of the black hole \cite{Liu:2014dva}
\begin{eqnarray} \label{temperaturegeneral}
	T=\frac{1}{4\pi}\lim_{u\to1}\sqrt{\frac{\dot{g}_{tt}(u)}{\dot{g}_{rr}(u)}}.
\end{eqnarray}
For the metric \eqref{r52-1}, this expression is simply
\begin{eqnarray} \label{Tofr}
	T=\frac{r_{0}}{\pi}\sqrt{\frac{\beta-2}{3-4\beta}}\ .
\end{eqnarray}
It is important to mention that expression \eqref{Tofr} is obtained by approximating the metric coefficients near the horizon, i.e. $g_{tt}(u=1)\approx g^{(0)}_{tt}(u=1)+g^{(1)}_{tt}(u=1)(u-1)+\ldots$, and similarly for $g_{uu}$. Fig. \ref{fig:1} illustrates Eq. (\ref{Tofr}) as a function of $\beta$.
\noindent {The deformed black brane temperature diverges at $\beta\to 3/4$, having imaginary values for either $\beta < 3/4$ or $\beta > 2$. As the deformed black brane temperature cannot attain divergent values or imaginary ones, the analysis of the deformed black brane temperature constrains the $\beta$ parameter in the open range $\beta\in(3/4,2)$. }
One can invert Eq. (\ref{Tofr}) to express $r_0$ as
\begin{eqnarray} \label{rofT}
	r_{0}=\pi\sqrt{\frac{3-4\beta}{\beta-2}}T\ .
\end{eqnarray} 
Finally, the free energy can be read off, when Eq. \eqref{rofT} is replaced into \eqref{partfnc}, yielding 
\begin{align}
		\label{freenergyofT}
		\!\!\!\!\!\!\!F=\frac{\pi^{3}V}{8G}\left(\frac{11-15\beta+3\beta^{2}}{2}\right)\left(\frac{3-4\beta}{\beta-2}\right)^{2}T^{4}\ .
\end{align}

\par The state functions can now be computed using standard statistical mechanics in the canonical ensemble
\begin{eqnarray} \label{entropyofT}
		\!\!\!\!\!\!\!\!\!\!\!\!\!\!s&=&-\frac{1}{V}\frac{\partial F}{\partial T}\!=\!-\frac{\pi^{3}}{2G}\!\left(\frac{11\!-\!15\beta\!+\!3\beta^{2}}{2}\right)\!\!\left(\frac{3-4\beta}{\beta-2}\right)^{\!\!2}\!T^{3},\\
		\!\!\!\!\!\!\!\!P&\!=\!&\!-\frac{\partial F}{\partial V}\!=\!-\frac{\pi^{3}}{8G}\!\left(\frac{11\!-\!15\beta\!+\!3\beta^{2}}{2}\right)\left(\frac{3\!-\!4\beta}{\beta-2}\right)^{2}\!T^{4}\ ,\label{pofT}\\
		\!\!\!\!\!\!\varepsilon &\!=\!& \frac{F}{V}\!-\!Ts \!=\!\frac{5\pi^{3}}{8G}\!\left(\frac{11\!-\!15\beta\!+\!3\beta^{2}}{2}\right)\!\left(\frac{3-4\beta}{\beta-2}\right)^{2}\!T^{4}\label{espsofT}.
\end{eqnarray}
Despite the negative sign in front of entropy and pressure, these quantities are positive in the range of $\beta$ to be considered in the analysis to come in the next session.

For a perfect fluid, the energy-momentum tensor reads
	\begin{equation}
		T^{ab}=\left(\varepsilon+P\right)u^{a}u^{b}+Pg^{ab}.\label{emt}
	\end{equation}
	From Eqs. (\ref{pofT}, \ref{espsofT}), evaluated at the
	boundary,  the trace of the energy-momentum tensor (\ref{emt})  
	is given by 
\begin{equation}
 g_{\mu\nu}T^{\mu\nu}=-\varepsilon+3P=-\frac{\pi^{3}}{G}\left(\frac{11-15\beta+3\beta^{2}}{2}\right)\!\left(\frac{3-4\beta}{\beta-2}\right)^{\!\!2}T^{4}.\label{trEMT}
\end{equation}

\par For future reference, changing $T$ to $r_0$ using \eqref{rofT}, the entropy density in Eq. \eqref{entropyofT} can be written as, 
\begin{eqnarray} \label{sofr}
		s=-\frac{r_{0}^{3}}{2G}\left(\frac{11-15\beta+3\beta^{2}}{2}\right)\left(\frac{3-4\beta}{\beta-2}\right)^{1/2} \,.
\end{eqnarray}
As the entropy of a black hole obtained from Einstein's equations is proportional to its area, in the particular case of metric \eqref{r52-1} we have a deformation of a Schwarzschild black hole that is asymptotically AdS. This deformation breaks the spherical symmetry of our problem, and we have just used the AdS/CFT correspondence to compute the surface area of the black hole, i.e. $A=4GsV$.

\subsection{Calculation of $\eta$ and KSS bound}
\par As metric \eqref{r52-1} arises from a deformation of the AdS$_5$--Schwarzschild \cite{Casadio:2001jg}, the same action-dependent results may be applied.  The metric determinant,  $g$, is such that $
\sqrt{-g} = \frac{r_0^4}{u^5} \sqrt{\frac{N}{A}}$, where, from now on, $N$ and $A$  refer respectively to $N(u)$ and $A(u)$.

Consider a bulk perturbation $h_{xy}$ such that:
\begin{equation}\label{ref12}
	ds^2 = ds^2_{{\rm AdS}_5-{SD}} + 2h_{xy} \mathrm{d}x \mathrm{d}y \ ,
\end{equation}
\noindent where $ds^2_{{\rm AdS}_5-{SD}}$ denotes the AdS$_5$--Schwarzschild deformed  black brane metric, Eq. \eqref{r52-1}. We can quickly show that the field associated to the perturbation propagates with the speed of light. In what follows we make $r_0=1$, and the metric \eqref{r52-1} is written in coordinates $\{t,r,x,y,z\}$, where $r=u^{-1}$ according to the present convention.

\par As we have already mentioned in Sect. \ref{eta-calc}, $h_{xy}$ can be considered as a field on its own, hence we define $\varphi=g^{xx}h_{xy}$. We now identify the action as $S\sim S_0+S_2$, where $S_0$ does not have any contribution from $\varphi$, i.e. it is the action as studied in the previous section, whereas $S_2$ contains contributions of $\varphi$ and its derivatives. Let
\begin{equation} \label{ap:pertbulk}
	\varphi=\int dk\Phi(r)e^{-i\omega t+ikr+iqz}\ ,
\end{equation}
where $dk=\frac{d\omega dq dk}{\left(2\pi\right)^{3}}$, so that $S_2\propto \int\mathcal{L}(\Phi,\Phi^{\prime},\Phi^{\prime \prime})$, the proportionality factor is discarded. The Lagrangian reads
\begin{align}
	\begin{aligned} \label{eqB2}
	\sqrt{\frac{A}{N}}\mathcal{L}&=\Phi^{2}\left[-6r^{5}+2rA+r^{2}\left(ikA+2A^{\prime}+\frac{2AN^{\prime}}{N}\right)+\frac{7}{2}r^{3}\left(-q^{2}-Ak^{2}+\frac{\omega^{2}}{N}\right)\right.\\&+\left.r^{3}\left(ikA^{\prime}+ikA\frac{N^{\prime}}{N}+\frac{AN^{\prime\prime}}{2N}+\frac{A^{\prime}N^{\prime}}{4N}-\frac{AN^{\prime2}}{4N^{2}}\right)\right]\\&+\Phi\Phi^{\prime}\left[r^{2}\left(8+7ikr\right)A+A^{\prime}+\frac{AN^{\prime}}{N}\right]+\frac{3}{2}\Phi^{\prime2}A+2A\Phi\Phi^{\prime\prime}\ .
	\end{aligned}
\end{align}
For an action dependent on a single field up to its second derivative one can show immediately that
\begin{equation}\label{eqB3}
	\delta S=\delta S_{bdy}+\int dr\delta\Phi\left[\left(\frac{\partial\mathcal{L}}{\partial\Phi^{\prime\prime}}\right)^{\prime\prime}-\left(\frac{\partial\mathcal{L}}{\partial\Phi^{\prime}}\right)^{\prime}+\frac{\partial\mathcal{L}}{\partial\Phi}\right]\ ,
\end{equation}
$\delta S_{bdy}$ are surface terms, while the factor inside the integral is the equation of motion.

\par The momentum vector is $k^{\mu}=(\omega,k,0,0,q)$. Evaluating the EOM from \eqref{eqB3} using Lagrangian \eqref{eqB2} we obtain, in the limit $k^{\mu}\to\infty$, the following
\begin{equation}
	k_{\mu}k^{\mu}=0\ ,
\end{equation}
i.e. the EOM for a light-ray. This shows that the graviton -- field associated with the perturbation \eqref{ref12} -- propagates with the speed of light.


\par For $h_{xy}^{(0)}$ being the perturbation added to the boundary theory, which is asymptotically related to $h_{xy}$, the bulk perturbation, by\footnote{We are now using the $u$ coordinate, instead of $r$.}
\begin{equation}
	g^{xx}h_{xy} \sim h_{xy}^{(0)} \left ( 1 + h_{xy}^{(1)} u^4 \right ) \ ,
	\label{eq:perturb_asym_def}
\end{equation}
according to Eq. \eqref{etas-37}. Notice that one can directly use the results for a massless scalar field, as  $g^{xx}h_{xy}$ obeys the EOM for a massless scalar field \cite{Son:2009zzc}. Besides, the deformed AdS$_5$--Schwarzschild  black brane has the same asymptotic behavior of the AdS$_5$--Schwarzschild black brane (namely, Eq. \eqref{etas-13}). One can identify $g^{xx}h_{xy}$ as the bulk field, $\varphi$, which plays the role of an external source of a boundary operator, in this case $\tau^{xy}$. Therefore, one can directly obtain the response $\delta \left \langle \tau^{xy} \right \rangle$, from Eq. \eqref{etas-15}, 
\begin{equation}
	\delta \left \langle \tau^{xy} \right \rangle = \frac{r_0^4}{16 \pi G}4 h_{xy}^{(1)}h_{xy}^{(0)} \ , 
	\label{eq:response_2}
\end{equation}
\noindent where it is now convenient to reintroduce the $1/16 \pi G$ factor. Comparing Eqs. \eqref{etas-11} and \eqref{eq:response_2} yields 
\begin{equation} \label{eq:eta_quase}
	i \omega \eta = \frac{r_0^4}{4\pi G} h_{xy}^{(1)} .
\end{equation}
Taking the ratio between Eq. \eqref{eq:eta_quase} and the entropy \eqref{sofr} we find
\begin{eqnarray}
		\!\!\!\!\!\!\!\!\!\!\!\!\!\!\frac{\eta}{s}=-\frac{r_{0}}{\pi}\left[\left(\frac{1}{11-15\beta+3\beta^{2}}\right)\left(\frac{\beta-2}{3-4\beta}\right)^{1/2}\right]\frac{h_{xy}^{\left(1\right)}}{i\omega},
		\label{eq:eta_s_geral}
\end{eqnarray}
where $h_{xy}^{(1)}$ is the solution of the EOM for the perturbation $g^{xx}h_{xy} \equiv \varphi$, which is that of a massless scalar field \cite{Son:2009zzc}
\begin{equation}
	\nabla_M \left( \sqrt{-g} g^{MN} \nabla_N \varphi \right) = 0\ .
\end{equation}
Considering a stationary perturbation, given by the form $\varphi(u,t) = \phi(u) e^{-i\omega t}$,  the perturbation equation reduces to a second-order ODE for $\phi(u)$,
\begin{equation}
	\ddot\phi + \frac12\left ( \frac{\dot{N}A}{2 N}+\frac{N\dot{A}}{A} - \frac{3}{u}\right ) \dot\phi + \frac{1}{NA} \frac{\omega^2}{r_0^2} \phi = 0 \ .
	\label{eq:pertu_edo2}
\end{equation}
To derive the solution of Eq. \eqref{eq:pertu_edo2}, two boundary conditions are imposed: the incoming wave boundary condition in the near-horizon region, corresponding to $u \rightarrow 1$, and a Dirichlet boundary condition at the AdS boundary, $\phi (u\rightarrow 0) = \phi^{(0)}$, where $h_{xy}^{(0)} =  \phi^{(0)} e^{-i \omega t}$. 

\par The incoming wave boundary condition near the horizon is obtained by solving Eq. \eqref{eq:pertu_edo2} in the limit $u \rightarrow 1$. After a straightforward computation, one finds the following
\begin{eqnarray}
	\phi \propto \exp \left (\pm i \frac{\omega}{r_0} \sqrt{\frac{4\beta -3}{\beta - 1}}\sqrt{1-u}\right).
\end{eqnarray}
This solution has a natural interpretation using tortoise coordinates, allowing one to identify it as a plane wave \cite{Natsuume:2014sfa}, c.f. Sect. \ref{eta-calc}. The positive exponent represents an outgoing wave, whereas the negative one describes the wave incoming to the horizon, which, according to the near-horizon boundary condition, allows us to fix
\begin{equation}
	\phi \approx \exp \left (- i \frac{\omega}{r_0} \sqrt{\frac{4\beta -3}{\beta - 1}}\sqrt{1-u}\right).
	\label{eq:sol_nh}
\end{equation}

\par Next we solve Eq. \eqref{eq:pertu_edo2} for all $u\in[0,1]$ as a power series in $\omega$. As we are interested in the hydrodynamic limit of this solution, i.e. $\omega \rightarrow 0$, it is sufficient to keep the series up to linear order:
\begin{equation}
	\phi(u) = \Phi_0(u) + \omega \Phi_1(u) \ .
	\label{eq:sol_omega_power}
\end{equation}
Since the second term in Eq. \eqref{eq:pertu_edo2} is of order $\omega^2$, it can be neglected. By direct integration, the solution reads
\begin{equation}
	\Phi_i = C_i + K_i \int \frac{u^3}{\sqrt{N(u)A(u)}}\, \mathrm{d} u\ , 
\end{equation}
for $C_i$ and $K_i$ the integration constants and $i=0,1$. Thus, according to Eq. \eqref{eq:sol_omega_power}, we have 
\begin{equation}
	\phi = \left (C_0 + \omega C_1 \right ) + \left ( K_0 + \omega K_1 \right ) \int \frac{u^3}{\sqrt{N(u)A(u)}}\, \mathrm{d} u \ .
	\label{eq:sol_omega_power_general}
\end{equation}
In order to impose the boundary conditions we expand the integral (\ref{eq:sol_omega_power_general}) around $u\rightarrow 0$ and $u\rightarrow 1$. It yields, up to leading order in the respective expansions,
\begin{eqnarray}
	\frac{u^3}{\sqrt{NA}}\, \mathrm{d} u  \!&=&\!
	\begin{cases} \frac{u^4}{4} \ ,&\;\;\;\text{for}\;\;u \rightarrow 0,\\
		\frac{3-4\beta}{\beta-1}\sqrt{\frac{\beta-1}{3-4\beta}}\sqrt{1-u} \ ,&\;\;\;\text{for}\;\;u \rightarrow 1.\end{cases}
\end{eqnarray}
The first pair of integration constants is fixed by the Dirichlet boundary condition
\begin{eqnarray}
	 \lim_{u\rightarrow0} \phi =\left (C_0 + \omega C_1 \right ) + \left ( K_0 + \omega K_1 \right ) \lim_{u\rightarrow0} \frac{u^4}{4} =  \phi^{(0)},
\end{eqnarray} 
implying that $\left (C_0 + \omega C_1 \right ) = \phi^{(0)}$. Near the horizon one has 
\begin{equation}
	\phi \approx \phi^{(0)} - \left ( K_0 + \omega K_1 \right ) \frac{(4\beta-3)}{\beta-1}\sqrt{\frac{\beta-1}{4\beta-3}}\sqrt{1-u}\ .
	\label{eq:sol_general_nh}
\end{equation}
Expanding Eq. \eqref{eq:sol_nh} up to $\mathcal{O}(\omega)$ yields
\begin{equation}
	\phi \propto 1 - i \frac{\omega}{r_0}\sqrt{\frac{4\beta-3}{\beta-1}}\sqrt{1-u}\ .
\end{equation}
It is straightforward to see that Eq. \eqref{eq:sol_general_nh} fixes the proportionality according to
\begin{equation}
	\phi \approx \phi^{(0)} - i \phi^{(0)}\frac{\omega}{r_0}\sqrt{\frac{4\beta-3}{\beta-1}}\sqrt{1-u}.
	\label{eq:sol_nh_power_omega}
\end{equation}
Comparison between Eqs.\eqref{eq:sol_general_nh} and \eqref{eq:sol_nh_power_omega} immediately fixes the second pair of integration constants:
\begin{eqnarray}
	\left ( K_0 + \omega K_1 \right ) = i \phi^{(0)}\frac{\omega}{r_0} \left (\frac{\beta -1}{4\beta-3} \right ) \frac{|4\beta - 3|}{|\beta-1|}\ .
\end{eqnarray}
\noindent Then the full solution reads 
\begin{equation}
	\phi = \phi^{(0)} \left ( 1 + i \frac{\omega}{r_0} \left (\frac{\beta -1}{4\beta-3} \right ) \frac{|4\beta - 3|}{|\beta-1|} \int \!\frac{u^3}{\sqrt{NA}}\, \mathrm{d} u \right ) \ . 
\end{equation}

\par Accordingly, the full time-dependent perturbation 
\begin{equation}
	\varphi = g^{xx}h_{xy} = \phi(u) e^{-i\omega t}\ ,
\end{equation}
\textcolor{black}{is asymptotically given by:}
\begin{equation}
	g^{xx}h_{xy} \sim  e^{-i\omega t}\phi^{(0)} \left ( 1 + i \frac{\omega}{r_0} \left (\frac{\beta -1}{4\beta-3} \right ) \frac{|4\beta - 3|}{|\beta-1|} \frac{u^4}{4}\right ) \ .
	\label{eq:perturb_asym_solution}
\end{equation}
Eqs. (\ref{eq:perturb_asym_def}, \ref{eq:perturb_asym_solution}) yield 
	\begin{equation}
		h_{xy}^{(1)} = \frac{i\omega}{4r_0} \left (\frac{\beta -1}{4\beta-3} \right ) \frac{|4\beta - 3|}{|\beta-1|} \ ,
		\label{eq:key}
	\end{equation}
where $h_{xy}^{(0)} =  \phi^{(0)} e^{-i \omega t}$. The term multiplying $\frac{i\omega}{4r_0}$ in Eq. \eqref{eq:key} can be visualized in the following plot:
\begin{figure}[h!]
	\centering\includegraphics[width=9.6cm]{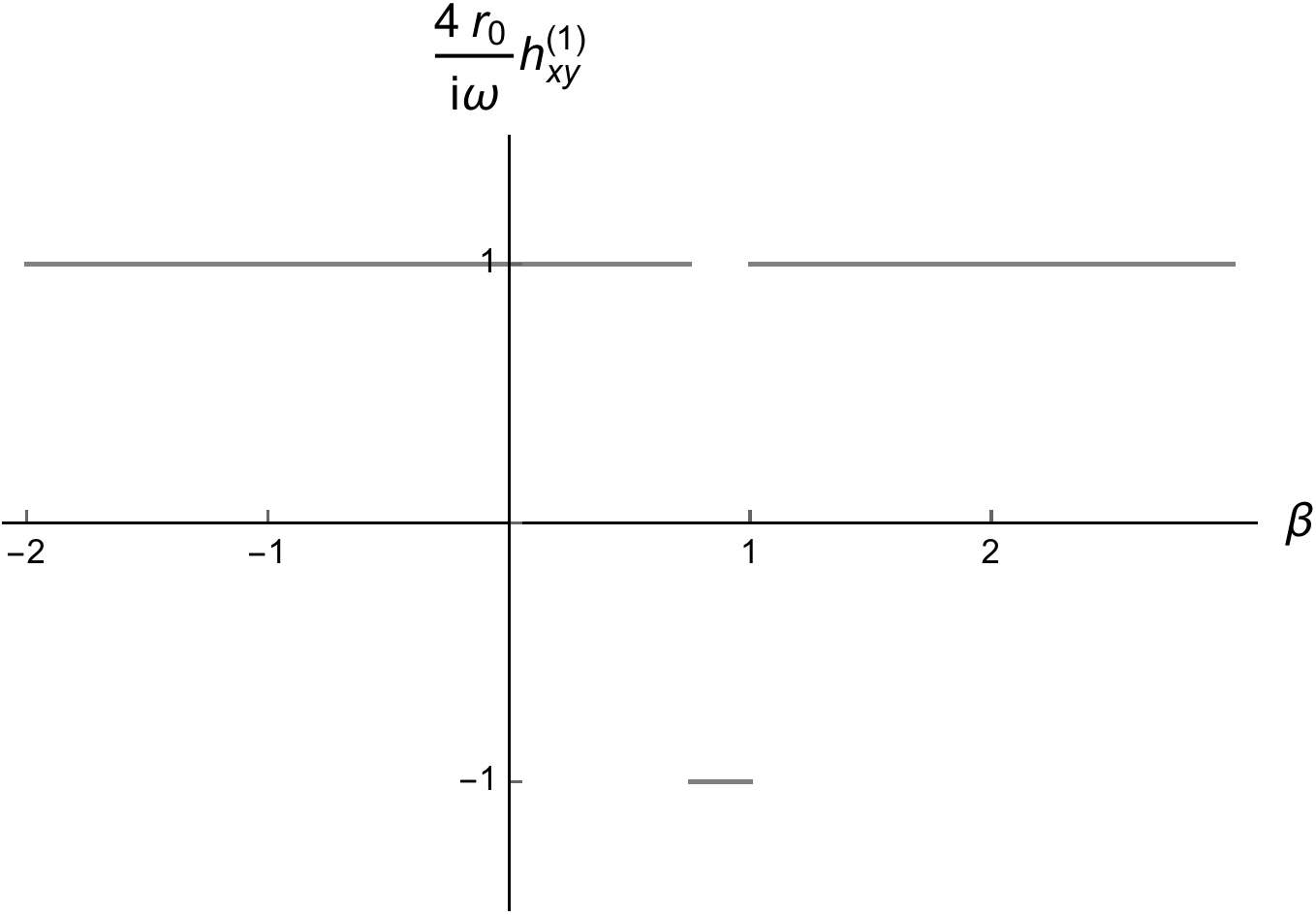}
	\caption{$4r_0 \frac{h_{xy}^{(1)}}{i\omega}$ as a function of $\beta$.}
	\label{fig}
\end{figure}
Therefore we have different signs depending on the value of $\beta$, 
\begin{equation}\label{ank}
	\begin{cases} 
		h_{xy}^{(1)}=-\frac{i\omega}{4r_{0}}\ , & \frac{3}{4}<\beta<1 \ , \\
		h_{xy}^{(1)}=\frac{i\omega}{4r_{0}}\ , & \beta<\frac{3}{4}\text{ or }\beta>1 \ . 
	\end{cases} 
\end{equation}
A negative value for $h_{xy}^{(1)}$, without further constraints, would imply a negative value of $\frac{\eta}{s}$, i.e., a negative viscosity or entropy density, which would violate the second law of thermodynamics. Therefore, demanding thermodynamic consistency leads to the following first bound in the deformation parameter: either $\beta < \frac{3}{4} \ \text{or} \ \beta > 1 $.

\par
\begin{figure}[h!]
	\centering\includegraphics[width=9.6cm]{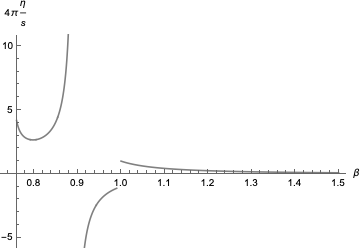}
	\caption{$\frac{\eta}{s}$ ratio of the deformed black brane, as a function of $\beta$.}
	\label{fi2}
\end{figure}

 Now, substituting \eqref{eq:key}  in Eq. \eqref{eq:eta_s_geral} yields
\begin{equation} \label{eq:etaSfinal}
		\frac{\eta}{s}=\begin{cases}
			-\frac{1}{4\pi}\left(\frac{1}{11-15\beta+3\beta^{2}}\right)\left(\frac{\beta-2}{3-4\beta}\right)^{1/2}\!\!, & \beta>1\\
			\frac{1}{4\pi}\left(\frac{1}{11-15\beta+3\beta^{2}}\right)\left(\frac{\beta-2}{3-4\beta}\right)^{1/2}\!\!, & \beta<1
		\end{cases} \ .
\end{equation}
Fig. \ref{fi2} illustrates Eq. (\ref{eq:etaSfinal}) as a function of $\beta$.
For the precise value $\beta =1$, the deformed black brane $\frac{\eta}{s}$ ratio is exactly $\frac{1}{4\pi}$, recovering the KSS result for the AdS$_5$--Schwarzschild   black brane. Besides, Fig. \ref{fi2} shows the divergence of $\frac{\eta}{s}$ for $\beta \approxeq 0.9$ as well as the vanishing of the $\frac{\eta}{s}$ ratio, for $\beta = 2$.

\par Therefore, \emph{a priori} the deformation parameter can attain the ranges 
		\begin{equation}\label{ra1}
			0.75< \beta < 0.9\;\;\;\text{and}\quad 1 < \beta \leq 2.
		\end{equation}
The value $\beta \leq 2$ is seen from \eqref{eq:etaSfinal}, since $\beta=2$ makes that quantity equal to zero, whereas the range  $0.9\leq\beta < 1$ imply $\frac{\eta}{s}<0$, which has no physical significance. The saturation $\frac{\eta}{s} = \frac{1}{4\pi}$, corresponding to  to the infinite 't Hooft coupling limit \cite{Cremonini:2011iq}, then implies $\beta = 1$. This result  has been expected, as this case recovers the AdS$_5$--Schwarzschild black brane (\ref{eq:2-14}). However, an additional consistence test must take into account Eq. (\ref{eq:Nu}), that defines the deformed AdS$_5$--Schwarzschild black brane event horizon. In fact, let us call by $u_\beta = 1/r_\beta$ the solution of the algebraic equation $N(u)=0$, in (\ref{eq:Nu}). The first consistence test must regard the choice of $\beta$ in such a way that it produces a real event horizon\footnote{Equivalently, that the algebraic equation $N(u)=0$, in (\ref{eq:Nu}) does not have only complex solutions.}. Therefore, this restricts more the possible range for $\beta$, from $1 < \beta \leq 2$ to $1 < \beta \leq 1.384$. A second consistence test involves the fact that the  $r_0=\lim_{\beta\to1} r_\beta$ horizon, corresponding to the standard AdS$_5$--Schwarzschild black brane event horizon,  is of Killing type. Along our previous calculations, the horizon is assumed to be at $r_0$. For it to be a good approximation in the proposed ranges of $\beta$, in such a way that $\left|r_0-r_\beta\right|\lesssim 10^{-2}$,
we must restrict a little more the allowed range to $1 < \beta \lesssim  1.2$, since for the another range $0.75< \beta < 0.9$ the condition $\left|r_0-r_\beta\right|\lesssim 10^{-2}$ already holds. Hence, the $\beta$ parameter is restricted into the ranges
\begin{equation}\label{ra2}
		0.75< \beta < 0.9\;\;\;\text{and}\quad 1 < \beta \lesssim 1.2.
\end{equation}

\par We now present a comparison between results obtained with metric \eqref{r52-1}, and the conventional AdS$_5-$Schwarzschild, which also gives us insight on the effect of the parameter $\beta$. Denoting $T_S$, $s_S$ and $\left(\frac{\eta}{s}\right)_{S}$ the temperature, entropy density and shear viscosity-to-entropy density ratio of the standard AdS$_5-$Schwarzschild spacetime, respectively, one can check that the corresponding {positive} quantities for fixed $\beta=1.05$ are

\begin{align}
	\begin{aligned} \label{compar1}
		\textcolor{black}{\!\!\!\!\!T=0.89T_{S}\ ,\qquad s=1.82s_{S}\ ,\qquad \frac{\eta}{s}=0.54\left(\frac{\eta}{s}\right)_{S}}\ .
	\end{aligned}
\end{align}
For instance, if $\beta=1.2$ one finds
\begin{align}
		\begin{aligned} \label{compar2}
			\!T=0.67T_{S}\ ,\qquad s=6.03s_{S}\ ,\qquad \frac{\eta}{s}=0.17\left(\frac{\eta}{s}\right)_{S}\ .
		\end{aligned}
\end{align}

\par Considering the results \eqref{compar1} and \eqref{compar2}, the effects of the deformation in the metric are clear,  changing thermodynamics and hydrodynamics by a numerical factor. In the range $1<\beta\leq1.2$, there is a violation of the KSS bound. One can speculate that the violation comes from the fact that the solution under investigation does not obey Einstein's equations of GR, since it was obtained via an embedding in a higher dimensional space-time, whose evolution is governed by an equation that has the Einstein's field equations as a certain limit, c.f. Eq. \eqref{adssch-hamconstraint}. Fig. \ref{fi2} illustrates that the range $0.75<\beta < 0.9$ is formally allowed, wherein the deformation parameter makes the KSS bound not to be violated. The existence of a range where the KSS bound is violated, namely $1<\beta\lesssim 1.2$, but no pathologies in the causality of space-time or thermodynamic functions can be seen, is also one of the main results of this work. We emphasize that it is a free constant parameter, generating a family of deformed AdS$_5$--Schwarzschild black branes, which has been constrained for different reasons. We have imposed compliance with the second law of thermodynamics, thus discarding the ranges which would yield negative values of $\frac{\eta}{s}$. Therefore, the family of solutions obtained with the allowed values of $\beta$ can be an interesting result worthy of further investigation, mainly in the AdS/QCD correspondence. The embedding bulk scenario and ADM procedure, in which the deformed AdS${}_5$--Schwarzschild black brane was obtained, provides one more counterexample setup to the KSS bound conjecture. Besides, these results can play a relevant role in the QGP, whose measured viscosity is close to the KSS bound, possibly violating the bound \cite{Cherman:2007fj}. In the next section, we also address a possible scenario that corroborates to the violation of the KSS bound in the range $1<\beta\lesssim 1.2$.

\subsection{Scrutinizing the $\beta$ parameter} \label{beta-sch}
This section is devoted to clarify aspects of the $\beta$ parameter.
If one considers AdS/CFT in the braneworld, it relates the electric part of the Weyl tensor $\Psi_{\mu\nu}$ in Eq. \eqref{eq:1.3.2-1}, that represents (classical) gravitational waves in the bulk, to the expectation value $\langle T_{\mu\nu}\rangle$ of the (renormalized) energy-momentum tensor of conformal fields on the brane\footnote{The large $N$ limit expansion of the CFT requires $N\sim 1/(\sigma\ell_p)^2\gg1$. In the original Randall--Sundrum braneworld models, the Planck length, $\ell_p$ (for $8\pi G_4 = \ell_p^2$, where $G_4$ is the 4D Newton constant), is related to the 5D fundamental gravitational length $\ell_5$ by $\ell^2_p = \sigma \ell^3_5$ \cite{Randall:1999vf,ssm2}, where $\sigma$ is the brane tension.} \cite{Shiromizu:2001jm,Kanno:2002iaa}. Besides, the presence of the brane introduces a normalizable 4D graviton and an ultraviolet (UV) cut-off in the CFT, proportional to $\sigma^{-1}$. The general-relativistic limit requires $\sigma\to\infty$, corresponding to a geometric rigid  brane with infinite tension.  In the AdS/CFT setup, 
$\Psi^\nu_{\;\mu}\sim\ell_p^2 \langle T_{\;\mu}^{\nu}\rangle$. 
Since the electric part of the Weyl tensor is traceless, such a correspondence would imply that $\langle T_{\;\mu}^{\mu}\rangle\equiv \langle T \rangle=0$. In other words, it would hold in the case where the conformal
symmetry is not anomalous.  Eq. (\ref{trEMT}) therefore 
indicates a conformal anomaly due to the quantum corrections induced by $\beta$.
Eq. (\ref{trEMT}) yields $ \langle T \rangle\neq0$ for any value of $\beta$ but $\beta=1$. It is in full compliance with the fact that if $ \langle T \rangle=0$, then the UV cut-off would be required to be much shorter than any physical length scale involved.
Besides, $  \langle T \rangle=0$ for any value of $\beta$ would also demand the absence of any intrinsic
4D length associated with the background, otherwise
the CFT is affected by that scale.
For the deformed AdS$_5$--Schwarzschild black brane, the horizon radius  $r_0$ is a natural length scale, and one, therefore, expects that only CFT modes with
wavelengths much shorter than $r_0$, that are much larger than $\sigma^{-1}$, can propagate freely.  Bulk perturbations at the boundary work as sources of the CFT fields, and can produce  $\langle   T \rangle=0$.

\par Of course, this requires that the UV cut-off be much shorter
than any physical length scale in the system.
For a black hole, the horizon  radius is a natural
length scale and one, therefore, expect that only CFT modes with
wavelengths much shorter than $r_{\rm h}$, that are still much larger than $\sigma^{-1}$, propagate
freely  \cite{Casadio:2003jc}.

\par Besides, for the deformed AdS$_5$--Schwarzschild black brane, one can emulate the holographic computation of the 
Weyl anomaly \cite{Henningson:1998gx}. In fact, denoting $a$ and $c$ central charges of the conformal gauge theory, according to Eq. (24) of Ref. \cite{Cremonini:2011iq}, 
\begin{equation}
	\label{wa}
\!\!\!\!\!\!\!\! \langle T^\mu_{\;\,\mu}\rangle_{\rm CFT}= \frac{c}{16\pi^2}\left(R_{\mu\nu\rho\sigma}R^{\mu\nu\rho\sigma}-2R_{\mu\nu}R^{\mu\nu}+\frac13 R^2\right)\nonumber\\
 -\frac{a}{16\pi^2}\left(R_{\mu\nu\rho\sigma}R^{\mu\nu\rho\sigma}\!-\!4R_{\mu\nu}R^{\mu\nu}\!+\!R^2\right),
\end{equation}
where the terms in parentheses are, respectively, the Euler density and the square of the Weyl curvature. 

\par It is worth mentioning the splitting of the allowed range of $\beta$ into $0.75<\beta<0.9$ and $1<\beta\lesssim 1.2$.
Firstly, considering the range $1<\beta\lesssim 1.2$, Ref.  \cite{Kats:2007mq} studied an effective 5D bulk gravity dual and showed that  the KSS bound is violated,  
whenever the central charges in the Weyl anomaly (\ref{wa}) satisfy $|c - a|/c \ll 1$. In this way, the inequality
$c>a$ yields the KSS bound to be violated \cite{Buchel:2008vz,viol1}. Ref. \cite{Kats:2007mq} showed that, as an effect of curvature squared corrections in the AdS bulk, the 
shear viscosity-to-entropy density ratio can be expressed as $\frac{\eta}{s}= \frac{1}{4\pi}\frac{a}{c}+\mathcal{O}(1/N^2)$.
Therefore, in the large $N$ limit, the equality $\frac{\eta}{s}\approxeq \frac{1}{4\pi}\frac{a}{c}$ holds, and the central charges ratio drive the KSS bound violation, whenever $c\neq a$. In fact, the well-known $\mathcal{N}=4$, SU($N$)  super-Yang--Mills theory implies $a = c$, however nothing precludes that $c\neq a$ in other cases \cite{Kats:2007mq}.

\par Secondly, now considering the allowed range $0.75<\beta<0.9$, the deformed AdS$_5$--Schwarzschild black brane, on the boundary $u\to0$, 
the square of the Weyl curvature can be expanded as
\begin{equation}
	\label{430}
\!\!\!\!\!\!\!N^2\left(\frac{40}{3}+\frac{32}{3} (\beta -1) u^4+8 (\beta -1) u^6\right)+\mathcal{O}\left(u^7\right),
\end{equation}
and  the Euler density as
\begin{equation}
	\label{120}
\!\!\!\!\!\!\!N^2\left(120+96 (\beta -1) u^4+72 (\beta -1) u^6\right)+\mathcal{O}\left(u^7\right),
\end{equation}
 where $N^2 = \pi L^3/2G$.
One notices in Eqs. (\ref{430}, \ref{120}) that the leading-order terms contain
factors $(\beta-1)u^p$, for $p=4,6$. Therefore, the limits $\beta\to 1$, corresponding to the standard AdS$_5$--Schwarzschild black brane, and the boundary $u\to0$ limit, are indistinguishable. Hence, the limit $u\to0$ yields 
\begin{equation}
\langle T^\mu_{\;\,\mu}\rangle_{\rm CFT}=\frac{520 N^2}9\ ,
\end{equation}
having the same result of the standard AdS$_5$--Schwarzschild black brane.

\par It is worth to compare an already known result about $\frac{\eta}{s}$ in presence of quantum corrections. In fact, Ref. \cite{Myers:2008yi} discusses quantum corrections to the $\frac{\eta}{s}$ ratio, by including higher derivative terms with the 5-form RR flux to the calculation. Corrections are implemented as inverse powers of the colour number $N$, and the leading $1/N^2$ correction adds two corrections terms to entropy density, $s$, modifying $\frac{\eta}{s}$ in QCD strongly coupled QGP. Its original value, $\frac1{4\pi}$, is increased by approximately 37\%, roughly 22\% due to the first correction term and 15\% due to the second. As discussed in this section, our setup yields corrections that can be interpreted as quantum ones, induced by $\beta$, as expressed in Eq. (\ref{trEMT}). For $\beta = 0.75$, consisting of a lower bound for $\beta$, the $\frac{\eta}{s}$ ratio increases $\sim4.1$ times the original $\frac{\eta}{s}=\frac1{4\pi}$ value. In the range $0.75<\beta < 0.9$, there is a minimum at $\beta\approx 0.8$, for which the shear viscosity-to-entropy ratio equals $2.5$ the KSS bound.	In the range $1< \beta \leq 1.2$, we showed that the KSS bound is violated. For example, as analyzed in Eq. (\ref{compar1}, \ref{compar2}), the value $\beta=1.05$ yields $\frac{\eta}{s}=0.54\left(\frac{\eta}{s}\right)_{S}$, whereas taking $\beta=1.2$ implies that $\frac{\eta}{s}=0.17\left(\frac{\eta}{s}\right)_{S}$.

\section{Transport coefficients associated with fermionic sector}
\par To lead the fluid/gravity correspondence -- essentially based on bosonic fields -- further, one aims to include fermionic modes into the description. To accomplish so, one refers to supersymmetry in the bulk and analyzes its effect in the boundary, describing supersymmetric hydrodynamics \cite{Kovtun:2003vj}. This setup indeed leads to predictions \cite{SSLee:2009,HLiu:2011,MCubrovic:2009,TFaulkner:2011,Iqbal:2009} and the quest which concerns us in this work is related to the problem of whether a quantity similar to the shear viscosity-to-entropy density ratio, associated to fermionic sectors, exists. In Ref. \cite{Policastro:2008cx} the sound diffusion constant was first calculated in a supersymmetric holographic background and indicated that this quantity is the obvious candidate for the task, which was investigated and asserted later by \cite{Erdmenger:2013thg}.

\par The sound diffusion constant is related to the supercurrent, which turns out to be the superpartner of the energy-momentum tensor. When one considers a field theory with $T>0$, supersymmetry is spontaneously broken and the emergence of a collective fermionic excitation called phonino arises in very general circumstances \cite{Lebedev:1989rz}. The sound diffusion is associated with the damping of this mode, i. e., the imaginary part of the dispersion relation. It was computed analytically for $\mathcal{N}=4$ supersymmetry. In the holographic setting, the EOM for the gravitino in AdS$_5$ background was solved to first order in the frequency and momentum using the retarded green function of the dual supersymmetric current, from where the dispersion relation was read off.

\par Classical spinor fields can be characterized by their properties with respect to the groups of symmetry characterizing rotations, in  Euclidean spacetimes,  or pseudo-rotations, regarding pseudo-Euclidean spacetimes. In a 4-dimensional Minkowski spacetime, spinor fields enclose two irreducible unitary representations of the Lorentz group,  comprising the Weyl and Majorana spinors, besides a unitary and reducible one, consisting of the Dirac spinors. Employing these spinor fields, one can construct bilinear covariants, which thoroughly designate the spinor fields and do satisfy the Pauli-Fierz-Kofink identities. On the other hand, if one reverses the protocol and assumes that the properties that define spinors consist of the Fierz identities, new unexpected and surprising possibilities do emerge. Fierz identities were used by Lounesto to classify spinor fields in Minkowski spacetime, regarding bilinear covariants, splitting the set of spinor fields into six disjoint classes, that embrace spinor fields in 4-dimensional Minkowski spacetime.  The first three spinor fields classes consist of regular spinor fields, as the scalar and pseudoscalar bilinear covariants differ from zero. The other additional three classes of singular spinors are called flag-dipole, flagpole, and dipole spinor fields. Notwithstanding these additional spinor field classes do include Weyl and Majorana spinor fields as particular cases of dipole and flagpole spinor fields,  respectively, these new classes further encompass new spinor fields with peculiar dynamics governing them. For example, other flagpole spinor fields in these classes are eigenspinors of the charge conjugation operator with dual helicity and may be prime candidates for dark matter \cite{daRocha:2011yr,Bernardini:2012sc,daRocha:2016bil,Cavalcanti:2015nna,daRocha:2007pz}. Also, flag-dipole spinor fields were found to satisfy the Dirac equation in f(R)-gravity \cite{daRocha:2013qhu}. A complete survey of this classification with relevant applications in quantum field theory and gravitation as well can be found in Ref. \cite{HoffdaSilva:2012uke}. This matter has been further explored in the context of black hole thermodynamics, as a special type of flagpoles \cite{daRocha:2014dla}. Driven by new possibilities concerning  such prominent recently discovered new spinor fields, higher-dimensional analog formulations have been explored, comprising new fermionic solutions in supergravity, string theory, and AdS/CFT \cite{Bonora:2015ppa,deBrito:2016qzl,Bonora:2014dfa,Cremmer:1978km,Englert:1983qe,Duff:1986hr,Lopes:2018cvu,Yanes:2018krn,Dantas:2015mfi}, based on generalized Fierz identities.  in particular for what concerns compactifications of 11-dimensional supergravity. The case involving black hole backgrounds was studied in Ref. \cite{DaRocha:2020oju}.

\par The second-quantized setup involving this classification was constructed in Ref. \cite{Bonora:2017oyb}. In the last two decades, new physical prospects, beyond the usual Dirac, Majorana, and Weyl spinor fields, have been proposed, scrutinized, and applied to QFT and gravity. Spinor field classifications have been also extended beyond the usual U(1) symmetry of quantum electrodynamics, encompassing also other compact gauge groups, in particular,  the electroweak gauge theory, in Refs. \cite{Fabbri:2017lvu,Arcodia:2019flm,daRocha:2005ti,HoffDaSilva:2020uov,HoffdaSilva:2019xvd}.

\par Flagpoles and flag-dipoles are types of the so-called singular spinor fields in Lounesto's U(1) gauge  classification. Flagpoles encompass neutral Majorana, and Elko, spinor fields, as well as charged spinor fields satisfying specific Dirac equations \cite{daRocha:2005ti}. Flag-dipoles are very rare in the literature, being their first appearance in Ref. \cite{esk}. The emergence of flagpole and flag-dipole singular spinor fields in the context of fermionic sectors of fluid/gravity correspondence is here scrutinized, exploring the duality between the graviton, in a supergravity bulk setup, and the phonino, in the boundary supersymmetric hydrodynamics. These spinor fields emerge when generalized black branes are considered, whose particular case in the AdS--Schwarzschild black brane for a very particular choice of parameter. This parameter appearing in the generalized black branes shall be shown to drive the flipping that takes regular into  singular spinors fields, as solutions of the equations of motion for the gravitino.

\subsection{General bilinear covariants and spinor classes} \label{Lounesto}
\par A classical spinor field $\uppsi$ is an object of the irreducible representation space of the Spin group. In $1+3$ dimensions, the isomorphism 
Spin($1,3)\simeq$ SL(2,$\mathbb{C}$) means that a classical spinor field then carries the 
representations of the SL(2,$\mathbb{C}$) Lorentz group. The bilinear covariants components, defined at each point $x$ on a 4D spacetime, with cotangent basis $\{e^\mu\}$  read
\begin{subequations}
	\begin{eqnarray}
		\Upsigma(x) &=& \overline{\uppsi}(x)\uppsi(x)\,,\label{sigma}\\
		J_{\mu }(x) &=&\overline{\uppsi}(x)\upgamma _{\mu }\uppsi(x)\,,\label{J}\\
		S_{\mu \nu }(x) &=&i\overline{\uppsi}(x)\upgamma
		_{\mu
			\nu }\uppsi(x),\label{S}\\
		K_{\mu}(x) &=&i\overline{\uppsi}(x)\upgamma_{0123}\upgamma _{\mu }\uppsi(x)\,,\label{K}\\\Upomega(x)&=&-\overline{\uppsi}(x)\upgamma_{0123}\uppsi(x)\,,  \label{fierz}
\end{eqnarray}\end{subequations}
where $\overline\uppsi=\uppsi^\dagger\upgamma_0$ is the spinor conjugate, $\upgamma
_{\mu
	\nu }=\frac{i}{2}[\upgamma_\mu, \upgamma_\nu]$, 
$\upgamma_5=i\upgamma_{0123}=i\upgamma_0\upgamma_1\upgamma_2\upgamma_3$ is the volume element, and $\upgamma_{\mu }\upgamma _{\nu
}+\upgamma _{\nu }\upgamma_{\mu }=2\eta_{\mu \nu }\mathbb{I}_{4\times 4}$, being the $\upgamma_\mu$  the Dirac gamma matrices and the $\eta_{\mu\nu}$ are the Minkowski metric components.  The form fields ${\bf J}(x) = J_\mu(x)\,e^\mu$,   ${\bf K}(x) = K_\mu(x)\,e^\mu$, and ${\bf S}(x) = \tfrac12 S_{\mu\nu}(x)\,e^\mu\wedge e^\nu$ are defined, where $\alpha\wedge\beta$ denotes the exterior product of form fields $\alpha, \beta$. Exclusively in the Dirac electron theory, the 1-form ${\bf J}$ represents a U(1) conserved current density. More precisely, in natural units, the time component $J_0$ is well known to regard the charge density, and the spatial components  $J_i$ typifies the electric current density. 
The spatial components $S_{jk}$ represent the magnetic dipole moment density, whereas the mixed components, $S_{i0}$, denote the electric dipole moment density. The $K_\mu$ denotes the chiral current density, that is solely conserved  in the massless case. The  scalar $\Upsigma$, responsible for the mass term in a fermionic  Lagrangian, and the pseudoscalar $\Upomega$, that is capable to probe CP symmetries, can be composed as $\Upsigma^2+\Upomega^2$ to be interpreted as a probability density. These  interpretations hold for the more common and usual cases where, for instance, the spinor describes the electron in the Dirac theory. Further cases can borrow similar interpretations for at least some of the bilinear covariants if the Dirac equation is satisfied for the given spinor field \cite{CoronadoVillalobos:2015mns}. 

\par The U(1) classification of spinor fields is described by the following  classes \cite{Lounesto:2001zz},
\begin{subequations}
	\begin{align}
	1)\;\;&\,\; \Upsigma(x)\neq0\neq\Upomega(x),\;\;\;\mathbf{S}(x)\neq 0\neq \mathbf{K}(x),\label{tipo1}\\
	 2)\;\;&\,\; \Upsigma(x)\neq0,\;\;\;
	\Upomega(x) = 0,\;\;\;\mathbf{S}(x)\neq 0\neq\mathbf{K}(x),\label{tipo2}\\
	 3)\;\;&\,\;\Upsigma(x)= 0, \;\;\;\Upomega(x) \neq0,\;\;\;\mathbf{S}(x)\neq 0\neq\mathbf{K}(x),\label{tipo3}\\
	 4)\;\;&\,\;\Upsigma(x)=0=\Upomega(x),\;\;\;\mathbf{S}(x)\neq 0\neq\mathbf{K}(x),
	\label{tipo4}\\
	 5)\;\;&\,\;\Upsigma(x)=0=\Upomega(x),\;\;\;\mathbf{S}(x)\neq0,\;\;\;
	\mathbf{K}(x)=0,
	\label{tipo5}\\
	6)\;\;&\,\;\Upsigma(x)=0=\Upomega(x),\;\;\; \mathbf{S}(x)=0,
	\;\;\; \mathbf{K}(x)\neq0.\label{tipo6}
\end{align} 	\end{subequations}
When both the scalar and the pseudoscalar vanish, a spinor field is called singular, otherwise, it is said to be regular. The objects in Eqs. \eqref{J} and \eqref{K}, being 1-form fields,  are named poles. Since spinor fields in the class 4, \eqref{tipo4}, have non vanishing {\bf K} and {\bf J}, 
spinor fields in this class are called flag-dipoles, because  ${\bf S}\neq 0$ is a 2-form field, identified by a flag, according to Penrose.
Besides, spinor fields in class 5, \eqref{tipo5}, have a vanishing pole, ${\bf K} = 0$, a non null pole, ${\bf J}\neq0$, and a non null flag, ${\bf S}\neq 0$, being  flagpoles. Spinor fields in class 6, \eqref{tipo6}, present two poles, ${\bf J}\neq0$ and ${\bf K}\neq0$, and a null flag, ${\bf S}=0$, corresponding therefore to a flag-dipole.  Flag-dipole spinor fields were shown to be a legitimate solution of the Dirac field equation in a torsional setup \cite{esk,Fabbri:2010pk,Vignolo:2011qt}, whereas Elko \cite{Ahluwalia:2009rh} and Majorana uncharged spinor fields represent type-5 spinors \cite{daRocha:2005ti}, although 
a recent example of a charged flagpole spinor has been shown to be a solution of the Dirac equation. Additional spinor fields classes, upwards of the Lounesto's classification, 
complete all the formal possibilities, including ghost fields  \cite{CoronadoVillalobos:2015mns}. 
The standard, textbook, Dirac spinor field is an element of the set of regular spinors in class 1, \eqref{tipo1}. Besides, chiral spinor fields were shown to 
correspond to be elements of class 6, \eqref{tipo6}, of (dipole) spinor fields. Chiral spinor fields governed by the Weyl equation are Weyl spinor fields. However, class 6 of dipole spinor fields further  allocates  
mass dimension one spinors, whose dynamics, of course, are not ruled by the Weyl equation, as well as flagpole spinor fields in class 5, that are not neutral and satisfy the Dirac equation \cite{CoronadoVillalobos:2015mns,Cavalcanti:2014uta}. The spinor field class 5, still, is also composed of mass dimension one spinor fields \cite{Ahluwalia:2009rh,daRocha:2005ti,Fabbri:2010wsa}. 
The Lounesto's spinor field classification was also explored in the lattice approach to quantum gravity \cite{Ablamowicz:2014rpa}. Flipping between regular and singular spinor fields was scrutinized in Ref. \cite{Cavalcanti:2014uta}, for very special cases.

\par The classification of spinor fields, according to the bilinear covariants, must not be restricted to the U(1) gauge symmetry of quantum electrodynamics. In fact, a more general classification, based on the ${\rm SU}(2)\times {\rm U}(1)$ gauge symmetry, embraces multiplets and provides new fermionic possibilities in the electroweak setup  \cite{Fabbri:2017lvu}. 

\par Regular spinor fields satisfy the Fierz identities, 
\begin{subequations} 
	\begin{eqnarray}\label{fifi}
		\!\!\!\!\!\!\!\!\!\!\! \Upomega S_{\mu\nu}+\Upsigma \epsilon_{\mu\nu}^{\quad\rho\sigma}S_{\rho\sigma}&=&K_\mu{J}_\nu,\\
		\Upomega^{2}+\Upsigma^{2} &=&{J}^\mu J_\mu,\\{J}^\mu{K}_\mu 
		&=&0={K}_\mu K^\mu +J_\mu J^\mu,\label{3c}
	\end{eqnarray}
\end{subequations}
what does not hold, in general, for singular spinor fields. Notwithstanding, a multivector field,  constructed upon the bilinear covariants, 
\begin{equation}
	\mathfrak{Z}(x)=\Sigma(x)+\mathbf{J}(x)+i\mathbf{S}(x)-\upgamma_{5}\mathbf{K}(x)+\upgamma_{5}\Omega(x),
	\label{boom1}
\end{equation}
is a {Fierz aggregate} if the listed bilinear covariants obey the Fierz identities (\ref{fifi} -- \ref{3c}). Besides,  Fierz aggregates,  that are  self-adjoint under the Dirac conjugation, are named    {boomerangs} \cite{Lounesto:2001zz}.
For singular spinor fields, the Fierz identities (\ref{fifi}) are promoted to the generalized ones,  \begin{subequations}
\begin{align}
		\label{nilp}\mathfrak{Z}^{2}(x)  &=4\Sigma(x) \mathfrak{Z}(x),\\
	\mathfrak{Z}(x)\upgamma_{\mu}\mathfrak{Z}(x)&=4J_{\mu}(x)\mathfrak{Z}(x),\\
	\mathfrak{Z}(x)i\upgamma_{\mu\nu}\mathfrak{Z}(x)&=4S_{\mu\nu}(x)\mathfrak{Z}(x),\\
	\mathfrak{Z}(x)\upgamma_{5}\upgamma_{\mu}\mathfrak{Z}(x)  &=4K_{\mu}(x)\mathfrak{Z}(x),\\
	\mathfrak{Z}(x)\upgamma_{5}\mathfrak{Z}(x)&=-4i\Omega(x) \mathfrak{Z}(x),\label{nilp1}
\end{align}
	\end{subequations}
that are satisfied for all the spinor fields in the Lounesto's classification. Moreover, spinor fields may be reconstructed from 
bilinear covariants, resulting in a classification  of spinor fields that is mutual to the Lounesto's one \cite{Cavalcanti:2014wia,fabbri}. In fact, given a spinor field $\upxi$
satisfying $\overline\upxi\uppsi\neq0$, then $
\uppsi=\frac{1}{4a}e^{-i\upalpha}\mathfrak{Z}\upxi$,  where $4a^2={\overline\upxi\mathfrak{Z}\upxi}$ and
$e^{-i\upalpha}=\frac{1}{a}\overline\upxi\psi$ \cite{Lounesto:2001zz}.

\subsection{Black hole absorption cross-sections and fermionic sectors}
\par Hydrodynamics plays the role of an effective account of quantum field theories (QFTs) in the  long
wavelength regime~\cite{Bu:2014ena},  regulated by the local fluid variables that are near the equilibrium. 
Transport coefficients, encompassing viscosities and conductivities, drive perturbations propagation and  can be experimentally measured.
One of the most remarkable predictions of AdS/CFT and fluid/gravity correspondence
is the shear viscosity-to-entropy density ratio, which is universal for a large class
of isotropic, strongly coupled, plasmas~\cite{Bu:2014ena}.
The fact that the shear viscosity-to-entropy density ratio is universal occupies a featured role in gauge theories that are dual to
certain gravitational backgrounds~\cite{Policastro:2001yc,Kovtun:2003wp}. The universality demonstrations of the KSS result,  
$ \frac{\eta}{s}=\frac{1}{4\pi}$,  
illustrate how the shear viscosity, $\eta$,  of the hydrodynamic limit of the QFT, with energy momentum tensor $T_{\mu\nu}$, is related to the low-energy absorption cross section $\sigma(\omega=0)$ of a transverse bulk graviton $h_{12}$ by a black brane \cite{Buchel:2003tz,Kovtun:2004de,Son:2007vk}. Comparing the Kubo formula, 
\begin{equation}
	\eta = \lim_{\omega\rightarrow 0} \frac{1}{2\omega}\int  \left\langle \left[T_{12}(x),T_{12}(0)\right]\right\rangle\,e^{i\omega t}\,d^4 x\ ,
\end{equation}
to the low-energy absorption cross section,  $\sigma(\omega)=-\frac{2\kappa^2}{\omega}  \Im  G^R(\omega)$~\cite{Klebanov:1997kc,Gubser:1997yh}, where $\kappa^2 = 8\pi G$ denotes the gravitational coupling constant, yields
\begin{equation}
	\sigma(\omega)=\frac{\kappa^2}{\omega}\int   \left\langle \left[T_{12}(x),T_{12}(0)\right]\right\rangle\,e^{i\omega t}\,d^4 x.\label{scs}
\end{equation}
The entropy, $S=\frac{A}{4G}$, of a black brane is only dependent on its area at the horizon. Denoting by $s$ and $a$ the respective entropic and areal densities, one may employ the Klein--Gordon equation of a massless scalar, as an equation of motion for the low-energy absorption cross-section associated with $h_{12}$, yielding the horizon area density $\sigma(0)=a$ \cite{Erdmenger:2013thg}. 

\par Black branes in a 5D bulk have near-horizon geometry  
\begin{equation}\label{3bbb}
	ds^2=-N(r)dt^2 +A(r) dr^2+ \frac{r^2}{\ell^2} d\vec{x}_3^2\ ,
\end{equation} 
where ${\ell}$ denotes the AdS$_5$ radius, which shall be considered unity herein, for the sake of conciseness. 
At the strong coupling regime $g_s N_c \gg 1$, the branes considerably  
curve the background bulk, sourcing the  geometry of the generalized black brane \cite{Casadio:2016zhu},
\begin{subequations}
\begin{eqnarray}
		\label{cfm1}
	N(r) &=& r^2\left(1-\frac{r_+^4}{r^4}\right)
	,\\
	A^{-1}(r) &=& r^2\left[\left(1-\frac{r_+^4}{r^4}\right)\frac{\left(1-\frac{5r_+^4}{2\,r^4}\right)}{1-(4\beta+1)\frac{r_+^4}{2r^4}}\right],\label{cfm2}
\end{eqnarray}
\end{subequations}
where $r_+$ denotes the black brane horizon. 
These generalized 
black branes can be equivalently obtained in two ways. The first one consists of deforming,  under the ADM formalism, the AdS--Schwarzschild black brane \cite{Casadio:2016zhu}. The second manner to derive them is as an analytical black brane solution of quadratic Ricci gravity 
with Lee--Wick terms. 
Clearly $\lim_{\beta\to 1} A(r) = N^{-1}(r)$, corresponding to the well known AdS--Schwarzschild black brane. 

\par The cross-section $ \sigma(0)=a$ was  previously proved  for 5D black brane metrics of type \cite{Das:1996we} 
\begin{equation}
	ds^2=-N(r)dt^2 + B(r)\left(dr^2 + r^2 d\Omega_3^2\right)\,.\label{cros1}
\end{equation}
In order to use the solutions (\ref{cfm1}, \ref{cfm2}), 
one needs to transform the metric \eqref{cros1}
into the standard one
\begin{equation}
	ds^2=-N(r)dt^2   + A(r)dr^2+ r^2 d\Omega_3^2\ ,
\end{equation}
by introducing the variable $\mathring{r} = rB^{1/2}(r)$, yielding the expression
\begin{eqnarray}
	A(\mathring{r}) = \left(\frac{B(r)}{B(r)+\frac{r}{2}B'(r)}\right)^2\ .
\end{eqnarray}
Therefore, one obtains 
	\begin{eqnarray}\label{mtcal}
B(r)=\frac{\left(1\sqrt{\frac{2r^{8}-5r^{4}r_{+}^{4}}{\left(r^{4}-r_{+}^{4}\right)\left(r^{4}-(4\beta+1)r_{+}^{4}\right)}}-\sqrt{2}\sqrt{r^{2}-\frac{r_{+}^{4}}{r^{2}}}\right)}{r^{3}\sqrt{r^{2}-\frac{r_{+}^{4}}{2r^{2}}(\beta+1)}\sqrt{\frac{2r^{8}-5r^{4}r_{+}^{4}+1}{\left(r^{4}-r_{+}^{4}\right)\left(r^{4}-(4\beta+1)r_{+}^{4}\right)}}}e^{\left({\frac{2r^{2}}{r_{+}^{2}}F\left(\left.i\sinh^{-1}\left(r\sqrt{-\frac{1}{r_{+}^{2}}}\right)\right|-1\right)}\right)},
	\end{eqnarray}
where $F(\;\cdot\;|\;\cdot\;)$ denotes the elliptic  integral of the first kind. 
Ref.~\cite{Das:1996we} derived an analogue result for the low-energy absorption cross section of a massless,  minimally coupled, fermion by the black brane \eqref{cros1},
\begin{equation}
	\sigma_{\frac12}(0)=2\, (B(r_+))^{-3/2}\,a\,,\label{crossf}
\end{equation}
which is solely dependent on the generalized black brane event horizon \cite{Erdmenger:2013thg}. 

\par Transport coefficients in the usual fluid/gravity correspondence can be emulated by the fermionic sector of the theory. In fact,  the cross section can be  computed by the standard QFT rate of decaying particles,
$
\sigma_{\frac12}=({2\omega})^{-1}\int  |\mathcal{M}|^2\,d\Uppi\,,
$ 
where the measure $d\Uppi$ accounts for the momentum space of final state particles. One then considers a spinor field, describing an AdS$_{5}$ bulk fermion $\Uppsi$, whose kinetic part in the action reads $
\int  \overline{\Uppsi}\upgamma^A{D}_A\Uppsi\,\sqrt{-g}\, d^{5}x,$ 
up to a constant, where $A=0,\ldots,4$, $\upgamma^A$ is a set of gamma matrices, and the $D_A$ stand for the covariant derivative.  The notation $x^4=r$ shall be used to denote 
the AdS radial coordinate. 
Denoting by $\Uppsi_b$ the fermionic fields at the boundary, one considers their coupling to a spinorial  boundary operator, $S$, that accounts the transverse component of the supersymmetric current \cite{Antoniadis:1985az,Henningson:1998cd,Iqbal:2009fd}, by $
\int \left(\overline{S}P^-\Uppsi_b + \overline{\Uppsi}_bP^+S\right)\,d^4x $ \cite{Erdmenger:2013thg}. The chiral projector $P^\pm =\frac{1}{2}\left(1\pm\upgamma^{4}\right)$ is employed, where  $\upgamma^4$ denotes the gamma matrix corresponding to the radial AdS$_5$ coordinate.  Taking a fermion at rest in the boundary  implies that 
\begin{align}
	\sigma_{\frac12}(\omega)&=
	\frac{\kappa^2}{{\rm tr}_0}\text{ Tr}\left(-\upgamma^0 \upzeta\right), 
\end{align}
where $\upzeta =  \Im\int \left\langle \,P^-P^+ S(x) \overline{S}(0)  \,\right\rangle\,e^{i \omega t}\,d^4x$ and, hereon, we use the notation ${\rm tr}_0\equiv\text{ Tr}\left(-\upgamma^0 \upgamma^0\right)$. 

\par The cross-section can be, therefore, associated with the Kubo formul\ae \ for  coefficients of transport in the boundary  CFT \cite{Erdmenger:2013thg}. 
The supersymmetry current  $S^i$ is associated with the supercharge density, $\uprho = S^0$, by the so-called constitutive relation
\cite{Erdmenger:2013thg}
\begin{equation}
	S^j = -\frac{P}{\upepsilon} \upgamma^0\upgamma^j \uprho +\left( \frac{{\rm D}_\upsigma}{2} [\upgamma^{j},\upgamma^{i}] - {\rm D}_{\rm s}\delta^{ij}\right)\nabla_i \uprho\,,\label{eq1230}
\end{equation}
where $\upepsilon$ and $P$ are the energy density and pressure of the fluid, respectively. Ref. \cite{Hoyos:2012dh} interprets $\uprho$ as a sound-like excitation, the phonino. The quantities  ${\rm D}_{\rm s}$ and ${\rm D}_\upsigma$ play the role of transport coefficients that govern the phonino, that has a speed dissipation given by $v_{\rm s}=\frac{P}{\upepsilon}$~\cite{Lebedev:1989rz,Leigh:1995jw}.  
Eq. (\ref{eq1230}) can be rewritten, taking into account to the spin-representations of O(3), 
\begin{equation}
	\!\!\!\!   S^i \!=\!-\frac{P}{\upepsilon} \upgamma^0 \upgamma^i \uprho  \!-\!\left[ D_{\frac12} \upgamma^i\upgamma_j\!+\!D_{\frac32} \left(\delta^i_j\!-\!\frac{1}{3} \upgamma^i \upgamma_j\right)\right]\nabla^j \uprho\,,
\end{equation}
where $D_{\frac12} = \frac{1}{3} \left({\rm D}_{\rm s} -  {\rm D}_\upsigma\right)$ and $D_{\frac32} = {\rm D}_{\rm s} + \frac{1}{2} {\rm D}_\upsigma$, respectively, denote transport coefficients associated with the spin-$\frac12$ and the spin-$\frac32$  components of $\nabla^j \uprho$ \cite{Erdmenger:2013thg}. It emulates the splitting of the stress-energy tensor  into the shear,  $\eta$, and the bulk, $\zeta$, viscosities \cite{Policastro:2002tn,Herzog:2003ke}. In the conformal setup, it reads 
\begin{equation}
	{\rm D}_{\rm s} = \frac{2}{3} D_{\frac32} \quad\text{and}\quad D_{\frac12} =0\,.\label{constantesdifusao}
\end{equation}
Using the Kubo formul\ae\, yields \cite{Kontoudi:2012mu,Erdmenger:2013thg}
\begin{align}
	D_{\frac32}&=\frac{1}{\upepsilon\,{\rm tr}_0}\lim_{\omega,k\rightarrow 0}\text{ Tr}\left(-\upgamma^0 \upzeta_1\right)\,. \label{Kubo}
\end{align} One has $\upzeta_1=\Im\int \langle \,P^+ \d{S}^i(x) \overline{\d {S}}^i(0)P^- \,\rangle\,e^{i \omega t}\,d^4x$. The supersymmetric (transverse)  current  \begin{equation}
	\d {\rm S}^i\equiv\left(\delta^i_j - \frac{1}{3}\upgamma^i\upgamma_j\right)S^j,
\end{equation}
and the equation of conservation, $\partial_\mu S^\mu=0$, auxiliates to derive the correlator $\left\langle \uprho \overline{\uprho}\right\rangle$ as \cite{Erdmenger:2013thg}
\begin{equation}
	\Im\,\left(\frac{k_i}{k^\mu k_\mu}\,\left\langle \d {\rm S}^i \overline{\uprho}\right\rangle\right)=-\frac23D_{\frac32}\,\Re\, \left\langle \uprho \overline{\uprho}\right\rangle\,.
\end{equation}
Therefore the gravitino can be coupled to the boundary supersymmetric current. The AdS/CFT correspondence 
associates the gravitino absorption cross section with the dual operator Green's function \cite{Gubser:1997yh}. Similar to the way that one considers transverse metric perturbations $h_{12}$ for the shear viscosity-to-entropy ratio, one takes into account the gravitino modes that have spin-$\frac32$, namely, $\upupsilon_i=P_{ij}\Uppsi_j$, where the $P_{ij}$ operator   
is responsible to project onto the $\upgamma^i$ component a quantity transversal to $\vec{k}$. The transverse gravitino components then satisfy equations of motion associated with fermions. One may, hence,   associate the absorption cross section  to the Kubo formul\ae\,~\eqref{Kubo}, \begin{equation}
	D_{\frac32} = \frac{2}{ {\upepsilon}\,\kappa^2}\,\sigma_{\frac12}(0)\,,\label{kfsi}
\end{equation}
characterizing the fermionic version of the viscosity $
\eta = \frac{1}{2\kappa^2}\,\sigma(0)$.

\par The cross-section \eqref{crossf} 
can be further tuned, by considering a  regular fermion, \cite{Das:1996we,Erdmenger:2013thg} satisfying the Dirac equation $
\left(\upgamma^\mu\nabla_\mu  - m\right)\Uppsi =0\,.$ 
It is worth mentioning that there are examples of 
regular and singular spinor fields, allocated in at least five of the six Lounesto's classes, that satisfy the Dirac equation.
Ref. \cite{Erdmenger:2013thg} studied the Dirac equation   
in the AdS--Schwarzschild background geometry. Taking into account the  generalized black brane geometry \eqref{cros1}, the Dirac equation yields 
\begin{equation}
	\left[k\upgamma^4\left(d_r\!+\!\frac{3}{2r}\right)\!+\!\frac{k}{r}\upgamma^j\text{\d{$\nabla$}}_j-\!m\sqrt{N}-\!i\omega \upgamma^0\right] \upxi\!=\!0\ ,
\end{equation}
where $d_r\equiv\frac{d}{dr}$, $k(r)=\sqrt{N(r)/B(r)}$, and one suitably scales 
the fermionic field $\Uppsi$ by $\upxi =  \sqrt[4]{NB^{3}}\,\Uppsi$, 
for easily solving the subsequent equations. Using an eigenspinor basis  of the operators $\{\upgamma^4,\upgamma^0\}$, namely, $\upgamma^{4}\lambda^\pm_n=\pm \lambda^\pm_n$ and $\upgamma^{0}\lambda^\pm_n=\mp \lambda^\mp_n$, the  expansion 
$
\upxi =\sum_{n=0}^\infty F_n^\pm(r) \lambda_n^\pm, 
$
can be then employed \cite{Erdmenger:2013thg},  yielding  
\begin{align}
	k\left(d_r + \frac{f_{n}^\pm}{r}\right)F_n^\mp \pm m \sqrt{N} F_n^\mp &= i \omega F_n^\pm\,,  \label{mde}
\end{align}
denoting $f_{n}^+=n+3$, $f_{n}^-=-n$. 
Their $n=0$ fermionic mode  satisfies
\begin{equation}
	\!\!\!\!\! \left[k\left(d_r+\frac{4}{r}+m\sqrt{B}\right)k\left(d_r - m\sqrt{B}\right) + \omega^2\right] F_0 =0\,.
\end{equation} 
Defining {\it x}-coordinates implicitly by $\frac{d}{dx} = k(r) \rho(r)r^3 \frac{d}{dr}$ \cite{Erdmenger:2013thg}, such that $d_r \rho=2m\rho\sqrt{B}$, implies that $\lim_{r\rightarrow \infty}\rho=  1$. Therefore, the $n=0$ fermionic mode can be rescaled as  $\mathfrak{F}_0 = e^{-m\int \sqrt{B}\,dr}F_0$, yielding 
\begin{equation}
	\left(\partial_x^2+(\omega \rho r^{3})^2 \right)\mathfrak{F}_0=0\,.
\end{equation}
Hence, the absorption cross section (per areal density) for a  massive spin-$\frac12$ fermion reads \begin{equation}
	\frac{\sigma_{\frac12}(0)}{a} = (B(r_+))^{-3/2} \exp\left(2m\int^{r_+}_\infty \sqrt{B}\,dr\right)\,.
\end{equation}
This result is led to the standard one derived in the context of the AdS--Schwarzschild black brane, in Refs. \cite{Das:1996we,Erdmenger:2013thg}, implemented in the limit  $\beta\to1$, in Eqs. (\ref{cfm1}, \ref{cfm2}, \ref{mtcal}). 
The supersound diffusion constant ${\rm D}_{\rm s}$, for the generalized black branes~\eqref{3bbb}, Eqs. \eqref{constantesdifusao} and~\eqref{kfsi}, then yields \cite{Policastro:2008cx,Kontoudi:2012mu}
\begin{equation}
	2 \, \pi T {\rm D}_{\rm s} =\frac{4\sqrt{2}}{9}\,, \label{sgen}
\end{equation}
where $T$ is the black brane temperature.

\subsection{Supersound diffusion constant from the transverse gravitino}\label{transverse}
\par Hereon the supersound diffusion constant
shall be studied with respect to the generalized black brane (\ref{cfm1}, \ref{cfm2}), also emulating the results for the AdS--Schwarzschild black brane in Ref. \cite{Erdmenger:2013thg} and the ones in Ref. \cite{Policastro:2008cx} for very particular limits. 
The bulk action for the gravitino,  
\begin{equation}
	S=\int \overline{\Uppsi}_\mu \left(\Gamma^{\mu\nu\rho}D_\nu-m\Gamma^{\mu\rho}\right)\Uppsi_\rho\,\sqrt{-g}\ , d^{5}x,\label{bulk}
\end{equation} 
is employed, for $\mu,\nu,\rho=0,\ldots,4$,
The covariant derivative acts on spinors as $D_\mu = \partial_\mu + \frac{1}{4}\omega_\mu^{bc}\upgamma_{bc}$, where the $\omega_\mu^{bc}$ denotes the spin-connection, as usual. Employing the usual gauge condition $\Gamma^\mu \Uppsi_\mu=0$,  the Rarita--Schwinger equation reads  $
\left(\upgamma^\mu {D}_\mu+m\mathbb{I}_{4\times 4}\right) \Uppsi_\mu =0\,.$
Supposing a boundary plane wave dependence $e^{-i \omega t + i k x}$ in the gravitino wave function, the projetor $P_{ij}$  is again used, to select the  gravitino components, $\upupsilon_i =\Uppsi_i - \frac{1}{2}\upgamma_i \upgamma^j \Uppsi_j$ ($i,j\neq 1$). Hence,  the equations of motion, in the background~(\ref{3bbb}, \ref{cfm1}, \ref{cfm2}),  read 
\begin{equation}
	\!\! \!\!\frac{\upupsilon^\prime}{\upupsilon} + \frac{\upgamma^5}{\sqrt{A}}\left(\frac{i k}{r} \upgamma^1\!-\!\frac{i \omega}{\sqrt{N}}  \upgamma^0\!-\! m\right)   \!+\!\frac{N^\prime}{4 \sqrt{NA}} \!+\! \frac{3}{2r}  \!=\!0\,.\label{eom_eta}
\end{equation}
Now, one writes \cite{Erdmenger:2013thg}
\begin{align}
	\upupsilon = \upupsilon^{\upalpha+} \upalpha^++ \upupsilon^{\upalpha-} \upalpha^- +\upupsilon^{\vartheta+} \vartheta^++\upupsilon^{\vartheta-} \vartheta^-\,,\label{esplit}
\end{align}
where the basis of chiral ($\upalpha^+, \vartheta^+)$ and anti-chiral ($\upalpha^-, \vartheta^-)$ eigenspinors of the $i\upgamma^{1}\upgamma^2$ operator, respectively with $+1$ and $-1$ eigenvalues, concomitantly satisfying 
\begin{subequations}
\begin{eqnarray}	
	&\upgamma^0 \upalpha^\pm=\pm \upalpha^\mp\,,\quad\upgamma^0 \vartheta^\pm = \mp \vartheta^\mp\,,\label{espin}\\&\upgamma^1 \upalpha^\pm = \pm i \, \vartheta^\mp\,,\quad\upgamma^1 \vartheta^\pm = \pm i\, \upalpha^\mp\,.
\end{eqnarray}
\end{subequations}
The correlator arises out of Eq.~\eqref{eom_eta} \cite{Son:2002sd,Policastro:2002se}, looking at the singular component  of Eq.~\eqref{eom_eta} near the event horizon, where the solutions $\upupsilon^{\upalpha\pm}$ are demanded to be of type \cite{Erdmenger:2013thg} 
\begin{equation}\label{solucoes}
	\upupsilon^{\upalpha\pm}\propto \left(r-r_+\right)^{-\frac{1}{4}\left(1- i\omega/\pi T\right)}\upupsilon^{\upalpha}_{0}\ ,
\end{equation}\vspace*{-1.5cm}
\par The supersound diffusion constant can be obtained from the Kubo formul\ae\, \eqref{Kubo}, used in the regime $\omega,k\to0$
into~\eqref{eom_eta}, in the black brane background brane (\ref{cfm1}, \ref{cfm2}), yielding 
\begin{equation}
	\upupsilon^\pm =c^\pm \sqrt[8]{AN} \sqrt[3]{r^{-2}} \sqrt[4]{\sqrt{r}+r^2\sqrt{1-\frac{r^4}{r_+^4}}}^{\;\pm3}\,.\label{sol}
\end{equation} The phonino dispersion relation 
$
\omega = v_{\rm s} k - i {\rm D}_{\rm s} k^2
$  can be then obtained from the pole of the longitudinal supersymmetry current correlator. The Rarita-Schwinger  equations read 
\begin{subequations}
	\label{eom2}
	\begin{align}
		&\!\!\!\!\!\!\!\upgamma^5 \Uppsi_0^\prime \!+\!\left(\!\frac{N^\prime}{4 \sqrt{A}} \upgamma^5\!-\! \frac{i \omega}{\sqrt{A}} \upgamma^0   \!+\! \frac{i k}{r} \upgamma^1 \!+\! \frac{3\sqrt{N}}{2 r} \upgamma^5 \!+\!{m\mathbb{I}_{4\times 4}}\right)\!\frac{\Uppsi_0}{\sqrt{N}}- \frac{N^\prime}{2 \sqrt{AN}} \upgamma^0 \Uppsi_4 =0 \,,\\	
		&\!\!\!\!\!\!\!\upgamma^5 \Uppsi_4^\prime \!+\!\left[\left(\frac{N^\prime}{4 \sqrt{AN}}+\!\frac{5}{2r}\right) \upgamma^5 \!+\! \frac{ik}{r\sqrt{N}} \upgamma^1\! \!-\! \frac{i \omega}{\sqrt{AN}} \upgamma^0\!\left.\quad+\frac{m\mathbb{I}_{4\times 4}}{\sqrt{N}} \right] \Uppsi_4  \right.\nonumber\\&\hspace{7cm}-\left(\frac{N^\prime}{2 \sqrt{AN}} \upgamma^0+ \frac{1}{r}\upgamma^0\right) \Uppsi_0  =0\,,\\
		&\!\!\!\!\!\!\!\upgamma^5 \Uppsi_j^\prime +\left(\frac{N^\prime}{4 \sqrt{AN}} \upgamma^5- \frac{i \omega}{\sqrt{AN}} \upgamma^0+ \frac{i k {}}{r\sqrt{N}} \upgamma^1 \right.\left.+ \frac{3}{2r} \upgamma^5+ \frac{m\mathbb{I}_{4\times 4}}{\sqrt{N}}\right)\Uppsi_j + \frac{1}{r} \upgamma^j \Uppsi_4=0\,.
	\end{align}
\end{subequations}
The gauge condition $\slash\!\!\!\Uppsi_\mu=0$ is again used in the above equations, yielding \cite{Erdmenger:2013thg}
\begin{eqnarray}
	&&\left(\frac{N^\prime}{2 \sqrt{NA}} \upgamma^5 \!-\! \frac{2 i \omega}{\sqrt{NA}} \upgamma^0 \!+\! \frac{2 i k {}}{r \sqrt{N}} \upgamma^1 \!-\! \frac{2 m\mathbb{I}_{4\times 4}}{\sqrt{N}}\!+\!\frac{2}{r}\upgamma^5\right)\upgamma^5 \Uppsi_4 \nonumber\\&&\qquad\qquad + \frac{2 i k {}}{r \sqrt{N}} \Uppsi_1\!+\! \left[\left(\frac{N^\prime}{2 \sqrt{NA}} \!-\!\frac{1}{r}\upgamma^5\right) \upgamma^5 \!-\! \frac{2 i \omega}{\sqrt{NA}} \upgamma^0 \right] \Uppsi_0 \! =\! 0\,.
\label{ccee}
\end{eqnarray}
Taking into account the  hydrodynamical regime, the gravitino can be expanded to the first order, with respect to $\omega$ and $k$, 
\begin{equation}\label{grav11}
	\Uppsi_\mu = \uppsi_\mu + k\uptau_\mu+ \omega\upvarphi_\mu   \,.
\end{equation}
To the  lowest order terms,  $\omega=0=k$,  in Eq.~\eqref{eom2} and~\eqref{ccee}, it reads 
\begin{equation}
	\uppsi_4^\prime +\left(\frac{3N^\prime}{4 \sqrt{NA}}+ \frac{9}{2r} - \frac{m}{\sqrt{N}}\upgamma^5\right) \uppsi_4 =0\,.
\end{equation}
Analogously, 
\begin{align}
	&	\!\!\!\!\!\Uppsi_0^\prime \!+\!\left(\frac{N^\prime}{4 \sqrt{NA}}\!+\! \frac{3}{2r} \!+\! \frac{m}{\sqrt{N}} \upgamma^5\!\right) \Uppsi_0 \!=\! -\frac{N^\prime}{2 \sqrt{NA}}\upgamma^0 \upgamma^5 \uppsi_4\,,\\
	&	\uppsi_1^\prime +\left(\frac{N^\prime}{4 \sqrt{NA}} + \frac{3}{2r} + \frac{m}{\sqrt{N}} \upgamma^5 \right) \uppsi_1 = \frac{1}{r} \upgamma^1 \upgamma^5 \uppsi_4\,.
\end{align}
The equation regarding the $\uppsi_4$ component was integrated in Ref. \cite{Erdmenger:2013thg}, after splitting it similarly to~\eqref{esplit}. The other components were obtained and derived in Ref. \cite{Erdmenger:2013thg}, employing the solution for $\uppsi_4$ component. Hereupon the  parameters $a_k, b_k, c_k$ and $d_k$ stand for (integration) constants that arise from the integration of the parameters in Eq. (\ref{esplit}), with respect to the gravitino component $\uppsi_k$ \cite{Erdmenger:2013thg}. 
In the near-horizon regime, one finds 
\begin{align} \label{fl0}
	\uppsi_4=\frac{r_+^{-\frac14}}{2^{3/2} }\begin{pmatrix} \beta a_4 \\ \frac{\beta+1}{2}a_4 \\ \beta c_4 \\ -\frac{\beta+1}{2}c_4 \end{pmatrix}  (r-r_+)^{-\frac34}
	\overset{\beta \to1}{=}\frac{r_+^{-\frac14}}{2^{3/2}}\begin{pmatrix} a_4 \\ a_4 \\ c_4 \\ -c_4 \end{pmatrix}  (r-r_+)^{-\frac34}.
\end{align}
On the CFT boundary, ingoing conditions 
impose that the solutions approach $\left(r-r_+\right)^{-\frac{i \omega}{4 \pi T}}$ at the horizon. It implies that $a_4 = b_4$ and $c_4 = - d_4$, in the limit $\beta\to1$. Besides, \begin{align}
	\Uppsi_0&=\frac{r_+^{-1}}{(4r_+)^{3/4} }\begin{pmatrix} \frac{4\beta+1}{5}a_0 \\ \frac{4\beta+1}{5}a_0 \\ \beta c_0 \\ -\beta c_0 \end{pmatrix}  (r-r_+)^{-\frac34}
	\overset{\beta \to1}{=}\frac{r_+^{-\frac74}}{2^{3/2} }\begin{pmatrix} a_0 \\ a_0 \\ c_0 \\ -c_0 \end{pmatrix}  (r-r_+)^{-\frac34}\,,\label{fl3}\\
	\!\!\!\uppsi_1\!&=\!-\frac{3ir_+^{-\frac{17}{4}}}{2^{3/2}}\!\begin{pmatrix} 3\,{}\beta\, r_+^2 c_4\!-\!2r_+^4c \\ 3{}\beta r_+^2 c_4\!-\!2 r_+^4 c \\ -3{} r_+^2 a_4\!+\!2r_+^4 a \\ 3{} r_+^2 a_4\!-\!2r_+^4 a \end{pmatrix}  (r-r_+)^{-\frac14}\ 	\nonumber	\\ 
	&\overset{\beta \to1}{=}\!-\frac{3ir_+^{-\frac{17}{4}}}{2^{3/2}}\!\begin{pmatrix} 3r_+^2 c_4\!-\!2r_+^4c \\ 3 r_+^2 c_4\!-\!2 r_+^4 c \\ -3{} r_+^2 a_4\!+\!2r_+^4 a \\ 3{} r_+^2 a_4\!-\!2r_+^4 a \end{pmatrix}  (r-r_+)^{-\frac14}\,.\label{fl4}
\end{align}
Now, we can calculate, for each spinor field, their associated bilinear covariants (\ref{sigma} -- \ref{fierz}) and, subsequently, attribute the spinor field type in the Lounesto's classification (\ref{tipo1} -- \ref{tipo6}). The solutions \eqref{fl0}, \eqref{fl3}, \eqref{fl4} compose the gravitino field (\ref{grav11}). Eq. (\ref{fl0}) describes a regular spinor field that flips into a singular, flag-dipole, spinor field, in the $\beta\to1$ limit described by Eq. (\ref{fl0}). On the other hand, the solution (\ref{fl3}) is already a flag-dipole spinor field solution, for any value of $\beta$, according to the protocol in Ref. \cite{Cavalcanti:2014wia}. In particular, the AdS--Schwarzschild limit $\beta\to1$ leads to Eq. (\ref{fl3}). 
In the same way, the spinor field solution (\ref{fl3}) is also a flag-dipole solution, irrespectively of the value of $\beta$. 

\par The AdS boundary behavior reads  
\begin{align}
	\!\!\!(\Uppsi_0^{\upalpha-}\!\!, \Uppsi_0^{\vartheta-}\!\!, \uppsi_1^{\upalpha-}\!\!, \uppsi_1^{\vartheta-})^\intercal
	=\frac{2^{4/3}}{3}i  (0,0,c,a)^\intercal r^{-\frac12}\,.
	\label{erre}
\end{align}
Defining $	\Uppsi_{\upzeta} =2\upgamma^1 \Uppsi_1 -\upgamma^{3}\Uppsi_{3}- \upgamma^2 \Uppsi_2$, and looking at the equations of motion~\eqref{eom2}, one realizes  that it partly decouples from the other gravitino components:
\begin{eqnarray}
	\Uppsi_{\upzeta}^\prime+\frac{i\omega}{N}\upgamma^0\upgamma^5\Uppsi_{\upzeta}+\frac{N^\prime}{4 \sqrt{AN}} \Uppsi_{\upzeta} -\frac{i k {}}{r\sqrt{A}}\upgamma^5\left(6\Uppsi_1-\upgamma^1\Uppsi_{\upzeta}\right)+\frac{3}{2r} \Uppsi_{\upzeta}-\frac{m}{(NA)^{1/4}}\upgamma^5 \Uppsi_{\upzeta}=0\ .
\end{eqnarray}
The gauge condition $\slash\!\!\!\Uppsi_\mu=0$ can be then employed to solve the  $\Uppsi_1$ components, 
\begin{equation}
	\Uppsi_1 =\frac{1}{3}\left(\upgamma^1 \Uppsi_{\upzeta} -\upgamma^1\upgamma^0\Uppsi_0 -\upgamma^1 \upgamma^5 \Uppsi_4\right)\,.
	\label{fgh}
\end{equation}
The equations of motion for the $\upvarphi_\mu$ component reads \begin{align}
	&\upvarphi_4^\prime\! +\!\left(\frac{3 N^\prime}{4\sqrt{AN}}+\frac{6}{2r}-\frac{m}{\sqrt{N}} \upgamma^5 \right)\upvarphi_4 =\frac{i}{\sqrt{AN}} \upgamma^0\upgamma^5 \uppsi_4-\frac{2i}{\sqrt{AN}}\Uppsi_0\,,\\
	&\!\!\!\!\!\upvarphi_\zeta^\prime\!+\!\left(\frac{N^\prime}{4 \sqrt{NA}}  \!+\!\frac{3}{2r} \!-\!\frac{m}{\sqrt{N}}\upgamma^5\! \right)\!\upvarphi_\zeta\!=\!-\frac{i}{\sqrt{NA}}\upgamma^0\upgamma^5\uppsi_\zeta\,.
\end{align}
The anti-chiral boundary values are given by 
\begin{align}\label{flagp}
	\begin{pmatrix} \upvarphi_0^{\upalpha} \\ \upvarphi_0^{\vartheta} \\ \upvarphi_1^{\upalpha} \\ \upvarphi_1^{\vartheta}\end{pmatrix}
	=- 2^{-\frac34} \, r_+^{-3}\begin{pmatrix}  -i a_1 \\   i c_1\\  c_1\\  a_1  \end{pmatrix}  r^{-\frac12}\,.
\end{align}
The spinor field in Eq. (\ref{flagp}) is 
a flagpole, when $a_1$ and $c_1$ are real constants. 

\par Now, the $\uptau_\mu$ gravitino components are, analogously, given by 
\begin{align}
	\begin{pmatrix} \uptau_0^{\upalpha} \\ \uptau_0^{\vartheta} \\ \uptau_1^{\upalpha} \\ \uptau_1^{\vartheta}\end{pmatrix}
	={2^{1/4}r_+^{-3}}\begin{pmatrix}
		(1{}- 
		r_+^{-2}/3) \,c_\upzeta\\
		(1{} - r_+^{-2}/3)\, a_\upzeta\\
		-3i{}  a\\
		3i{}  c
	\end{pmatrix}  r^{-\frac12}\,.\label{fp1}
\end{align}
This is clearly a regular spinor field.
If $c=c_\upzeta$, $a=a_\upzeta$,  and $r_+=\sqrt{3}$, then Eq. (\ref{fp1}) is a flagpole singular spinor field, as explicitly 
verified to satisfy the conditions (\ref{tipo5}).
Besides, if either $a_\zeta=0=c_\zeta$ or 
$a=0=c$, then Eq. (\ref{fp1}) satisfies the conditions (\ref{tipo6}), hence regarding a dipole spinor field. 

\par Ref. \cite{Erdmenger:2013thg} shows that the components $\Uppsi_0, \Uppsi_1$ have anti-chiral  boundary values given by 
\begin{align}
	(\Uppsi_0^{\upalpha}, \Uppsi_0^{\vartheta}, \Uppsi_1^{\upalpha},  \Uppsi_1^{\vartheta})^\intercal
	=\mathfrak{B} (a_4,a,c_4,c)^\intercal r^{-\frac12}\,,
\end{align}
for a matrix $\mathfrak{B}$. Solutions can be evaluated at the poles of the boundary values, being therefore interpreted as  phonino modes \cite{Erdmenger:2013thg}. Computing the  $\det\mathfrak{B}$ and substituting the dispersion relation, 
$
\omega = v_{\rm s} k - i {\rm D}_{\rm s} k^2
$, and solving for $v_{\rm s}$ and ${\rm D}_{\rm s}$,  yields $
v_{\rm s}=\frac{1}{3}$ and $2 \pi T {\rm D}_{\rm s} = \frac{4\sqrt{2}}{9},$ 
being equal to Eq. (\ref{sgen}), obtained in another context. 

\subsection{Discussion and further analysis}
\par The occurrence of flag-dipole fermions 
in physics is very rare,  comprising  
features that approach regular spinor fields, although being singular ones, in  the Lounesto's classification (\ref{tipo1} -- \ref{tipo6}) according to the bilinear covariants. Besides the two previous 
examples in the literature, here flag-dipole solutions corresponding to the spinor part of the gravitino field  (\ref{grav11}) were obtained, together with flagpole spinor fields. They were derived 
as solutions of the bulk action for the gravitino field,  in the 
background of black brane solutions that generalize the AdS--Schwarzschild one.
Besides, the generalized black brane
has a free parameter driving the singular spinor field solutions, which can flip between regular and singular spinor fields.
The relation between hydrodynamic transport coefficients and the universal absorption cross-sections in the corresponding gravity dual was also studied in the fermionic sectors of the fluid/gravity correspondence. The Kubo formul\ae\, was employed to derive the transport coefficients for the gravitino and it is dual, the phonino. In the limit where the generalized black brane parameter tends to the unit, these results are in full compliance with the ones in Ref. \cite{Erdmenger:2013thg} for the AdS--Schwarzschild black brane. The supersound diffusion constants were also discussed through the solutions of the equations of motion for the gravitino. 
\section{Gravitational decoupling, hairy black holes and conformal anomalies} \label{anomalies}
\par Gravitational decoupling methods comprise established successful protocols used to generate analytical solutions of Einstein's effective field equations \cite{Ovalle:2017fgl,Ovalle:2019qyi,Ovalle:2020kpd,Casadio:2012rf,Ovalle:2017wqi,Antoniadis:1998ig,Cavalcanti:2016mbe,Casadio:2015gea}. 
The gravitational decoupling and some extensions were studied in Refs. \cite{covalle2,Ovalle:2014uwa,Ovalle:2016pwp,Casadio:2013uma,Ovalle:2013vna,Ovalle:2013xla,Ovalle:2018vmg,Casadio:2012pu,Casadio:2015jva,Casadio:2016aum,DaRocha:2019fjr,daRocha:2017lqj,Casadio:2017sze,daRocha:2020jdj,daRocha:2021sqd,Meert:2021khi,daRocha:2021aww,daRocha:2020gee,Ovalle:2018ans} and have been applied to kernel solutions of general relativity to construct new physically realistic solutions that describe stellar distributions, including anisotropic ones \cite{daRocha:2020rda,Fernandes-Silva:2017nec,Contreras:2018gzd,Ovalle:2007bn,Sharif:2018tiz,Sharif:2019mzv,Morales:2018urp,Rincon:2019jal,Hensh:2019rtb,Ovalle:2019lbs,Gabbanelli:2019txr,Tello-Ortiz2020,Gabbanelli:2018bhs,Panotopoulos:2018law,Heras:2018cpz,Contreras:2018vph,Tello-Ortiz:2020euy}. Refs. \cite{Fernandes-Silva:2019fez,daRocha:2017cxu,Fernandes-Silva:2018abr} derived accurate physical constraints on the parameters in gravitational decoupled solutions, using the WMAP, eLISA and LIGO.  
The gravitational decoupling procedure iteratively constructs, upon a given isotropic source of the gravitational field, anisotropic compact sources of gravity, that are weakly coupled. One starts with a perfect fluid, then couples it to more elaborated stress-energy-momentum tensors that underlie realistic compact configurations \cite{Maurya:2019kzu,PerezGraterol:2018eut,Morales:2018nmq,Contreras:2019iwm,Contreras:2019fbk,Singh:2019ktp,Tello-Ortiz:2019gcl,Maurya:2019xcx,Cedeno:2019qkf,Sharif:2018toc,Estrada:2018zbh,Torres:2019mee,Abellan:2020jjl,Estrada:2018vrl,Leon:2019abq,Casadio:2019usg,Sharif:2019mjn,Abellan:2020wjw,Rincon:2020izv,Sharif:2020arn,Maurya:2020gjw}.

\par Any action related to a classical conformal theory is invariant under Weyl transformations. Since the variation of the action with respect to the background metric is proportional to the stress-energy-momentum tensor, then the variation of the action with respect to a conformal rescaling is proportional to the trace of the stress-energy-momentum tensor, which vanishes for conformally invariant theories. However, upon quantization, conformal invariance under Weyl rescalings may be broken and conformal anomalies set in \cite{Capper:1973mv}. In this case, the trace of the stress-energy-momentum tensor may achieve a non-null expectation value and, thus, a conformal anomaly regards a trace anomaly \cite{Duff:1993wm,Henningson:1998gx,Kuntz:2017pjd,Bonora:1983ff,Bonora:2014qla,Kuntz:2019omq}.
In the context of the gravitational decoupling procedure, comparing the holographic Weyl anomaly to the trace anomaly of the energy-momentum tensor from 4D field theory leads to a quantity that can probe and measure the source of the gravitational decoupling \cite{Casadio:2003jc}. Hence, the calculation of the trace anomaly-to-holographic Weyl anomaly ratio makes one capable to place the gravitational decoupling, in the context of three possible metrics describing hairy black holes, as a reliable AdS/CFT realization.

\subsection{Gravitational decoupling and hairy black holes}	\label{Sgd}
\par The gravitational decoupling procedure can be straightforwardly introduced when kernel solutions of Einstein's effective field equations can be used to decouple any intricate stress-energy-momentum
tensor into manageable pieces \cite{Ovalle:2017wqi,Ovalle:2019qyi}, including the case of hairy black holes \cite{Ovalle:2020kpd}. When one regards Einstein's field equations,
\begin{equation}	\label{corr2}
	G_{\mu\nu}:=
	R_{\mu\nu}-\frac{1}{2}R g_{\mu\nu}
	=
	\upkappa^2\,\mathring{T}_{\mu\nu}\ ,
\end{equation}
where the stress-energy-momentum tensor, satisfying the conservation equation $
\nabla_\mu\,\mathring{T}^{\mu\nu}=0$, can be split as 
\begin{equation}	\label{emt}
	\mathring{T}_{\mu\nu}
	=
	\mathsf{T}^{\rm}_{\mu\nu}
	+
	\upalpha\,\Uptheta_{\mu\nu}\ ,
\end{equation}
for $\mathsf{T}_{\mu\nu}$ being a general-relativistic solution and $\Uptheta_{\mu\nu}$ encoding additional sources in the gravitational sector, for $\upalpha$ being an arbitrary decoupling parameter that is not perturbative, in general.  
One considers static, spherically symmetric, stellar distributions described by the metric  
\begin{equation} 	\label{metric}
	ds^{2}
	=
	e^{\upnu (r)}dt^{2}-e^{\uplambda (r)}dr^{2}
	-r^{2}d\Omega^2\ ,
\end{equation}
where $d\Omega^2$ denotes the solid angle element.
The Einstein's field equations~(\ref{corr2}) are equivalently written as 
\begin{subequations}
	\begin{eqnarray}
		\label{ec1}
		\upkappa^2
		\left(
		\mathsf{T}_0^{\ 0}+\Uptheta_0^{\ 0}
		\right)
		&=&
		\frac 1{r^2}
		-
		e^{-\uplambda }\left( \frac1{r^2}-\frac{\uplambda'}r\right),
		\\
		\label{ec2}
		\upkappa^2
		\left(\mathsf{T}_1^{\ 1}+\Uptheta_1^{\ 1}\right)
		&=&
		\frac 1{r^2}
		-
		e^{-\uplambda }\left( \frac 1{r^2}+\frac{\upnu'}r\right),
		\\
		\label{ec3}
		\upkappa^2
		\left(\mathsf{T}_2^{\ 2}+\Uptheta_2^{\ 2}\right)
		&=&
		-\frac {e^{-\uplambda }}{4}
		\left(2\upnu''+\upnu'^2-\uplambda'\upnu'
		+2\,\frac{\upnu'-\uplambda'}r\right)\ ,
	\end{eqnarray}
\end{subequations}
where the prime denotes the derivative with respect to the variable $r$.
Eqs. (\ref{ec1} -- \ref{ec3}) regard the effective density, and the effective radial and tangential pressures, respectively given by \cite{Ovalle:2017wqi,Ovalle:2019qyi}
\begin{subequations}
\begin{align}
			\mathring{\rho}
		&=
		\rho+
		\upalpha\Uptheta_0^{\ 0}\ ,\label{efecden}\\
		\mathring{p}_{r}
		&=
		p
		-\upalpha\Uptheta_1^{\ 1}\ ,
		\label{efecprera}\\
		\mathring{p}_{t}
		&=
		p
		-\upalpha\Uptheta_2^{\ 2}\ , 
		\label{efecpretan}
\end{align}
\end{subequations}
with anisotropy 
\begin{equation}
	\Updelta = 
	\mathring{p}_{t}-\mathring{p}_{r}\ .
\end{equation}
\par A solution to Einstein's field equations \eqref{corr2} for the single kernel source $\mathsf{T}_{\mu\nu}$ was considered \cite{Ovalle:2017wqi,Ovalle:2020kpd}, 
\begin{equation}	\label{pfmetric}
	ds^{2}
	=
	e^{\upxi (r)}dt^{2}
	-e^{\upmu (r)}dr^{2}
	-
	r^{2}d\Omega^2
	\ ,
\end{equation}
where 
\begin{equation}	\label{standardGR}
	e^{-\upmu(r)}
	\equiv
	1-\frac{\upkappa^2}{r}\int_0^r x^2\,\mathsf{T}_0^{\ 0}(x)\, dx
	=
	1-\frac{2m(r)}{r}\ ,
\end{equation}
is the Misner--Sharp--Hernandez function.
The additional source $\Uptheta_{\mu\nu}$ drives the gravitational decoupling of the kernel metric~\eqref{pfmetric}, implemented by the mappings 
\begin{subequations}
	\begin{eqnarray}
		\label{gd1}
		\upxi(r)
		&\mapsto &
		\upnu(r)=\upxi(r)+\upalpha g(r)\ ,
		\\
		\label{gd2}
		e^{-\upmu(r)} 
		&\mapsto &
		e^{-\uplambda(r)}=e^{-\upmu(r)}+\upalpha f(r)
		\ , 
	\end{eqnarray}
\end{subequations}
where $f(r)$ [$g(r)$] is the geometric deformation for the radial [temporal] metric
component.
Eqs.~(\ref{gd1}, \ref{gd2}) split the Einstein's field equations~(\ref{ec1}) -- (\ref{ec3}) into two distinct arrays. 
The first one encodes Einstein's field equations for $\mathsf{T}_{\mu\nu}$, solved by the kernel metric~(\ref{pfmetric}). The second one is associated to $\Uptheta_{\mu\nu}$ and reads
\begin{subequations}
	\begin{eqnarray}
		\label{ec1d}
		\upkappa^2\,\Uptheta_0^{\ 0}
		&=&
		-\upalpha\left(\frac{f}{r^2}+\frac{f'}{r}\right)\ ,
		\\
		\label{ec2d}
		\upkappa^2\,\Uptheta_1^{\ 1}
		+\upalpha\,\frac{e^{-\upmu}\,g'}{r}
		&=&
		-\upalpha\,f\left(\frac{1}{r^2}+\frac{\upnu'}{r}\right)\ ,
		\\
		\label{ec3d}
		\upkappa^2\Uptheta_2^{\ 2}\!+\!\upalpha{f}\left(2\upnu''\!+\!\upnu'^2\!+\!\frac{2\upnu'}{r}\right)\!&=&\!-\upalpha\frac{f'}{4}\!\left(\upnu'\!+\!\frac{2}{r}\right)\!+\!V\ ,
	\end{eqnarray}
\end{subequations}
where \cite{Ovalle:2017fgl}
\begin{equation}
	V(r) = \upalpha e^{-\upmu}\left(2g''+g'^2+\frac{2\,g'}{r}+2\upxi'\,g'-\upmu'g'\right)\ .
\end{equation}
The tensor-vacuum, defined for $\Uptheta_{\mu\nu}\neq 0$ and $\mathsf{T}_{\mu\nu}=0$, leads to hairy black hole solutions \cite{Ovalle:2018umz}. 
Eqs. (\ref{ec1}) -- (\ref{ec2}) then yield a negative radial pressure, 
\begin{equation}
	\mathring{p}_{r}
	=
	-\mathring{\rho}\ ,
	\label{schwcon}
\end{equation}
and, together with the Schwarzschild solution, it implies that 
\begin{equation}
	\label{fg}
	\upalpha\,f(r)
	=
	\left(1-\frac{2M}{r}\right)\left(e^{\upalpha\,g(r)}-1\right)
	,
\end{equation}
so that the line element~\eqref{metric} becomes
\begin{equation}		\label{hairyBH}
	ds^{2}
	=\
	\left(1-\frac{2M}{r}\right)
	e^{\upalpha g(r)}
	dt^{2}
	\!-\!\left(1-\frac{2M}{r}\right)^{-1}
	e^{-\upalpha\,g(r)}
	dr^2-r^{2}\,d\Omega^2
	.
\end{equation}
In the radial range $r\geq 2M$, the tensor-vacuum is given by expressing $\Uptheta_0^{\ 0}$ by the most general linear combination of the radial and tangential components of the stress-energy-momentum tensor, as 
	\begin{equation}
		\Uptheta_0^{\ 0}
		=
		a\,\Uptheta_1^{\ 1}+b\,\Uptheta_2^{\ 2},
	\end{equation}
	with $a,b\in\mathbb{R}$ denoting the coefficients of the linear combination. 
Eqs.~(\ref{ec1d}) -- (\ref{ec3d}) then yield  
\begin{equation}		\label{master}
	b\,r\,(r-2M)\,h''+2\,\left[(a+b-1)\,r-2\,(a-1)\,M\right]
	h'+2\,(a-1)\,h=2\,(a-1)\ 	,
\end{equation}
for 
$ 
h(r)
=
e^{\upalpha\,g(r)}$. A trivial deformation corresponding to the standard Schwarzschild solution can be yielded when $a = 1$.
The solution of Eq. (\ref{master}) can be written as
\begin{equation}
	\label{master2}
	e^{\upalpha\,g(r)}
	=
	1+\frac{1}{r-2M}
	\left[\ell_0+r\left(\frac{\ell}{r}\right)^{n}
	\right]
	,
\end{equation}
where $\ell_0=\upalpha\ell$ is a primary hair charge, whereas
\begin{equation}\label{nnn}
	n
	=
	\frac{2\left(a-1\right)}{b}	\ ,
\end{equation}
with $n>1$ for asymptotic flatness. 

\par In the tensor-vacuum background, this line element is produced by the effective density, the radial, and tangential pressures, respectively, 
\begin{subequations}
\begin{align}
			\mathring{\rho}
		&=
		\Uptheta_0^{\ 0}
		=
		\upalpha\,\frac{(n-1)\,\ell^n}{\upkappa^2\,r^{n+2}},
		\label{efecdenx}\\
		\mathring{p}_{r}
		&=
		-\Uptheta_1^{\ 1}
		=
		-\mathring{\rho},\\
		\mathring{p}_{t}
		&=
		-\Uptheta_2^{\ 2}
		=
		\frac{n}{2}\,\mathring{\rho}
		.
		\label{efecptanx}
\end{align}
\end{subequations}
On the other hand, the dominant energy conditions,
\begin{equation}
	\mathring{\rho}\geq |\mathring{p}_r|,\quad \mathring{\rho}\geq|\mathring{p}_t|,\label{dec0}
\end{equation}
yield $n\le2$ \cite{Ovalle:2017wqi,Ovalle:2019qyi,Ovalle:2020kpd}.
Besides, the strong energy conditions, 
\begin{subequations}
	\begin{eqnarray}
	\mathring{\rho}+\mathring{p}_r+2\,\mathring{p}_t
	\geq
	0, 
	\label{strong01}\\
	\mathring{\rho}+\mathring{p}_r
	\geq
	0,\\
	\mathring{\rho}+\mathring{p}_t
	\geq
	0,
\end{eqnarray}
\end{subequations}
make Eq.~\eqref{schwcon} to read 
	\begin{equation}		\label{t001}
	-\Uptheta_0^{\ 0}\leq\Uptheta_2^{\ 2}\leq0.
	\end{equation}
Therefore, together with Eqs.~\eqref{ec1d} and~\eqref{ec3d}, Eq. (\ref{t001}) can be written as
\begin{subequations}
	\begin{eqnarray}
	\label{strong5}
	\!\!\!\!\!\!\!\!\!\!\!\!\!\!\!\!\!\!G_1(r)&:=&{h''(r-2M)+2h'}
	\geq
	0,\\
	\label{strong51}
	\!\!\!\!\!\!\!\!\!\!\!\!\!\!\!\!\!\!G_2(r)&:=&
	h''r(r-2M)
	+4h'M
	-2h+2
	\geq
	0.
\end{eqnarray} 
\end{subequations}
The mapping 
\begin{equation}
	\label{gauge}
	h(r)
	\mapsto
	h(r)-\frac{\ell_0}{r-2M}\ ,
\end{equation}
leaves $G_1(r)$ and $G_2(r)$ invariant. 
Solutions with a proper horizon at $r\sim 2M$, which also
behave approximately like the Schwarzschild metric for $r\gg 2M$, yield $
G_1(r)=0$. Hence, solving Eq.~\eqref{strong5} implies that  
\begin{equation}
	\label{strongg}
	h(r)
	=
	c_1
	-
	\upalpha\,\frac{\ell-r\,e^{-r/M}}{r-2M}\ .
\end{equation}
Also, Eq.~\eqref{strongg} is also contrained to \eqref{strong51}. Replacing~\eqref{strongg} in ~\eqref{hairyBH} implies the metric 
\begin{equation}
	\label{strongBH}
	e^{\upnu}
	=
	e^{-\uplambda}
	=
	1-\frac{2{\cal M}}{r}+\upalpha\,e^{-r/({\cal M}-\upalpha\,\ell/2)}
	,
\end{equation}
to represent a hairy black hole, 
where ${\cal M}=M+\upalpha\,\ell/2$.

\par Now, the strong energy conditions are consistent with $\ell
\geq
2M/e^{2}$, whose extremal case $\ell=2M/e^{2}$ leads to 
\begin{eqnarray}
	\label{strongh2M}
	e^{\upnu}
	=
	e^{-\uplambda}
	=
	1-\frac{2M}{r}+\upalpha\,\left(e^{-r/M}-\frac{2M}{e^2\,r}\right)
	.
\end{eqnarray}
which has the horizon at $r_{\scalebox{.56}{\textsc{hor}}} = 2M$. 
The dominant energy conditions, 
\begin{subequations}
	\begin{eqnarray}
	\mathring{\rho}
	\geq
	|\mathring{p}_r|,\\
	\mathring{\rho}
	\geq
	|\mathring{p}_t|,
	\label{dom2} 
\end{eqnarray}
\end{subequations}
in terms of \eqref{efecden} and~\eqref{efecpretan} are respectively equivalent to 
\begin{subequations}
	\begin{eqnarray}
	\label{dom6}
	-r(r-2M)h''
	-4(r-M)h'
	-2h+2
	&\geq&
	0,
	\qquad
	\\
	\label{dom61}
	r\,(r-2M)\,h''
	+4\,M\,h'
	-2\,h+2
	&\geq&
	0.
\end{eqnarray} 
\end{subequations}
Solving~\eqref{dom6} for $r\sim
2M$ and $r
\gg
M$ yields \cite{Ovalle:2018umz} 
\begin{equation}
	\label{dominantg}
	h(r)
	=
	1
	-
	\frac{1}{r-2M}
	\left(
	\upalpha\,\ell
	+\upalpha\,M\,e^{-r/M}
	-\frac{Q^2}{r}
	\right)
	,
\end{equation}
where the charge $Q=Q(\upalpha)$ encompasses also tidal charges generated by additional gravitational sectors.
Eq.~\eqref{dominantg} also has to satisfy \eqref{dom61},
which reads
\begin{equation}
	\frac{4\,Q^2}{r^2}
	\ge
	\frac{\upalpha}{M}\,(r+2M)\,e^{-r/M}
	.
\end{equation}
Using~\eqref{dominantg} into the line element~\eqref{hairyBH}, yields \begin{equation}
	\label{dominantBH}
	e^{\upnu}
	=
	e^{-\uplambda}
	=1-\frac{2M+\upalpha\,\ell}{r}
	+\frac{Q^2}{r^2}
	-\frac{\upalpha\,M\,e^{-r/M}}{r},
\end{equation}
such that 
\begin{equation}
	\mathring{\rho}
	=
	\Uptheta_0^{\ 0}
	=
	-\mathring{p}_r
	=
	\frac{Q^2}{\upkappa^2\,r^4}
	-\frac{\upalpha\,e^{-r/M}}{\upkappa^2\,r^2}\ .
	\label{dendom}
\end{equation}
The metric~\eqref{dominantBH} also represents hairy black holes, where $Q$ and $\ell_0=\upalpha\,\ell$
comprise charges generating primary hair.

\par The horizon radii $r_{\scalebox{.56}{\textsc{hor}}}$ are given by solutions of
\begin{equation}
	\label{horizondom}
	\upalpha\,\ell
	=
	r_{\scalebox{.56}{\textsc{hor}}}
	-2M
	+\frac{Q^2}{r_{\scalebox{.56}{\textsc{hor}}}}
	-\upalpha\,M\,e^{-r_{\scalebox{.56}{\textsc{hor}}}/M}
	,
\end{equation}
which allows us to write the metric functions~\eqref{dominantBH} as
\begin{eqnarray}
	\label{dominantBHH}
	e^{\upnu}
	=
	e^{-\uplambda}
	=
	1-\frac{r_{\scalebox{.56}{\textsc{hor}}}}{r}
	\left(1+\frac{Q^2}{r_{\scalebox{.56}{\textsc{hor}}}^2}-\frac{\upalpha\,M}{r_{\scalebox{.56}{\textsc{hor}}}}\,e^{-r_{\scalebox{.56}{\textsc{hor}}}/M}\right)+\frac{Q^2}{r^2}-\frac{\upalpha\,M}{r}\,e^{-r/M}.
\end{eqnarray}

\par To find analytical solutions to $r_{\scalebox{.56}{\textsc{hor}}}$, appropriate values
	of the parameters $\upalpha$, $Q$, and $\ell$ must be chosen.
	However, since the dominant energy conditions demand $r_{\scalebox{.56}{\textsc{hor}}}\geq 2\,M$,
	the choice of these values cannot be arbitrary. 
	Evaluating the effective density~\eqref{dendom} at the event horizon, and making use of Eq. \eqref{horizondom} imply that  
	\begin{equation}
		\label{condi}
		Q^2
		\geq
		\frac{4\upalpha M^2}{e^2}, \qquad
		\quad
		\ell
		\geq
		\frac{M}{e^2}.
	\end{equation}
	The physical interpretation of $Q$ encompasses the case of an electric charge, but not only restricted to it, but also encodes the possibility of hidden gauge charges, tidal charge, and eventually, Kaluza--Klein stringy effects \cite{daRocha:2017cxu}, or any other source. In the case where $Q$ represents an electric charge, the electrovacuum generated by the Reissner--Nordstr\"om solution additionally accommodates a tensor-vacuum that is proportional to $\upalpha$ in Eq.~\eqref{dendom}.
	One must emphasize that the Reissner--Nordstr\"{o}m metric has an event horizon 
	\begin{equation}
		\label{domhor}
		r_{{\scalebox{.56}{\textsc{RN}}}}
		=
		M+\sqrt{M^2-Q^2}
		<
		2\,M,
	\end{equation}
	and also an inner Cauchy horizon, 
	\begin{equation}
		\label{cauchy}
		r_{\scalebox{.56}{\textsc{C}}}
		\equiv
		M-\sqrt{M^2-Q^2}
		<
		r_{\scalebox{.56}{\textsc{hor}}}.
	\end{equation}
	The solution~\eqref{dominantBH} can thus yield three ramifications wherein the event horizon $r_{\scalebox{.56}{\textsc{hor}}}$ has straightforward analytical formul\ae\, and the dominant energy conditions are satisfied.
	Similarly to the Reissner--Nordstr\"{o}m solution, the three cases to be studied have an inner Cauchy horizon $r_{\scalebox{.56}{\textsc{C}}}<r_{\scalebox{.56}{\textsc{hor}}}$. 

\par If the event horizon is made proportional to the mass, as $r_{\scalebox{.56}{\textsc{hor}}}=k\,M$,
	as long as $k\ge 2$, to satisfy the dominant energy conditions, the metric components are given by 
	\begin{equation}
		\label{dominantBHf}
		e^{\nu}
		=
		e^{-\lambda}
		=
		1-\frac{2{\cal M}}{r}
		+\frac{Q^2}{r^2}
		-\frac{\upalpha r_{\scalebox{.56}{\textsc{hor}}}}{kr}e^{-kr/r_{\scalebox{.56}{\textsc{hor}}}},
	\end{equation}
	where the Reissner--Nordstr\"om-like event horizon reads 
	\begin{equation}
		\label{domhorf}
		r_{\scalebox{.56}{\textsc{hor}}}
		=
		\tilde{\cal M}
		+\sqrt{\tilde{\cal M}^2-\tilde{Q}^2},
	\end{equation}
	where $\tilde{\cal M}={\cal M}/\upbeta$ and $\tilde{Q}^2={Q}^2/\upbeta$ with
	\begin{equation}
		\upbeta=1-\upalpha\,\frac{e^{-k}}{k}
		\ .
	\end{equation}
	Therefore the metric components~\eqref{dominantBHf} describe a black hole solution arising from nonlinear electrodynamics, with event horizon 
	\begin{eqnarray}
		r_{\scalebox{.56}{\textsc{hor}}}
		=
		\frac{r_{\scalebox{.56}{\textsc{RN}}}}{\upbeta}
		\geq
		{r_{\scalebox{.56}{\textsc{RN}}}},
	\end{eqnarray}
	as $\upbeta\leq 1$. The nonlinear electrodynamics is obtained when one identifies
	\begin{eqnarray}
		\label{tmunu}
		\Uptheta_{\mu\nu}
		=
		-\mathcal{L}(F)g_{\mu\nu}-\mathcal{L}_{F}F_{\mu}^{\;\rho}F_{\rho\nu},
	\end{eqnarray}
	where 
	\begin{equation}
		F
		=
		\frac{1}{4}F_{\rho\sigma}F^{\rho\sigma},
		\quad
		\quad 
		\mathcal{L}_{\scalebox{.56}{\textsc{$F$}}}=\frac{\partial\mathcal{L}}{\partial F}.
	\end{equation}
	When static, spherically symmetric, stellar distributions described by the metric (\ref{metric}) are regarded, 
	the field strength reads 
	\begin{equation}
		F_{\mu\nu}(r)
		=
		\left(\updelta^0_\mu\updelta^1_\nu-\updelta^1_\mu\updelta^0_\nu\right)E(r),
	\end{equation}
	where the electric field is given by 
	\begin{eqnarray}
		E(r)
		=
		\frac{Q}{r^2}
		-\frac{\upalpha\, e^{-\frac{kr}{r_{\scalebox{.56}{\textsc{hor}}}}} }
		{4r_{\scalebox{.56}{\textsc{hor}}}Q}(kr+2r_{\scalebox{.56}{\textsc{hor}}}).
	\end{eqnarray}
	Introducing the field 
	\begin{equation}
		P
		=\mathcal{L}_{\scalebox{.56}{\textsc{$F$}}}^2 F_{\rho\sigma}F^{\rho\sigma},
	\end{equation}
	the underlying nonlinear electrodynamics can be thus placed into the
	$P$-dual framework \cite{Salazar:1987ap,Ovalle:2017fgl}, described by the Lagrangian 
	\begin{eqnarray}
		\label{Lp}
		\!\!\!\!\!\mathcal{L}(P)
		\!=\!
		-4 \pi P
		\!-\!\frac{\upalpha k({-2P})^{\frac14}}
		{4\sqrt{\pi Q} \,r_{\scalebox{.56}{\textsc{hor}}}}\exp\left[\frac{k\sqrt{Q}}{2\, \sqrt{\pi }({-2P})^{\frac14}r_{\scalebox{.56}{\textsc{hor}}}}\right],
	\end{eqnarray}
	where
	\begin{eqnarray}
		\mathfrak{G}(P)
		=
		-\frac{k\sqrt{Q}}{2\, \sqrt{\pi }r_{\scalebox{.56}{\textsc{hor}}}({-2P})^{\frac14}}.
	\end{eqnarray}
	One can read off the relation $Q^2\sim\upalpha$,
	yielding the Schwarzschild standard solution whenever $\upalpha\to 0$.
	When $Q$ represents an electric charge, one can state that the Reissner--Nordstr\"om electrovacuum is permeated by a tensor-vacuum governed by \eqref{Lp}.

\par As concrete examples, one can saturate the inequalities~\eqref{condi},
and \eqref{dominantBHH} become, defining $\mathsf{M}=M\left(1+\frac{\upalpha}{2\,e^2}\right)$, 
\begin{equation}
	\label{98}
	e^{\upnu_{\scalebox{.62}{(I)}}}
	=
	e^{-\uplambda_{\scalebox{.62}{(I)}}}
	=
	1-\frac{2\mathsf{M}}{r}
	+\frac{Q^2}{r^2}
	-\frac{\sqrt{\upalpha}\,Q}{2\,r}\,e^{1-2\sqrt{\upalpha}\,r/e\,Q}
	,
\end{equation}
which can be interpreted as nonlinear electrodynamics coupled with gravity, similarly to the content in the last paragraph \cite{Ovalle:2020kpd}. 
The event horizons are placed at $r_{\scalebox{.56}{\textsc{hor}}}=2M$, and $
r_{\scalebox{.56}{\textsc{hor}}}
=
\frac{e}{\sqrt{\upalpha}}\,Q.$

\par In the second case, the relation 
\begin{equation}
	\label{qcase2}
	Q^2
	=
	\upalpha\,\ell\,M
	\left(2+\upalpha\,e^{-{\upalpha\,\ell}/{M}}\right)
	,
\end{equation}
leads to
\begin{eqnarray}
	\label{100}
	e^{\upnu_{\scalebox{.62}{(II)}}}
	=
	e^{-\uplambda_{\scalebox{.62}{(II)}}}
	=
	1-\frac{2M+\upalpha\,\ell}{r}
	+\frac{2\,\upalpha\,\ell\,M}{r^2}
	-\frac{\upalpha\,M}{r^2}\,e^{-r/M}\left(r-\upalpha\,\ell\,e^{\frac{r-\upalpha\,\ell}{M}}\right)
	.
\end{eqnarray}
The event horizon is now at $r_{\scalebox{.56}{\textsc{hor}}}=\upalpha\,\ell\geq 2M$.
As $\upalpha\,\ell\sim\,M$, Eq. \eqref{100} can be also realized as a solution in nonlinear electrodynamics coupled with gravity.  

\par Finally, when 
\begin{equation}
	Q^2
	=
	\upalpha\,M\left(2M+\upalpha\,\ell\right)
	e^{-\frac{(2M+\upalpha\,\ell)}{M}}
	,
\end{equation}
the metric components $e^{\upnu_{\scalebox{.62}{(III)}}}
=
e^{-\uplambda_{\scalebox{.62}{(III)}}}
$ read
\begin{eqnarray}
	\label{102}
	e^{\upnu_{\scalebox{.62}{(III)}}}
	&=&1
	-\frac{\upalpha M}{r^2}\,e^{-r/M}\left[r-\left(2M+\upalpha\ell\right)e^{\frac{r-\left(2M+\upalpha\ell\right)}{M}}\right]
	\nonumber
	\\
	&&\qquad\qquad\qquad\qquad\qquad-\frac{2M+\upalpha\,\ell}{r}.
\end{eqnarray}
The event horizon is at $r_{\scalebox{.56}{\textsc{hor}}}=2M+\upalpha\,\ell=2{\cal M}\geq 2M$.

\subsection{Weyl and trace anomalies of gravitational decoupled hairy black holes}	\label{dsfb}
\par The holographic Weyl anomaly is reminiscent of the regularization process applied to the gravitational part of the action under conformal transformations \cite{Henningson:1998gx}. In general, the anomaly can be expressed by  
\begin{equation}
	\mathcal{A}\propto E_{\left(d\right)}+I_{\left(d\right)},
\end{equation}
where $E_{(d)}$ is the $d$-dimensional Euler density, and $I_{(d)}$ denotes a conformal invariant\footnote{For $d=4$ this invariant is unique, being given by the contraction of the Weyl tensor to itself \cite{Henningson:1998gx}.}. In four dimensions the Euler density takes the form
\begin{equation}
	E_{\left(4\right)}=\frac{1}{64}\left(K-4R^{\mu\nu}R_{\mu\nu}+R^{2}\right).
\end{equation}
Up to a multiplicative constant, the holographic Weyl anomaly becomes
\begin{equation}\label{TCFT}
	\left\langle T\right\rangle _{\scalebox{.62}{CFT}}\sim \left(R^{\mu\nu}R_{\mu\nu}-\frac{1}{3}R^{2}\right).
\end{equation}

\par On the other hand, the trace anomaly is a function of the matter content on a curved background together with its geometric aspects \cite{Birrell:1982ix} \begin{equation} \label{T4D}
	\left\langle T\right\rangle _{\scalebox{.62}{4D}}\sim \left({K}-R^{\mu\nu}R_{\mu\nu}-\Box R\right).
\end{equation}
This anomaly quantifies the deviation from conformal invariance, i. e., the vanishing of this particular quantity indicates that the associated dual theory preserves conformal symmetry.

\par Trace anomalies from the field theory side can be compared to the one found in the CFT using the coefficient \cite{Casadio:2003jc}
\begin{equation}\label{gammaCFT}
	\Upgamma=\left|1-\frac{\left\langle T\right\rangle _{\scalebox{.62}{4D}}}{\left\langle T\right\rangle _{\scalebox{.62}{CFT}}}\right|,
\end{equation}
where the definitions given in Eqs. (\ref{TCFT}, \ref{T4D}) have been applied. 
The quantity $K = R_{\mu\nu\rho\sigma}R^{\mu\nu\rho\sigma}$ denotes the Kretschmann scalar ${K}$. 
This result holds in the context of asymptotic AdS backgrounds \cite{Henningson:1998gx} and here we discuss the possibility of emulating this result to the gravitational decoupling of hairy black holes. 

\par The coefficient \eqref{gammaCFT} can quantify how AdS/CFT is reliable in the context where the metrics (\ref{98}, \ref{100}, \ref{102}) of gravitational decoupled hairy black holes are taken into account. It measures the trace anomalies associated with them and how the additional sources backreact, in the gravitational decoupling setup. The coefficient $\Upgamma$ can formally run from $0$ to infinity, and AdS/CFT can underlie this setup 
for values that are close to the unit \cite{Casadio:2003jc,Meert:2020sqv}.
In fact, for the Schwarzschild case, $\left\langle T\right\rangle _{\scalebox{.62}{4D}}\propto K$ and $\left\langle T\right\rangle _{\scalebox{.62}{CFT}}=0$ yielding $\Upgamma\to\infty$, what compromises AdS/CFT in the general-relativistic case \cite{Casadio:2003jc}. 
\par On braneworld scenarios, one seeks for spacetimes where $\Upgamma \ll 1$, where the quantum conformal field theory on the brane and the classical gravity on the bulk descriptions are dual and equivalent \cite{Casadio:2003jc,Shiromizu:2001ve}. The case when $\Upgamma=1$ was obtained evaluating the coefficient \eqref{gammaCFT} for a braneworld black hole solution \cite{Casadio:2003jc}. One expects other solutions to be more intricate, as in the case of gravitational decoupled hairy black holes (\ref{98}, \ref{100}, \ref{102}). 

\par
\begin{figure}[h]
	\centering
	\includegraphics[width=8.6cm]{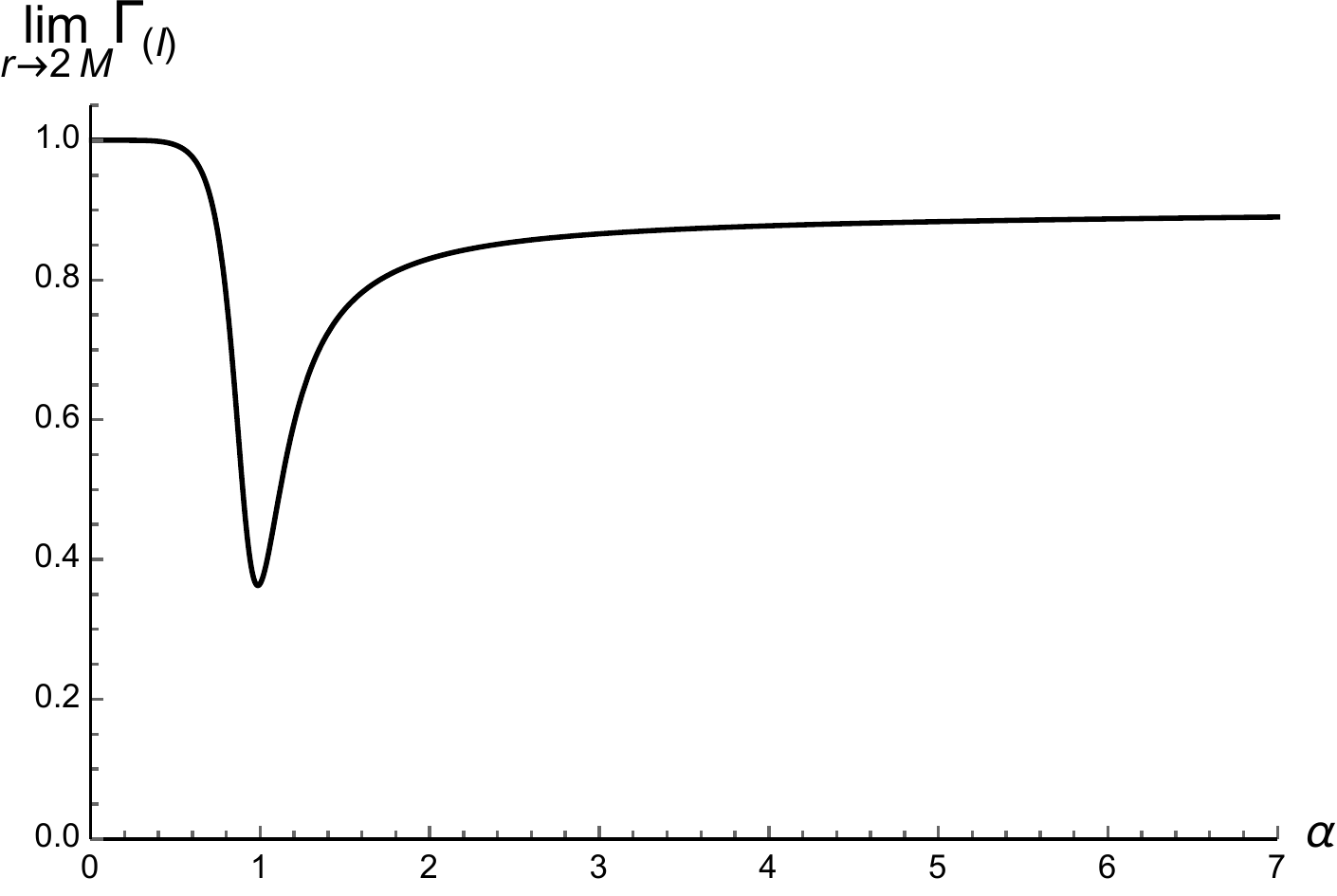}
	\caption{Plot of $\lim_{r\to2M}\Upgamma_{\scalebox{.62}{(I)}}$ as a function of the decoupling parameter $\upalpha$.}
	\label{f1}
\end{figure}

\begin{figure}[h]
	\centering
	\includegraphics[width=8.6cm]{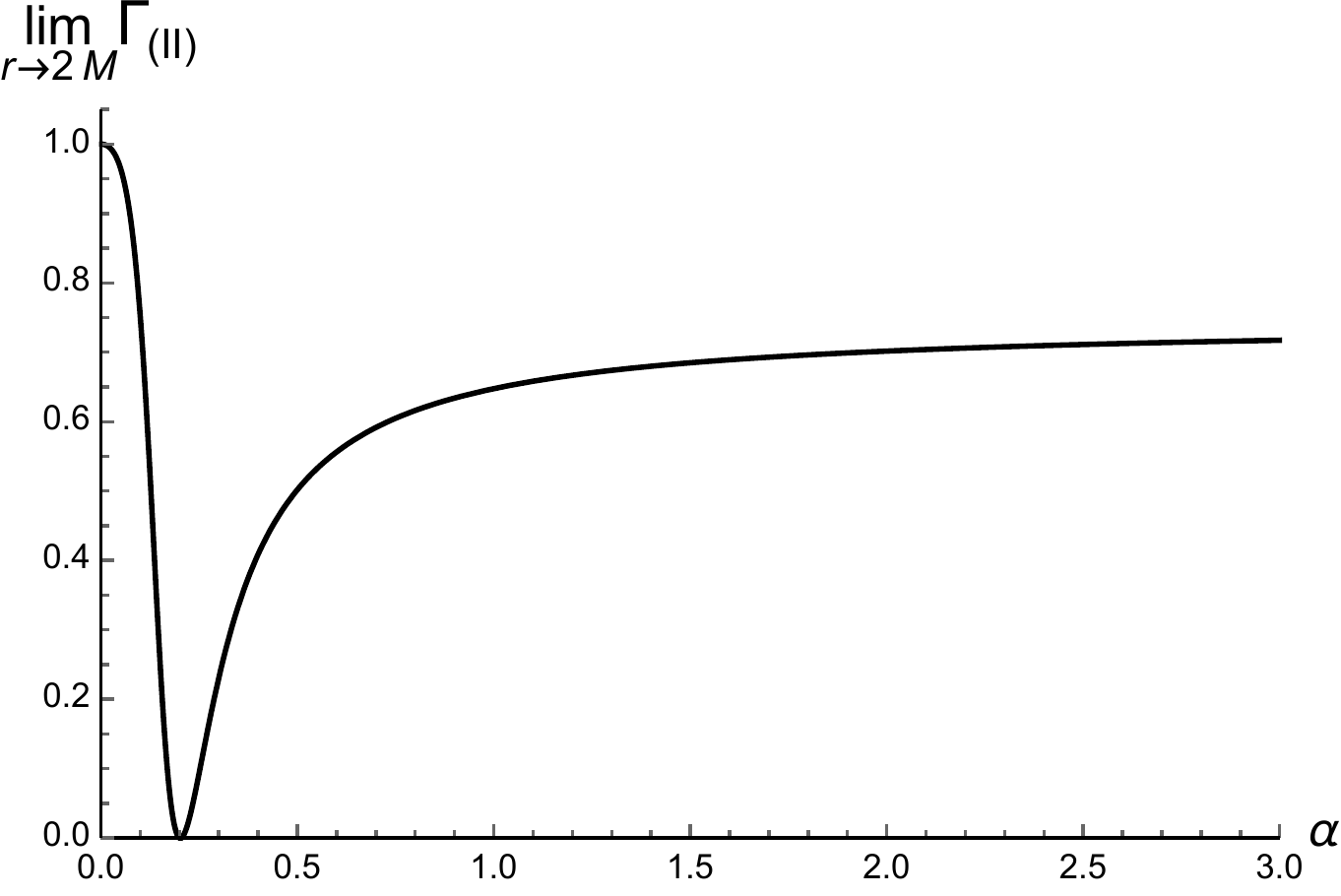}
	\caption{Plot of $\lim_{r\to2M}\Upgamma_{\scalebox{.62}{(II)}}$ as a function of the decoupling parameter $\upalpha$.}
	\label{f2}
\end{figure}

\begin{figure}[h!]
	\centering
	\includegraphics[width=8.6cm]{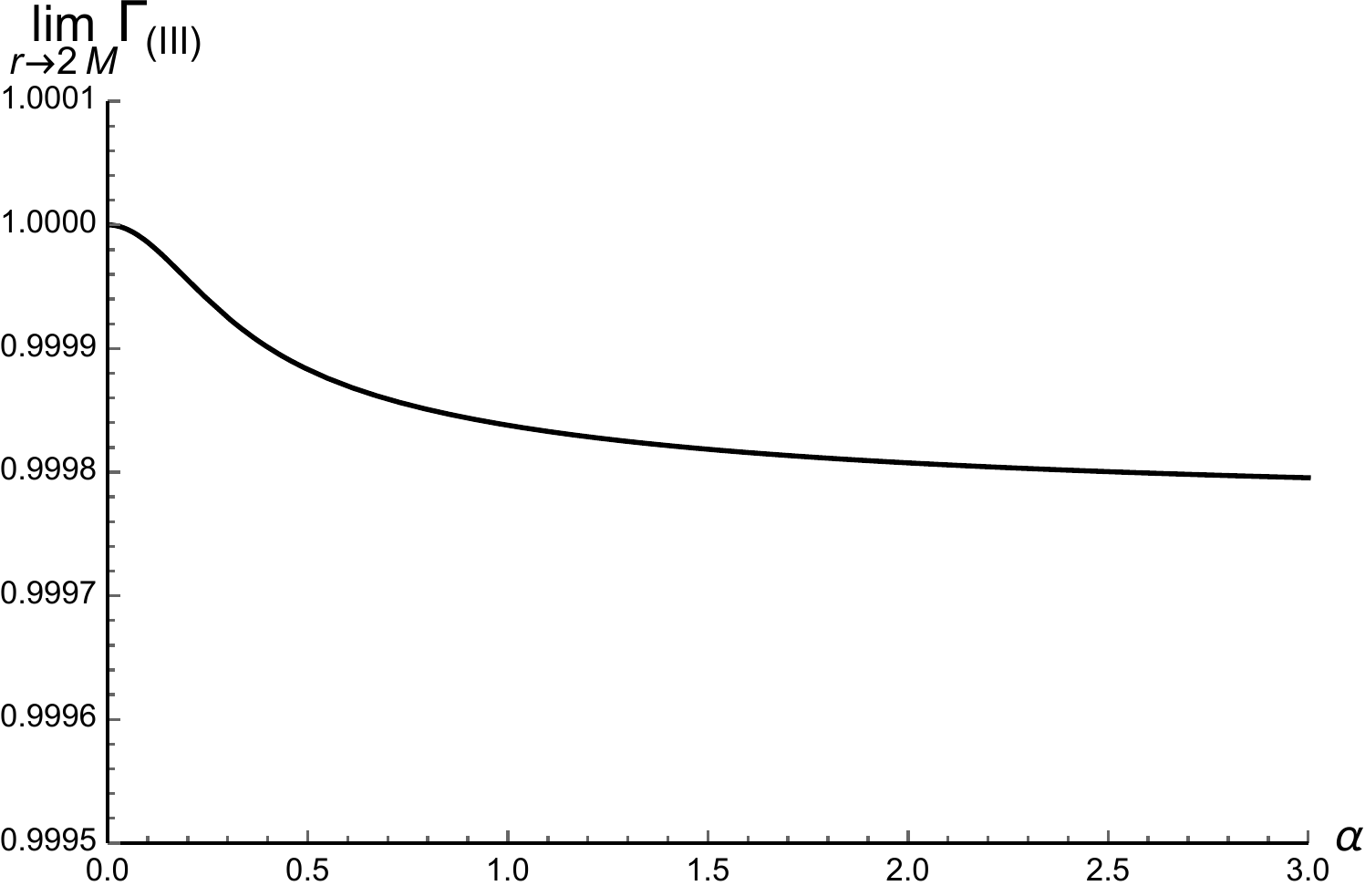}
	\caption{Plot of $\lim_{r\to2M}\Upgamma_{\scalebox{.62}{(III)}}$ as a function of the decoupling parameter $\upalpha$.}
	\label{f3}
\end{figure}

 Notice that for large values of $r$, all three coefficients associated with  (\ref{98}, \ref{100}, \ref{102}) have near unit values, 
\begin{align}
\lim_{r\to\infty}\Upgamma^{\scalebox{.63}{(I)}}\approx 1,\\
\lim_{r\to\infty}\Upgamma^{\scalebox{.63}{(II)}}\approx 1,\\
\lim_{r\to\infty}\Upgamma^{\scalebox{.63}{(III)}}\approx 1.
\end{align}
This is interesting and relevant, as such behavior is different of the Schwarzschild kernel metric used to derive these solutions using the gravitational decoupling method. Next, the limiting expressions for $r\to2M$ are analyzed in Figs. \ref{f1} -- \ref{f3}, which display $\Upgamma|_{r\to 2M}$ as a function of $\upalpha$. Since $\upalpha$ is not a perturbation parameter, at least technically it can assume any value. Thus the coefficient $\Upgamma_{\scalebox{.62}{CFT}}$ can be determined in the $\upalpha \to \infty$ limit, 

\begin{align} \label{g1}
\lim_{\substack{\upalpha\to\infty \\ r \to 2M}}\Upgamma_{\scalebox{.62}{(I)}} &= 0.923-\mathcal{O}\left(\frac{1}{\upalpha}\right), \\	\label{g2}
\lim_{\substack{\upalpha\to\infty \\ r \to 2M}}\Upgamma_{\scalebox{.62}{(II)}}&=1+\mathcal{O}\left(\frac{1}{\upalpha}\right), \\ \label{g3}\lim_{\substack{\upalpha\to\infty \\ r \to 2M}}\Upgamma_{\scalebox{.62}{(III)}}&=0.
\end{align}
Considering these limits along with the plots in Figs. \ref{f1} -- \ref{f3}, one can safely state that $\Upgamma\leq 1$ at the regions of interest, namely, for the concomitant limits $r\to \infty$ and $r \to 2M$.

\par In Figs. \ref{f1} and \ref{f2} one can clearly see a bump, where the value of $\Upgamma$ is minimum. These values are $\Upgamma_{\scalebox{.62}{(I)}} \approx 0.367$, when $\upalpha \approx 0.988$, as $\Upgamma_{\scalebox{.62}{(II)}} \approx 0$, for $\upalpha \approx 0.202$. In fact, the tiny range $0.2015 < \upalpha < 0.2032$ yields $\Upgamma_{\scalebox{.62}{(II)}} < 3\times 10^{-5}$.


\par As the limiting values of $\Upgamma$ for $r\to \infty$ do not depend on the parameter $\upalpha$, we can use the values mentioned above on the metrics to conclude that, if the AdS/CFT correspondence holds for these particular solutions obtained via gravitational decoupling, then the best agreement between classical gravity and the associated field theory is then implemented.

\subsection{Relation to AdS/CFT correspondence} \label{newSect}
\par Having these results, it is important to point out how the solutions can be used in the context of AdS/CFT correspondence, given the compliance between gravity and the associated boundary theory from the computation of $\Upgamma$. For this connection to be established, one recalls Einstein's equations in their full form
	\begin{equation} \label{n1}
		R_{\mu\nu}-\frac{1}{2}Rg_{\mu\nu}=\Lambda_{4}g_{\mu\nu}+\overline{T}_{\mu\nu}, \\
	\end{equation}
	where $\Lambda_{4}$ is the cosmological constant in 4-dimensions, and $\overline{T}_{\mu\nu}$ is the energy-momentum tensor describing all other fields except the cosmological constant one. In Sec. \ref{Sgd} the solutions studied were derived requiring that $T_{\mu\nu}=0$ and $\Theta_{\mu\nu}\neq0$. From the explicit expressions of the solutions, c.f. Eqs. (\ref{98}, \ref{100}, \ref{102}), one can check that the limit $r\to\infty$ leads to the Minkowski spacetime. However, the MGD has a strong connection to brane-world scenarios \cite{Ovalle:2009xk}, which can be employed to establish the connection.

\par In the brane-world setup, the stress-energy-momentum tensor associated with the brane has the most general form \cite{GCGR}
	\begin{equation}\label{n2}
		\overline{T}_{\mu\nu}=T_{\mu\nu}+E_{\mu\nu}+\gamma^{-1}S_{\mu\nu}+L_{\mu\nu}+P_{\mu\nu},
	\end{equation}
	where $\gamma$ denotes the brane tension. 
	The first term is the ordinary energy-momentum tensor from Einstein's equations, as already pointed out $T_{\mu\nu}=0$. The third term contains corrections of second-order on the energy-momentum tensor $T_{\mu\nu}$, such that $S_{\mu\nu}\propto\mathcal{O}\left(T_{\mu\nu}^{2}\right)$, and therefore vanishes as well. The remaining terms in Eq. \eqref{n2} carry non-local effects and also affect the energy-momentum tensor depending on the geometric procedure one uses to embed the brane in the bulk. Specifically, $L_{\mu\nu}$ accounts for the embedding and is associated with the bending of the brane concerning the codimension-1 bulk. $P_{\mu\nu}$ contains possible stringy effects living in the bulk. $E_{\mu\nu}$ describes a Weyl fluid in the bulk and is responsible for non-local effects and anisotropies. Explicit expressions for these terms can be found in \cite{GCGR,maartens,Antoniadis2011}.

\par Notice that this description is valid on a context where the solutions describe a brane, which is embedded in a higher-dimensional space-time, and therefore the quantities appearing on \eqref{n2} are related to the higher dimensional bulk. It is important to remark that the bulk is governed by its own Einstein's equations, such as Eq. \eqref{n1} in one extra dimension, where other matter fields can be defined, and it has its own cosmological constant. The cosmological constants in the bulk and on the brane are related to each other by fine-tuning with the brane tension $\gamma$, \cite{Randall:1999ee}
	\begin{equation}\label{n:3}
		\Lambda_{4}=\frac{\kappa_{5}^{2}}{2}\left(\Lambda_{5}+\frac{1}{6}\kappa_{5}^{2}\gamma^{2}\right),
	\end{equation}
	where $\kappa_{5}=8\pi G_5$ and $G_5$ is the Newton constant in five dimensions, which is related to the 4 dimensional Newton constant by the Planck length $\ell_p$ as $G_5=\ell_p G_4$. The brane tension cannot be arbitrarily defined and has a predicted lower bound for its value $\gamma\geq2.8131\times10^{-6}$ \cite{Fernandes-Silva:2019fez}. To prevent matter fields living in the bulk to interact with matter fields in the brane one has to fine-tune the cosmological constant in Eq. \eqref{n:3} such that $\Lambda_{4}=0$ \cite{Ovalle:2013vna}. Given the lower bound on the brane tension, one finds immediately that
	\begin{equation}
		\Lambda_{5}=-\frac{1}{6}\kappa_{5}^{2}\gamma^{2}\\ ,
	\end{equation}
	Implying that the bulk where the brane is located is an AdS space-time. Considering the overall setup described in Sec. \ref{Sgd}, one can therefore identify $\overline{T}_{\mu\nu}=\alpha\Theta_{\mu\nu}$, therefore
	\begin{equation} \label{n:4}
		\alpha\Theta_{\mu\nu}=E_{\mu\nu}+L_{\mu\nu}+P_{\mu\nu}.
	\end{equation}
	From this connection, the AdS/CFT conjecture can be applied to the described metrics following the prescription of \cite{Soda:2010si}. Ref. \cite{Ovalle:2017fgl} established the way how the general gravitational decoupling can be led to the membrane paradigm of AdS/CFT.

\subsection{Discussion and further analysis} \label{4}
\par The gravitational decoupling of hairy black holes was utilized and inspected with the apparatus provided by trace and Weyl anomalies. The gravitational decoupling was shown to be a trustworthy model, in the context of AdS/CFT. Since the value of the $\Upgamma$ coefficient, for the gravitational decoupling case, was shown to be near the unit, it means that the gravitational decoupling solutions may occupy a privileged place and can play a prominent role in emulating AdS/CFT on gravitational decoupled solutions. 
The $\upalpha\to\infty$ limit in Eqs. (\ref{g1} -- \ref{g3}) can be seen as a regime where the stress-energy-momentum tensor (\ref{emt}) has only the additional source contribution, in the gravitational sector, being the general-relativistic source negligible. This characterizes, in fact, the tensor vacuum regime. The coefficient (\ref{gammaCFT}) quantifies the excitation of gravitationally decoupled matter fields and estimates the signatures of gravitational waves beyond the general-relativistic setup, measuring fluctuations of the decoupled source. Hence, the anomaly coefficient \eqref{gammaCFT} brings useful information about placing gravitational decoupled hairy black holes in the AdS/CFT framework, implementing a method to quantify trace anomalies in this context, besides also quantifying backreaction of gravitationally decoupled additional sources, driven by the parameter $\upalpha$. 
\newpage

%% file: chapters/conclusions.tex
\chapter{Conclusions and Perspectives}
\par Having discussed the results and how we obtained them for each of the cases, we will now present some thoughts on the meaning of these results, possible consequences, implications and perspectives on how one could possibly improve them, as well as further developments and possibilities. Each section of this chapter is dedicated to the specific result mentioned in the previous chapter, and we close with general thoughts about the work as a whole.

\section{AdS-Reissner-Nordstr\"om}

\par In this work, we explored the viability of extending a charged black hole with a planar horizon, obtained by applying a decomposition similar to the ADM method on Einstein's equations projected onto the brane. The criteria used are the ratio of shear viscosity-to-entropy density and the KSS lower bound. We propose a new horizon from the identification of a coordinate singularity, and demand that this surface, called $r_{\beta}$, be a horizon by satisfying a constraint associated with the time-like Killing vector. We are then able to calculate entropy associated with this surface, as well as apply the GKPW formula to the metric and, considering the horizon radius as $r_{\beta}$ associate a shear viscosity to it.

\par In doing so we are able to narrow down the value of the free parameter to the range of $0 < \beta \leq 1$, where $\beta = 1$ corresponds to the AdS-Reissner-Nordstr\"om black hole. Our results are presented in terms of the tidal charge $Q$, but notice that in Eq. \eqref{adsrn-13} the deformation parameter $\beta$ is itself determined by $Q$. Moreover, in Figure \ref{eta-s-plot1} we show that the max value for $Q$, established by the KSS value, gives $\beta = 1$. 

\par Preliminary numerical analysis provides us with  deformations of the AdS$_4$--RN black brane \eqref{eq:adscft-129}, \eqref{eq:adscft-130} without AdS$_5$ bulk embedding, as an exact solution of a Lee--Wick-like action of gravity, and then more members of the family of deformed AdS$_4$-RN  black branes might be taken into account, with free parameter given by Eq. \eqref{adsrn-14}. In order to implement it, the value of $\eta/s$ cannot be conjectured to be equal to $1/4\pi$, and must be derived for the Lee--Wick-like action. However, up to now, we have  not concluded these computations, as the machining time employed is awkwardly long. If these calculations can be finally implemented, and if the $\eta/s$ ratio allows a value $\beta\neq 1$, one can therefore apply the deformed AdS$_4$-RN black brane in the context of the AdS/CMT correspondence. In fact, the standard AdS$_4$-Reissner-Nordstr\"om black brane is already known to describe the  strange metals in the holographic duality setup. By promoting the spacetime metric from the standard AdS$_4$--RN black brane, corresponding to the particular value in Eq. \eqref{adsrn-13},  to the deformed family \eqref{adsrn-9} studied here, with the $\beta$ parameter given by Eq. \eqref{adsrn-14}, we expect to gain freedom in describing such materials. Hence this family of deformed black branes can model a wider class of strange metals and the $\beta$ parameter in Eq. \eqref{adsrn-9} can be then used to compute other transport coefficients, or fine-tune quantities already known, like the electric conductivity and the thermal conductivity, for example.

\section{AdS-Schwarzschild}
\par The ADM procedure was used to derive a family of AdS$_5$--Schwarzschild deformed gravitational backgrounds, involving a free parameter, $\beta$, in the black brane metric  (\ref{r52-1}, \ref{eq:Nu}, \ref{eq:Au}). Computing the $\frac{\eta}{s}$ ratio for this family provided two possible values to $\beta$. The first one, $\beta=1$, was physically expected, corresponding to the AdS$_5$--Schwarzschild black brane. Besides the importance of the result itself, in particular for the membrane paradigm of AdS/CFT, it has a good potential for relevant applications, mainly in AdS/QCD. Taking into account the thermodynamics that underlies the family of deformed black branes solutions, arising from the Einstein--Hilbert action in the bulk, with a Gibbons--Hawking term and a counter-term that eliminates divergences, yields 
the deformed black brane temperature \eqref{Tofr}. This expression, together with the fact that  the event horizon of the deformed AdS$_5$--Schwarzschild black brane must assume real values, constrain the range of the free parameter $\beta$ in the range (\ref{ra2}). 

\par Although we have derived our results using the ADM formalism, in a bulk embedding scenario, the KSS bound violation in the range $1<\beta\leq 2$ represents, as a matter of speculation, a possible smoking gun towards the fact that the deformed AdS$_5$--Schwarzschild black brane \eqref{r52-1}, with metric coefficients (\ref{eq:Nu}, \ref{eq:Au}), might be, alternatively, derived from an action with higher curvature terms. However, up to our knowledge, no result has been obtained in this aspect, yet.

\par The  family of AdS$_5$--Schwarzschild deformed black branes,  here derived using the ADM formalism, is also not the first example in the literature of a setup that violates the KSS bound \emph{and} does not involve higher derivative theories of gravity, in the gauge/gravity correspondence. In fact, strongly coupled $\mathcal{N} = 4$ super-Yang--Mills plasmas can describe pre-equilibrium stages of the quark-gluon plasma (QGP) in heavy-ion collisions. In this setup, the shear viscosity, transverse to the direction of anisotropy, was shown to  saturate the KSS viscosity bound \cite{Rebhan:2011vd}. Besides, anisotropy in the shear viscosity induced by external magnetic fields in a strongly coupled plasma also provided violation in the KSS bound \cite{Critelli:2014kra}. Theories with higher-order curvature terms in the action, in general, comprise attempts of describing quantum gravity.
Hence, one is restricted to consider CFT for which  the central charges  satisfy $|c - a|/c \ll 1$ and $c>a$, in such a way that still $c\sim a \gg 1$, also yielding violation of the KSS bound \cite{Buchel:2008vz,viol1}.   Up to now,  the equations of motion for 5-dimensional actions with higher curvature terms up to third order are already established in the literature, but it has been not possible to obtain  the deformed AdS$_5$--Schwarzschild black brane \eqref{r52-1} yet as an exact solution to any of them. We keep trying to compute higher-curvature terms, including  fourth-order terms, and we have not exhausted all the possibilities, yet.  Any effective action is expected to contain curvature terms of  higher-order, each one of them accompanying their respective coefficients. To derive a sensible derivative expansion, one should restrict to the classes of CFTs wherein these coefficients are proportional to inverse powers of the central charge $c$ \cite{Buchel:2008vz}.

\par As large-$N_c$ gauge theories considered by AdS/CFT are good approximations to QCD, one could expect that the result $\frac{\eta}{s}=\frac{1}{4\pi}$ may be applied to the QGP, which is a natural phenomenon in QCD, when at high enough temperature the quarks and gluons are deconfined from protons and neutrons to form the QGP  \cite{qgpexp2}. In fact, experiments in the RHIC have shown that the QGP behaves like a viscous fluid with very small viscosity, which implies that the QGP is strongly-coupled, thus discarding the possibility of using perturbative QCD in the study of the plasma. Therefore, the new AdS$_5$-Schwarzschild deformed  black brane  \eqref{r52-1} can be widely used to probe additional properties in the AdS/QCD approach.  As in  the holographic soft-wall AdS/QCD, the AdS$_5$-Schwarzschild black brane provides a reasonable description of mesons at finite temperature \cite{BoschiFilho:2004ci}, we can test if using the AdS$_5$-Schwarzschild deformed  black brane derives a more reliable meson mass spectra for the mesonic states and their resonances, better matching experimental results. Besides,  the new AdS$_5$-Schwarzschild deformed  black brane can be also explored in the context of the Hawking--Page transition and information entropy \cite{Bernardini:2016hvx,Braga:2016wzx}.

\section{Fermionic sectors of gauge/gravity duality}
\par We considered a black brane solution \cite{Casadio:2016zhu}, and apply a procedure first proposed at \cite{Erdmenger:2013thg}. It is an analog of the shear viscosity for fermionic sectors, where the formalism to compute transport coefficients is applied to the fermionic sector of the action, rather than the gravitational part, thus the association with another field -- in this case the gravitino. Also, notice that the method applied in this case makes use of the cross-section associated with the fields, rather than applying perturbation as we presented in Sect. \ref{eta-calc}. This method leads to the same results though, as one can check that the ratio of $\frac{\eta}{s}=\frac{1}{4\pi}$ can be obtained using both methods, c.f. \cite{Son:2009zzc}.

\par What is interesting in the case upon consideration in this work is the nature of the spinor fields that emerge from the computations. In particular, the fact that the deformation parameter defined in Eqs. \eqref{cfm1} and \eqref{cfm2} determines whether these spinor fields are regular or singular. These conclusions are based on the Lounesto's Classification, c.f. Sect. \ref{Lounesto}, that gives the criteria to evaluate the nature of the spinor fields. 

\par The main result here to be emphasized consists of observing that the free parameter, regulating the generalized black brane solution, controls a flipping phenomenon, mapping regular to singular fermioninc field solutions appearing in the equations of motion for the gravitino. Except for a result in Ref. \cite{HoffdaSilva:2017vic}, but in a completely different context, there is no similar result in the literature considering the flipping between Lounesto's spinor field classes. Besides, our results encompass and generalize the results by Erdemenger and Steinfurt in Ref. \cite{Erdmenger:2013thg}, as the free parameter governing the generalized black brane solution yields the flag-dipole singular spinor fields, which does not occur in Ref. \cite{Erdmenger:2013thg}. Since Ref. \cite{Bonora:2017oyb} introduced the fermionic spinor field classification in a second quantized setup, with a perturbative approach to the bilinear covariants and the associated Feynman--Dyson propagators, the flag-pole and flag-dipole fermionic fields here obtained can be employed to construct new second-quantized quantum fields with self-interaction in the Rarita--Schwinger equation, governing the gravitino.

\section{Anomalies}
\par We employed a method presented on \cite{Casadio:2003jc}, where a coefficient is defined as the ratio between the conformal and Weyl anomaly. From this ratio, we are able to establish if the AdS/CFT correspondence would give an accurate outcome when the metrics considered were used.

\par The solutions used were derived on \cite{Ovalle:2017wqi,Ovalle:2020kpd} and we simply reproduce the calculations. The MGD method provides a way to extend solutions of the Einstein's equations by adding an unknown matter term, in a way similar to how it is done in linear theories such as electrodynamics. The drawback of this procedure is that one cannot know what the matter term actually represents, and even harder it is to find an action that reproduces the source -- making it very difficult to actually apply the AdS/CFT correspondence, since it relies heavily on the knowledge of the action.

\par However, we are able to find regions of the parameter $\alpha$ where one can expect the AdS/CFT correspondence to provide good results, given that the coefficient approaches one as we showed in the results, c.f. Sect. \ref{dsfb}. We point out that the solutions analyzed are not asymptotically AdS, but the AdS/CFT correspondence as presented in \cite{Soda:2010si} allows for the application since the bulk where the solution is embedded is pure AdS -- so at least technically the correspondence can be realized.

\section{General overview and perspectives}
\par In all of the studies presented in this work, we have considered some kind of extension to familiar solutions from General Relativity, namely, black branes and minimal geometric deformation. We were able to apply the AdS/CFT correspondence to compute the shear viscosity associated with these solutions, as well as compute the thermodynamic state functions and, from there, draw conclusions about the parameter governing the deformation of the solutions.

\par In the case of the AdS-Reissner-Nordstr\"om black brane we intend to apply a development similar to the one presented in Sect. \ref{conductivities}. In fact this was the goal when we first started to study the system, but technical limitations at the time ended up taking a lot of effort. Mainly the computations related to conductivity rely heavily on numerical methods to extract the properties, and very few can be solved analytically. The deformed AdS-RN black brane is very complex and will require a lot of effort in this direction in order to produce worthy results. The idea is to compare the conductivity associated with the deformed solution to the already known results reviewed in Sect. \ref{conductivities}.

\par  For the AdS-Schwarzschild black brane, we mentioned some speculations about the nature of $\beta$ parameter, c.f. Sect. \ref{beta-sch}. It is interesting to consider higher-curvature actions, and check if this solution can be found as solving an equation of motion, rather than an ADM-like constraint as we have done here. This would lead to a better understanding of the deformation parameter, and also enable a better interpretation of its origin.

\par In the work related to the gauge/gravity duality applied to the fermionic sector, one way to improve the model is to consider the coupling of fermionic fields to gauge fields, this is pointed out at \cite{Erdmenger:2013thg}. Another point that might be worthy of further investigation is the spinors of the fifth class in Lounesto's classification, the flag-poles. This class contains Majorana and so-called ELKO, which are dark matter candidates because their mass dimension is equal to $1$, and therefore do not couple to gauge fields. If a theory contains only fields of this type only, in light of the AdS/CFT correspondence this would mean that the chemical potential vanishes.

\par Finally, for the study regarding the conformal anomalies, we are looking into methods to implement the AdS/CFT correspondence as proposed at \cite{Soda:2010si}. This could provide interesting results, especially considering the ranges of $\alpha$ where the calculations in Sect. \ref{dsfb} suggest that one can expect good results for the correspondence. 

%% file: chapters/app-admtex.tex
\chapter{ADM formalism} \label{adm-app}
\par The standard treatment of GR as a field theory is based on the Lagrangian approach, c.f. Eq. \eqref{eq:28}. However this is a matter of choice, and one can formulate the same theory by adopting the Hamiltonian description, which is sometimes called Canonical Formalism. The Hamiltonian formulation of GR was pioneered by Arnowitt, Deser, and Misner \cite{Arnowitt:1959ah}, hence the name ADM formalism. Usually, the ADM method is employed in numerical relativity, or in a context where quantization of the gravitational field is relevant. The ADM method is presented here to make the analogy mentioned in Sect. \ref{blackbranes} clear, as well as present some key concepts appearing in that section. The crucial point is to understand where the constraints come from and how they are related to the equations of motion.

\par Recall that the Hamiltonian formulation of particle mechanics makes an explicit distinction between time and space, this becomes very clear once one recalls that the momenta are defined by the time derivative, and one cannot generalize this to include all the coordinates in the definition of momenta. When fields are considered, instead of a particle, the time coordinate must be defined everywhere, not only along the trajectory, but the splitting between time and space must still occur such that a Hamiltonian can be properly defined. This is done by foliating the space-time in time slices, a well-known technique in differential geometry. Although Einstein's equations are no longer manifestly covariant, the choice of the time coordinate, as well as the coordinate system, remains totally arbitrary - i.e. diffeomorphism invariance is not lost -, and the results obtained are completely equivalent to the ones obtained from the Lagrangian approach. 

\par Suppose the space-time of interest is described by a manifold $M$ equipped with metric, it is usual to refer to such a manifold simply by the pair $\left(M,g\right)$. Now consider the diffeomorphism
\begin{equation}\label{eq:adm-1}
	\varphi:M\rightarrow\mathbb{R}\times S\ ,
\end{equation}
where $S$ is a space-like surface and $t\in\mathbb{R}$ is the time coordinate. A splitting such as \eqref{eq:adm-1} is, in principle, arbitrary, as long as the roles of being space-like, for $S$, and time-like, for $\mathbb{R}$, are kept. Let $n^{\mu}\in M$ be a future pointing time-like vector, i.e. $ n^{\mu}n_{\mu}=-1$. The induced metric on $S$ is
\begin{equation}\label{eq:adm-2}
	\gamma_{\mu\nu}=g_{\mu\nu}-n_{\mu}n_{\mu}\ .
\end{equation}
One can check that $\gamma$ is purely spatial by verifying that $n^{\mu}\gamma_{\mu\nu}=0$. The space-like foliation of $M$ is denoted by the pair $\left(S,\gamma\right)$. So, if $D$ is the dimension of $M$, $S$ is a hyper-surface with dimension $D-1$.

\par Next, define the time-evolution vector field $t=t^{\mu}\nabla_{\mu}$, this vector field defines the direction of time derivatives, and is normalized according to $t^{\mu}\nabla_{\mu}t=1$. The shift vector $N^{\mu}=\gamma^{\mu\nu}t_{\nu}$, and lapse function $N=-n_{\mu}t^{\mu}$ are introduced in order to decompose the time-evolution vector field, $t^{\mu}$, in normal and spatial parts as
\begin{equation}\label{eq:adm-3}
	t^{\mu}=Nn^{\mu}+N^{\mu}\ .
\end{equation}
From Eq. \eqref{eq:adm-2} one can write $g_{\mu\nu}$ in terms of $\gamma_{\mu\nu}$ and $n_{\mu}$, and from Eq. \eqref{eq:adm-3} the normal vector $n_{\mu}$ can be written in terms of the time-evolution vector field, the shift vector and the lapse function. Putting these all together, and considering a coordinate basis $x^{\mu}$ such that $t^{\mu}\nabla_{\mu}=\partial_{t}$ one can check that
\begin{equation}\label{eq:adm-4}
	ds^{2}=g_{\mu\nu}dx^{\mu}dx^{\nu}=-N^{2}dt^{2}+\gamma_{\mu\nu}\left(dx^{\mu}+N^{\mu}dt\right)\left(dx^{\nu}+N^{\nu}dt\right)\ .
\end{equation}
This allows the description of space-time geometry by the spatial slicing, $\gamma_{\mu\nu}$, and the deformation of neighbouring slices with respect to each other described by $N$ and $N^{\mu}$.

\par What is aimed here is to describe Einstein's equations with respect to the foliation $\left(S,\gamma_{\mu\nu}\right)$ and the normal vector $n^{\mu}$. This is necessary to re-write the Einstein-Hilbert action as the so-called ADM action, which allows one to obtain solutions that are not easily seen in the conventional formulation. For a complete discussion, one is referred to the original work by ADM \cite{Arnowitt:1959ah}, as well as a pedagogical exposition of the subject in the book by Baez \cite{Baez:1995sj}.

\subsection*{Some notions of geometry} \label{subsec:geometry}
\par The spatial metric $ \gamma_{\mu\nu}$ is an intrinsic quantity, meaning that this metric itself defines the geometry of a manifold. Start by defining the covariant derivative, denoted by $D_{\mu}$, of a tensor $F_{\text{ }\rho}^{\lambda}$
\begin{equation}\label{eq:adm-5}
	D_{\mu}F_{\text{ }\rho}^{\lambda}=\gamma_{\mu}^{\text{ }\alpha}\gamma_{\beta}^{\text{ }\lambda}\gamma_{\rho}^{\text{ }\delta}\nabla_{\alpha}T_{\text{ }\delta}^{\beta}\ .
\end{equation}
The derivative $\nabla_{\mu}$ is the usual covariant derivative on$\left(M,g_{\mu\nu}\right)$. Notice how the metric $\gamma_{\mu}^{\text{ }\nu}$ acts as a transformation matrix - in fact this is the projection of a quantity on $M$ to $S$.

\par Recall that the Riemann tensor is defined as the symmetric part of the commutator between two covariant derivatives, c.f. Eq. \eqref{eq:7}. In the same way, the Riemann tensor on $\left(S,\gamma_{\mu\nu}\right)$ is defined as
\begin{equation}\label{eq:adm-6}
	^{\left(D-1\right)}R_{\mu\nu\lambda}^{\text{ }\text{ }\text{ }\text{ }\text{ }\sigma}V_{\sigma}=D_{\mu}D_{\nu}V_{\lambda}-D_{\nu}D_{\mu}V_{\lambda}\text{ },
\end{equation}
given that $V_{\sigma}n^{\sigma}=0$, i.e. $V_{\sigma}$ is a purely spatial vector. From Eq. \eqref{eq:adm-6} the other quantities such as Ricci tensor and scalar curvature can be obtained.

\par Since $\left(S,\gamma_{\mu\nu}\right)$ is embedded in $\left(M,g_{\mu\nu}\right)$, the extrinsic curvature plays an important role. Roughly, it measures the bending of $S$ when viewed from a higher dimension - hence, extrinsic. It can be shown that
\begin{equation}\label{eq:adm-7}
	K_{\mu\nu}=D_{\mu}n_{\nu}=\gamma_{\mu}^{\text{ }\alpha}\gamma_{\nu}^{\text{ }\beta}\nabla_{\alpha}n_{\beta}\text{ },
\end{equation}
where the last equality follows from the consideration in the last paragraph, as well as the definition of the covariant derivative on $\left(S,\gamma_{\mu\nu}\right)$. It should be clear at this point that quantities defined on the hyper-surface $S$ have their indices lowered/raised with the metric $\gamma_{\mu\nu}$, as well as inner products of vectors and contractions of various tensors. The trace of the extrinsic curvature is simply
\begin{equation}\label{eq:adm-8}
	\gamma^{\mu\nu}K_{\mu\nu}=K\text{ }.
\end{equation}

\par From these definitions, one is able to write the scalar curvature in terms of the scalar curvature on $S$ and the extrinsic curvature, $K_{\mu\nu}$. This can be found in the literature \cite{Baez:1995sj} and will be skipped, once one goes through all computations should arrive in the so-called Gauss-Codazzi equations, c.f. Eqs. \eqref{1.3.1-2} and \eqref{1.3.1-3}, basically, the equations describing the projections of intrinsic quantities to $\left(M,g_{\mu\nu}\right)$ in terms of quantities defined on $\left(S,\gamma_{\mu\nu}\right)$. 

\subsection*{ADM equations}
The Einstein-Hilbert action is\footnote{The cosmological constant is not considered here because it can be regarded as a matter term. Moreover, for the applications considered here, it is always constant.}
\begin{equation}\label{eq:adm-10}
	S=\frac{1}{2\kappa_{D}^{2}}\int d^{D}x\sqrt{-g}R\text{ },
\end{equation}
rewriting this action using Eq. \eqref{1.3.1-6} (the equation for scalar curvature in terms of extrinsic curvature) is straightforward, notice that $\sqrt{-g}=\sqrt{-N^{2}\gamma}$, from Eq. \eqref{eq:adm-4}, where $\gamma=\det\gamma_{\mu\nu}$. So
\begin{equation}\label{eq:adm-11}
	S=\frac{1}{2\kappa_{D}^{2}}\int d^{D}x\sqrt{-N^{2}\gamma}\left(^{\left(D-1\right)}R+K_{\mu\nu}K^{\mu\nu}-K^{2}\right)\text{ },
\end{equation}
notice that the last term on Eq. \eqref{1.3.1-6} vanishes upon integration - its actual contribution would be a surface term but, as usual, it vanishes on the limits. Action \eqref{eq:adm-11} shows that the components of the metric $g_{\mu\nu}$ are replaced by the components of $\gamma_{\mu\nu}$, the shift vector $N^{\mu}$ and the lapse function $N$ - one can check that the number of degrees of freedom is not changed.

\par To write the Hamiltonian associated to Eq. \eqref{eq:adm-11} one needs to compute the conjugate momenta to $\gamma_{\mu\nu},\text{ }N^{\mu}$ and $N$. Recall that in classical mechanics the momenta are defined as
\begin{equation}\label{eq:adm-12}
	p_{a}=\frac{\partial L}{\partial\dot{q}^{a}}\text{ },
\end{equation}
where $L$ is the Lagrangian describing the system, and $\dot{q}=\partial_{t}q$ the velocities. Generalizing Eq. \eqref{eq:adm-12} one has
\begin{align}\label{eq:adm-13}
	p^{\mu\nu}=\frac{\delta L_{ADM}}{\delta\dot{\gamma}_{\mu\nu}},		\ \ \ \
	&p_{\mu}=\frac{\delta L_{ADM}}{\delta\dot{N}^{a}},		\ \ \ \
	p_{N}=\frac{\delta L_{ADM}}{\delta\dot{N}}\ ,
\end{align}
with $L_{ADM}$ defined from Eq. \eqref{eq:adm-11}:
\begin{equation}\label{eq:adm-14}
	L_{ADM}=\frac{1}{2\kappa_{D}^{2}}\int d^{D-1}x\sqrt{-N^{2}\gamma}\left(^{\left(D-1\right)}R+K_{\mu\nu}K^{\mu\nu}-K^{2}\right)\text{ },
\end{equation}
i.e. the action \eqref{eq:adm-11} is equivalent to $S=\int dtL_{ADM}$. The computation of Eqs. \eqref{eq:adm-13} is rather technical, it is enough to state here that, when the foliation \eqref{eq:adm-1} is considered, tensor fields on $M$ are regarded as time-dependent tensor fields, from the point of view of $S$. So the evolution of tensor fields on $S$ is given by the Lie derivative of such tensor fields with respect to the time-evolution vector field, Eq. \eqref{eq:adm-3}. Given this, the extrinsic curvature can be written as \cite{Baez:1995sj}
\begin{equation}\label{eq:adm-15}
	K_{\mu\nu}=\frac{1}{2N}\left(\dot{\gamma}_{\mu\nu}-2D_{(\mu}N_{\nu)}\right)\text{ },
\end{equation}
with $\dot{\gamma}_{\mu\nu}=\gamma_{\mu}^{\text{ }\alpha}\gamma_{\nu}^{\text{ }\beta}\mathfrak{L}_{t}\gamma_{\alpha\beta}$, $\mathfrak{L}_{t}$ being the Lie derivative with respect to the time-evolution vector field. From this, Eqs. \eqref{eq:adm-13} read
\begin{align}\label{eq:adm-16}
	p^{\mu\nu}=\frac{\sqrt{\gamma}}{2\kappa_{D}^{2}}\left(K^{\mu\nu}-K\gamma^{\mu\nu}\right), \ \ \ \
	p_{\mu}=0,\ \ \ \
	p_{N}=0,
\end{align}
the last two are zero because $L_{ADM}$ does not depend on time derivatives of the shift vector or the laps function - $p_{\mu}$ and $p_{N}$ are, in fact, constraints on the phase space. The first of Eqs. \eqref{eq:adm-16} can be inverted for $\dot{\gamma}_{\mu\nu}$
\begin{equation}\label{eq:adm-17}
	\dot{\gamma}_{\mu\nu}=\frac{N\kappa_{D}^{2}}{\sqrt{\gamma}}\left(2p_{\mu\nu}-p_{\lambda}^{\text{ }\lambda}\gamma_{\mu\nu}\right)+2D_{(\mu}N_{\nu)}\text{ },
\end{equation}
so the total Hamiltonian is
\begin{align}
	\begin{aligned}\label{eq:adm-18}
		H&=\int d^{D-1}\left(\dot{\gamma}_{\mu\nu}p^{\mu\nu}+\lambda p_{N}+\mu^{\mu}p_{\mu}\right)-L_{ADM} \ ,	\\
		&=\int d^{D-1}\left\{ \frac{N\kappa_{D}^{2}}{\sqrt{\gamma}}\left[p_{\mu\nu}p^{\mu\nu}-\frac{1}{2}\left(p_{\lambda}^{\text{ }\lambda}\right)^{2}\right]+2p^{\mu\nu}D_{\mu}N_{\nu}-\frac{N\sqrt{\gamma}}{\kappa_{D}^{2}}\text{ }^{\left(3\right)}R+\lambda p_{N}+\mu^{\mu}p_{\mu}\right\} \ .
	\end{aligned}
\end{align}
In Eq. \eqref{eq:adm-18} $\lambda=\dot{N}$ and $\mu^{\mu}=\dot{N}^{\mu}$. From constraints in Eqs. \eqref{eq:adm-16} one finds the secondary constraints from the Poisson brackets
\begin{align}\label{eq:adm-19}
	0=\dot{p}_{N}=\left\{ p_{N},H\right\} =\zeta, \ \ \ \
	0=\dot{p}_{\mu}=\left\{ p_{\mu},H\right\} =\zeta_{\mu},
\end{align}
which give the Hamiltonian constraint
\begin{equation}\label{eq:adm-20}
	\frac{N\kappa_{D}^{2}}{\sqrt{\gamma}}\left[p_{\mu\nu}p^{\mu\nu}-\frac{1}{2}\left(p_{\lambda}^{\text{ }\lambda}\right)^{2}\right]-\frac{N\sqrt{\gamma}}{\kappa_{D}^{2}}\text{ }^{\left(3\right)}R\approx0\text{ },
\end{equation}
and the momentum constraint (sometimes this is called diffeomorphism constraint)
\begin{equation}\label{eq:adm-21}
	-2D_{\mu}p_{\nu}^{\text{ }\mu}\approx0\text{ }.
\end{equation}
The sign $\approx$ here means that the equality holds on the surface $S$.

\par Einstein's equations, i.e. equations of motion, are obtained by considering the Poisson brackets of the metric and the Hamiltonian, as well as its momentum conjugate
\begin{align}\label{eq:adm-22}
	\left\{ \gamma_{\mu\nu},H\right\}=\dot{\gamma}_{\mu\nu},\ \ \ \
	\left\{ p^{\mu\nu},H\right\}=\dot{p}^{\mu\nu}\ .
\end{align}
These are equivalent to $G_{\mu\nu}=0$ for the metric $\gamma_{\mu\nu}$. Notice how constraints \eqref{eq:adm-19} and Eqs. \eqref{eq:adm-22} form a system with the same amount of equations as Einstein's equations in one dimension higher.

%% file: chapters/app-electricweyl.tex
\chapter{Electric part of the Weyl tensor} \label{app:1}
\par In Chapter \ref{ch-2} we claimed that the term $\Psi_{\alpha\beta}$ equals the electric part of the Weyl tensor. In this appendix, we will explicitly show this, and also discuss some other properties related to this quantity. Throughout the calculations, we are going to use capital letters for quantities defined in the 5 dimensions bulk, whereas greek letters are for quantities defined on the brane. We will not be using the indicators for the dimension of the quantities since no use of the Gauss equation is made. 

\par Recall the definition of $\tilde{\Psi}_{\alpha\beta}$
\begin{align}
	\begin{aligned}
		\tilde{\Psi}_{\alpha\beta}&=R_{MNP}^{\text{ }\text{ }\text{ }\text{ }Q}n_{Q}n^{N}q_{\alpha}^{\text{ }M}q_{\beta}^{\text{ }P}\ ,\\
		&=\left(R_{MNPR}^{\text{ }\text{ }\text{ }\text{ }}g^{RQ}\right)n_{Q}n^{N}q_{\alpha}^{\text{ }M}q_{\beta}^{\text{ }P}\ ,\\
		&=R_{MNPR}^{\text{ }\text{ }\text{ }\text{ }}n^{R}n^{N}q_{\alpha}^{\text{ }M}q_{\beta}^{\text{ }P}\ .
	\end{aligned}
\end{align}
Where we used the metric to write the Riemann tensor with all indices down and rearranged the contraction with the normal vector. Next, recall the definition of decomposition of the Riemann tensor
\begin{equation}
	R_{MNPR}=C_{MNPR}+\frac{2}{3}\left(g_{M[P}R_{R]N}-g_{N[P}R_{R]M}\right)-\frac{2}{12}Rg_{M[P}g_{R]N}\ ,
\end{equation}
where we keep the factor 2 in the expression to simplify the calculations later. We can now introduce the electric part of the Weyl tensor
\begin{equation}
	\Psi_{\alpha\beta}=C_{MNPR}n^{R}n^{N}q_{\alpha}^{\text{ }M}q_{\beta}^{\text{ }P}\ .
\end{equation}

\par The other terms in the contraction can be simplified using the relation between the Ricci tensor and the energy-momentum tensor. Consider the term
\begin{align}
	\begin{aligned}
		g_{M[P}R_{R]N}n^{R}n^{N}q_{\alpha}^{\text{ }M}q_{\beta}^{\text{ }P}&=\left(g_{MP}R_{RN}-g_{MR}R_{PN}\right)n^{R}n^{N}q_{\alpha}^{\text{ }M}q_{\beta}^{\text{ }P}\ ,\\ &=R_{RN}n^{R}n^{N}\left(g_{MP}q_{\alpha}^{\text{ }M}q_{\beta}^{\text{ }P}\right)-R_{PN}n^{R}n^{N}\left(g_{MR}q_{\alpha}^{\text{ }M}\right)q_{\beta}^{\text{ }P}\ ,\\	&=R_{RN}n^{R}n^{N}q_{\alpha\beta}-R_{PN}n^{R}n^{N}q_{\alpha R}q_{\beta}^{\text{ }P}\ .
	\end{aligned}
\end{align}
By definition we have
\begin{equation}
	n^{R}q_{\alpha R}=0\ ,
\end{equation}
therefore the second term vanishes immediately. Also, by \eqref{1.3.1-9}
\begin{equation}
 R_{RN}=8\pi G_{5}\left(T_{RN}-\frac{T}{3}g_{RN}\right)\ ,
\end{equation}
such that
\begin{equation}
	g_{M[P}R_{R]N}n^{R}n^{N}q_{\alpha}^{\text{ }M}q_{\beta}^{\text{ }P}=8\pi G_{5}\left(T_{RN}-\frac{T}{3}g_{RN}\right)n^{R}n^{N}q_{\alpha\beta}\ .
\end{equation}
Similarly, 
\begin{align}
	\begin{aligned}
		g_{N[P}R_{R]M}n^{R}n^{N}q_{\alpha}^{\text{ }M}q_{\beta}^{\text{ }P}&=\left(g_{NP}R_{RM}-g_{NR}R_{PM}\right)n^{R}n^{N}q_{\alpha}^{\text{ }M}q_{\beta}^{\text{ }P}\ ,\\	&=R_{RM}n^{R}n^{N}q_{\alpha}^{\text{ }M}\left(g_{NP}q_{\beta}^{\text{ }P}\right)-R_{PM}\left(n^{R}n^{N}g_{NR}\right)q_{\alpha}^{\text{ }M}q_{\beta}^{\text{ }P}\ ,\\	&=R_{RM}n^{R}q_{\alpha}^{\text{ }M}\left(n^{N}q_{N\beta}\right)-R_{PM}q_{\alpha}^{\text{ }M}q_{\beta}^{\text{ }P}\ ,\\	&=-8\pi G_{5}\left(T_{PM}-\frac{T}{3}q_{PM}\right)q_{\alpha}^{\text{ }M}q_{\beta}^{\text{ }P}\ .
	\end{aligned}
\end{align}
Finally, for the last term
\begin{align}
	\begin{aligned}
		g_{M[P}g_{R]N}n^{R}n^{N}q_{\alpha}^{\text{ }M}q_{\beta}^{\text{ }P}&=\left(g_{MP}g_{RN}-g_{MR}g_{PN}\right)n^{R}n^{N}q_{\alpha}^{\text{ }M}q_{\beta}^{\text{ }P}\ ,\\	&=\left(g_{RN}n^{R}n^{N}\right)\left(g_{MP}q_{\alpha}^{\text{ }M}q_{\beta}^{\text{ }P}\right)-\left(n^{R}g_{MR}q_{\alpha}^{\text{ }M}\right)\left(n^{N}g_{PN}q_{\beta}^{\text{ }P}\right)\ ,\\&=q_{\alpha\beta}\ .
	\end{aligned}
\end{align}
From \eqref{1.3.1-8} 
\begin{equation}
	R=-\frac{16\pi G_{5}}{3}T\ ,
\end{equation}
and therefore
\begin{equation}
	g_{M[P}g_{R]N}n^{R}n^{N}q_{\alpha}^{\text{ }M}q_{\beta}^{\text{ }P}=-\frac{16\pi G_{5}}{3}Tq_{\alpha\beta}\ .
\end{equation}
Plugging these expressions in the decomposition of the Riemann tensor
\begin{align}
	\begin{aligned}
		\tilde{\Psi}_{\alpha\beta}&=R_{MNPR}^{\text{ }\text{ }\text{ }\text{ }}n^{R}n^{N}q_{\alpha}^{\text{ }M}q_{\beta}^{\text{ }P}\ ,\\	&=\left(C_{MNPR}+\frac{2}{3}\left(g_{M[P}R_{R]N}-g_{N[P}R_{R]M}\right)-\frac{2}{12}Rg_{M[P}g_{R]N}\right)n^{R}n^{N}q_{\alpha}^{\text{ }M}q_{\beta}^{\text{ }P}\ ,\\	&=\Psi_{\alpha\beta}+\frac{1}{3}\left[8\pi G_{5}\left(T_{RN}-\frac{T}{3}g_{RN}\right)n^{R}n^{N}q_{\alpha\beta}-\left(-8\pi G_{5}\left(T_{PM}-\frac{T}{3}q_{PM}\right)q_{\alpha}^{\text{ }M}q_{\beta}^{\text{ }P}\right)\right]\\&-\frac{1}{12}\left(-\frac{16\pi G_{5}}{3}Tq_{\alpha\beta}\right)\ ,\\	&=\Psi_{\alpha\beta}+\frac{8\pi G_{5}}{3}\left[\left(T_{MN}n^{M}n^{N}-\frac{T}{2}\right)q_{\alpha\beta}+T_{MN}q_{\alpha}^{\text{ }M}q_{\beta}^{\text{ }N}\right]\ .
	\end{aligned}
\end{align}
Which leads to the Effective Einstein's Equations \eqref{1.3.1-13}, and $\Psi_{\alpha\beta}=C_{MNPR}n^{R}n^{N}q_{\alpha}^{\text{ }M}q_{\beta}^{\text{ }P}$ is the electric part of the Weyl tensor, as we claimed.

%% file: chapters/Frr0beta.tex
\chapter{Right hand-side of Eq. (1.134)} \label{app-C}
In Sect. \ref{adsschw-def}, in Eq. \eqref{adssch-hamconstraint} we make use of the following function on the right hand-side of the equation:
\begin{align*}
	\begin{aligned}
F\left(r,r_{0},\beta\right)=-\frac{1}{r^{10}}\Bigg\{-&\left(10(\beta-1)+r^{6}-3r^{2}r_{0}^{4}\right)\left(\beta+r^{6}-r^{2}r_{0}^{4}-1\right)\\&+\frac{4r^{8}\left(-2\beta+r^{6}+r^{2}r_{0}^{4}+2\right)^{2}}{\left(\beta+r^{6}-r^{2}r_{0}^{4}-1\right)^{2}}\\&+\frac{4r^{8}\left(4r^{12}+8(2-3\beta)r^{8}r_{0}^{4}+(20\beta-23)r^{4}r_{0}^{8}+3(4\beta-1)r_{0}^{12}\right)^{2}}{\left(2r^{8}-5r^{4}r_{0}^{4}+3r_{0}^{8}\right)^{2}\left(2r^{4}+(1-4\beta)r_{0}^{4}\right)^{2}}\\&-\frac{2r^{8}\left[8r^{16}-60r^{12}r_{0}^{4}+6\left(40\beta\left(2\beta-3\right)+67\right)r^{8}r_{0}^{8}\right]}{\left(2r^{8}-5r^{4}r_{0}^{4}+3r_{0}^{8}\right)\left(2r^{4}+\left(1-4\beta\right)r_{0}^{4}\right)^{2}}\\&-\frac{2r^{8}\left[\left(4\beta-1\right)\left(20\beta+43\right)r^{4}r_{0}^{12}-9\left(1-4\beta\right)^{2}r_{0}^{16}\right]}{\left(2r^{8}-5r^{4}r_{0}^{4}+3r_{0}^{8}\right)\left(2r^{4}+\left(1-4\beta\right)r_{0}^{4}\right)^{2}}\\&+\frac{1}{2r^{4}+\left(1-4\beta\right)r_{0}^{4}}\left[r^{2}\left(2r^{8}+2r^{6}-5r^{4}r_{0}^{4}\right.\right.\\&\left.\left.+\left(1-4\beta\right)r^{2}r_{0}^{4}+3r_{0}^{8}\right)\left(\beta+r^{6}-r^{4}-r^{2}r_{0}^{4}-1\right)\right]\\&+\frac{4r^{8}\left(r^{6}+r^{2}r_{0}^{4}+2-2\beta\right)\left(4r^{12}+8(2-3\beta)r^{8}r_{0}^{4}+3(4\beta-1)r_{0}^{12}\right)}{\left(2r^{4}-3r_{0}^{4}\right)\left(r^{4}-r_{0}^{4}\right)\left(2r^{4}+(1-4\beta)r_{0}^{4}\right)\left(\beta+r^{6}-r^{2}r_{0}^{4}-1\right)}\\&+2r^{8}\left(\frac{2r^{8}+5r^{4}r_{0}^{4}-9r_{0}^{8}}{2r^{8}-5r^{4}r_{0}^{4}+3r_{0}^{8}}-\frac{4r^{4}}{2r^{4}+(1-4\beta)r_{0}^{4}}+\frac{r^{2}\left(3r^{4}-r_{0}^{4}\right)}{\beta+r^{6}-r^{2}r_{0}^{4}-1}\right)\Bigg\}
	\end{aligned}
\end{align*}